\documentclass[twocolumn]{article}
\usepackage{amsthm}

\usepackage{style_licence/arxiv}

\usepackage[utf8]{inputenc} 
\usepackage[T1]{fontenc}    
\usepackage{hyperref}       
\usepackage{url}            
\usepackage{booktabs}       
\usepackage{amsfonts}       
\usepackage{nicefrac}       
\usepackage{microtype}      
\usepackage{lipsum}		
\usepackage{graphicx}
\usepackage{natbib}
\usepackage{doi}
\usepackage{amsmath}
\newtheorem{definition}{Definition}
\newtheorem{theorem}{Theorem}
\usepackage[dvipsnames]{xcolor}
\usepackage{titlesec}
\usepackage{comment}

\usepackage[strict]{changepage}
\usepackage{calc}

\def\eps{\epsilon}
\def\Tr{\,\hbox{Tr}\,}
\def\div{\,\hbox{div}\,}
\def\det{\,\hbox{det}}
\def\dis{\displaystyle}
\def\dt{\Delta t}

\def\eps{\epsilon}

\def\mod{\,\,\hbox{mod}\,}
\def\dis{\displaystyle}

\def\Tr{\,\hbox{Tr}\,}
\def\eps{\epsilon}

\def\idm#1{{\mbox{\scriptsize #1}}}

\def\mean#1{\langle{}#1{}\rangle}

\def\Tr{\hbox{Tr}\,} 

\DeclareMathAlphabet{\mathsfit}{\encodingdefault}{\sfdefault}{m}{sl}
\SetMathAlphabet{\mathsfit}{bold}{\encodingdefault}{\sfdefault}{bx}{sl}

\newcommand{\xb}{x_{\idm{1}}}

\newcommand{\lambdab}{\lambda_{\idm{1}}}

\newcount\exno

\newcommand\hide[1]{}

\def\K2{{\sc K2}}

\def\idm#1{{\mbox{\scriptsize #1}}}

\font\grassettogreco=cmmib10
\font\scriptgrassettogreco=cmmib7
\font\scriptscriptgrassettogreco=cmmib10 at 5 truept
\font\sansserif=cmss10
\font\scriptsansserif=cmss10 at 7 truept
\font\scriptscriptsansserif=cmss10 at 5 truept
\def\bgr{\fam=13}
\def\ssm{\fam=14}
\input amssym.def
\input amssym.tex
\def \ebf{{\bf e}}

\def \kbf{{\bf k}}

\def \pbf{{\bf p}}
\def \qbf{{\bf q}}
\def \rbf{{\bf r}}

\def \ubf{{\bf u}}
\def \vbf{{\bf v}}
\def \xbf{{\bf x}}
\def \ybf{{\bf y}}
\def \wbf{{\bf w}}

\def \Fbf{{\bf F}}

\def \Rbf{{\bf R}}

\def \Xbf{{\bf X}}

\def \Aop{{\mathchardef\alpha="710B \ssm \char'101}}
\def \Bop{{\mathchardef\alpha="710B \ssm \char'102}}

\def \Dop{{\mathchardef\alpha="710B \ssm \char'104}}

\def \Gop{{\mathchardef\alpha="710B \ssm \char'107}}

\def \Iop{{\mathchardef\alpha="710B \ssm \char'111}}
\def \Jop{{\mathchardef\alpha="710B \ssm \char'112}}

\def \Lop{{\mathchardef\alpha="710B \ssm \char'114}}
\def \Mop{{\mathchardef\alpha="710B \ssm \char'115}}
\def \Nop{{\mathchardef\alpha="710B \ssm \char'116}}

\def \Pop{{\mathchardef\alpha="710B \ssm \char'120}}
\def \Qop{{\mathchardef\alpha="710B \ssm \char'121}}
\def \Rop{{\mathchardef\alpha="710B \ssm \char'122}}
\def \Sop{{\mathchardef\alpha="710B \ssm \char'123}}
\def \Top{{\mathchardef\alpha="710B \ssm \char'124}}
\def \Uop{{\mathchardef\alpha="710B \ssm \char'125}}
\def \Vop{{\mathchardef\alpha="710B \ssm \char'126}}
\def \Wop{{\mathchardef\alpha="710B \ssm \char'127}}
\def \Xop{{\mathchardef\alpha="710B \ssm \char'130}}
\def \Yop{{\mathchardef\alpha="710B \ssm \char'131}}
\def \Zop{{\mathchardef\alpha="710B \ssm \char'132}}

\def\Xibf{{\mathchardef\Xi="7104 \bgr \Xi}}

\def\Phibf{{\mathchardef\Phi="7108 \bgr \Phi}}

\def\epsilonbf{{\mathchardef\epsilon="710F \bgr \epsilon}}

\def\etabf{{\mathchardef\eta="7111 \bgr \eta}}

\def\xibf{{\mathchardef\xi="7118 \bgr \xi}}

\def \Acal{{\cal A}}
\def \Bcal{{\cal B}}
\def \Ccal{{\cal C}}

\def \Ecal{{\cal E}}

\def \Pcal{{\cal P}}

\def \Rcal{{\cal R}}

\def \Wcal{{\cal W}}

\def \eps{\epsilon}

  2
  3
 4
 5

\def \spazio{\vskip .5	  truecm \noindent }

\def \spa{\vskip    .3	  truecm \noindent }

\def \pan {\par \noindent}

   \newcount\Silicon \global\Silicon=1
   \newcount\PC \global\PC=2
   \newcount\parziale
   \def\Colori#1{\global\parziale=#1
                \ifnum\parziale=\Silicon
                    \input colors
                    \gdef\Color##1{\Black{##1}}                    
                \else\ifnum\parziale=\PC
                    \input colordvi
                    \gdef\textRGB##1{\textColor{##1 0.}}
                    \gdef\GrayA##1{\textGray{##1}}
                    \gdef\GrayB##1{\textGray{##1}}
                    \gdef\GrayC##1{\textGray{##1}}
                    \gdef\GrayD##1{\textGray{##1}}
                    \gdef\GrayE##1{\textGray{##1}}
                    \gdef\GrayF##1{\textGray{##1}}
                    \gdef\GrayG##1{\textGray{##1}}
                    \gdef\GrayH##1{\textGray{##1}}\fi                   
                \fi}        
\def\mean#1{\langle \,#1\,\rangle}

\def \parton#1{\left({{#1}}\right)}
\def \parqua#1{\left[{{#1}}\right]}

\def \parbar#1{\left|{{#1}}\right|}

\def \parmean#1{\left\langle{{#1}}\right\rangle}
\def \derp#1#2{{\partial{#1} \over \partial{#2} }}

\font \rmsmm=cmr7
\def \hbix#1{\rmsmm \hbox{ {#1} } }

\def \underwrite#1{\mathop{\vtop{\ialign {##\crcr
$\hfil\displaystyle {#1}\hfil$\crcr\noalign{\kern3pt\nointerlineskip}
\crcr\noalign{\kern3pt}}}} \limits}

\def\Reali{\Bbb R}
\def\Interi{\Bbb Z}
\def\Toro{\Bbb T}

\def\sqr#1#2{{\vcenter{\hrule height.#2pt
     \hbox{\vrule width.#2pt height#1pt \hskip#1pt
       \vrule width.#2pt}
     \hrule height.#2pt}}}

\def\DAL{\hbox{\raise.250ex \hbox{$\sqr7{10}\,$}}} 

\def\bib#1.#2/#3/#4/#5/#6/#7.{\frenchspacing\item{[#1]}#2:\ {\it ``#3''}
~--~#4\ $\underline{\bf #5}$,\ #6 (#7)}
\def\diagramma#1#2#3#4#5#6#7#8{
 \vbox to 2.5cm{
       \hbox to 3cm{\hfil ${#1}$ \hfil}
       \hbox to 3cm{ ${#2}$\rightarrowfill ${#3}$ }
\hbox to 3cm{${#4} \Biggl\uparrow\hfill\Biggr\uparrow {#5}$}
       \hbox to 3cm{ ${#6}$\rightarrowfill ${#7}$ }
       \hbox to 3cm{\hfil ${#8}$ \hfil}   }    }
\def\sopra#1#2{{\raise 0.8 ex
\hbox{$
{{\scriptstyle \,{#2}}	\atop \displaystyle{#1}}
$}}
}

\def\figuraps#1#2#3#4#5{
\par
\midinsert
\centerline{\bf #4}
\vbox to #3 truecm{
\vskip #3 truecm
\ifnum #1 = 0	
\special {ps: plotfile #2}
\else		
\special {#2 0 0 moveto 16} \fi
}
\centerline{#5}
\endinsert}

\font\cofon = cmr6
\font\cobfon = cmbx6
\font\copi = cmr9
\def\codlib{{\copi\copyright}{\cofon 88-08- }{\cobfon 9820}}
\def\riga{\vskip .1  truecm   \hrule \vskip .2	    truecm \noindent }
\def\HeadLinea#1#2{	\headline={\vbox to 0pt{\vss\noindent
{\ifnum \pageno=1  \hfill {\bf \folio}		 
\else {\ifodd \pageno			   
{\noindent     \hfill  {\it     #2} \quad {\bf \folio}
 }\riga
\else				 
{\noindent {\bf\folio} \quad  {\it  #1} \hfill 	}
\riga
\fi } \fi }			}}
}
\def\testatina#1#2{	\headline={\vbox to 0pt{\vss\noindent
{\ifnum \pageno=1  \hfill {\bf \folio}		 
\else {\ifodd \pageno			   
{\noindent  \codlib   \hfill  {\it     #2} \quad {\bf \folio}
 }\riga
\else				 
{\noindent {\bf\folio} \quad  {\it  #1} \hfill \codlib	}
\riga
\fi } \fi }			}}
}
\def\testatinacap#1#2#3{	\headline={\vbox to 0pt{\vss\noindent
{\ifnum \pageno=#3  \hfill {\bf \folio}		 
\else {\ifodd \pageno			   
{\noindent  \codlib   \hfill  {\it     #2} \quad {\bf \folio}
 }\riga
\else				 
{\noindent {\bf\folio} \quad  {\it  #1} \hfill \codlib	}
\riga
\fi } \fi }			}}
}
\def\oggi{\number\day\space\ifcase\month
   \or gennaio\or febbraio\or marzo\or aprile\or maggio\or giugno\or
   luglio\or agosto\or settembre\or ottobre\or novembre\or dicembre
   \fi\space\number\year}
\def\today{\number\day\space\ifcase\month
   \or January\or February\or March\or April\or May\or June\or
   July\or August\or September\or October\or November \or December
   \fi\space\number\year}   
\def\frame#1{\ifmmode\dframe{#1}\else\leavevmode\lower 2.4 pt
    \hbox{\vrule\unskip\vbox{\hrule\kern 1.5 pt\hbox{\kern
    1.5 pt{#1}\kern 0.5 pt}\kern 2 pt\hrule}\unskip\vrule}\fi}

\def\dframe#1{\hbox{\vrule\unskip$\vcenter{\hrule\kern 3 pt\hbox
    {\kern 3 pt$\displaystyle{#1}$\kern3pt}\kern 3 pt\hrule}$\vrule}}

\font \rmsmm=cmr7

\title{Lyapunov and Reversibility Error invariant indicators}

\author{ 
	\href{https://orcid.org/0000-0001-6261-7928}{\hspace{1mm}Federico Panichi} \\
	31, Bramwell Drive, M139YB\\
	  Manchester, United Kingdom\\
	\texttt{federico.panichi@studio.unibo.it} \\
    \And
\href{https://orcid.org/0000-0001-9481-0490}{\hspace{1mm}Giorgio Turchetti} \\
	Department of Physics and Astronomy, \\
    Alma Mater Studiorum, University of Bologna, \\
    Viale Berti Pichat 6/2, 40-127 Bologna, Italy \\
 	\texttt{giorgio.turchetti@bo.infn.it} \\
}

\renewcommand{\shorttitle}{Lyapunov and Reversibility Error}

\begin{document}
\maketitle

\twocolumn[
  \begin{@twocolumnfalse}
    \begin{abstract}

      In this review, we present a survey of the Lyapunov Error and Reversibility Error (\cite{Faranda2012}), and we propose a generalization to make them invariant to the choice of initial conditions.
      We first define a process as the evolution in time of a a map or a flow, we then introduce the covariance matrix of a given process, and use their trace to compute LE and RE. The determinant of the covariance matrices is used to compute all invariant indicators of higher order.
      In this way, two set of invariant indicators are proposed within the framework introduced here, one for the Reversibility and one for the Lyapunov Errors, respectively. LE and RE which have been used in the literature, are the \textit{first-order} invariant indicators in their respective sets.   
      The new sets of invariant indicators have the same fundamental meaning, the set for RE is used to characterize the dynamical evolution of a continuous small perturbation over an orbit, while LE is used to study the evolution of an initial displacement between two nearby orbits. 
      We obtain again also the Reversibility Error Method (REM), which is a particular case of RE, where the additive noise is replaced by the round-off errors. 
      REM has been proved to be a practical, accurate, and fast dynamical indicator and the results are comparable with the RE induced by a random noise of given amplitude. the behaviour of RE invariants  depends on the sum of the positive Lyapunov exponents. We showcase the accuracy and reliability of those indicators on two well studied Hamiltonian.
      Within this new framework, a full dynamical characterization of a Hamiltonian system can be obtained. For instance, the Gibbs entropy, its asymptotic behaviour and a method to compute the fidelity over the perturbed orbit, are provided.

    \end{abstract}
 \end{@twocolumnfalse}
]

\newpage
\tableofcontents


%
%
\section {Introduction}   
\label{S:Introduction}
\spazio
The stability theory of Hamiltonian systems is mathematically well established
     \cite{Arnold1961},  \cite{Oseledets1961} and algorithms to compute Lyapunov exponents are available (see, for example, \cite{Benettin, Benettin2,  Skokos2010b, Politi2016}).
However, since the computation time required to calculate accurately the Lyapunov exponents is often difficult to achieve numerically, finite-time stability indicators
were introduced.  The basic idea is to compare the sensitivity to initial conditions
at a fixed time $t$ for all orbits whose initial points belong to a compact
low-dimensional subset of the phase space, rather than looking at the behavior of single orbits when $t$ increases.
        The Fast Lyapunov indicator (FLI)  was first proposed in
        \cite{Froeschle1997,Lega2001,Froeschle2000b}
        and was followed by several finite-time variational indicators such as the Relative Lyapunov indicator (RLI) \cite{Sandor2004}, the Orthogonal Fast Lyapunov Indicator (OFLI) \cite{Barrio2016},
the Smaller ALignment Index  (SALI)  \cite{Skokos2001} and its generalization (GALI) \cite{Skokos2015}, just to cite some of them.
Extensive research has been done to filter out short-frequency oscillations, for example, by proposing methods based on the double time average (MEGNO) \cite{Cincotta2000} or Birkhoff averaging \cite{Yorke2018}.
A comparison between these indicators can be found in \cite{Maffione2011, Darriba2012}. 
Spectral and geometric chaos indicators were also introduced:
frequency map analysis (FMA) \cite{Laskar1990,Laskar1992,Laskar1993,Nesvorny1996}, power
spectrum of the autocorrelation function \cite{Anischenko2004,Melbourne2007}, digital indicator ( $0,1$) \cite{Gottwald2004,Gottwald2005, Nicol}, and curvature of constant energy manifolds
\cite{Pettini1996},  \cite{Pettini2008}, \cite{Pettini1993}.

 In this paper, we review two dynamical indicators, namely, the Lyapunov error (LE) and the reversibility error (RE) indicators, previously proposed for Hamiltonian systems \cite{Turchetti2017}. We recall that RE has been formally defined in \cite{Turchetti2019} and applied to beam dynamics in \cite{Gradoni2021}, while REM has been defined in \cite{Panichi2016} and mostly used in celestial mechanics \cite{Panichi2017}.
 We propose a framework to generalize the above indicators to a from that is independent to the choice of initial conditions, and define them as \textit{invariant} indicators. Within this framework, LE and RE are the \textit{first-order} invariant indicators on a set composed by $n$ degrees of freedom.
 
  To this end, we follow a similar process as in \cite{Skokos2008}. We first introduce the following concepts.

  \begin{definition}
      Given   the vector field $\Phibf(\xbf,t)$   where 
$\xbf$ and $\Phibf$ belong to  $\Reali^{d}$, the \texttt{semi-group of solutions} $ S_{t,t_0}(\xbf)$ is defined as a continuous time map, and is obtained by solving
the following
\begin{equation}
 \frac{d}{dt}\,\xbf(t)=\Phibf(\xbf(t),t)
 \label{eq_2_1}
 \end{equation}
with initial condition $\xbf$ for $t=t_0$.
\label{def:S}
  \end{definition}

We also define the forward evolution $S_t$ according to 
\begin{equation}
  S_t(\xbf)\equiv S_{t,0}(\xbf) \qquad \quad     S_{t,\,t_0}(\xbf)=
  S_t\circ S^{-1}_{t_0}(\xbf)
 \label{eq_2_2}
 \end{equation}
and re-write $\xbf(t)=S_t(\xbf)$ choosing from now on the initial condition $\xbf$ at $t=0$.
Letting $\etabf$ be a unit vector  we consider a nearby initial condition
$\ybf=\xbf+\eps\,\etabf$ and the corresponding solution $\ybf(t)= S_t(\ybf)$.
To the initial deviation  $\ybf-\xbf=\eps\,\etabf$   corresponds at time $t$ the deviation
$\ybf(t)-\xbf(t)= \eps\,\etabf(t) +O(\eps^2)$.  The remainder depends on $t$ and usually is bounded by
$\eps^2 e^{\lambda t}$ with $\lambda >0$ and therefore is small with respect to $\eps$ only
for $t< \lambda^{-1}\log (1/\eps)$. A time independent bound $M\eps^2$
to the remainder can be given only on phase space
subsets which exclude a neighborhood of the unstable manifolds. 

\begin{definition}
  The \texttt{tangent map} $\Lop(\xbf,t)$ is defined as a $ d\times d$
matrix whose entries are the partial derivatives of the continuous time map $S_t(\xbf)$ (from \ref{def:S})
\begin{equation}
  (L)_{ij}(\xbf)= \derp{}{x_j} \Bigl(S_t(\xbf)\Bigr)_i  \quad \longrightarrow\quad \Lop(\xbf,t)= DS_t(\xbf)
  \label{eq_2_4}
\end{equation}
\label{def:1-tagentmap}
\end{definition}

\begin{definition}
  Given an initial deviation at time $t$, we define the \texttt{displacement vector} $\etabf(\xbf,t)$ as
\begin{equation}
  \etabf(\xbf,t)= 
  \lim_{\eps \to 0} \frac{ S_t(\xbf+\eps\etabf)-S_t(\xbf)}{\eps}
 \label{eq_2_3}
 \end{equation}

\label{def:dispVect}
\end{definition}
\begin{definition}
We define the \texttt{linear response} the matrix $\Lop(\xbf,t)\,\etabf$, which can be calculated in any phase space region where $\Lop(\xbf,t)$ is defined and is given the tangent map of $S_t(\xbf)$.
\label{def:LinResponse}
\end{definition}

\begin{theorem}
The matrix $\Lop(\xbf,t)$ is the \textbf{fundamental solution} of the linear equation satisfying to
\begin{equation}
  \begin{split}
    {\frac{d}{dt}}\Lop(\xbf,t) = D\Phibf\bigl(S_t(\xbf),\,t\bigr)\,\Lop(\xbf,t)   \qquad \Lop(0)=\Iop
    \end{split}
 \label{eq_2_5}
 \end{equation}
 \label{def:FondMatrix}
 \end{theorem}
 We have two alternatives to prove that $\Lop(\xbf,t)$ satisfies equation (\ref{eq_2_5}).

 \begin{proof}
    We can interchange the space and time  derivatives $D$ ad $d/dt$
according to 
\begin{equation}
  \begin{split}
     {d\over dt}\Lop(\xbf,t) & = {d\over dt}\,D\,S_t(\xbf) = D  {d\over dt}\,S_t(\xbf)=
    D\,\Bigl( \Phi(S_t(\xbf)\bigr) \Bigr)= \\ \\
   & = D \,\Phi\Bigl(S_t(\xbf)\Bigr) \, D\,S_t(\xbf)=
    D\,\Phi\bigl( S_t(\xbf)\bigr)\,\Lop(\xbf,t)
    \end{split}
 \label{eq_2_6}
 \end{equation}
The interchange is allowed in any phase space-time region where $S_t(\xbf)$ is $C^2$,
according to Schwarz's theorem. If $\Phibf(\xbf,t)$
is  holomorphic in $\xbf$ and $t$  then $S_t(\xbf)$ is  analytic in a ball centered at the origin
and can be analytically continued.  Approaching the singularities, the $C^2$ property is lost.
\end{proof}

\begin{proof}
The second  alternative consists  in  proving  that the time derivative of (\ref{eq_2_3})
  can be interchanged with the  $\eps \to 0 $ limit.
In this case,  we should prove that the limit $\eps\to 0$ is uniform with
respect to $\xbf$ and $t$.
The  space time regions on which   $S_t(\xbf)$ is $C^2$  are the same on which  
one can prove that the limit $\eps \to 0$ is uniform.
For a map the problem of the 
limits interchange is not present since the differential equation  is
replaced by a recurrence.
\end{proof}

We note that the linear response $\Lop(\xbf,t)\,\etabf$ depends on the direction of the initial deviation
and can be used to compute the Fast Lyapunov Indicator (FLI, \cite{Froeschle1997}), in fact, we can compute FLI as the logarithm of norm of the linear response.  

\begin{definition}[Lyapunov Error]
\label{pythagorean}
The Lyapunov error (LE), we have previously introduced in \cite{} as $E_L(\xbf,t)$, is independent from $\etabf$. It is defined by
considering the linear response $\ebf_i(\xbf,t)= \Lop(\xbf,t)\,\etabf_i$
to  all the  unit vectors $\ebf_i$ of an orthonormal base. The definition  is
\begin{equation}
  \begin{split}
   E_L(\xbf,t)= \sqrt{\sum_{i=1}^d  \, \Vert  \ebf_i(\xbf,t)\Vert^2 }  = \sqrt{\Tr\bigl (\Lop^T(\xbf,t)\Lop(xbf,t)\, \bigr)}
    \end{split}
 \label{eq_2_7}
 \end{equation}
\end{definition}
The relation is evident if  $(\ebf_i)_j=\delta_{ij}$   and  is still valid if  we change
$\ebf_i$ into $\Rop \ebf_i$ where  $\Rop$ is an orthogonal matrix, namely for an arbitrary orthonormal base.
 The dependence on the initial condition $\xbf$ will be often dropped unless this
 may cause ambiguities.

\begin{definition}
We define an indicator as \texttt{invariant} if and only if it does not depend on the direction and amplitude of the initial deviation. 
\label{def:invariant}
\end{definition}
\spa
We recall that $LE$ as introduced in \cite{Turchetti2019} is an invariant indicator. The geometrical meaning of \ref{def:invariant} is that an invariant indicator does not depend on the initial choice of the deviation vectors
or on the chosen orthogonal reference frame and also have a simple geometric interpretation, as discussed in \cite{Skokos2008}. 

To summarize, we first consider the linear response of a flow in $\Reali^d$ (or any compact subset) to a small deviation at time $t$, from the initial point $\xbf$ and the corresponding covariance matrix $\Sigma^2_L(\xbf,t)$. Then, we compute the Lyapunov indicator as the square root of the trace of the product of $\Lop^T\Lop$. This is in line with what has been reviewed in \cite{Skokos2008}.


 The linear response  to a small additive noise after its forward
 evolution from time 0 up to time $t>0$, is a random vector with  covariance matrix
  $\Sigma^2_F(\xbf,t)$ and the square root of its trace defines the Forward Error (FE). 
 Also in this case,  we compute  all the   invariants  $I^{(j)}_F$  of the
 covariance matrix. 
 The Gibbs entropy for the F process is defined as the log of the square root  of the last invariant  $I^{(d)}_F$.
 The asymptotic behaviour   of the invariant  $I^{(j)}_F$  is  determined by the sum of the positive
 Lyapunov exponents among the first $j$. 
  \\
 The  linear response  to the forward evolution with a small additive  noise up to time $t$ followed by
 the backward evolution  up to time $2t$  is a random vector  with  covariance matrix
 $\Sigma^2_{BF}(\xbf,t)$. The BF reversibility error (RE) is the square root of the trace of this matrix.  Also in this case we  consider  the     invariants  $I^{(j)}_{BF}$  of the covariance matrix.
 The log of the square root of the last invariant $I^{(d)}_{BF}$   is the Gibbs entropy for the BF process.
 The  asymptotic  behaviour of   $I^{(j)}_{BF}$  is determined 
 by the sum of the absolute values of the negative
 Lyapunov exponents among the last $j$.
 For Hamiltonian systems the  asymptotic behaviour of the  Gibbs entropy   for  BF processes
 is determined by the sum of positive Lyapunov exponents, just as the Kolmogorov-Sinai  entropy  \cite{Sinai1959}
 according to Pesin theorem \cite{Pesin1977}, see also \cite{Eckman1985,Ginelli2012}
 \\
 The Lyapunov, Forward and Reversibility error invariants can be computed for  invertible maps. 
 The covariance matrices  can be written   in a  closed form involving the tangent map
 and the procedure to obtain them from the linear response  is  rigorously justified.
 The asymptotic behaviour of the invariants is determined by the Lyapunov exponents
 just as   for the flows.
 \\
 For invertible maps  we have previously  introduced the  reversibility error method (REM). 
 In this case the  round off replaces the noise  during  the  F evolution,   given 
 the fist $n$ iterations of the map. The round off affects also  the B evolution, 
 given by    $n$ iterations  of the  inverse map.
 The basic idea is that the round off error is a pseudo-random noise. If we compute
 the BF reversibility error  with a   noise   during   F  evolution and   the
 B evolution as well, 
 by introducing  a  factor 1/2 in the definition of the covariance matrix, we obtain the same result as
 when the noise is present only in  the F evolution. As a consequence it is reasonable
 to expect a correspondence between REM and RE. The latter is defined as the square root  of the trace
 of $\Sigma^2_{BF}$.  A good correspondence was systematically found  between RE and REM
 in all the analyzed Hamiltonian models.
  Since  RE is obtained by averaging over a  random process,
  whereas REM corresponds to a single realization, the  latter  exhibits  fluctuations when
  $n$ varies for a given initial point $\xbf$ in phase space,  or when  $n$ is fixed and $\xbf$ varies.
  In addition REM is computed for a finite round off amplitude $\eps$,   whereas RE is  obtained
  from the linear response corresponding to the $\eps\to 0$ limit. The systematically good agreement found
  between  REM and RE, implies that the $O(\eps)$ remainder of the linear approximation
  (REM) w.r.t. the linear response (RE), is negligible for $\eps \sim 10^{-16}$, which is the  round off amplitude in the 8 bytes
  representation of real numbers.   The   $O(\eps)$ remainder can be neglected when
  the linear approximation is valid. As a consequence, we expect that the linear approximation
  can be used when the  reversibility error RE  is computed with a noise amplitude $\eps$
  equal to the round off amplitude. Heuristic arguments suggest that for regions where the difference is negligible are at the border with the   unstable invariant  manifolds. 
  To conclude we point out that the key differences between RE and REM  is that the former provides
  the whole set of invariants,  the latter just the first one. In addition, RE is the result
  of an average on the random process whereas REM corresponds to a unique realization  and exhibits
  significant fluctuations.
  Finally we notice that for the Hamiltonian systems the square of RE
  divided by $t$ is the time average of the square of LE. As a consequence the oscillations
  occurring during a quadratic or exponential time growth are averaged  and the use of a filter like
  MEGNO is not needed when RE is computed.
  \\
 The sensitivity to a small additive noise  is measured by  the 
 random fidelity decay rate \cite{Liverani2007}.
 We introduce a local  random fidelity defined as the autocorrelation
 for the BF noisy reverse  process   which  can be easily computed on a torus defined
 as a cube with center
 $\xbf$ and identified opposite faces,  if we let the length  $\ell$ of its sides  
 go to zero jointly with the noise amplitude $\eps$ while keeping their ratio  $\rho=\eps/\ell$ fixed.
 In this case, the linear response applies and  the  transition  probability density is a Gaussian
 on the torus with covariance matrix $\Sigma^2_{R}(\xbf,t)$. 
 The decay  rate to the asymptotic limit for $t\to \infty$  can be easily determined and depends just
 on the covariance matrix.
 \\
The reversibility error  RE  and REM have been previously  computed   for  Hamiltonian flows. 
In that case, the inverse of the Lyapunov matrix $\Lop^T\Lop $   whose time  integral defines 
covariance matrix of RE can be replaced by the matrix itself,   since the invariants do not change. 
Even though  LE and RE are fast indicators   we notice that, letting $M$ be the overflow 
(equal to $10^{308}$  in standard double precision) and $\lambda_1>0$ the maximum 
Lyapunov exponent,   they can   be evaluated  even for a long time period,
provided that the time does not exceed   $t_M \sim  \lambda_1^{-1}\, \log M$. 
When REM is evaluated for orbits which belong to an invariant subset  whose diameter 
is of order 1, letting $\eps$ be  the numerical  accuracy  (equal to $10^{-16}$  in standard double precision) 
then the computation can be extended to times not exceeding    $t_\eps \sim  \lambda_1^{-1}\, \log \eps^{-1}$.
For $t<t_\eps$ it was found that RE and REM are comparable and  the agreement improves  after  a
running average of REM  which smooths its fluctuations. Similar considerations hold for the
the higher order LE and RE errors taking into account   that  asymptotic behaviour  for order $k$
depends on the sum of the first $k$ Lyapunov exponents and the sum of the positive exponents
among the first $k$, respectively.  We stress that REM  is  the indicator whose implementation is 
the easiest one and it discriminates regions of regular motion, where it grows as a power law, from regions of chaotic motions
where  it grows following an exponential law. 
The higher order LE and RE  provide more detailed information  related to higher  Lyapunov exponents.
\spa
Numerical applications were developed for systems of celestial mechanics,
beam dynamics, mechanical systems and rays propagation an cavities and waveguides.
A list of references  is found in  Sect. \ref{S:twoSimpleModels} where we 
 present the numerical results for the above indicators 
 for   two canonical  Hamiltonian systems,  a modulated
 pendulum coupled with a  standard pendulum and   two   coupled standard maps.
 \\
The paper is organized as follows. 
In section \ref{S:TheLyapunovError}, we  introduce  the Lyapunov error invariants.
In section \ref{S:LyapunovInterpretation}, we analyze   their asymptotic behaviour.
In section \ref{S:TheReversibility Error} we introduce the reversibility  error invariants.
In section \ref{S:Asymptotic} we examine  their asymptotic behaviour.
In section \ref{S:twoSimpleModels} present the results for two Hamiltonian models.
Section \ref{S:conclusions} is devoted to the  summary and  conclusions.
The appendices are devoted to numerical procedures and technical aspects.
%
%

\section{ Lyapunov error invariant indicators }
\label{S:TheLyapunovError}

In this section, we review and extend the properties and theorems used to compute the Lyapunov Error (LE). In \ref{def:dispVect}, we choose an arbitrary vector. This can cause problems related to the choice of the initial deviations and the emergence of spurious structures is well documented in the literature (see, for instance \cite{Barrio2016}). This issue can be avoided by considering the linear response to a random initial deviation. 

We start by replacing  $\etabf$ in \ref{eq_2_3} with a random vector $\xibf$ having zero mean and unit covariance matrix  $\mean{\xibf\,\xibf^T}=\Iop$. With this assumption, we can extend the definition of displacement vector in  \ref{def:dispVect} as follow.

\begin{definition}
The displacement vector of the initial random  deviation  $\ybf-\xbf =\eps\,\xibf$ is given by
\begin{equation}
  \begin{split}
    \Xibf(\xbf,t) & = \lim_{\eps \to 0} { S_t(\xbf+\eps\xibf)-S_t(\xbf)\over \eps}
    \end{split}
 \label{eq_2_8}
 \end{equation}
 \end{definition}
\begin{definition}
We define the \texttt{covariance matrix} of $ \Xibf(\xbf,t)$ as $\Sigma^2_L(\xbf,t)$. 
    \label{def:CovMatLy}
\end{definition}

Such a matrix is given by $\Lop\Lop^T$ and has the same  spectrum as the
Lyapunov matrix $\Lop^T\Lop$ \cite{Politi2016, Skokos2008}.

\begin{theorem}
As a consequence, we can extend the definition of $E^2_L$ given in \ref{eq_2_7} using the property of $\Sigma^2_L$ and $E^2_L$
\begin{equation}
  \begin{split}
    \Sigma^2_L(\xbf,t)  &= \mean{\Xibf(\xbf,t)\,\Xibf^T(\xbf,t)}= \Lop(\xbf,t)\,\Lop^T(\xbf,t)  \\ \\
    E^2_L(\xbf,t)& = \Tr\bigl(\Lop^T(\xbf,t)\,\Lop(\xbf,t)\bigr)= 
    \Tr\bigl(\Sigma^2_L(\xbf,t)\bigr) 
 \end{split}
\label{eq_2_9}
\end{equation}
\end{theorem}
\space
With this extended definition, LE is an invariant indicator, satisfying the definition given in \ref{def:invariant}. This is not the case for FLI and GALI$^{(j)}$ indicators. 
Moreover, we can observe that $E^2_L$ is the first invariant   of $\Sigma^2_L$  and for this reason, we can naturally extend the computation of all the next invariants.
\begin{definition}
We define a \texttt{spectrum} of invariant indicators, the group of $I^{(k)}_L$  for $1\le k\le d$, indicators, where $d$ is the dimension of the system.
    \label{def:spect}
\end{definition}
\space

The following theorem is used to compute all invariants for the LE other than the first one:
\begin{theorem}
Given any   matrix $\Aop$,  whose eigenvalues $\mu_j$, we suppose to be all distinct, 
the characteristic polynomial is  $P(z)=\det \,(\Aop -z\,\Iop)$, which is calculated as follow 
\begin{equation}
 P(z) =\det \,(\Aop -z\,\Iop)=  \prod_{j=1}^d(\mu_j-z)=  \sum_{j=0}^d I^{(d-j)}\,(-z)^j  
\label{eq_2_10}
\end{equation}
where $I^{(0)}=1$. 
\end{theorem}

By using \ref{eq_2_10}, we can consider all the invariants $I^{(k)}_L$ of the covariance matrix, for $k=1,\ldots,d$. For instance, the first two invariants and the last one are given by 
\begin{equation}
  \begin{split}
      &I^{(1)} = \sum_{j=1}^d \,\mu_j= \Tr(\Aop) \qquad \qquad 
      I^{(d)}=\det(\Sigma^2) \\ 
      & I^{(2)}= \sum_{j_1<j_2}^d \,\mu_{j_1}\,\mu_{j_2}= {1\over 2}\Bigl( \bigl(\Tr(\Aop)\,\bigr)^2-
      \Tr\bigr( (\Aop^2) \bigr)\,\Bigr)
      \end{split}
\label{eq_2_11}
\end{equation}
 Since the covariance matrix has the same spectrum as the Lyapunov matrix, the asymptotic behaviour of the invariant $I^{(j)}_L$ is determined by the sum of the first $j$ Lyapunov exponents, which has the same value on any invariant ergodic subset for Hamiltonian systems. 
 \space
 All the remaining invariants are  obtained from  the traces of  the powers of $\Aop$  
according to the Faddeev-Leverrier  recurrence  formula
\cite{Faddeev-Leverrier1952,Barnett1989}
\def\dis{\displaystyle }
\begin{equation}
   \begin{split}
     & \Bop_k = \Aop( I^{(k-1)}\,\Iop-\Bop_{k-1})   \\
      &I^{(k)}={\dis 1\over \dis k} \Tr(\Bop_k) \quad \, 1\le k\le d 
 \end{split}
 \label{eq_2_12}
 \end{equation} 
%
initialized by $I^{(0)}=\Iop$ and $\Bop_0=0$. The relation between the  invariants and the eigenvalues
is 
\begin{equation}
  \begin{split}
      I^{(k)}= \sum_{1\le j_1<j_2<\ldots<j_k}^d \, \mu_{j_1}\,\mu_{j_2}(t)\,\ldots\,\mu_{j_k}
      \end{split}
\label{eq_2_13}
\end{equation}
The numerical computation of  the invariants of a  positive  matrix   using the Faddeev-Leverrier
recurrence should be avoided since  the leading terms cancel and  accuracy is lost.
In appendix I, we describe the algorithms which are suited to compute the invariants.
According to \ref{eq_2_13} the invariant $I^{(k)}$ is the sum of all the  ${d\choose k}$ distinct products
of the eigenvalues of $\Lop^T\Lop$.  The  Jacobi method can be used to diagonalize
$\Lop^T\Lop$ but,  ordering  the eigenvalues in a decreasing sequence
$\mu_1(t)>\mu_2(t)>\ldots>\mu_d(t)$,  only the first $j$ eigenvalues can be determined  where
$\mu_j(t)/\mu_1(t)$ is sufficiently large with respect to the machine accuracy $\eps$.
In the standard IEEE 754 representation of reals  with 8 bytes, 52 bits are reserved
for the mantissa  the amplitude of the round-off error is $\eps=2^{-53}\simeq 10^{-16}$.
All the eigenvalues whose ratio
with the first one is below $\eps$ are lost.  In appendix I, we show how to use the QR decomposition to find the products of $k$ eigenvalues.
In appendix II, we discuss the validity conditions of the linear approximation.
\\
With the spectrum of indicators in \ref{def:spect}, it is possible to define the  Lyapunov  errors $E_L^{(k)} $ of order $k$ as their square root. 
%
%
%
\subsection{The case of  a map}
\begin{definition}
 Given a vector $\xbf$ and a map $M(\xbf,n)$, where  $\xbf \in \Reali^d$ and $n$ is an integer, which we assume to be analytic or $C^{\infty}$ in $\xbf$ for any $n$, we define the \texttt{orbital evolution} $\xbf_n(\xbf)$ by the recurrence.

\begin{equation}
  \begin{split}
  \xbf_n(\xbf) & = M(\xbf_{n-1},n-1) \equiv  M_n(\xbf)  \qquad  \xbf_0=\xbf \\  \\
  M_n(\xbf) & = M\bigl(M_{n-1}(\xbf),\,   n-1\bigr)
  \end{split}
 \label{eq_2_14}
 \end{equation}
\end{definition}

Similarly to what we introduced in Sect. \ref{S:Introduction}, given an initial displacement $\ybf= \xbf+\eps\,\xibf$  corresponding to a new
orbit $\ybf_n= M_n(\xbf+\eps \xibf)$, we then define the following:
\begin{definition}
The  \texttt{linear response} after  $n$ iterations is
given by
\begin{equation}
  \begin{split}
    \Xibf_n(\xbf) & = \lim_{\eps \to 0} {M_n(\xbf+\eps \xibf)-M_n(\xbf)  \over \eps} =
  \end{split}
 \label{eq_2_15}
 \end{equation} 
\end{definition}

From which it follows:
\begin{theorem}
    The tangent map  $\Lop_n(\xbf)\equiv DM_n(\xbf)$  satisfies the recurrence
\begin{equation}
    \begin{split}
  \Lop_n(\xbf) & =D \,M\bigl(M_{n-1}(\xbf),\,   n-1\bigr)= DM(\xbf_{n-1},n-1)\, \times  \\ \\
  & \times  DM_{n-1}(\xbf)=DM(\xbf_{n-1},n-1)\,\Lop_{n-1}(\xbf) 
    \end{split}
 \label{eq_2_16}
 \end{equation}
with initial condition $\Lop_0=\Iop$. 
\end{theorem}

The initial deviation is a random vector $\eps\,\xibf$ where $\xibf$ has a unit covariance  matrix
$\mean{\xibf \, \xibf^T}=\Iop$. 
The linear response  after $n$ iterations is $\Xibf_n(\xbf)=\Lop_n(\xbf)\,\xibf$ 
and its covariance matrix is $\Sigma^2_{L\,\,n}(\xbf) = \Lop_n(\xbf)\,\Lop_n^T(\xbf)$.
We denote with $I^{(k)}_{L\,\,n}(\xbf)$ the invariants of the covariance matrix and with 
$E_{L\,\,n}^{(k)}$ the Lyapunov errors of order $k$ given by their square root.
Whenever there is no ambiguity the Lyapunov invariants and errors will be denoted by $I^{(k)}_{n}(\xbf)$
and $E^{(k)}_{n}(\xbf)$.
%
%
%
%
%
%
%
%
%
%
%
%
\section{Geometrical  interpretation of  Lyapunov error  invariants}
\label{S:LyapunovInterpretation}
The  asymptotic  behaviour of  the  invariants of the 
covariance matrix    $\Sigma^2(\xbf,t)= \Lop(\xbf,t)\Lop^T(\xbf,t)$ is
governed by  the Lyapunov exponents.   In this section, we omit to write explicitly
the dependence on $\xbf$  of these matrices.
Notice that   $\Lop(t)\Lop^T(t)$ has the same spectrum as
the Lyapunov  matrix $\Lop^T(t)\Lop(t)$ but  their   eigenvectors are different.  

\def\diag{\hbox{diag}}

\begin{theorem}
    Any real matrix $\Lop(t)$ can be written according to the polar decomposition, called singular value decomposition (SVD)
in the  case of a real rectangular matrix.
\begin{equation}
\Lop(t)= \Rop(t) e^{t\Lambda(t)}\,\Wop^T(t) \qquad  \Lambda(t)=\diag(\lambda_1(t),\ldots,\lambda_d(t)) 
\label{eq_3_1}
\end{equation}
\end{theorem}

where $\Rop(t)$ and $\Wop(t)$ are two orthogonal matrices and the exponents  are ordered
in a non-increasing sequence $\lambda_1(t)\ge\lambda_2(t)\ge\ldots\ge\lambda_d(t)$. 

As a consequence of Eq. \ref{eq_3_1}, we have
\begin{equation}
  \begin{split}
  \Lop^T(t)\Lop(t)& = \Wop(t) e^{2t\Lambda(t)}\,\Wop^T(t) \\ \\
  \Lop^T(t)\Lop(t)&  =  \Rop(t) e^{2t\Lambda(t)}\,\Rop^T(t) \qquad \quad
  \end{split}
\label{eq_3_2}
\end{equation}
Oseledet theorem  \cite{Oseledets1961} states that $(\Lop^T(t)\Lop(t))^{1/2t}$ has a limit $\Wop\,e^{\Lambda}\,\Wop^T$ where
the  elements $\lambda_j$ of the diagonal matrix $\Lambda= \lim_{t\to \infty}\,\Lambda(t)$ are
the Lyapunov exponents while $\Wop=(\wbf_1,\ldots,\wbf_d)= \lim_{t\to\infty}\,\Wop(t)$ is an orthogonal matrix,  whose columns
$\wbf_k$ are the corresponding  Lyapunov vectors.
There is no convergence theorem for $\Rop(t)$. Indeed if  a couple  or all the eigenvalues
  have an asymptotic power law behaviour, implying that two or all the Lyapunov exponents vanish,
  the limit of $\Rop(t)$,  might not  exist.  Counterexamples,  showing  $\Rop(t)$  has no limit
  due to indefinite oscillations,  are  provided in  Sect. \ref{S:twoSimpleModels}
  where the case of an integrable  Hamiltonian  near  elliptic point
  is explicitly worked out. 
       \begin{theorem}
The asymptotic behaviour of the Lyapunov error invariants 
is determined by the Lyapunov exponents according to 
 \begin{equation}
    \begin{split}
      & \lim_{t\to \infty}\, {1\over 2t}   \log I^{(k)}(t)  =
      \lim_{t\to \infty}\, {1\over 2t} \log\Biggl(\quad \sum_{1\le j_1<\ldots < j_k}^d  \\
     &  \quad \exp     \bigl( 2t \lambda_{j_1}(t)+  
     \ldots + 2t \lambda_{j_k}(t)\,\bigr) \Biggr) =  
         \lambda_1+\ldots+\lambda_k
\label{eq_3_3}
\end{split}
\end{equation} 
\end{theorem}

Recalling that:
\begin{theorem}
    For  a Hamiltonian system in $\Reali^{2d}$, the matrix $\Lop$ 
is symplectic $\Lop^T\,\Jop\,\Lop=\Lop\,\Jop\,\Lop^T=\Jop$ where
$\Jop=\begin{pmatrix} 0 & \Iop \\ -\Iop& 0 \end{pmatrix}$, and the matrix $\Lop^T(t)\,\Lop(t)$
is  symplectic  too. 
\end{theorem}

As a consequence, the eigenvalues are pairwise opposite
$\lambda_{d+k}(t)=-\lambda_{d-k+1}$ for  $1\le k\le d$.

\subsection{The volumes of parallelotops}
\def\Vol{{\hbox{Vol}}}
The invariants of the covariance matrix have a  simple geometric meaning.
Let $\ebf_j$  the  orthonormal base in
$\Reali^d$ with $(\ebf_k)_j=\delta_{jk}$ and  let $\ebf_j(t)=\Lop(t)\ebf_j$
be their images at time $t$. The vectors $\ebf_j(t)$
 are the columns of the matrix $\Lop(t)$   namely 
 $\Lop(t)= \bigl(\ebf_1(t)\,\ebf_2(t),\ldots,\ebf_d(t)\bigr)$.
Written in a more explicit form the  vectors $\ebf_j(\xbf,t)$ read 
 \begin{equation}
   \begin{split}
     \ebf_i(t)= \Rop(t) \,\,\begin{pmatrix} e^{t\,\lambda_1(t)} \,\, \ebf_i\cdot \wbf_1(t) \cr
          \vdots \cr
         e^{t\,\lambda_d(t)} \,\, \ebf_i\cdot \wbf_d(t)
     \end{pmatrix}
\label{eq_3_4}
\end{split}
\end{equation}
 Consider  the parallelotopes
 $ \Pcal_{\ell_1\,\ell_2,\ldots,\,\ell_k}(t)$ whose sides are the vectors $\ebf_{\ell_1}(t),\ldots,\ebf_{\ell_k}(t))$, then there are   ${d\choose k}$  distinct  parallelotopes $ \Pcal_{\ell_1\,\ell_2,\ldots,\,\ell_k}(t)$
 with $k$ edges, corresponding to  all the possible  choices of   $k-$uples of  distinct  indices
 chosen from $1,2,\ldots,d$.  
 In the following, we introduce a series of theorems to prove that the asymptotic behavior of the Lyapunov Error's invariants $I^{(k)}(t)$, are governed by the Lyapunov exponents. 

 \begin{theorem}
 The  volume  squared of each parallelotope   is equal to
 the determinant of the   corresponding Graham   matrix $G_{j_1\,j_2,\ldots,\,j_k}(t)$
 whose entries are the scalar products   of these vectors.
 \begin{equation}
   \begin{split}
     &  \Bigl(\Gop_{\ell_1\,\ell_2,\ldots,\,\ell_k}(t)\Bigr)_{i\,j}\equiv  \ebf_{\ell_i}(t)\cdot \ebf_{\ell_j}(t) \quad \qquad
     i,j=1,\ldots,k  \\ \\
     &  \det \Bigl(\Gop_{\ell_1\,\ell_2,\ldots,\,\ell_k}(t)\Bigr)= \Vol^2\Bigl(\Pcal_{\ell_1\,\ell_2,\ldots,\,\ell_k}(t)\Bigr)
\label{eq3_5}
\end{split}
\end{equation}
\label{def:Gram}
\end{theorem}
 
 We can prove Theorem \ref{def:Gram} for $k=1$ and $k=2$.
 
 \begin{proof}
 Fr the case $k=1$, given a vector $\ubf_\ell$, we have $G_\ell=\ubf_\ell\cdot\ubf_\ell=
 \Vol^2(\Pcal_\ell)$   
where $\Pcal_\ell$ is the oriented  segment $\ubf_\ell$.
For $k=2$  given   two vectors $\ubf_{\ell_1},\ubf_{\ell_2}\in \Reali^d$ which form an angle, we denote
by $\theta$
 \begin{equation}
   \begin{split}
     \det(G_{\ell_1\,\ell_2})& = \ubf_{\ell_1}\cdot\ubf_{\ell_1}\, \,\,\ubf_{\ell_2}\cdot\ubf_{\ell_2}-
    ( \ubf_{\ell_1}\cdot\ubf_{\ell_2})^2= \\
     & =  \Vert \ubf_{\ell_1}\Vert^2\, \Vert \ubf_{\ell_2}\Vert^2     \,\sin^2\theta =
     \Vol^2\bigl(\Pcal_{\ell_1\,\ell_2}\bigr)
\label{eq_3_6}
\end{split}
\end{equation}
 where $\Pcal_{\ell_1\,\ell_2}$ is the parallelogram whose sides are $\ubf_{\ell_1}$ and $\ubf_{\ell_2}$. 
 Given $d$ vectors $\ubf_1,\ubf_2,\ldots,\ubf_d\in \Reali^d$
 let $\Uop=(\ubf_1,\ldots,\ubf_d)$ and $\Gop_{1\,2\,\ldots d}=\Uop^T\,\Uop$ be the Graham
 matrix whose entry $i,j$ is $\ubf_i\cdot\ubf_j$.
  \end{proof}
  
 It is well known that $\det(\Uop)=\Vol(\Pcal_{1\,2\ldots d})$ which implies
 $\det(\Gop_{1\,2\,\ldots d})= \Vol^2(\Pcal_{1\,2\ldots d})$.
As a consequence  chosen $k$ vectors, for instance $\ubf_1,\ubf_2,\ldots,\ubf_k$ it is natural to interpret
$\det (\Gop_{1\,\ldots\,k})$ as the square of the parallelotope $\Pcal_{1\,\ldots,k}$ whose edges are precisely
these vectors.
Let's now compute the second invariant $I^ {(2)}(t)$ by using (\ref{eq_2_11}) and taking into account that
$(\Lop^T\Lop)_{ij}=\ebf_i^T\ebf_j=\ebf_i\cdot \ebf_j$
 \begin{equation}
   \begin{split}
   &  I^{(2)}(t)= {1\over 2} \Bigl[ \Bigl( \Tr(\Lop^T(t)\Lop(t)\Bigr)^2 - \Tr\Bigl( \bigl(\Lop^T(t)\Lop(t)\bigr)^2\Bigr) \Bigr] = \\ 
   &\;=  {1\over 2}\,\sum_{i,j=1}^d\Bigl( (\ebf_i(t)\cdot\ebf_i(t)\,\,\ebf_j(t)\cdot\ebf_j(t)
     - \bigl(\ebf_i(t)\cdot\ebf_j(t)\bigr)^2\Bigr)=  \\ 
     & \;= {1\over 2} \,\sum_{i,j=1}^d \,\Vol^2\bigl(\Pcal_{ij}(t)\bigr) = \sum_{1\le i<j}^d \,\Vol^2\bigl(\Pcal_{ij}(t)\bigr)
\label{eq_3_7}
\end{split}
\end{equation} 
 since $\Vol(\Pcal_{i \,i})=0$ and $\Vol(\Pcal_{i \,j}(t))=\Vol(\Pcal_{j \,i}(t))$.
 More generally  according to a standard result of external algebra the volume of a  parallelotope 
 with $k$ edges   is given by
 \begin{equation}
   \begin{split}
     \hskip -.2 truecm \Vol\bigl(\Pcal_{\ell_1\,\ell_2\,\ldots \ell_k}(t)\bigr) =
     \Vert \ebf_{\ell_1}(t)\wedge\ebf_{\ell_1}(t) \wedge
     \ldots\wedge \ebf_{\ell_k}(t)\Vert 
\label{eq_3_8}
\end{split}
\end{equation} 
 The square of the volume of  this parallelotope is equal to the determinant of the Graham matrix
 $\Gop_{\ell_1,\ldots,\ell_k}(t)$. 
 
 We can now define the following theorem:
 \begin{theorem}
      The  invariant $ I^{(k)}(t)$ is the sum of the squared volumes of all the
 distinct ${d\choose k}$    parallelotopes 
 \begin{equation}
   \begin{split}
     & I^{(k)}(t)=  \sum_{ \ell_1<\ell_2\ldots<\ell_k}^d \,\Vol^2\bigl(\Pcal_{\ell_1\,\ell_2\,\ldots \ell_k}(t)\bigr) =  \\  
   & \quad =  \sum_{ \ell_1<\ell_2\ldots<\ell_k}^d  \exp\Bigl(  2t\,\bigl (\lambda_{\ell_1}(t)+\lambda_{\ell_2}(t)+\ldots+\lambda_{\ell_k}(t)\bigr) \Bigr)
\label{eq_3_9}
\end{split}
\end{equation} 
 From Oseledet  theorem \cite{Oseledets1961} the  limit of $\lambda_j(t)$ exists for $t\to \infty$
 and is given by the Lyapunov exponent $\lambda_j$. 
 \label{def:lyapVol}
  \end{theorem}

 As a consequence of Theorem \ref{def:lyapVol}, the invariants $I^{(k)}(t)$, which do not depend on  the  choice of the initial
 orthonormal  base vectors, have an asymptotic behaviour which is governed by the sum
 of the first $k$ Lyapunov exponents. We recall that the matrices $\Lop$ and $\Sigma^2_L$ depend on
 $t$ and the initial point $\xbf$,  even though we omitted to write
 this dependence explicitly. Below we recall the properties of the Lyapunov invariants:
\begin{itemize}
  \item  The  eigenvalues of $\Sigma^2_L(\xbf,t)=\Lop^T(\xbf,t)\Lop^T(\xbf,t)$  are $e^{2t\,\lambda_j(\xbf,t)}$ for $j=1,\ldots,d$ and   their invariants are $I^{(k)}_L(\xbf,t)$.
  \item The Lyapunov exponents   $\lambda_j(\xbf)$ are the limits of $\lambda_j(\xbf,t)$ for $t\to +\infty$ and depends on the initial conditions.
  \item  The limit of $(2t)^{-1}\log I^{(k)}(\xbf,t)$ for $t\to \infty$,  given by the sum of the first $k$ Lyapunov exponents, depends on $\xbf$ as well.
  \item If the system has ergodic invariant components, then the Lyapunov exponents  have the same value  for almost every  initial point $\xbf$ in each one of these  domains.
\end{itemize}
 
 The advantage of using the invariants is evident. For an initial displacement if the direction
 $\ebf_i$   corresponds to a linear response $\ebf_i(\xbf,t)= \Lop(\xbf,t)\,\ebf_i$   whose norm squared is
\begin{equation}
 \Vert\ebf_i(\xbf,t)\Vert^2 = \sum_{j=1}^d \,e^{2t \,\lambda_j(\xbf,t)} \,\bigl(\,\ebf_i\cdot \wbf_j(\xbf,t)\,\bigr)^2
\label{eq_3_10}
\end{equation} 
 If the direction $\ebf_i$  of the chosen initial displacement is orthogonal
 to the Lyapunov eigenvector $\wbf_1(\xbf)$  the leading behaviour corresponding to the largest Lyapunov exponent
 is missed. If $\ebf_i$ is almost
 orthogonal to $\wbf_1$ then from the norm of  $\ebf_i(\xbf,t)$ we recover the largest Lyapunov exponent
 only for $t\gg\tau$ where 
\begin{equation}
\begin{split}
  \tau= {1\over \lambda_1-\lambda_2}  \,\log\parbar{\ebf_i\cdot \wbf_2\over \ebf_i\cdot \wbf_1}
  \end{split}
\label{eq_3_11}
\end{equation} 
 Taking the sum over $i$ the leading term   $e^{2t\,\lambda_1(t)}$  has  a unit  coefficient
 and therefore the first Lyapunov exponent is correctly obtained.
 This is the key  difference between our Lyapunov indicators  and other  indicators which
 depend on the choice of the initial displacements.

%
%
%
%
\section{Forward and Reversibility error invariant indicators}
\label{S:TheReversibility Error}
In the previous section, we have analyzed the sensitivity to a random initial displacement
$\eps \xibf$  defined by the linear response limit $\eps\to 0$.
Here, we instead analyze the sensitivity to a small noise during the evolution
by considering the linear response when the $\eps\to 0$ limit of
the noise amplitude.
The validity conditions for the linear approximation are discussed in Appendix II.
Following the framework defined in \ref{S:Introduction}, we first study \textit{ noise} Forward evolution (hereafter defined as F) from $t=0$
up to $t>0$ and the corresponding linear response which is a random vector
of zero mean and  covariance matrix $\Sigma^2_F(\xbf,t)$. The trace of the covariance matrix
gives the square of the Forward Error (FE)
and its higher invariants characterize its spectrum.
We also consider the \textit{noisy} Backward evolution (hereafter defined as B) from $t$ to $2t$ by reversing the
vector field and changing $t'$ into $2t-t'$. In the absence of noise, at time $2t$ we are back to the initial condition. 

When applying both the F and B evolutions, the computation of the trace of the covariance matrix $\Sigma^2_{BF}(\xbf,t)$ provides the square of the Reversibility Error (RE) which, jointly with the higher
invariants, quantifies the reversibility violation in the
linear response limit.
Formally, a different value for RE could be computed, by applying first B and after the F evolution, for Hamiltonian systems the resulting RE is the same. 

The presence of noise during both the F and B evolutions
mimics the effect of round-off in numerical computations, we also note that the Reversibility Error is
the same as when the noise is present only in the F or B evolution (up to a $\sqrt{2}\,\,$
factor).  
\\
Without loss of generality, in the following text, the covariance matrix for the RE is denoted by $\Sigma^2_R$
and the corresponding invariants are denoted by $I^{(k)}_R$.
\spa
\subsection{The F process}   
\pan

\begin{definition}
We define a \texttt{white noise} by $\xibf(t')$ with $\mean{\xibf(t')}=0$ and
$\mean{\xibf(t')}\xibf^T(t'')=\Iop \,\,\delta(t'-t'')=0$.
\label{def:whiteNoise}
\end{definition}

The white noise might be replaced by a stationary
process, such as the Uhlenbeck process, so that the covariance matrix is
$\mean{\xibf(t')}\xibf^T(t'')=\Iop \,\, C(t'-t'')=0$. However, a delta-correlated noise simplifies the computation with respect to a process  for  which $C(t)$ has an exponential or power-law decay.
Using \ref{def:S} and \ref{def:whiteNoise}, we define the evolution  $\xbf(t')=S_{t'}(\xbf)$  
and the noisy evolution $\ybf(t')$  for $0\le t'\le t$ which satisfy the
following differential equations 
\begin{equation}
\begin{split}
  {d \over dt'} \xbf(t') &  =\Phibf(\xbf(t'), t')  \\ \\ 
  {d \over dt'} \ybf(t') &  = \Phibf(\xbf(t'), t') +\eps\,\xibf(t') 
  \end{split}
\label{eq_4_1}
\end{equation} 
The second equation in \ref{eq_4_1}, is equivalent
to $d\ybf(t')= \Phibf dt' + \eps d\wbf(t')$ where $\wbf(t')$ is the Wiener noise.

Similarly to \ref{def:LinResponse}, we can now introduce
\begin{definition}
the \texttt{linear response} for the F evolution $\Xibf(\xbf,t)$ is defined according to 
\begin{equation}
\begin{split}
  & \Xibf_F(\xbf,t') =  \lim_{\eps \to 0}   {\ybf(t')-\xbf(t')\over \eps}  \\ \\
  &   {d\over dt'} \Xibf_F(\xbf,t')= D\Phibf(S_{t'}(\xbf),t')\,\Xibf_F(\xbf,t') +\xibf(t')
  \end{split}
\label{eq_4_2}
\end{equation} 
\end{definition}
It is not straightforward to justify the second equation in \ref{eq_4_2}, obtained by inverting the
time derivative with the $ \eps\to 0$ limit. The problem disappears  when  the flow is replaced by a map, which is always the case
in numerical computations. The problem  in this case is to control the global truncation  error
by choosing a time step sufficiently small so that the  error can be neglected.

We can now expand \ref{def:FondMatrix} and \ref{def:CovMatLy} to the F process as follow
\begin{theorem}
Giving the random vector $\Xibf_F(\xbf,t)$, the \texttt{fundamental solution} $\Lop(\xbf,t')$ and the \texttt{covariance matrix} are given by
\begin{equation}
\begin{split}
 &  \Xibf_F(\xbf,t) =  \Lop(\xbf,t)\,\int_0^t \,\Lop^{-1}(\xbf,t')\,\xibf(t')\,dt' \\ \\
 &  \Sigma^2_F(\xbf,t)= \Lop(\xbf,t)\,\,\int_0^t \,\Bigl(\Lop^T(\xbf,t')\Lop(\xbf,t')\Bigr)^{-1}
 \,dt'\,\,\Lop^T(\xbf,t)
  \end{split}
\label{eq_4_3}
\end{equation} 
\label{def:CovForward}
\end{theorem}

By using \ref{def:CovForward}, we can then define the FE as
\begin{definition}
The \texttt{Forward Error} (FE) is defined as the square root of the trace of $\Sigma^2_F(\xbf,t)$
and the higher invariants of this matrix can be computed.  
\label{def:FE}
\end{definition}
\spa
\subsection{The BF process}   
\pan
In absence of noise, the BF evolution in the time interval $[0,2t]$ is given by
\begin{equation}
\begin{split}
 &  {d \over dt'} \xbf(t')  =\phantom{-}\Phibf(\xbf(t'), t')      
                       \qquad \qquad \qquad       0\le t'\le t  \\
 &     {d \over dt'} \xbf(t')  = -\Phibf(\xbf(t'), 2t-t')   
                       \qquad \qquad   t\le t'\le 2t \phantom{\Biggl(}
  \end{split}
\label{eq_4_4}
\end{equation} 
The solution of equation (\ref{eq_4_4})
has the time-reflection symmetry which
reads $\xbf(t')=\xbf(2t-t')$  for $0\le t'\le 2t$.
The reversibility condition $\xbf(2t)=\xbf(0)$ is broken if we introduce a small additive
noise. 
Similarly to how we defined the \textit{noisy} F evolution in \ref{eq_4_1}, the \texttt{noisy} solution  $\ybf(t')$  satisfies the  stochastic equation   
\begin{equation}
\begin{split}
  {d \over dt'} \ybf(t') & =\phantom{-}\Phibf(\ybf(t'), t')  +\eps\xibf(t')  
  \qquad  \qquad \;    0\le t'\le t
  \\ 
   {d \over dt'} \ybf(t') & = -\Phibf(\ybf(t'), 2t-t')  +\eps\xibf(t')  \qquad      t\le t'\le 2t  \phantom{\Biggr )}
  \end{split}
\label{eq_4_5}
\end{equation}    
We may write   $ \ybf(t')= \xbf(t') +\eps\,\Xibf_{R}(\xbf, t')  + \eps^2 \Rbf(\xbf,t')$ but it is not simple to control the remainder $\Rbf$. For this reason, we introduce the following definition of linear response. 
\begin{definition}
The \texttt{linear response} $\Xibf_{BF}(\xbf,t')$ is defined by the $\eps\to 0$ limit as
\begin{equation}
\begin{split}
  \Xibf_{BF}(\xbf,t') =\lim_{\eps \to 0}\,{\ybf(t')-\xbf(t') \over \eps} 
  \end{split}
\label{eq_4_6}
\end{equation}
For the linear response theory of nonlinear equations with noise see \cite{Zang2020}.
\end{definition}

Under suitable regularity conditions on the vector field and the noise, which allow
to interchange the time derivative with the $\eps\to 0$ limit, the following definition of linear equation can be provided: 
\begin{definition}
The stochastic process $\Xi_{BF}(\xbf,t')$ in the time interval $[0,2t]$ satisfies the \texttt{linear equation} defined as
\begin{equation}
\begin{split}
  &  {d \over dt'} \Xibf_{BF}(\xbf,t')  = \\ 
  & D\Phibf(\xbf(t'), t')\Xibf_{BF}(\xbf,t')+\xibf(t'), \hskip 1  truecm 0\le t'\le t  \\ 
&      {d \over dt'} \Xibf_{BF}(\xbf,t')  = \\ 
& -D\Phibf(\xbf(2t-t'), 2t-t')\Xibf_{BF}(\xbf,t')+\xibf(t') \\
& \hskip 1 truecm t\le t'\le 2t
  \end{split}
\label{eq_4_7}
\end{equation} 
\end{definition}
where the matrix $D\Phibf(\xbf,t)$ is  the matrix whose elements are
$(D\Phibf)_{ij}=\partial \Phi_i/\partial x_j$.
The solution of the linear equations
(\ref{eq_4_7})  is given by 
\begin{equation}
\begin{split}
 &  \Xibf_{BF}(\xbf,t') = \Lop(\xbf,t')\,\int_0^{t'}\,\Lop^{-1}(\xbf,s)\,\xibf(s)ds  \qquad   0\le t'\le t  \\ 
 &   \Xibf_{BF}(\xbf,t') = \Lop(\xbf,(2t-t'))\,\Lop^{-1}(\xbf,t) \Xibf_R(\xbf,t) + \phantom{\Bigl(} \\ 
 & \; + \Lop(\xbf, 2t-t') \,\int_t^{t'}\,\Lop^{-1}(\xbf,(2t-s)\,\xi(s)\, ds   \qquad t\le t'\le 2t  \phantom{\Bigl(}
  \end{split}
\label{eq_4_8}
\end{equation} 
The solution at time $2t$ for the linear response in Eq. \ref{eq_4_6} reads  
\begin{equation}
\begin{split}
  & \hskip -.5 truecm \Xibf_{BF}(\xbf,2t) = \int_0^t\,\Lop^{-1}(\xbf,s)\,\xibf(s) + \\
  & + \int_t^{2t }\,\Lop^{-1}(\xbf,2t-s)\,\xi(s)\, ds  = \\
  & \qquad \quad = \int_0^{t }\,\Lop^{-1}(\xbf,s)\,\bigl(\xi(s)+\xibf(2t-s)\bigr) \, ds
  \end{split}
\label{eq_4_9}
\end{equation} 
We notice that if the noise satisfied $\xibf(2t-t')=-\xibf(t')$ for $0\le t'\le t$ then we would have 
$\Xibf_{BF}(\xbf,2t-t')=\Xibf_{BF}(\xbf,t')$  implying  $\Xibf_{BF}(\xbf,2t)=\Xibf_{BF}(\xbf,0)=0$.
However, this is not the case since the noise for any $t'\in[0,2t]$ is uncorrelated.
Assuming the  noise components are stationary and uncorrelated the correlation
matrix for $\Xibf_{BF}(\xbf,2t)$ can be computed.
If $\xibf(t')$ is a white noise 
\begin{equation}
\begin{split}
  \mean{\xibf(t')\xibf^T(t'')}= \Iop \,\,\delta(t'-t'') \qquad t',\,t''>0 
  \end{split}
\label{eq_4_10}
\end{equation} 
which affects both the F and B evolution. 
\begin{theorem}
The \texttt{covariance matrix} for the BF process is defined as 
\begin{equation}
\begin{split}
 \Sigma^2_{BF}(\xbf,t) &  ={1\over 2}\,\mean{\Xibf_{BF}(\xbf,2t) \Xibf^T_{BF}(\xbf,2t)}= \\ 
  &\quad = \int _0^t \bigl(\Lop^T(\xbf,s)\,\Lop(\xbf,s)\bigr)^{-1} \,ds
  \end{split}
\label{eq_4_11}
\end{equation} 
\end{theorem}
The  invariants $I_{BF}^{(k)}(\xbf,t) $  of the covariance matrix (\ref{eq_4_11}) provide a
complete characterization of the BF process defined by (\ref{eq_4_9}).
The following properties hold when the linear approximation is valid for the BF process, then:
\begin{itemize}
  \item $\ybf(2t)=\xbf + \eps \Xibf_{BF}(\xbf,t) + O(\eps^2) $ is a random process whose average is $\xbf$.
  \item The covariance matrix is  $\eps^2\Sigma^2_{BF}(\xbf,t)$.
  \item The probability density is multivariate Gaussian.
  \item The transition probability density from $(\xbf,0)$ to $(\ybf,2t)$ is given by $G(\ybf-\xbf, \eps^2\Sigma^2_{BF}(\xbf,t))$.
\end{itemize}

If the noise is present only in the $F$ process, namely in the  $[0,t]$ time-interval,
then the random deviation at time $2t$  is given by
\begin{equation}
\begin{split}
  \Xibf_{BF}(x,t)=\Lop^{-1}(\xbf,t)\,\Xibf_F(\xbf,t)=
  \int_0^{t }\,\Lop^{-1}(\xbf,s)\,\,\xi(s) \, ds
  \end{split}
\label{eq_4_12}
\end{equation} 
We can compute the value for the covariance matrix as follow
\begin{theorem}
The \texttt{covariance matrix} when a random noise is applied only to the F process is defined by 
\begin{equation}
  \begin{split}
    \Sigma^2_{BF}(\xbf,t)& = \mean{\Xibf_{BF}(\xbf,2t) \Xibf^T_{BF}(\xbf,2t)}= \\ \\
    & = \int _0^t \bigl(\Lop^T(\xbf,s)\,\Lop(\xbf,s)\bigr)^{-1} \,ds
  \end{split}
\label{eq_4_13}
\end{equation} 
and the result is the same as the r.h.s. of equation (\ref{eq_4_11}).
\end{theorem}

%
%
%
%
\subsection{BF invariants for Hamiltonian systems }
For a Hamiltonian system the fundamental matrix $\Lop(\xbf,t)$ is symplectic
and the same property is inherited by $\Lop^T\Lop$ and $\Lop\Lop^T$. Letting $\xbf^T=(\qbf^T,\,\pbf^T)$
where $\qbf^T=(q_1,\ldots,q_d)$,  $\pbf^T=(p_1,\ldots,p_d)$ and $q_i$ ,$p_i$  denote  the coordinates and
conjugate moments  we have 
\begin{equation}
\begin{split}
 &  \Lop\Jop\Lop^T =\Lop^T\Jop\Lop=\Jop \qquad \Jop=\begin{pmatrix} 0  & & \Iop \\ \\ -\Iop & & 0 \end{pmatrix}  \\ 
 &  (\Lop^T \Lop)\Jop  (\Lop^T \Lop)= \Jop \qquad \quad  (\Lop \Lop^T)\Jop  (\Lop \Lop^T)= \Jop  \phantom{\Biggl(}
  \end{split}
\label{eq_4_23}
\end{equation}
From $\Lop^{-1}= \Jop\,\Lop^T\,\Jop^{-1}$ it follows that $\Tr(\Lop^{-1})= \Tr (\Lop^T)=\Tr (\Lop)$
and  that  the traces of $(\Lop^T\Lop)^{-1}$  and  $\Lop^T\Lop$ are equal.
In the following, we report

A consequence
the reversibility error  $E_{BF}(\xbf,t)= \parton{I^{(1)}_{BF}(\xbf,t)}^{1/2}$ is  related to  the Lyapunov error
by 
\begin{equation}
\begin{split}
  E_{BF}^2& (\xbf,t)= \Tr \int _0^t \,\bigl( \Lop^T(\xbf,s)\,\Lop(\xbf,s)\bigr)^{-1} \,ds = \\ 
 &= \int _0^t \Tr( \Lop^T(\xbf,s)\,\Lop(\xbf,s)\bigr)\,ds \equiv
  \int _0^t E^2_L(\xbf,s)\, \,ds
  \end{split}
\label{eq_4_24}
\end{equation}   
Concerning the higher invariants we notice that
\begin{equation}
\begin{split}
 &  \Tr(\Sigma_{BF}^{2k})  =\int _0^t ds_1\ldots ds_k \Tr\Bigl( \bigl(\Lop^T(\xbf,s_1) \Lop(\xbf,s_1)\bigr)^{-1}
  \cdots \\ 
   & \qquad  \cdots \bigl(\Lop^T(\xbf,s_k) \Lop(\xbf,s_k)\bigr)^{-1}\,\Bigr) 
  = \int _0^t ds_1\ldots ds_k    \\ 
  & \qquad  \Tr\bigl( \Lop^T(\xbf,s_k) \Lop(\xbf,s_k)\cdots   \Lop^T(\xbf,s_1) \Lop(\xbf,s_1) \bigr)
  \phantom{\int}
  \end{split}
\label{eq_4_25}
\end{equation} 
As a consequence the matrix   $\Sigma_{BF}^2(\xbf,t)$   has the same invariants as the matrix
$\overline{\Sigma}_{BF}^2(\xbf,t)$  obtained  by replacing
$(\Lop^T\Lop)^{-1}$ with $\Lop^T\Lop$ in the integral defining it
\begin{equation}
\begin{split}
\overline{\Sigma}_{BF}^2(\xbf,t) = \int _0^t \,\Lop^T(\xbf,s)\,\Lop(\xbf,s) \,ds 
  \end{split}
\label{eq_4_26}
\end{equation}   
The same procedure shows that  invariants  and the spectrum of  of $\Sigma_{FB}^2(\xbf,t)$ are
the same as for the matrix $\overline{\Sigma}_{BF}^2(\xbf,t)=\int_0^t \Lop^T(\xbf,-s) \Lop(\xbf,-s)\,ds$.
\\ \\
Letting   $\Iop_R=\begin{pmatrix} \Iop & 0 \\ 0 & -\Iop \end {pmatrix}$  a Hamiltonian system is time reversal
invariant  if  $S_{-t}(\xbf)= \Iop_R \,S_t(\Iop_R\xbf)$
so that $\Lop(\xbf,-t)= \Iop_R\Lop(\Iop_R\,\xbf,t)\,\Iop_R$. In this case we have
\begin{equation}
\begin{split}
  \Sigma^2_{FB}(\xbf,t)= \Iop_R \, \int_0^t \,dt' \Bigl(\Lop^T(\Iop_R\xbf,t)\,\Lop(\Iop_R\xbf,t) \Bigr)^{-1}\,\Iop_R
\end{split}
\label{eq_4_27}
\end{equation} 
An autonomus system is time reversal invariant  if  $H(\Iop_R\xbf)=H(\xbf)$,  a condition satisfied
for instance when  $H=T(\pbf)+V(\qbf)$  if  $T(-\pbf)=T(\pbf)$. In this case  invariants of the $FB$
covariance matrix  computed at $\xbf, t$ are the same as the invariants of $FB$ computed at $\Iop_R\xbf,\,t$.
This means the invariants are the same provided that we change $\pbf$ into $-\pbf$
\begin{equation}
\begin{split}
I_{FB}^{(k)}(\qbf,\pbf,t)= I_{BF}^{(k)}(\qbf,-\pbf,t)
\end{split}
\label{eq_4_28}
\end{equation} 
For more details see Appendix IX
%
%
%
%
\subsection{The BF  process for a map}
Given a  map  $M(\xbf,n)$  with a unique inverse $M^{-1}(\xbf,n)$  we consider the 
forward orbit $\xbf_{n'}$ 
for $1\le n'\le n$ and  the backward  orbit $\xbf_{n'}$  for  $  n+1\le n'\le 2n$ defined by
\begin{equation}
\begin{split}
  & \xbf_{n'}= M(\xbf_{n'-1},n'-1) \qquad  \qquad  1\le n'\le n \\ \\
  &\xbf_{n'}= M^{-1}(\xbf_{ n'-1},2n-n')\qquad n+1 \le n'\le 2n
\end{split}
\label{eq_4_29}
\end{equation} 
The inverse map being defined by $M^{-1}(M(\xbf,k),k)=\xbf$. The orbit has the property
$\xbf_{n+k}=\xbf_{n-k}$ for $1\le k\le n$ which implies the following symmetry
 $\xbf_{n'}= \xbf_{2n-n'}$ by setting $n'=n-k$. The end point is equal to the initial point  $\xbf_{2n}=\xbf$.
By adding a small random deviation at each iteration
we obtain the orbit $\ybf_{n'}$ with $\ybf_0=\xbf$ defined by 
\begin{equation}
 \begin{split}
   &\ybf_{n'}  = M(\ybf_{n'-1},n'-1) +\eps\xibf_{n'}\qquad\qquad  1\le n'\le n \\ \\ 
   &\ybf_{n'} = M^{-1}(\ybf_{n'-1}, 2n-n')+\eps  \xibf_{n' }  \qquad   n+1   \le n'\le 2n
  \end{split}
\label{eq_4_30}
\end{equation}
where $\xibf_{n'}$ has a unit covariance matrix $\mean{ \xibf_{n'}\,\xibf^T_{n''} }= \Iop\, \delta_{n'\,n''}$.
The linear response is defined by 
\begin{equation}
  \Xibf_{BF\,\,n'}(\xbf)= \lim_{\eps\to 0}\,\,{\ybf_{n'}-\xbf_{n'}\over \eps} \qquad 1\le n'\le 2n
\label{eq_4_31}
\end{equation}
and satisfies  the following linear recursion relation  
\begin{equation}
  \begin{split}
    & \Xibf_{BF\,\,n'}(\xbf) = DM(\xbf_{n'-1},n'-1)\, \Xibf_{BF\,\,n'-1}(\xbf) + \xibf_{n'}  \\
    & \hskip 6 truecm   1\le n'\le n \\ \phantom{\Biggl)} 
    & \Xibf_{BF\,\,n'}(\xbf) = DM^{-1}(\xbf_{2n-n'+1},2n-n')\, \Xibf_{BF\,\,n'-1}(\xbf)+ \\
    & \hskip 4 truecm + \xibf_{n'}\qquad   n+1 \le n'\le 2n
  \end{split}
\label{eq_4_32}
\end{equation}
Since  $M^{-1}(M(\xbf,k),k) =\xbf$   from
$\xbf_{k+1}=M(\xbf_k,k)$  we obtain  $\xbf_k=M^{-1}(\xbf_{k+1},k)$.    
From the recurrence of the tangent map, we get
\begin{equation}
 \begin{split}
  &  \Lop_k(\xbf) =  DM_k(\xbf)= \\ 
  & DM(M_{k-1}(\xbf),\,k-1) \,DM_{k-1}(\xbf)=  \\
   &\qquad \;= DM(M_{k-1}(\xbf),\,k-1)\, \Lop_{k-1}(\xbf)     \\ \\
   & DM(\xbf_{k-1},\,k-1) 
   = \Lop_k(\xbf)\,\Lop_{k-1}^{-1}(\xbf)
  \end{split}
\label{eq_4_33}
\end{equation}
Replacing (\ref{eq_4_32}) in the first recurrence (\ref{eq_4_31}) and  taking into account that
$\Xibf_0(\xbf)=0$, we obtain 
\begin{equation}
 \begin{split}
   \Xibf_{BF\,\,n'}(\xbf) & = \Lop_{n'}(\xbf)\,\Lop_{n'-1}^{-1}(\xbf)\,\Xibf_{BF\,\,n'-1}(\xbf)\,+ 
   \, \xibf_{n'}  \\   & \hskip 3.5 truecm  1\le n'\le n 
  \end{split}
\label{eq_4_34}
\end{equation}
For $n'=n$ the random vector  provides the linear response to the forward  process and we can write
\begin{equation}
 \begin{split} 
  \Xibf_{F\,\,n}(\xbf) \equiv \Xibf_{BF\,\,n}(\xbf)= \,\Lop_n(\xbf)\,\sum_{k=1}^n\,\Lop_k^{-1}(\xbf) \,\xibf_k   
\end{split}
\label{eq_4_35}
\end{equation}
The second recurrence of equation (\ref{eq_4_32})  can be written taking into account the
relation
\begin{equation}
 \begin{split}
  & DM^{-1}(\xbf_{2n-n' +1},2n-n') = \\ 
   & = \Bigl(DM(\xbf_{2n-n'},2n-n')\,\Bigr)^{-1}= \\
  & = \Lop_{2n-n'}(\xbf)\,\Lop^{-1}_{2n-n'+1}(\xbf)
  \end{split}
\label{eq_4_36}
\end{equation}
which follows from $DM^{-1}(M(\xbf,k),k)\,DM(\xbf,k)=\Iop$ and (\ref{eq_4_33}).
As a consequence, taking into account that the second recurrence is initialized by
$\Xibf_{BF\,\,n}(\xbf)$, we obtain 
\begin{equation}
 \begin{split}
   & \Xibf_{BF\,\,n'} (\xbf) = \Lop_{2n-n'}(\xbf)\,\Lop^{-1}_{2n-n'+1}(\xbf)\,
   \Xibf_{n'-1}^{BF}(\xbf)+  \xibf_{n'}\\
     & \hskip 1 truecm n+1 \le n'\le 2n  \\  
   & \Xibf_{BF\,\,2n}(\xbf)= \Lop_n^{-1}\,\Xibf_{BF\,\,n}(\xbf)+ \sum_{k=0}^{n-1}\,\Lop^{-1}_k(\xbf)\,
   \xibf_{2n-k}  \\
\end{split}
\label{eq_4_37}
\end{equation} 
If the backward iteration is performed without noise, we immediately obtain   
\begin{equation}
 \begin{split}
   & \Xibf_{BF\,2n} (\xbf) = \Lop_n^{-1}\,\Xibf_{BF\,\,n}(\xbf) \equiv 
   \Lop_n^{-1}\,\Xibf_{F\,\,n}(\xbf)= \\
   & = \sum_{k=1}^n\,\Lop_k^{-1}(\xbf) \,\xibf_k   
\end{split}
\label{eq_4_38}
\end{equation} 
If the noise is present also during  the B iterations we obtain 
\begin{equation}
 \begin{split}
   \Xibf_{BF\,2n} (\xbf) & =\sum_{k=1}^n\,\Lop_k^{-1}\,\xibf_k + \sum_{k=0}^{n-1}\,\Lop^{-1}_k \,\xibf_{2n-k}= \\
   &  = \sum_{k=1}^{n-1}\,\Lop_k^{-1}\,(\xibf_k+\xibf_{2n-k}) \,+
  \,\xibf_{2n}+ \Lop_n^{-1}(\xbf) \xibf_n
  \end{split}
\label{eq_4_39}
\end{equation}   
If the noise is present only in the F iterations, which amounts to choosing
$\xibf_{n'}=0$ for $n'\ge n+1$,      the covariance matrix
is given  by 
\begin{equation}
 \begin{split}
   \Sigma^{2}_{BF\,\,n}(\xbf) =\mean{\Xibf_{2n}(\xbf)\,\Xibf_{2n}(\xbf)} = 
        \sum_{n'=1}^{n} \,\bigl(\Lop^T_{n'}(\xbf)\Lop_{n'}(\xbf)\bigr)^{-1} 
  \end{split}
\label{eq_4_40}
\end{equation}  
If   the noise  is uncorrelated  and is present during the  F and B iteration  the  covariance matrix becomes
\begin{equation}
 \begin{split}
     \Sigma^{2}_{BF\,\,n}(\xbf)&  ={1\over 2} \mean{\Xibf_{2n}(\xbf)\,\Xibf_{2n}(\xbf)} = \\ 
    & =   {1\over 2}\Iop+ {1\over 2}\bigl(\Lop^T_n(\xbf)\Lop_n(\xbf)\bigr)^{-1}  
          +    \sum_{k=1}^{n-1} \,\bigl(\Lop^T_k(\xbf)\Lop_k(\xbf)\bigr)^{-1} 
  \end{split}
\label{eq_4_41}
\end{equation}  
For a symplectic map in $\Reali^{2d}$, the matrices  $\Lop_n$ and  $\Lop_n^T\Lop_n$ are simplectic  
and the first invariant, which is equal to the square of   the reversibility error $E_{BF}$,
is given by
\def\scr{\scriptstyle}
\begin{equation}
  \begin{split}
    E_{BF\,\,n}^2(\xbf) = \Tr\Bigl( \Sigma^2_{BF\,\,n}(\xbf)  \Bigr) = 
   \sum_{n'=1}^{n-1}\, E_{L\,n'}^2 + d +{\scr1\over \scr2}\,E_{L\,n}^2(\xbf)
  \end{split}
\label{eq_4_42}
\end{equation}
If the noise affects only  the B process 
then we have  $E_{BF\,\,n}^2 =\sum_{n'=1}^{n}\, E_{L\,n'}^2$

\subsection{REM. Round-off induced reversibility error method }
To  any map $M(\xbf,n)$, we associate another  map $M_{\eps_*}(\xbf,n)$ defined as  the result of its  numerical evaluation 
with accuracy $\eps_*$. To the  inverse $M^{-1}(\xbf,n)$ we associate $M^{-1}_{\eps_*}(\xbf,n)$, which is the  result
of a computation with finite accuracy.  Since $M_{\eps_*}^{-1}(M_{\eps_*}(\xbf,n),n)\not =\xbf$
we  consider  an alternative   
formulation, the Reversibility Error Method (REM),  based on the computation of
the normalized deviation from the initial point   induced  by round-off for the BF orbit. 
We  introduce the  matrix
\begin{equation}
  \begin{split}
  \Xop_{n}(\xbf)& = {1\over 2}\,  {\ybf_{2n} - \xbf \over
    \eps_*}\,\parton{\ybf_{2n} - \xbf \over
  \eps_*}^T
    \end{split}
  \label{eq_4_57}
\end{equation}  
where $\ybf_{2n}$ is the end point of the BF orbit computed with round-off
according to the recurrence 
\begin{equation}
  \begin{split}
    & \ybf_{n'} =M_{\eps_*}(\ybf_{n'-1},\,n'-1) \qquad \quad n'\le n \\
    &\ybf_{n'} =M_{\eps_*}^{-1}(\ybf_{n'-1},\,2n-n')\qquad n+1\le n'\le 2n \\ 
\end{split}
  \label{eq_4_58}
\end{equation}  
The matrix $\Xop_n$ has one postive eigenvalue, whose square root, 
denoted  $REM_n(\xbf)$,  defines the reversibility error induced by the rounf off.
The positive  eigenvalue,
whose eigenvector is $(\ybf_{2n}-\xbf)/\Vert \ybf_{2n}-\xbf\Vert$, 
is given by
\begin{equation}
  \begin{split}
   REM_n(\xbf) = {\Vert \ybf_{2n} - \xbf\Vert \over
    \sqrt{2}\, \eps_*}
\end{split}
  \label{eq_4_59}
\end{equation}  
The  matrix  $\Xop_n$ has the  eigenvalue 0 with multiplicity $d-1$ and eigenspace orthogonal
to the eigenvector of the unique  positive eigenvalue. In tis case no other invariants are provided
by the method.
For the eight bytes representation of reals, according to the standard IEE 754, in which 52
bits are reserved for the mantissa, the  amplitude of the round off error  is
$\eps_*= 2^{-53} \simeq 10^{-16}$.  The round off introduces pseudo random displacements from
  the exact orbit.  Uncorrelated random displacements  produce a similar result, 
  the key difference  being that  only one realization of the pseudo-random process is available.
As a consequence  $REM_n$  exhibits large fluctuations
when  $n$ varies or when we change the initial condition,
whereas  the reversibility error $E_{BF\,n}(\xbf)$ does not,  since 
it is the result of an average  over all possible  realizations of the noise
\cite{Hairer2006,Faranda2012,Turchetti2010b}. 
Finally, we notice that REM should be compared with the  BF  reversibility error  
computed for a finite noise amplitude  $\eps_*$, equal to the amplitude of round off,
rather than in the limit  of zero noise amplitude.  Indeed we should compare  REM,  defined by
 (\ref{eq_4_59})   where $\ybf_n$ is computed with round off,  with the random vector
$\Xibf_n=(\ybf_n-\xbf)/(\sqrt{2}\,\eps_*)$ where $\ybf_n$ is obtained  by iterating $M(\xbf,n')+ \eps_*\xibf_{n'}$
for $n'\le n$ and $M^{-1}(\xbf,2n-n')+ \eps_*\xibf_{n'}$ for $n+1\le n'\le 2n$  keeping
$\eps_*=10^{-16}$. The result is comparable with the $\eps_*\to 0$ limit
only if the linear approximation is valid. Taking the  square root of $\mean{ \Vert\Xibf_n\Vert^2}$
just  smoothes out the fluctuations when $n$ varies. 
The linear approximation differs from the linear response by a $O(\eps_*)$ remainder for which
it is  hard to provide rigorous bounds.
\\
\\
In Appendix II  we consider  the logistic equation   and the Lyapunov error
for and initial deviation $\eps \xi$ of finite amplitude $\eps$, 
were  $\xi$ is a random vector uniformly distributed with zero mean and unit variance,
and compare it with the linear response.
The linear approximation is valid for any $t\ge 0$  provided  that the initial condition is
is at a  finite distance from the origin   which is an unstable fixed point for the flow.
We choose  $x\ge \ell$ recalling that   for any  $t\ge 0$ the solution is bounded
for any  initial condition $x\ge 0$.  In this case the ratio $r$
between the averages of the absolute values of the  remainder $\eps^2R(x,t;\xi)$
and  of the linear approximation $\eps L(x,t) \xi$ is bounded by $\eps/\ell$ so that
the linear approximation is valid for $\eps/\ell\ll1$.
When the initial condition $x>0$  approaches  the origin the ratio $r$
is bounded by $\eps e^{\lambda\, t}$ so that
the linear approximation holds only for short times $\lambda t \ll \log(1/\eps)$
where $\lambda=1$ is the Lyapunov exponent at $x=0$.
 For  one dimensional
systems whose orbits are bounded for any $t$ and $x\in \Reali$  the linear approximation
holds provided  the initial condition is at a distance  greater than $\ell$ from any
unstable fixed point and $\eps/\ell\ll 1$. For plane integrable Hamiltonian systems
with  $H=p^2/2m+V(x)$ and a confining potential, the linear approximation
holds for initial conditions at  a  distance greater than $\ell$ from the  separatrices
provided that $\eps/\ell \ll 1$.  For higher dimensional systems it plausible that the
linear approximation holds  for any $t>0$ 
if the  initial condition is  a finite distance not less than $\ell$
from unstable fixed points and manifolds provided that $\eps/\ell \ll 1$. 
\\
\\
No estimates are available to justify the linear approximation when a small additive
noise in introduced to perturb the vector field even for one dimensional systems
or plane Hamiltonian systems.  However for a one dimensional system
we  believe  that  the linear
approximation is valid for any $t>0$  and any $x$   such that the orbit is bounded
provided that the distance of $x$ from any unstable point   is greater than $\ell$
and $\eps/\ell \ll 1$. For a  plane Hamiltonian system with bounded orbits
we believe that the linear
approximation is  justified for any initial condition whose distance from any
separatrix is greater  than $\ell$   and $\eps/\ell \ll 1$.
\\
The motivation comes from numerical simulations performed for a variety of models.
When we have a continuous time model the computation is performed by 
a sequence of   discrete times  $t_n=n\dt$    and the exact orbit   $S_{t_n}(\xbf)$
is approximated by  $M^{\circ n}(x)$ where $M(\xbf)$ is an integrator
of order $m\ge 1$ whose local error is $(\dt)^{m+1}$. 
If we add a white noise $\eps\xibf(t)$   to the  vector field $\Phibf(\xbf)$ generating the flow $S_t(\xbf)$
the REM flow at time $2t_n$ is approximated by $\ybf_{2n}$ defined by the linear approximation recurrence as
\begin{equation}
 \begin{split}
   \ybf_{n'}& =  M(\ybf_{n'-1})+ \eps\,\sqrt{\dt}\,\,\xibf_{n'}  \qquad  1\le n'\le n   \\ \\
    \ybf_{n'}& =  M^{-1}(\ybf_{n'-1})+ \eps\,\sqrt{\dt}\,\,\xibf_{n'}  \qquad  n+1\le n'\le 2n
   \end{split}
\label{eq_4_60}
\end{equation}  
initialized by $\ybf_0=\xbf$.   In this case, we may
identify $\eps_*$ with $\eps\,\sqrt{\Delta t}$  and consequently the error
for the orbit  $y_{2n}$ obtained for a single realization of noise given  by (\ref{eq_4_58})
is  
\begin{equation}
 \begin{split}
{\Vert \ybf_{2n}-\xbf \Vert \over \sqrt{2}\,\, \eps} = {\Vert \ybf_{2n}-\xbf \Vert \over \sqrt{2}\,\,\eps_*}\,\sqrt{\dt}
\simeq REM_n \,\sqrt{\dt}  
  \end{split}
\label{eq_4_61}
\end{equation}  
Replacing $\Vert \ybf_{2n}-\xbf \Vert$ with  $\mean{\Vert \ybf_{2n}-\xbf \Vert^2}^{1/2}$
the fluctuations of the   stochastic process are damped and taking the $\eps\to 0$
limit one obtains the Reversibility Error. Choosing a finite value of the noise amplitude
$\eps=\eps_*/\Delta t$  the results for  RE  and REM are comparable if the linear approximation
is valid. The fluctuations of REM can be smoothed with a running average.
In the absence of a rigorous justification of the linear approximation
the claim that REM is close to RE  after smoothing its fluctuations comes
from empirical evidence provided by the numerical results on a variety
of models, most of which Hamiltonian.
To this end we remark that RE and REM  for one dimensional systems are  typically computed  at $N_g$
points,  which are the centers of $N_g$ equal sub-intervals partitioning an interval of length $L$.
Choosing  $\ell=L/N_g^2$ the linear approximations holds if $\eps/\ell\ll 1$. As a consequence
the linear approximation holds provided that  $\eps/\ll 1$ where 
\begin{equation}
 \begin{split}
 {\eps\over \ell}={\eps_*\over \sqrt{\dt}} { N_g^2\over L}= {10 ^{-16}  N_g^2\over \sqrt{\dt}} {1\over L}
  \end{split}
\label{eq_4_62}
\end{equation}  
For $L=1,\; N_g=10^3, \,\dt=10^{-4}$ the ratio is $10^{-6}$ so that the condition is satisfied.
The  probability  that in the interval to which an unstable fixed belongs the center   has a  distance
less than $\ell$ from the fixed point  is $1/N_g$. Of course if we make a magnification by  decreasing   $L$
the accuracy of the linear approximation decreases.
For a plane Hamiltonian system we choose a grid of $N_g\times N_g$  cells and compute the
reversibility error at their centers. Letting $\ell=L/N_g^2$  the probability
that in any  cell intersecting  the separatrix the distance of the  the center from the separatrix
is less than $\ell$  is still $1/N_g$.  As a consequence  it is likely that our claim is correct.
We notice that  the standard analysis is performed on a rectangular domain in a phase plane
and the REM  fluctuations, which appear   when $\xbf$ varies keeping $n$ fixed, do not
prevent to obtain a satisfactory   stability analysis.  Since only the inverse map is needed
to compute REM,  and for symmetric integrators based on splitting,
the inverse  map is obtained by changing $\Delta t$ into $-\Delta t$,
this  analysis is recommended for the first stability screening of a given dynamical system.
The trivial parallelization for a grid of initial points allows the
analysis to be performed interactively  for simple models.   
%
%
%
%
%
\section{Asymptotic behaviour of Lyapunov and reversibility error  invariants} 
\label{S:Asymptotic}
The asymptotic behaviour of the Lyapunov  and reversibility  error invariants
for Hamiltonian systems is governed by the  Lyapunov exponents, which are
pairwise opposite.
If the Lyapunov exponents vanish then the asymptotic growth follows  a power law.
For autonomous time reversal invariant Hamiltonian systems the BF and FB invariants  have the same
asymptotic behaviour.  Here we analyze the plane Hamiltonian systems in
the linear and nonlinear case, where the results are obtained using the
normal coordinates.
The normal forms are a basic tool to investigate  systems of celestial mechanics
and were extended to analyze the behaviour of symplectic maps describing non linear
effects in particle accelerators \cite{Bazzani1988,Yellow1994}

The extension to non autonomous plane systems  or to autonomous 
maps defined in a phase  plane  is possible in the neighborhood of fixed points, but
the the non integrable nature of these systems requires the use of normal normal
forms and Nekhoroshev remainder estimates, which involve  a considerable technical work.
Higher dimensional extensions require similar procedures \cite{Marmi1990}
\spa
We analyze first the linear systems   for which the vector field is $\Aop\xbf$
with $\xbf\in \Reali^d$ and the evolution matrix is $\Lop(t)=e^{\Aop t}$.
If  $\Aop$ has real distinct eigenvalues $\Lambda=\diag(\lambda_1,\ldots,\lambda_d)$
and $e^{\Aop t}=\Top\,e^{\Lambda t}\,\Top^{-1}$   where  $\det(\Top)=1$ we introduce 
two positive auxiliary matrices $\Xop(t)$ and $\Yop(t)$ defined by  
\begin{equation}
  \begin{split}    
    & \Lop^T(t)\Lop(t)=\Top \,\Xop(t)\,\Top^{-1}   \phantom{\int} \qquad   \Vop=\Top^T\,\Top  \\
    &  \Xop(t)= \Vop^{-1}\,e^{\Lambda  t}\,\Vop\,e^{\Lambda  t}  \\
    & \Yop(t) =\int_0^t \Xop(t') \,dt' \equiv \int_0^t \,\Vop^{-1} \,e^{\Lambda\,t'}\, \Vop\,
       e^{\Lambda\,t'}\,dt'
  \end{split}
\label{eq_5_1}
\end{equation} 
where $\Vop$ is a positive matrix with unit determinant. The BF covariance matrix is 
\begin{equation}
  \begin{split}       
   &  \Sigma^2_{BF} (t)=  \int_0^t \,\bigl(\Lop^T(t')\,\Lop(t')\,\bigr)^{-1}\,dt'=  \\
     & \quad   = \bigl(\Top^T\bigr)^{-1} \,\,
    \parton{  \,\,\int_0^t \,\Vop \,e^{-\Lambda t'}\,\Vop^{-1} \,e^{-\Lambda t'}\,dt'\,\,} \,\,\Top^T
  \end{split}
\label{eq_5_2}
\end{equation}
The FB covariance matrix is  obtained BF covariance matrix  by changing $\Lop(t') $ into $\Lop(-t')$ in
the integral defining them (see \ref{eq_4_13}).
%
%
The Lyapunov matrix $\Lop^T\Lop$ is conjugated to $\Xop$,
the covariance matrix  $\Sigma^2_{BF}$ is conjugated to a matrix  obtained from $\Yop(t)$
by interchanging $\Vop$  with  $\Vop^{-1}$ and changing  $\Lambda$ into $-\Lambda$.
If the matrix $\Aop$ has complex eigenvalues it is convenient to introduce a real normal form
for $\Lop$ rather than a complex diagonal normal form.
   If  $d=2$  and  the eigenvalues of  $\Aop$
   are $\lambda\pm i\omega $  and we write   $e^{\Aop t}=e^{\lambda t}\Top\,\Rop(\omega t)\,\Top^{-1} $.
   In this case   $\Xop=e^{2\lambda t}\Vop^{-1}\,\Rop(-\omega t) \Vop \Rop(\omega t) $ and $\Yop(t)$
   is still defined as the integral of $\Xop(t')$ in $[0,t]$. The matrix  $\Sigma^2_{BF}(t)$ is conjugated 
   to  a mtarix obtained from $\Yop(t)$ by interchanging $\Vop$  with  $\Vop^{-1}$ and   $\lambda$  with  $-\lambda$.
%
%
%
\def\scr{\scriptstyle}
\subsection{Lyapunov error for plane Hamiltonian systems}  
 \spa
 For a plane linear system  the   Hamiltonian is the   quadratic
 form $H(\xbf)= {1\over 2}\,\xbf\cdot \Bop\xbf$  where  $\xbf=(q,p)^T$. The  phase flow is 
 $\xbf(t)= \Lop(t) \xbf$ where  $\Lop(t)=e^{\Aop t}$  with $\Aop=\Jop\Bop$.
 If $\Bop$  has  positive or negative  eigenvalues    the origin is an elliptic  point,
 if $\Bop$ has eigenvalues of opposite sign the origin is an hyperbolic  point,
 The matrix $\Lop(t)$  takes the simplest possible form in normal coordinates.
 \\
 For an elliptic point the normal form is $H={1\over 2}\,\omega (q^2+p^2)$  so  that 
 $\Lop(t)= e^{\Jop\,\omega t}= \Rop(\omega t)$.
 For an hyperbolic point $H=\lambda\,qp$
 so that   $\Lop(t)=e^{\Lambda t}$ where $\Lambda= \diag(\lambda ,\,-\lambda )$.
 If $\Bop$ has one vanishing  eigenvalue the normal form is    $H={1\over 2}\alpha p^2$   and
 $\Lop(t)= e^{\Nop\,\alpha t}= \Iop+\alpha t\,\Nop$
 where $\Nop=\Bigl(\begin{matrix} {\scr 0} & {\scr 1} \\ {\scr 0} & {\scr 0} \end{matrix} \Bigr )$.
 Notice that $\omega$ and $\alpha$ can be positive or negative. 
 \\
 \\
 Given a generic quadratic Hamiltonian $H={1\over 2}\xbf\cdot \Bop \xbf$  we introduce the normal
 coordinates $\xbf'=(q',p')^T$  by an orthogonal transformation $\xbf'= \Top^{-1} \xbf$ where  $\Top^T\Top=\Iop$.
 such that the the Hamiltonian $H$ takes the normal form described above.
 The matrix $\Lop(t)$ becomes 
  $\Lop(t)= \Top \Rop(\omega t)\Top^{-1}$  for an elliptic fixed point,  $\Lop(t)= \Top \, e^{\Lambda t}\Top^{-1}$
  for an hyperbolic fixed  point.
  \\
  \\
 For a  generic Hamiltonian the evolution in  the neighborhood of a fixed point  takes a simple form in
 nonlinear normal coordinates.  For an elliptic fixed  point    $H=H\bigl({1\over 2}(q^2+p^2) \bigr)$, 
  for an hyperbolic fixed  point $H=H(qp)$  and  
  $H=H\bigl({1\over 2}p^2 \bigr)$   when  Hamiltonian for the linarized system is $H={1\over 2}\alpha p^2$.
 The  asymptotic behaviour of the Lyapunov invariants  is easy to obtain. 
 If the coordinates are not normal one should  consider the normalizing transformation 
 $\xbf'=\psi(\xbf)$  tangent to the  identity  which brings the Hamiltonian to  its normal form.  
 \\
 \\
\spa
  {\bf  A1. Linear  system with elliptic fixed  point} 
  \spa
  The tangent map is $\Rop(\omega t)$ in normal coordinates 
   $\xbf'=\Top^{-1}\xbf$  and in the given  coordinates $\xbf$ it becomes
  $\Lop(t)= \Top\,\Rop(\omega t) \Top^{-1}  $ 
  Introducing the  symplectic matrix $\Vop=\Top^T\Top$  we  notice that  the spectrum of
  $\Lop^T\Lop$ is the same as the matrix $\Xop(t)$ defined by  
\begin{equation}
  \begin{split}
    & \Xop(t) 
    =  \Vop^{-1}\Rop(-\omega t)\Vop \Rop(\omega t)  \\ \\
   &\Vop\equiv \Top^T\Top = \begin{pmatrix} a & & b \\ \\ b & & c
     \end{pmatrix}
  \quad \det(\Vop)=ac-b^2=1  \\ \\
  \end{split}
\label{eq_5_3}
\end{equation} 
From $ac=1+b^2\ge 1$ follows that $a$ and $c$ have the same sign and we can choose them positive
since $\Xop$  is invariant for $\Vop\to -\Vop$.   From $a>0,\,c>0$ and $ac\ge 1$ follows that 
\begin{equation}
  \begin{split}
  {a+c\over 2}\ge {a+a^{-1}\over 2} \ge 1 
  \end{split}
\label{eq_5_4}    
\end{equation} 
It is convenient to write the matrix $\Xop(t)$ in a different form setting $C(t)=\cos(\omega t)$
and $S(t)=\sin(\omega t)$  so that $\Rop(\omega t)=C(t) \Iop+S(t)\Jop$
\begin{equation}
  \begin{split}
     \Xop(t) & =\Vop^{-1}\bigl(C(t) \Iop-S(t)\Jop\bigr)  \Vop \bigl(C(t) \Iop+S(t)\Jop\bigr) \\ 
     \Xop(t)& = \Iop +\Uop \bigl(-\Iop + C_2(t) \Iop +S_2(t) \Jop \bigr) \\ 
     & \Uop = {1\over 2}\bigl(\Iop+\Vop^{-1}\Jop\Vop\Jop\bigr)
  \end{split}
\label{eq_5_5}    
\end{equation} 
where $C_2(t)=\cos(2\omega t)$ and $S_2(t)=\sin(2\omega t)$
%
%
The matrix $\Uop$ explicitly reads 
\begin{equation}
 \begin{split} 
   \Uop(t)={1\over 2}\begin{pmatrix} 2 -ac -c^2 & ab+bc \\ \\ ab+bc & 2-ac -a^2 \end{pmatrix}
\end{split}
\label{eq_5_6}    
\end{equation}   
so that its trace and determinant are given by 
\begin{equation}
 \begin{split} 
   \Tr(\Uop)= 
   2\parqua{1-\parton{a+c\over 2}^2} \quad   
   \det(\Uop)=     1-\parton{a+c\over 2}^2
\end{split}
\label{eq_5_7}    
\end{equation}  
The trace of $\Xop(t)$ is  obtained noticing that $\Tr(\Uop\Jop)=0$ and reads
\begin{equation}
  \begin{split}
  \Tr(\Xop(t))   =   2\,\parton{a+c\over 2}^2  -  2 \cos(2\omega t)   
     \parqua{ \parton{a+c\over 2}^2 -1}\, 
 \end{split}
\label{eq_5_8}    
\end{equation}
The trace of $\Xop(t)$ can be written in a more convenient form  
\begin{equation}
  \begin{split}
    {1\over 2} \Tr(\Xop(t))   = 1+ 2\sin^2(\omega t)\,\parqua{ \parton{a+c\over 2}^2 -1} \ge 1
 \end{split}
\label{eq_5_9}    
\end{equation}
which shows that the  first invariant $I^{(1)}=\Tr(\Xop)$ oscillates  between 2 and
$ 4\bigl({a+c\over 2}\bigr)^2-2$  with a mean value  $2\bigl({a+c\over 2}\bigr)^2\ge 2$
due to (\ref{eq_5_4}).
One can check that $\det(\Xop)=1$.  Indeed setting $\Xop=\Iop+\Aop$  where
$\Aop=\Uop\,\bigl(-\Iop+\Rop(2\omega t)\bigr)$
we use the relation $\det(\Iop+\Aop)= 1+\det(\Aop)+\Tr(\Aop)$.   From  $\det(\Aop)=
4\sin^2(\omega t) \det \Uop$ and $\Tr(\Aop)= \Tr(\Xop-\Iop)= \Tr(\Xop)-2$  from
(\ref{eq_5_9}) we obtain $\Tr(\Aop)=-\det(\Aop)$ and the check is complete.
\\
The second invariant is $I^{(2)}= \det(\Xop)=1$.
From (\ref{eq_5_9}) follows that   $\Delta = \parton{{1\over2}\Tr(\Xop)}^2-1>0$ so that
we verify that the eigenvalues of $\Xop$ are real and positive.
If $(a+c)/2$ close to 1 we obtain the following first order approximation to the eigenvalues $\mu_{j}^2$ of
$\Xop$ by setting    $((a+c)/2)^2 = (1-\eps)^{-1}= 1+\eps+O(\eps^2)$. The result is
%
%
\begin{equation}
  \begin{split}
   \mu_1^2&=1+   2\sqrt{\eps} \sin(\omega t) +2\eps\,\sin^2(\omega t)+O(\eps^{3/2}) \\
   \mu_2^2&={1\over \mu_1^2}
 \end{split}
\label{eq_5_10}    
\end{equation}
The eigenvalues oscillate around 1 with an amplitude $2\sqrt{\eps}$ whereas the first invariant
oscillates with an amplitude $4\eps$.
\\
\spa
  {\bf A2. Linear system with hyperbolic fixed  point} 
  \spa
In the hyperbolic case we have $\Lop(t)  = \Top\,e^{\Lambda t} \Top^{-1} $
where $\Lambda=\diag(\lambda,\,-\lambda)$ . If the coordinates are normal then $\Top=\Iop$
and  $\Lop^T(t)\,\Lop(t)= e^{2\Lambda \,t}$.
We compare the $\log $ of the
Lyapunov error with FLI for an initial displacement $\etabf=(\eta_1,\,\eta_1)^T$
of unit norm
\begin{equation}
  \begin{split}
   &\log E_L(t)= {1\over 2}\log \bigl(e^{2\lambda t}+e^{-2\lambda t}\bigr) \\  
    & FLI(t)=
    {1\over 2} \,\log \bigl(\eta_ 1^2\,e^{2\lambda t}+    \eta_ 2^2\,e^{-2\lambda t}\bigr)
  \end{split}
\label{eq_5_11}   
\end{equation}
The first indicator  has a linear growth with $t$ whereas  if $|\eta_1|<|\eta_2|$  the second
indicator FLI  is decreases  for $t\le t_*= (2\lambda)^{-1}\log(|\eta_1|/|\eta_2|)$ and increases for $t>t_*$,
after reaching the minimum for $t=t_*$. The linear growth of FLI is observed only for $t\gg t_*$.
Moreover, the first indicator does not change for $t \to -t$.
\spa
When  the coordinates are not normal the fundamental matrix is
$\Lop(t)= \Top e^{\Lambda t}\Top^{-1} $, where $\Lambda=\diag(\lambda,-\lambda)$.
In this case  spectrum of $\Lop^T\Lop$ is the same as for the matrix
$\Xop(t)=\Vop^{-1} e^{\Lambda t} \Vop e^{\Lambda t}$ which explicitly reads 
\begin{equation}
 \begin{split}   
   \Xop(t)= \begin{pmatrix}  ac\, e^{2\lambda t} -b^2 & \qquad 
         bc\,(1-e^{-2\lambda t}) \\ \\ ab\,(1-e^{2\lambda t})&  \qquad ac \,e^{-2\lambda t} -b^2
         \end{pmatrix}
  \end{split}
\label{eq_5_12}   
\end{equation}
The invariants  of this matrix are
\def\ch{\hbox{ch}}
\def\sh{\hbox{sh}}
\begin{equation}
 \begin{split}   
  &  \Tr\bigl(\Xop(t)\bigr)= 2 + 2ac\bigl (\ch(2\lambda t) -1 \bigr) \\
  &  \det \bigl(\Xop(t)\bigr)=1
  \end{split}
\label{eq_5_13}   
\end{equation}
We notice that ${1\over 2}\Tr(\Xop)=1+2ac\,\, \sh^2(\lambda t)>1$ so that
$\Delta = \parton{{1\over2}\Tr(\Xop)}^2-1>0$.
The asymptotic expression of the eigenvalues is given by 
\def\sh{\hbox{sh}}
\begin{equation}
 \begin{split}   
   \mu_1^2(t)= ac e^{2\lambda t} \,\,\bigl( 1+O(e^{-2\lambda t}) \bigr)   \quad 
   \mu_2^2(t)= {1\over \mu_1^2(t)}
  \end{split}
\label{eq_5_14}    
\end{equation}

\spa
  {\bf A3.  Nonlinear system with an elliptic fixed point}
\spa
If the system is nonlinear but integrable near an elliptic fixed point the Hamiltonian written in normal coordinates becomes
$H=H(J)$ where $J={1\over 2}\xbf\cdot\xbf$ and $\xbf=\sqrt{2J}(\cos\theta,-\sin\theta)^T$.
  The orbits are still circles as in the linear
case when $a=b=1$ but the frequency $\Omega(J)=H'(J)$ varies. Since the rotations
are asynchronous, the Lyapunov error is no longer constant but grows linearly with $t$.
Since   $S_t(\xbf)= \Rop(\Omega(J)t)\xbf$   the matrix $\Lop(\xbf,t)$ is no longer a rotation matrix 
\begin{equation}
 \begin{split}
    \qquad    
   \Lop(\xbf,t) = \Rop(\Omega t)+\Omega' t\,\Rop'(\Omega t)\,\xbf\xbf^T 
  \end{split}
\label{eq_5_15}   
\end{equation}
We notice that that  $\Rop'=\Rop \Jop$ so that
$\Rop^T\Rop'=\Jop$  which implies ${\Rop'}^T\,\Rop'=\Iop$.
Letting $\alpha=2J\Omega'(J) \,t$ and $\xbf\xbf^T= 2J\, \Zop$ where
$\Zop^2=\Zop$ we write 
\begin{equation}
 \begin{split} 
  & \Lop= \Rop\,(\Iop+\alpha \Jop\,\Zop) \qquad 
   \Zop= \begin{pmatrix} \cos^2\theta & -\sin \theta \cos\theta \\ \\
     -\sin \theta \cos\theta & \sin^2\theta \end {pmatrix} \\ \\
  & \Xop\equiv \Lop^T \Lop=(\Iop-\alpha\,\Zop\,\Jop)(\Iop+\alpha\,\Jop\,\Zop)= \\
  & = \Iop +\alpha^2\,\Zop+\alpha(\Jop\Zop-\Zop\Jop) \\ \\
  &  \Lop\Lop^T= \Rop(\Iop+\alpha\Jop\Zop)(\Iop-\alpha\Zop\Jop)\Rop^T \qquad \alpha=2J\,\Omega'(J)\,t
  \end{split}
\label{eq_5_16}   
\end{equation} 
The trace of $\Xop$ is obtained  taking into account that  $\Tr(\Zop)=1$. The  Lyapunov error
has a linear asymptotic growth in $t$ and the  asymptotic estimates of the eigenvalues $\mu_j(t)$  of
$\Xop(t)$  are a immediately obtained 
\begin{equation}
 \begin{split} 
   & \Tr(\Xop)=2+\alpha^2  \qquad  \qquad E_L(\xbf,t) = \bigl(2 + \alpha^2\bigr)^{1/2}  \\ \\
   & \mu_1^2= 2+\alpha^2 +O(\alpha^{-2})  \qquad \mu_2^2=1/\mu_1^2 
  \end{split}
\label{eq_5_17}   
\end{equation}
The   eigenvector
of $\Lop^T\Lop$ corresponding to $\mu_1^2$ is given by $\wbf_1=(\cos\theta,\, -\sin \theta)^T  +O(\alpha^{-1})$  the
eigenvector corresponding to $\mu_2$ is 
$\wbf_2=(\sin\theta, \,  \cos \theta)^T  +O(\alpha^{-1})$. As a consequence the diagonalizing
matrix is $\Wop(t)=(\wbf_1,\,\wbf_2)= \Rop(\theta)+ O(\alpha^{-1})$  whose $t\to \infty$ limit is
$\Wop=\Rop(\theta)$.
\\
\\
We notice that $\Lop\Lop^T=\Rop(\Omega t)(\Iop+\alpha\Jop\Zop)(\Iop-\alpha\Zop\Jop)\Rop(-\Omega t)$.
The eigenvalues of $(\Iop+\alpha\Jop\Zop)(\Iop-\alpha\Zop\Jop)$ are the same as $\Xop$
and its eigenvectors are  $\rbf_1=\Jop^{-1}\wbf_1 +O(\alpha^{-1})=(\sin \theta,\cos\theta)^T+O(\alpha^{-1})$ and
$\rbf_2=\Jop^{-1}\wbf_2 +O(\alpha^{-1})=(-\cos \theta,\sin\theta)^T+ O(\alpha^{-1})$.
Finally the eigenvectors matrix of $(\Iop+\alpha\Jop\Zop)(\Iop-\alpha\Zop\Jop)$  is given by
$(\rbf_1,\rbf_2)= \Rop(\theta-\pi/2)+O(\alpha^{-1})$.  Finally we  
we obtain  the following asymptotic estimates 
\begin{equation}
  \begin{split}
   &   \Xop(\alpha)\equiv \Lop^T(t)\Lop(t)= \Wop(t)\,\diag\bigl(\mu_1^2,\,1/\mu_1^2)^{-1}\bigr)\,\Wop^T(t) \\ \\
   &  \qquad \Wop(t)=\Rop(\theta) +O(\alpha^{-1}) \\ \\ 
   &    \Lop(t)\Lop^T(t)= \Rop_L(t) \,\diag\bigl(\mu_1^2,\,1/\mu_1^2)^{-1}\bigr)\,\Rop_L^T(t) \\ \\
   &    \qquad   \Rop_L(t)=\Rop(\Omega t) \,(\rbf_1,\rbf_2)= \Rop(\Omega t+\theta-\pi/2) +O(\alpha^{-1})
  \end{split}
\label{eq_5_18}   
\end{equation} 
The Lyapunov eigenvectors matrix  $\Wop(t)$ has a limit $\Rop(\theta)$ for $t\to \infty$
in agreement with Oseledet theory,  and  the Lyapunov exponents are zero.
The matrix  $\Rop_L(t)$ does not have a limit for $t\to \infty$.
We recall that $\Lop(t)= \Rop_L(t) \,\diag(\mu_1(t)\,,\,1/\mu_1(t))\,\Wop^T(t)$.
 The first eigenvector
 corresponding to the largest  linearly growing eigenvalue  is asymptotically
 equal to  $\wbf_1=\xbf/\Vert \xbf\Vert$,
the second eigenvectors corresponding to the  smallest  eigenvalue  is orthogonal to it
and asymptotically is tangent to the orbit at the initial point.
Conversely the eigenvectors of  $\Lop\Lop^T$  rotate and have no asymptotic  limit.
\\
If the coordinates are not normal   the square of the Lyapunov error 
has quadratic  growth   with a periodic modulation, introduced by the breakup-up of the  rotation invariance
$E_L^2(t) \sim (2J\Omega'\,t)^2\bigl(1+f(t)\bigr)$ where $f(t)$ is a periodic function with zero mean and period $2\pi/\Omega$.
The MEGNO filter, consisting in a double average with respect to $t$
of $d\log E^2_L(\xbf,t)/d\log t$, was introduced to wash out  these oscillations.  The
square of the BF reversibility error is given by
\begin{equation}
  \begin{split}
      E^2_{BF}(t) & =\int_0^t\,E_L^2(t')\,dt' \sim \\
  & \sim (2J\Omega')^2 {t^3\over 3} \parton{ 1+ {g(t)\over t} + O(t^{-2}) }
  \end{split}
\label{eq_5_19}   
\end{equation} 
where $g(t)$ is a periodic function.  Since the  amplitude of oscillations decreases as $t^{-1} $
they  disappear for large $t$ and a filter like MEGNO is not really needed.
\spa
  {\bf A4.  Nonlinear   system  with hyperbolic fixed   point} 
  \spa
  The Hamiltonian  for  with a hyperbolic fixed point at the origin in normal coordinates
  is given by  $H(qp)=\lambda_0 qp+{1\over 2}\lambda_1 (qp)^2 +\ldots$ so that the trajectories
  are the same as for the linear system. 
  Letting  $\lambda(qp)= H'(qp)=\lambda_0+\lambda_1 \,qp +\ldots$  the solution is $q(t)= e^{\lambda(qp)\, t}q$
and $p(t)= e^{-\lambda(qp)\, t}p$, the  $p$ and $q$ axis  are the stable and unstable manifolds.
The  matrix $\Lop(\xbf,t)$ is given by  
\def\ch{\hbox{ch}\,}
\def\sh{\hbox{sh}\,}
\begin{equation}
 \begin{split} 
   \Lop(\xbf,t)= \begin{pmatrix} e^{\lambda\,t} & 0 \\ \\ 0 &  e^{-\lambda\, t}  \end{pmatrix}
   \begin{pmatrix} 1 + qp\,\lambda'\,t &   q^2\, \lambda'\,t \\ \\
  -p^2\, \lambda'\,t & 1 - qp\,\lambda'\,t \end{pmatrix} 
  \end{split}
\label{eq_5_20}    
\end{equation}    
The  square of Lyapunov error defined by  $\Tr(\Lop^T\Lop)$ reads 
\begin{equation}
 \begin{split} 
  &  E_L^2(\xbf,t) = 
   2\ch(2\lambda t) + 4qp\,\lambda' t \,\sh(2\lambda t) + 2 q^2p^2 (\lambda' t)^2 \times \\ 
  & \qquad  \qquad \times \ch(2\lambda t) +  
  q^4(\lambda' t)^2  \,e^{2\lambda t}+  p^4(\lambda' t)^2  \,e^{-2\lambda t}
  \end{split}
\label{eq_5_21}     
\end{equation}
If  the intial condition is chosen on the stable mainifold $q=0$ we have 
\begin{equation}
 \begin{split} 
  E_L^2(0,p,,t) =  e^{2\lambda_0 t}+ e^{-2\lambda_0 t}\bigl(1+ (p^2\lambda_1 t)^2\bigr)
  \end{split}
\label{eq_5_22}    
\end{equation}   
The first term which grows  exponentially  appears because the Lyapunov error
is an  average of the displacements on all  the directions.
If  the initial displacement is not random  but   is chosen  starting from the 
stable manifold $q=0$   and directed along it   $\etabf=(0,1)^T$,    the error
decreases exponentially    $\Vert \Lop(0,p,t) \etabf \Vert= e^{-\lambda_0 t}$.
If  the initial condition  is on the unstable manifold $p=0$  we have 
\begin{equation}
 \begin{split} 
  E_L^2(q,0,t) =  e^{-2\lambda_0 t}+ e^{2\lambda_0 t}\bigl(1+ (q^2\lambda_1 t)^2\bigr)
  \end{split}
\label{eq_5_23}    
\end{equation}
Starting from the unstable manifold $p=0$ and choosing the displacement along it 
$\etabf=(1,0)^T$  we have   $\Vert \Lop(x,0,t) \etabf \Vert= e^{\lambda_0 t}$.
The asymptotic diagonal form  of  $\Lop^T\Lop$ can be written just as we did in the elliptic case.
\\
The diagonalization of $\Lop^T\Lop$ and $\Lop\Lop^T$  is carried out by introducing the invariant
$\rho=pq$ and a angle like parameter  $\theta\in \Reali$. Choosing  $\rho>0$ and $\theta\in \Reali$
we map the points of the first  sector $q>0$ and $p>0$ excluding the half axis $p=0$ and
$q=0$.
\begin{equation}
 \begin{split} 
   & q  =\sqrt{\rho}\,e^\theta \qquad p=\sqrt{\rho}\,e^{-\theta}  \qquad \qquad \alpha(t)= \rho\lambda'(\rho)\,t  \\ \\
   & \Lop(\xbf,t)=\begin{pmatrix} e^{\lambda\,t} & 0 \\ \\ 0 &  e^{-\lambda\, t}  \end{pmatrix}
   \begin{pmatrix} 1+\alpha & \alpha\,e^{2\theta}\\ \\ -\alpha \,e^{-2\theta} &  1-\alpha \end{pmatrix}
  \end{split}
\label{eq_5_24}   
\end{equation}
We compute now $\Lop^T\Lop$ 
\begin{equation}
 \begin{split} 
   (\Lop^T\Lop)_{11} &=  (1+\alpha)^2e^{2\lambda t} +\alpha^2 e^{-2(\lambda t+2\theta)}  \\
   (\Lop^T\Lop)_{12} &=  \alpha (1+\alpha)e^{2(\lambda t+\theta)}-\alpha(1-\alpha) e^{-2(\lambda t+\theta)} \\  
   (\Lop^T\Lop)_{22} &= \alpha^2 e^{2(\lambda t+2\theta)}+ (1-\alpha)^2e^{-2\lambda t}
    \end{split}
\label{eq_5_25}    
\end{equation}
Since the determinant  of $\Lop^T\Lop$ is 1 its eigenvalues are determined by the  trace 
\begin{equation}
 \begin{split} 
  & \mu_1^2= e^{2\lambda t}\,\Bigl( (1+\alpha)^2 + \alpha^2 e^{4\theta} \Bigr)  + O\bigl(e^{-2\lambda t}\bigr) \\
   & \mu_2^2=1/\mu_1^2
    \end{split}
\label{eq_5_26}     
\end{equation}
The eigenvectors  $\wbf_j(t)$ and  the eigenvectors matrix
$\Wop(t)=(\wbf_1(t),\,\wbf_2(t))$, recalling that  $\alpha= \rho\, \lambda'(\rho)\,t$,
for $t$  large or more precisely $t\gg (\rho\,|\lambda'|)^{-1}$ we have   
\begin{equation}
 \begin{split} 
   & \wbf_1(t)= N_1 \begin{pmatrix} (1+\alpha) e^{-2\theta} \\ \\ \alpha \end{pmatrix} \\ \\
   & \wbf_2(t)= N_2\begin{pmatrix}  - \alpha e^{2\theta} \\ \\ (1+\alpha) \end{pmatrix}  \\ \\ 
   & \Wop= \lim_{t\to\infty}\,\Wop(t) = N \begin{pmatrix} e^{-\theta}  &  -e^{\theta}\\ \\
   e^{\theta} & e^{-\theta}\end{pmatrix} \\ 
   & N={1\over \sqrt{2\ch(2\theta)}}\,       
    \end{split}
\label{rq_5_27}     
\end{equation}
where $N_j$ denote the normalization factors of the eigenvectors. We notice that the  second Lyapunov vector
$\wbf_2=N(-e^\theta,e^{-\theta})^T$ is tangent to the trajectory  defined by
$\xbf(\theta)=\rho(e^{\theta}, e^{-\theta})^T$ for any given  $\rho$.
Indeed  the tangent vector,  given by  $d\xbf(\theta)/d\theta=\rho(e^\theta,\,-e^{-\theta})^T$ divided by its
norm,  is equal to $-\wbf_2(t)$ for $t \to \infty$.
The Lyapunov vector $\wbf_1$ is normal to the previous one and its
direction corresponds to $\hbox{grad}(H)$.
As in the elliptic case the  first Lyapunov vector is normal
to the trajectory at the initial point.
\spa
The matrix $\Lop\Lop^T$ has  has the same spectum but the eigenvectors differ. In this case we have 
\begin{equation}
 \begin{split} 
   (\Lop\Lop^T)_{11} &=  (1+\alpha)^2e^{2\lambda t} +\alpha^2 e^{2(\lambda t+2\theta)}  \\
   (\Lop\Lop^T)_{12} &= \alpha(1-\alpha) e^{2\theta} - \alpha(1+\alpha) e^{-2\theta}  \\
   (\Lop\Lop^T)_{22} &=  (1-\alpha)^2e^{-2\lambda t} +\alpha^2 e^{-2(\lambda t+2\theta)}+
    \end{split}
\label{eq_5_28}      
\end{equation}
For the eigenvectors  $\rbf_j(t)$ of $\Lop\Lop^T$  the asymptotic limit is easily obtained and reads  
\begin{equation}
 \begin{split} 
   & \rbf_1(t)= N_1 \begin{pmatrix} (  \alpha(1-\alpha) e^{2\theta}  -\alpha(1+\alpha) e^{-2\theta} \\ \\
     O(e^{-2\lambda t})  \end{pmatrix} \to
    \begin{pmatrix} 1   \\ \\ 0  \end{pmatrix}     \\ \\
 & \rbf_2(t)= N_2 \begin{pmatrix}  \alpha (1+\alpha) e^{-2\theta} - \alpha(1-\alpha) e^{2\theta}  \\ \\
     e^{2\lambda t} \,\bigl(\,\, (1+\alpha)^2+\alpha^2 e^{4\theta} \,\,\bigr)   \end{pmatrix} \to
   \begin{pmatrix} 0   \\ \\ 1  \end{pmatrix}  
    \end{split}
\label{eq_5_29}     
\end{equation}
Finally, unlikely the elliptic case,  the matrix  $\Rop_L(t)= (\rbf_1(t),\rbf_2(t))$  has an asymptotic
limit for $t\to +\infty$ given by  the identity matrix $\Rop_L(t)\to \Iop$. 
%
%
%
%
%
%
%
%
%
%
\subsection{Reversibility Error invariants for  Hamiltonian systems}
The  invariants of the BF and FB Reversibility Error matrices are  simply related to the invariants
of the matrix $\Yop(t)$ previously introduced. We consider first  the  plane linear
and plane nonlinear Hamiltonian systems as we did in the previous subsection.
\spazio
\spazio
  {\bf B1. Linear  system with elliptic fixed  point} 
\spa
We first compute the matrix $\Yop=\int_0^t \,\Xop(t')dt'$ where $\Xop(t)$ is given by (\ref{eq_5_15}).
The covariance matrix $\Sigma^2_{BF}(t)$ is conjugated to a matrix obtained from  $\Yop$
by interchanging $\Vop$  with  $\Vop^{-1}$ .   This change leaves the trace and
determinant of $\Uop$ invariant. As a consequence, we   can show that
$\Tr( \Sigma^2_{BF}(t))= \Tr (\Yop(t)) $ where  
\begin{equation}
 \begin{split} 
   &  \Tr(\Yop(t)) =    2t\, \parton{a+c\over 2}^2 \,+\,{\sin(2\omega t)\over \omega}
   \parton{1-\parton{a+c\over 2}^2 }  
\end{split}
\label{eq_5_30}      
\end{equation}  
The determinant of $\Yop(t)$ is no longer equal to 1  but using repeatedly
$\det(\Iop+\Aop)= 1 +\Tr(\Aop)+\det(\Aop)$ and the expression for $\det \Uop= {1\over 2}\,\Tr(\Uop)$
jointly with $\Tr(\Uop\Jop)=0$ we finally can prove that
\begin{equation}
 \begin{split} 
     \det(\Yop(t))  =  t^2 \, \parton{a+c\over 2}^2 - {\sin^2\omega t\over \omega^2}
     \parton{\parton{a+c\over 2}^2-1 }
\end{split}
\label{eq_5_31}     
\end{equation}  
and this is equal to $\det(\Sigma^2_{BF}(t))$.
The invariants for the FB matrix are identical to the invariants for the BF matrix.
%
%
\spa
  {\bf  B2. Linear  system with  hyperbolic  fixed  point} 
  \spa
In order to compute the BF invariants we first compute $\Yop(t)=\int_0^t \,\Xop(t')\,dt'$ starting from
(\ref{eq_5_12}).   The results  reads
\begin{equation}
  \begin{split}
    & \Yop(t)= \\
    & = \begin{pmatrix} ac\,e^{\lambda t}\,\sh(\lambda t)/\lambda  -b^2 t & bc \,\bigl ( t- e^{-\lambda t}\,\sh(\lambda t)/\lambda\bgr) \\ \\
     ab \,\bigl ( t- e^{\lambda t}\,\sh(\lambda t)/\lambda\bgr) & ac\,e^{-\lambda t}\,\sh(\lambda t)/\lambda -b^2 t
    \end{pmatrix}
    \label{eq_5_32}    
    \end{split}
\end{equation}    
The invariants of $\Yop(t)$ are given by 
\begin{equation}
  \begin{split}
    \Tr(\Yop(t))& =  ac{\sh(2\lambda t)\over \lambda }\,-2b^2 t \\ \\
    \det(\Yop(t))& = ac\,{\sh^2(\lambda t)\over \lambda^2}-b^2 t^2 
    \label{eq_5_33}    
    \end{split}
\end{equation}    
Since $ac=\Vop_{11}\,(\Vop^{-1})_{11}$ and $ac-b^2=1$, the interchange $\Vop$ with $\Vop^{-1}$ and $\lambda$ with
$-\lambda$  does not affect the invariants of $\Yop(t)$ and therefore the invariants of $\Yop(t)$
are the invariants of  $\Sigma^2_{BF}(t)$. In this case also the invariants of $\Sigma^2_{FB}(t)$  are the same as
$\Yop(t)$. 
Notice that for $t\to 0$ we have $\Tr(\Yop(t)\simeq  2t+{4\over 3}\,ac\,\lambda^2 t^3 +O(t^5)$ and $\det(\Yop(t))
=  t^2+{1\over 3}\,ac\,\lambda^2 t^4 +O(t^6)$
whereas for $t\to \infty$ the invariants are $\Tr(\Yop(t)\simeq ac e^{2\lambda t}/(2\lambda)$ and
$\det(\Yop(t)\simeq ac e^{2\lambda t}/(4\lambda^2)$

%
\spazio
  {\bf  B3. Non linear  system with elliptic fixed  point} 
\spa
We compute the matrix $\Xop\equiv \Lop^T\Lop$  for  $p=0$ which corresponds to $\theta=0$
and $\Zop= \begin{pmatrix} 1 & 0 \\ 0 & 0 \end{pmatrix} $. The  inverse of $\Lop^T(t)\Lop(t)$ and its 
integral  equal to  $\Sigma^2_{BF}(t)$ are given by
\begin{equation}
  \begin{split}
    & \Xop^{-1}(t)\equiv(\Lop^T(t)\Lop(t))^{-1} =\begin{pmatrix}  1 & \alpha \\ \\ \alpha & 1+\alpha^2  \end{pmatrix} \quad \;
    \alpha=2J\,\Omega'(J) \,t\\ \\   
    & \Sigma^2_{BF}(t)= \int_0^t \Xop^{-1}(t')\,dt'= \begin{pmatrix}  t    & {1\over 2} t^2 (2J\Omega') \\ \\
   {1\over 2} t^2 (2J\Omega')   & t +{1\over 3}\,t^3 (2J\Omega')^2 \end{pmatrix}
    \label{eq_5_34}    
    \end{split}
\end{equation}    
As a consequence the invariants of $\Sigma^2_{BF}(t)$ which are the same as $ \overline{\Yop}(t) $
and $\Sigma^2_{FB}(t)$   are given by
\begin{equation}
  \begin{split}
    \Tr(\Sigma^2_{BF}(t))& = 2t + {t^3\over 3}(2J\Omega')^2 \\ \\
   \det(\Sigma^2_{BF}(t))& = t^2 + {t^4\over 12}(2J\Omega')^2 
  \end{split}
    \label{eq_5_35}   
\end{equation}    
%
The largest eigenvalue grows asymptotically as $t^3$ the smallest as $t$ to be compared with the eigenvalues
of the Lyapunov matrix which behave as $t^2$ and $t^{-2}$ respectively.

\spa
  {\bf B4.   Non linear  system with hyperbolic fixed  point} 
\spa
We  compute first the matrix $ \Lop^T\Lop$ on the stable manifold $q=0$. Its inverse is given by 
\begin{equation}
  \begin{split}
    \bigl(\Lop^T\Lop\bigr)^{-1} = \begin{pmatrix}  e^{-2\lambda_0 t} & \quad  p^2\lambda_1 t\,
      e^{-2\lambda_0 t} \\ \\
p^2\lambda_1 t\,  e^{-2\lambda_0 t} & \quad  e^{2\lambda_0 t}  + (p^2\,\lambda_1\,t)^2  e^{-2\lambda_0 t} \end{pmatrix} 
    \label{eq_5_36}    
    \end{split}
\end{equation}    
The  covariance matrix $\Sigma^2_{BF}(t)$ is given by the integral of $(\Lop^T(t')\Lop(t'))^{-1}$
which explicitly reads 
\begin{equation}
  \begin{split}
    & \bigl(\Sigma^2_{BF}\bigr)_{11}= {1-e^{-2\lambda_0 t}\over 2\lambda_0} \\ 
    & \bigl(\Sigma^2_{BF}\bigr)_{12}={p^2\lambda_1\over 4\lambda_0^2}\,\Bigl( 1- e^{-2\lambda_0 t}(1+2\lambda_0 \,t)\,\Bigr) \\ 
    & \hskip -.25 truecm \bigl(\Sigma^2_{R}\bigr)_{22}={e^{2\lambda_0 t} -1 \over 2\lambda_0}+
    {p^4\lambda_1^2 \over 4\lambda_0^3}\Bigl(1 -
    e^{-2\lambda_0 t}\,(2\lambda_0^2t^2 +2\lambda_0 t +1)\,\Bigr)
    \label{eq_5_37}    
    \end{split}
\end{equation}    
The invariants of $\Sigma^2_{BF}(t)$ are given by  $\Tr(\Sigma^2_{BF}(t))= e^{2\lambda_0 t}/(2\lambda_0) +O(1)$
and $\det(\Sigma^2_{BF}(t))= e^{2\lambda_0 t}/(4\lambda_0^2) +O(1)$ 

%
%
%
%


\subsection{ Lyapunov, Forward and  Reversibility Error
  invariants for linear and nonlinear flows in \texorpdfstring{$\Reali^d$}{Rd}} 
The vector field of an autonomous linear system is  $\Phibf(\xbf)=\Aop\xbf$  and the flow 
is $\xbf(t)= e^{\Aop t}\,\xbf$  where  $\Aop$   is a real matrix whose eigenvalues $\lambda_j$
we assume to be real,  distinct and ordered in  a decreasing sequence  
\begin{equation}
  \begin{split}
    \lambda_1>\lambda_2>\ldots\lambda_{n_+}>0>\lambda_{n_++1}>\ldots>\lambda_{d} \qquad   \end{split}
    \label{eq_5_39}     
\end{equation}       
where $n_+$  is the number of positive eigenvalues and $n_-=d-n_+$ is the number of negative eigenvalues.
\spa
{\bf C1 Linear flows}
\spa
The Lyapunov matrix  can be  written as $\Lop^T(t)\Lop(t)= \Wop(t) e^{2t\,\Lambda(t)}\Wop^T(t)$
and its eigenvalues are $\mu_j^2(t)=e^{2t\,\lambda_j(t)}$.
We have seen that $\Lop^T(t)\Lop(t)$ is conjugated to the matrix $\Xop(t)$ and that
the reversibility error matrix $\Sigma^2_{BF}(t)$ is conjugated to the  matrix 
$\Yop(t)=\int_0^t\,\Xop(t')\,dt'$  after  trivial changes. 
\spa
For the invariants of the  Lyapunov error matrix we have the following limit 
\begin{equation}
   \begin{split}
     & \lim_{t\to +\infty}\,{1\over 2t}\log (I^{(k)}_L)  =\lambda_1+\ldots+\lambda_k   \\ \\
  \end{split}
   \label{e_5_40}    
\end{equation}
in agreement with the Oseledet theorem.
For the invariants of the  BF reversibility error the asymptotic limit is expected to be given by
\begin{equation}
  \begin{split}
     \lim_{t\to +\infty } {1\over 2t} \log (I^{(k)}_{BF}) =
   \begin{cases}   0    \hskip 3.3 truecm   n_-=0  \phantom{\biggr)}  \\
     |\lambda_d|+ \ldots+ |\lambda_{d-k+1}| \\
     for \quad 1\le k\le n_- \phantom{\biggr)} \\
      |\lambda_d|+ \ldots+ |\lambda_{d-n_-+1}| \\
      for \quad n_- \le k\le d
    \end{cases}   
  \end{split}
   \label{eq_5_41}    
\end{equation}     
For the forward error invariants the asymptotic limit is expected to be  given by  
\begin{equation}
   \begin{split}
     \lim_{t\to +\infty } {1\over 2t}\,\log (I^{(k)}_F) =
     \begin{cases}   0   \qquad \qquad\qquad \qquad n_+=0 \phantom{\biggr)} \\
       \lambda_1+ \ldots+ \lambda_k, \quad 1 \le k\le n_+   \phantom{\biggr)}  \\
      \lambda_1+ \ldots+ \lambda_{n_+} \quad  n_+ \le k\le d
    \end{cases}   
  \end{split}
   \label{eq_5_42}    
\end{equation}     
The same limit as (\ref{eq_5_42})  is expected for FB Reversibility Error invariants.
In the previous subsection we have analyzed the invariants  of the Lyapunov and
reversibility error matrices of linear Hamiltonian systems in $\Reali^2$.
The previous relations  (\ref{eq_5_41}) and (\ref{eq_5_42})  are easily proved if the matrix $\Aop$ is symmetric
since in this case the Lyapunov matrix is given by $\Lop^T\Lop(t)= e^{2\Aop t}= \Wop e^{2\Lambda t}\,\Wop^T$
and the integral of its inverse is immediately obtained.
\\
If the matrix $\Aop$ is not symmetric the asymptotic behaviour of
the integral of $(\Lop^T(t')\Lop(t'))^{-1} $  is harder to prove.
\\  \\
In  Appendix IV    we perform  the analysis  for a real matrix $\Aop$ when $d=2$
obtaining the asymptotic behaviour of  the  invariants, the eigenvalues and the eigenvectors
of  $\Lop^T(t)\Lop(t)$. 
We show that the asymptotic behaviour of the eigenvalues  of the Lyapunov matrix   is given by
$\mu_j^2(t)=c_j^2 e^{2\lambda_j\,t}\,(1+O(\eta))$ where $\eta$ decreases exponentially
so  that  $\lambda_j(t)= \lambda_j + t^{-1}\,\log c_j+O(\eta)$.
For the invariants  of  the Lyapunov matrix we show that  $(2t)^{-1}\,\log I_L^{(1)}(t)= \lambda_1 + t^{-1}\,\log c_1+O(\eta)$
and $(2t)^{-1}\,\log I_L^{(2)}(t)= \lambda_1+\lambda_2 + t^{-1}\,\log (c_1 c_2)+O(\eta)$. 
The asymptotic behaviour for the invariants of the BF  Reversibility Error in $\Reali^2$ 
is obtained  and provides a rigorous proof  of  (\ref{eq_5_41})
for $d=2$.
For $d\ge 3$  the asymptotic limit (\ref{eq_5_41}) is not  supported by a
 rigorous analysis but   numerical analysis could be performed to support it 
\phantom{Obtain the asymptotic estimates  of  the invariants of $\Yop(t)$ written as
  $\int_0^t \Wop(t')\,e^{2t'\,\Lambda(t')}\,\Wop^T(t')\,dt'$  by far   more difficult}.
\\
\\
  {\bf C2.  Nonlinear flows}
  \\
  \\
  We consider a  nonlinear flow  in $\Reali^d$  generated by an analytic   vector field $\Phibf(\xbf)$
  whose linear part is $\Lambda \xbf$ where $\Lambda$ is a real diagonal matrix.
  The spectral decomposition of the Lyapunov matrix  is given by 
  $\Lop^T\Lop=  \Wop(\xbf,t)   \, e^{2t\,\Lambda(\xbf,t)}\,\Wop^T(\xbf,t)$
  and Oseledet system insures that the  diagonal matrix
  $\Lambda(\xbf,t)$ has a  limit $\Lambda(\xbf)$ and the orthogonal matrix $\Wop(\xbf,t)$
  has a limit $\Wop(\xbf)$.
  If $\Ecal$ is an invariant ergodic subset then $\Lambda(\xbf)$ has the same value for
  almost every $\xbf\in \Ecal$,   whereas $\Wop(\xbf)$ depends on $\xbf$.
  No  result is available on the convergence rate to $\Lambda$ and
  $\Wop$ except for  linear systems in $\Reali^2$, see Appendix III.  
  \\
  In the previous section, we  have analyzed the behaviour of the invariants of the Lyapunov
  and reversibility error matrices for Hamiltonian flows  in $\Reali^2$ written in  normal coordinates
  in the neighborhood of an elliptic or hyperbolic fixed point.   The   first Lyapunov invariant and
  the first two   reversibility error invariants for an  integrable Hamiltonian system in $\Reali^2$
  have a power law growth, with a coefficient which diverges while approaching the  separatrices,
  where the growth becomes exponential. This is the
  reason why the se\-pa\-ra\-tri\-ces can be  easily detected in numerical computations. 
  \\
  \\
  For generic nonlinear flows the existence of  a normalizing transformation, which linearizes the vector field in the neighborhood of a fixed point $\xbf_c$, is established by the  
  the Poincar\'e-Dulac theorem if the fixed point is   attractive and the vector field $\Phibf(\xbf)$
  is real analytic. 
  We suppose that  the linear part is  $\Lambda \,(\xbf-\xbf_c)$ where $\Lambda=\diag(\lambda_1,\ldots,\lambda_s)$
  with $0>\lambda_1>\ldots>\lambda_d$. We denote with  $\Xbf=\xbf_c +\psi(\xbf-\xbf_c)$
  the normalizing transformation tangent to the identity $D\psi({\bf 0})=\Iop$,
  and  with $T$   its inverse so that  $\xbf=\xbf_c+T(\Xbf-\xbf_c)$. The conjugation   $\xbf(t)=
  \xbf_c + T(e^{\Lambda t}\psi(\xbf-\xbf_c))$ implies that
\begin{equation}
  \begin{split}
   \Lop(\xbf,t)=D\,T (e^{\Lambda t}\psi(\xbf-\xbf_c))\,e^{\Lambda \,t}\,D\psi(\xbf-\xbf_c) 
  \end{split}
   \label{eq_5_43}    
\end{equation}    
Since $e^{\Lambda t}(\xbf-\xbf_c) \to {\bf 0}$ for  $t\to +\infty$ and $DT ({\bf 0})=\Iop$   
asymptotically for $t\gg 1$  we have
$\Lop(\xbf,t)= e^{\Lambda \,t}\,D\psi(\xbf) $ so that 
the Lyapunov matrix is asymptotically conjugated to $\Vop(\xbf)e^{2\Lambda t}$ where
$\Vop(\xbf)=D\psi(\xbf-\xbf_c)\,(D\psi(\xbf-\xbf_c))^T$ is a symmetric matrix.
The asymptotic behaviour of the invariants $I^{(k)}$ is given by $c_k(\xbf) \,\,e^{2t(\lambda_1+\ldots\lambda_k)}$ 
where $c_k$ is the determinant of the the $k\times k$ matrix given by the the first $k$
rows and columns of $\Vop$. As a consequence unless   $c_k(\xbf)=0$ vanishes at some point, we have
\begin{equation}
  \begin{split}
    \lim_{t\to +\infty} {1\over 2t}\log I_L^{(k)}(\xbf,t)
    =\lambda_1+\ldots\lambda_k
  \end{split}
   \label{eq_5_44}    
\end{equation}     
The covariance matrix matrix $\Sigma^2_F$ has a constant asymptotic limit and this is the case also for its invariants
so that  $(2 t)^{-1}\,\log  I_F^{(j)}(\xbf,t) \to 0$.  The BF reversibility error invariants are obtained from
the covariance matrix $\Sigma^2_{BF}(\xbf,t)$ which is the integral in $[0,t]$ of $(\Lop^T(\xbf,t')\Lop^T(\xbf,t'))^{-1}$.
The asymptotic expression of $\Lop^T\Lop$  does not allow  to estimate the integral, however  we expect that 
$(2 t)^{-1}\,\log  I_{BF}^{(j)}(\xbf,t) \to |\lambda_d|+\ldots+|\lambda_{d-j+1}|$ for $t\to +\infty$.
\\
For one dimensional systems $d=1$ the proof    $t^{-1}\,\log E_F^{(1)}(x,t) \to 0$  and
$t^{-1}\,\log E_{BF}^{(1)}(x,t) \to |\lambda|$ when $t\to +\infty$ 
for any $x$ in the basin of  an attractive fixed point $x_c$,  where the linearized  field $\Phi(x)$ 
becomes $\lambda(x-x_c)$ with $\lambda <0$,  can be obtained with  direct computation  
if the  field $\Phi(x)$ is a polynomial with simple real zeroes.
The procedure is straightforward  in the case of the  
logistic equation  $\Phi(x)=x(1-x)$ whose attractive fixed point is $x_c=1$  with $\lambda=-1$ 
and the Duffing equation $\Phi(x)=x(1-x^2)$ whose attractive fixed points are $x_c=\pm 1$ with $\lambda=-2$.
Indeed in this case we have
\begin{equation}
  \begin{split}
    & \Lop(x,t)=   {  e^{-t}  \over (x+(1-x)\,e^{-t})^2 }     \qquad \qquad \hbox {Logistic} \\
    & \Lop(x,t)={ e^{-t}\over (x^2+(1-x^2)\,e^{-t})^{3/2}} \qquad \hbox{ Duffing}
  \end{split}
   \label{eq_5_45}    
\end{equation}  
and the integrals defining $E_{BF}^2$  and $E_F^2$ can be  easily evaluated, see Appendix V
\\
\\
If $\xbf_c$ is an hyperbolic fixed point  of the flow generated by by the vector field
$\Phibf(\xbf)\in C^{m+1}$    
whose   linear part  is $\Lambda(\xbf-\xbf_c)$ with  $\Lambda=\diag(\lambda_1,\ldots,\lambda_d)$,
where the $\lambda_j$ are ordered as (\ref{eq_5_39}) whith $n_+\ge 1$ and $n_-=d-n_+ \ge 1$, 
 then the Hadamard-Perron theorem states  that  the flow has two  $C^m$ invariant  manifolds $\Wcal_s$ and $\Wcal_u$
  whose tangent spaces at $\xbf_c$ are $\Ecal_s$ and $\Ecal_u$. 
  For any initial condition $\xbf \in \Wcal_s$
  the trajectory tends exponentially fast to $\xbf_c$ for $t\to +\infty$ and for any  $\xbf \in \Wcal_u$
  exponentially fast to $\xbf_c$ for $t\to -\infty$.  The vector field is linearizable  for $\xbf\in \Wcal_s$
  close to $\xbf_c$ by a $C^m$ transformation close to the identity which maps $\Wcal_s$ into $\Ecal_s$
  and the flow is mapped  into $\xbf_c + e^{\Lambda t}(\Xbf-\xbf_c)$  which converges to $\xbf_c$
  for $t\to +\infty$.     If $\xbf\in \Wcal_u$ and is close to $\xbf_c$ then the flow is mapped into
  a linear flow which converges to $\xbf_c$ for $t\to -\infty$.
  If  $\Lambda$ is non resonant 
  ( $\sum_{i=1}^d\, k_i\,\lambda_i=0$  for $k_i$ non negative integers only if  all 
  $k_i$  vanishes identically), then  in a neighborhood of $\xbf_c$ there
  is  a transformation tangent to the identity that conjugates the flow with the linear part.
  However  only  if $\xbf \in \Wcal_s$  so that $\Xbf\in \Ecal_s$ we have $\xbf_c$ as a limit when
  $ t \to +\infty$    and   if $\xbf \in \Wcal_u$  so that $\Xbf\in \Ecal_u$  we have a
  limit $\xbf_c$ when $t\to +\infty$. Without restricting $\xbf$ to $\Wcal_s$ or $\Wcal_u$ we have 
  no asymptotic estimate even  though the field is linearizable in a neighborhood of $\xbf_c$.
  The asymptotic behaviour of the Lyapunov error invariants  is determined if  the initial point
  belongs to   $\Wcal_s$ or $\Wcal_u$. Even if the flow is linearizable in a neighborhood
  of $\xbf_c$ when  $\Lambda$ is not resonant, no asymptotic estimate for $t$ positive is available
  unless $\xbf$ belongs to $\Wcal_s$ because the the orbit leaves the neighborhood of the fixed point
  rather than approaching it.  For $t$ negative no estimete is available unless $\xbf\in \Wcal_u$.
  Similar considerations apply  to  the  reversibility  error invariants. 
%
%
%
%
%
%
%
%

%
%
\section{Two Hamiltonian  models}
\label{S:twoSimpleModels}
\begin{figure}[!ht]
\centerline{
\includegraphics[width=2.5 cm, height=3 cm]{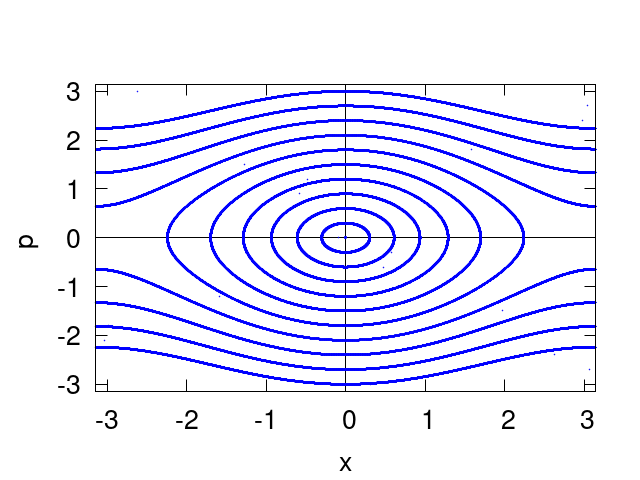}    
\includegraphics[width=2.5 cm, height=2.5 cm]{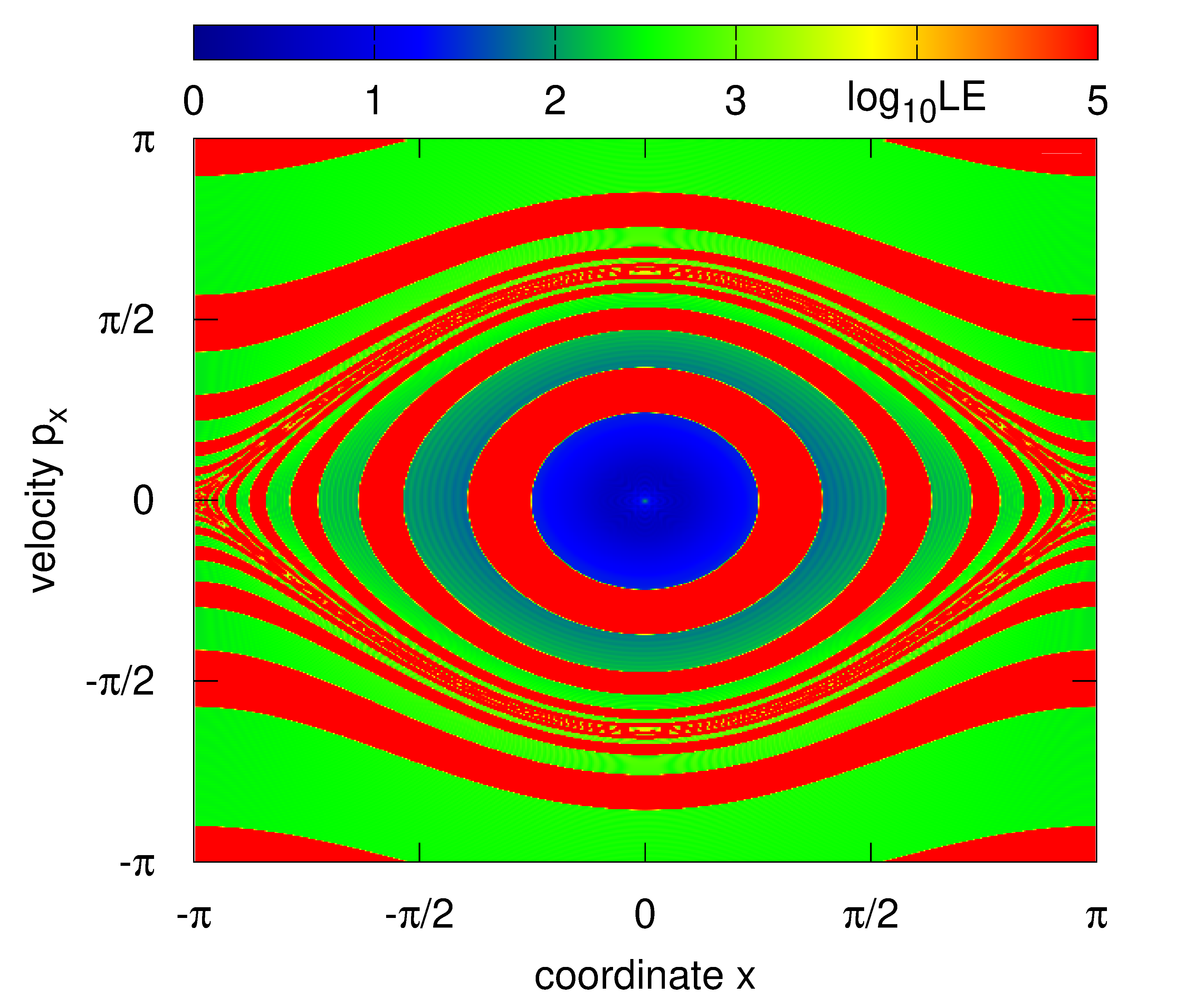}    
 \includegraphics[width=2.5 cm, height=2.5 cm]{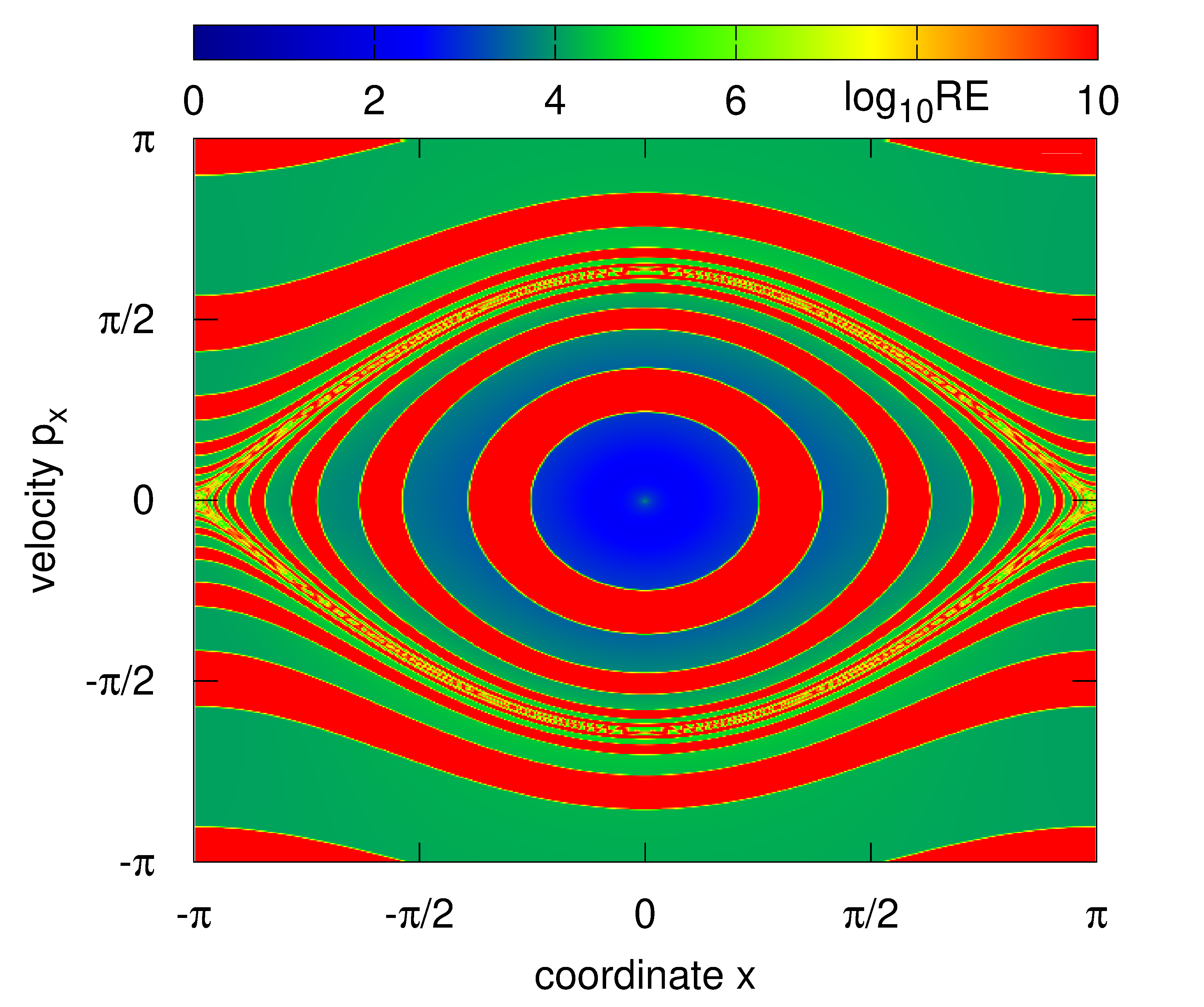}
}
\centerline{
\includegraphics[width=2.5 cm, height=2.5 cm]{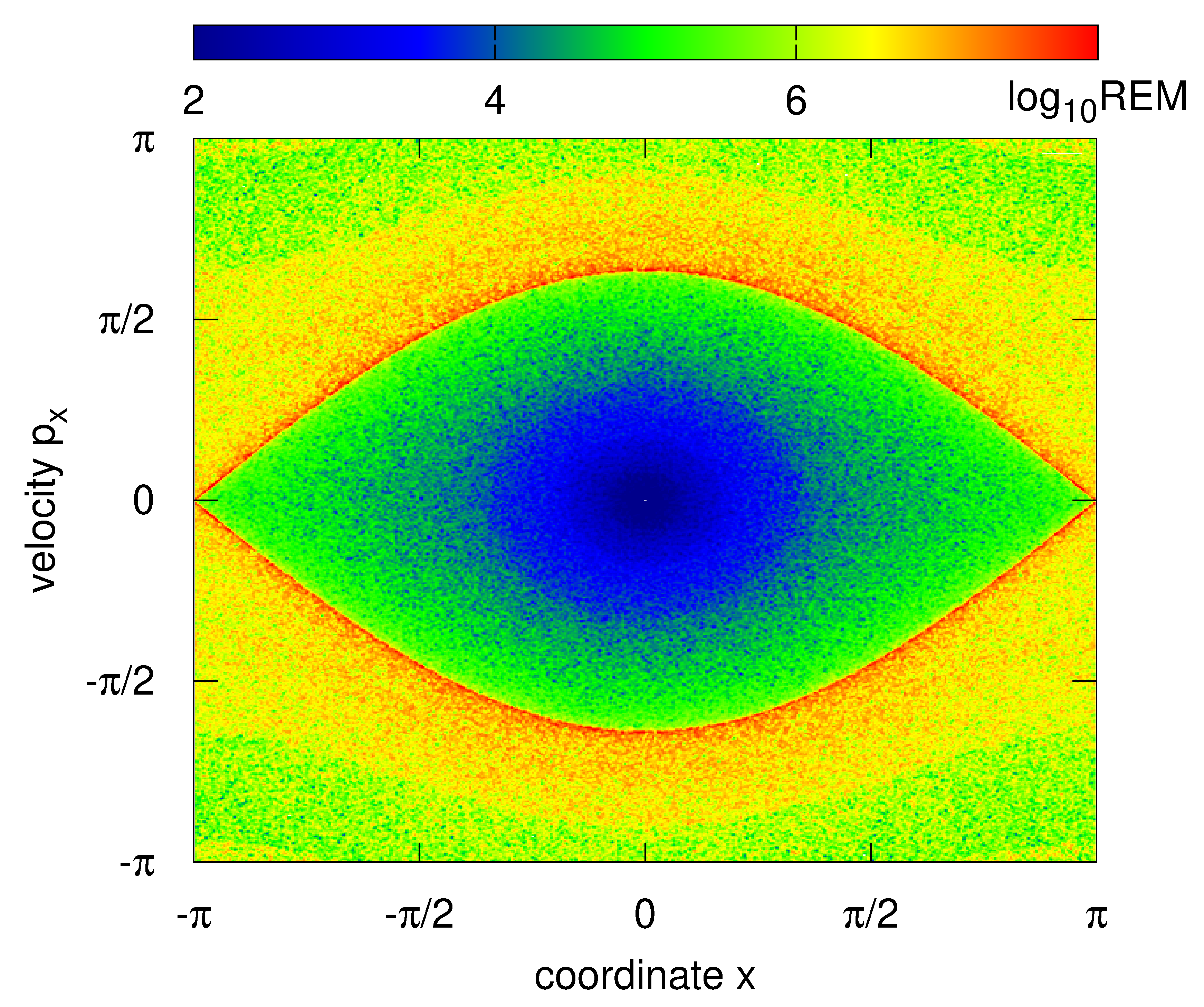}
\includegraphics[width=2.5 cm, height=2.5 cm]{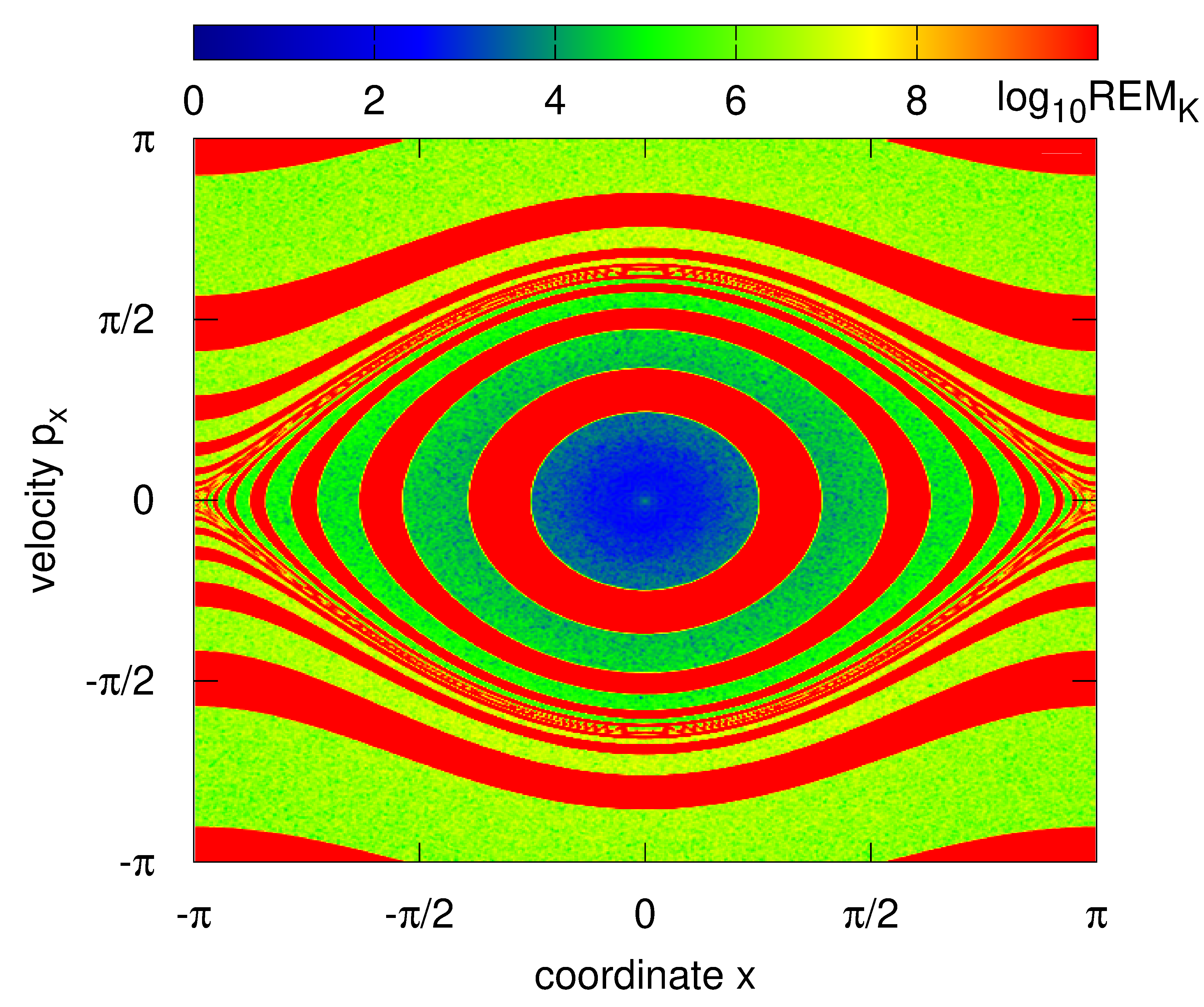}
}
\caption{ 
  First frame (leftmost):
  phase portrait the $(x ,p_x)$ phase plane
  with initial   conditions in the same phase plane
  for the  coupled pendulums without     time dependence $\eps=0$.
  The parameters are    $\omega_x^2=5,\,\omega_y^2=4,\,\mu=4$   the integration has
  been performed by the second order symplectic integrator described in  \cite{Yoshida1990}, with $\Delta t=2\pi/100$.
  Second frame: LE color plot at $t= 2\pi \times 50$.
  Third frame:  RE color plot at $t= 2\pi \times 50$.
  Fourth frame: REM  color plot at $t= 2\pi \times 50$.
  Fifth frame: REM with kick
  at $t= 2\pi \times 50$. }
\label{fig:fig_1} 
\end{figure}

The validity and efficiency of the proposed fast indicators have been checked on symplectic maps
\cite{Faranda2012,Turchetti2010b}  and Hamiltonian systems modeling  the dynamics of particle beams
\cite{Turchetti_beam,Montanari23}, of mechanical systems
\cite{Turchetti2017,Panichi2018}, planetary systems
\cite{Panichi2017,Panichi2016} and  rays propagation in waveguides
\cite{Gradoni2021, Gradoni2023}
%
Typically, for models of Celestial Mechanics, variational indicators like MEGNO
and spectral methods are used and compared to analyze the 
dynamical stability \cite{Gozdziewski01,Nesvorny1996}.
In this section  present the  application of the Lyapunov and Reversibility Error
indicators to a Hamiltonian system with two degrees
of freedom having  a periodic time dependence and  to a four dimensional symplectic
map. 
The first model is given by two coupled pendulums,  one of which
has a periodically varying linear frequency
\begin{equation}
 \hskip -.4 truecm    H = {p_x^2+p_y^2\over 2} - \omega_x^2 \,(1+\eps\cos(\Omega t))\,\cos x
    -\omega_y^2 \cos y  +\mu \cos x  \cos y 
\label{eq_7_1}        
\end{equation}
%
%
According to \cite{Barrio2009}, we consider  $\omega_x^2=1+\mu,\; \omega_y^2=\mu$ 
choosing  $\mu=4$ in our numerical simulations 
to investigate how the orbital stability depends on $\mu$ and $\epsilon$. 
The model without modulation that is with  $\eps=0$ was first proposed to investigate the
dependence of FLI and MEGNO
on the initial displacement  vector $\etabf_0$ in \cite{Barrio2009}.
Given an error $E(t)$  the following double time average
\begin{equation}
\begin{split}
  Y(E(t)) & = \mean{\mean{d\log E^2/d \log t }}= \\
 &  =\mean{\log E^2(t)} - \mean{\mean{\log E^2(t)}}
\end{split}
\label{eq_7_2}    
\end{equation}
is known as MEGNO in the special case in which $E(t)$ is replaced by 
$\Vert \Lop(\xbf,t)\,\etabf\Vert$
where $\etabf$ is a given unit vector.
From the  last expression in  (\ref{eq_7_2}), obtained with an integration
by parts,  we see that   that  MEGNO is twice  the  time average of FLI     
minus its double time  average   \cite{Mestre2011}.
MEGNO is a filter designed to remove the time  oscillations  for a given orbit ($\xbf$ fixed)
and can be applied to any indicator. 
It is interesting to compare the MEGNO filter applied to LE with the BF Reversibility error
for autonomous Hamiltonian systems
\begin{equation}
\begin{split}
 & Y(E_L(\xbf,t)) =  \,\mean{\log E_L^2(\xbf,t)} -
  \mean{\mean{\log E_L^2(\xbf t)}} \, \\ \\
 &  E_BF(\xbf,t) = \parton{\int_0^t E_L^2(x,t')\,dt'}^{1/2} = t^{1/2\,}\,\mean{\,E_L^2(\xbf,t)\,}^{1/2}
\end{split}
\label{eq_7_3}    
\end{equation}
With  $E_{BF}$ we denote the $BF$ Reversibility error, which in our case is equal to the FB error
wup to  a change of sign on the initial moments.
Since $E_{BF}$  involves a time average  of $E_L^2$ whereas  MEGNO applied to $E_L$  involves a time  average of
$\log E_L^2$, the oscillations present in $E_L$ are filtered when $E_{BF}$
is computed and no further averaging is needed.
On the contrary when $E_L$  is considered the MEGNO filter is needed to kill the oscillations.
The appearance of oscillations is due   to a loss of symmetry, which occurs when we change from normal
to ordinary coordinates in the neighborhood of an elliptic fixed point.
The same considerations
apply to simplectic maps,  with obvious changes (integrals are replaced by sums when the averages are
computed).
We notice that REM  filters the oscillations just as $E_{BF}$. However, it exhibits
fluctuations for  a  fixed initial condition $\xbf$  when the iteration number $n$ varies and when
$\xbf$ varies for $n$ fixed.  Since the computations are performed at a fixed value of
$n$ local phase space averages allow to filter the fluctuations.
\\
\\
The  time independent Hamiltonian with  $\eps=0$   was first  proposed  because
the  $(x \,p_x)$ plane is invariant
but any displacement out of it excites parametric instabilities.   The variational equation for $y$ is
$\ddot \eta_y= -\bigl (1-\mu\,\cos x(t)\bigr)\,\eta_y$   so that for initial conditions
in the $(x,p_x)$-plane the parametric instability affects $\eta_y$   when    $\eta_y(0)$ and or
$\dot\eta_y(0)$  are  not zero leading to an alternation of annular  unstable regions of decreasing thickness
when the separatrix is approached.
The BF reversibility error $E_{BF}$   and REM   were   computed abd compared  .
Since the  Hamiltonian has  an even time dependece on $t$  if we compute $E_L$  for$-t$
the result is the same provided that we change $\pbf$ into $-\pbf$ due to the time reversal
invariance. As a consequence th eReversibility error RE  and REM computed for the FB process
are the same as for BF process provided that   $\pbf$ is changed into $-\pbf$.
This is  the case also for the next model which is an autonomous symplectic map.
\\
In Fig.~\ref{fig:fig_1}, we compare the  fast indicators for the coupled pendulums model without
time dependence, $\epsilon=0$.
We have integrated the Hamiltonian using a second and fourth order symplectic
integrators \cite{Yoshida1990}, and we computed the corresponding tangent map
according to \cite{Skokos2010}.
Choosing the initial conditions in the $(x, p_x)$ plane,  the  change of stability is  analyzed  with
$E_L$  and  $E_R$  which provide  almost exactly the same   pattern.  The parametric instability
is not felt by REM because the round-off error does not bring the orbit out of the invariant
$(x, p_x)$-plane. However, a small   kick  applied  before reversing the motion brings the orbit
out of the plane rendering the REM pattern similar to RE,
even though significant fluctuations are  present.

\begin{figure}[!ht]
\centerline{
\includegraphics[width=2.5 cm, height=3 cm]{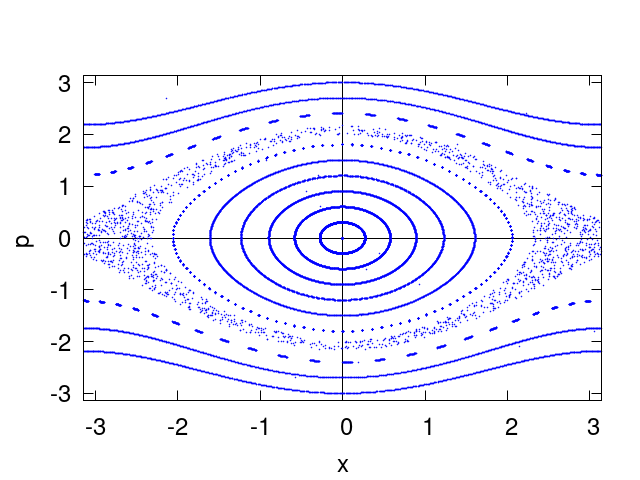}   
\includegraphics[width=2.5 cm, height=2.5 cm]{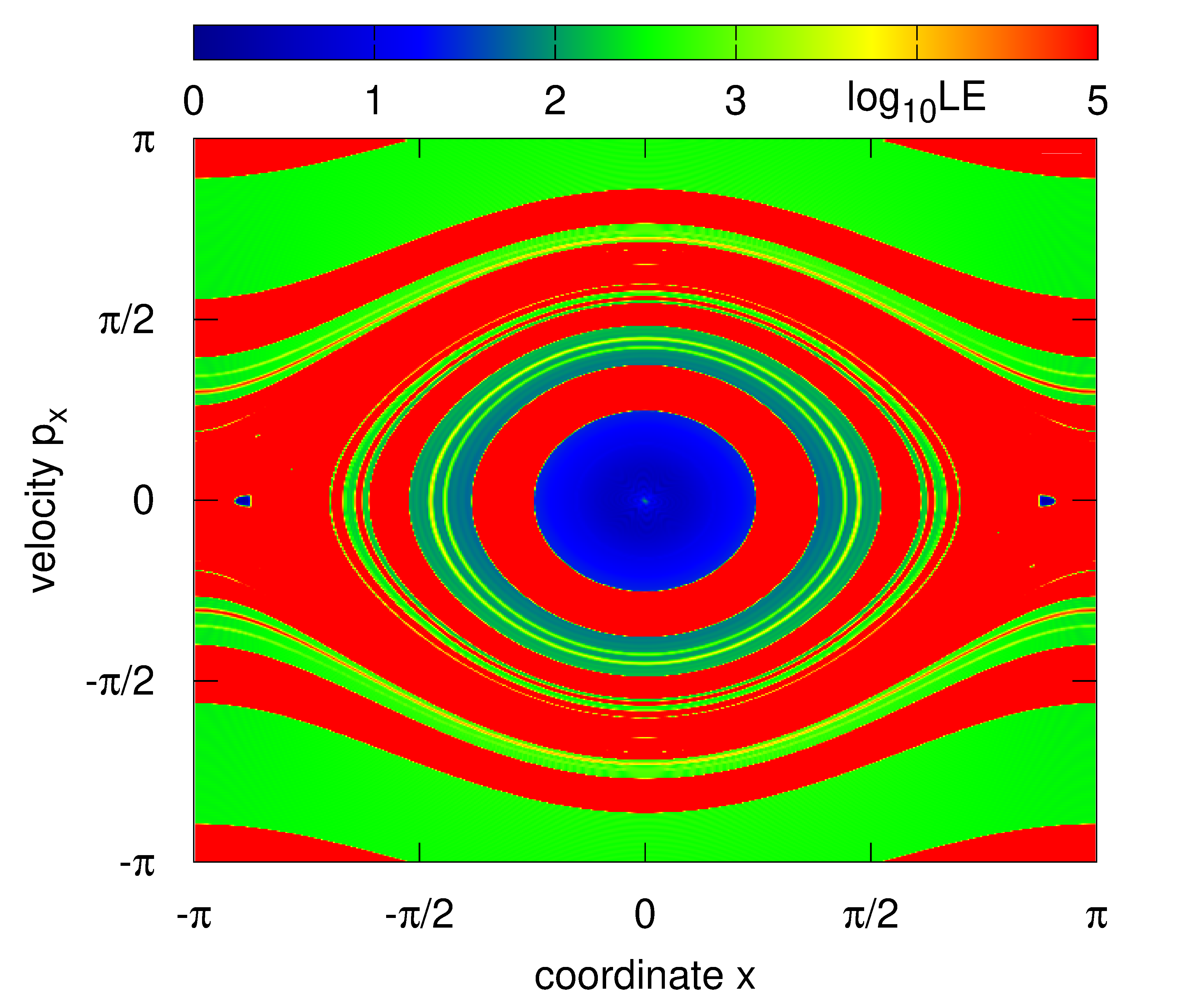}    
\includegraphics[width=2.5 cm, height=2.5 cm]{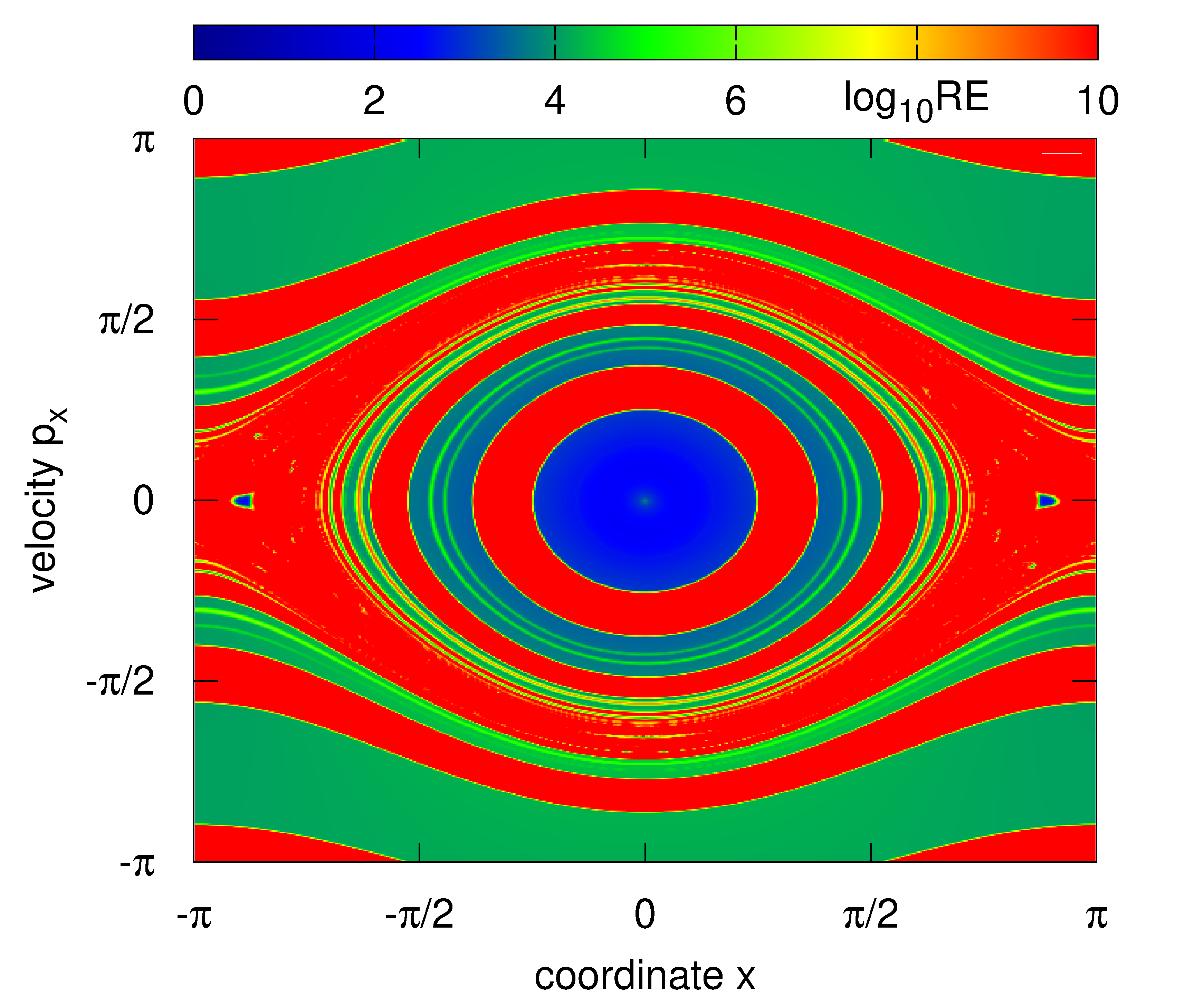}   }
 \centerline{
\includegraphics[width=2.5 cm, height=2.5 cm]{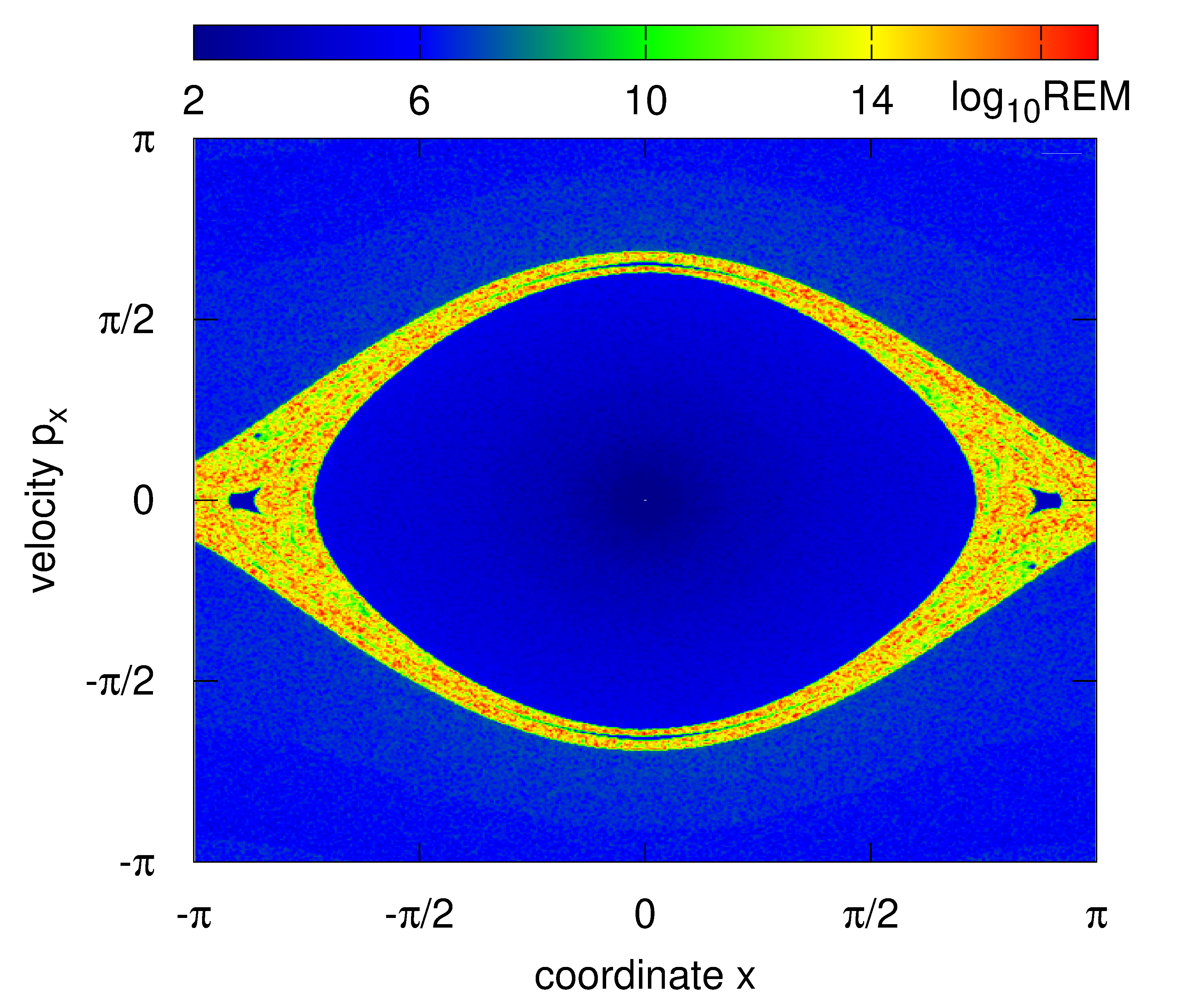}
\includegraphics[width=2.5 cm, height=2.5 cm]{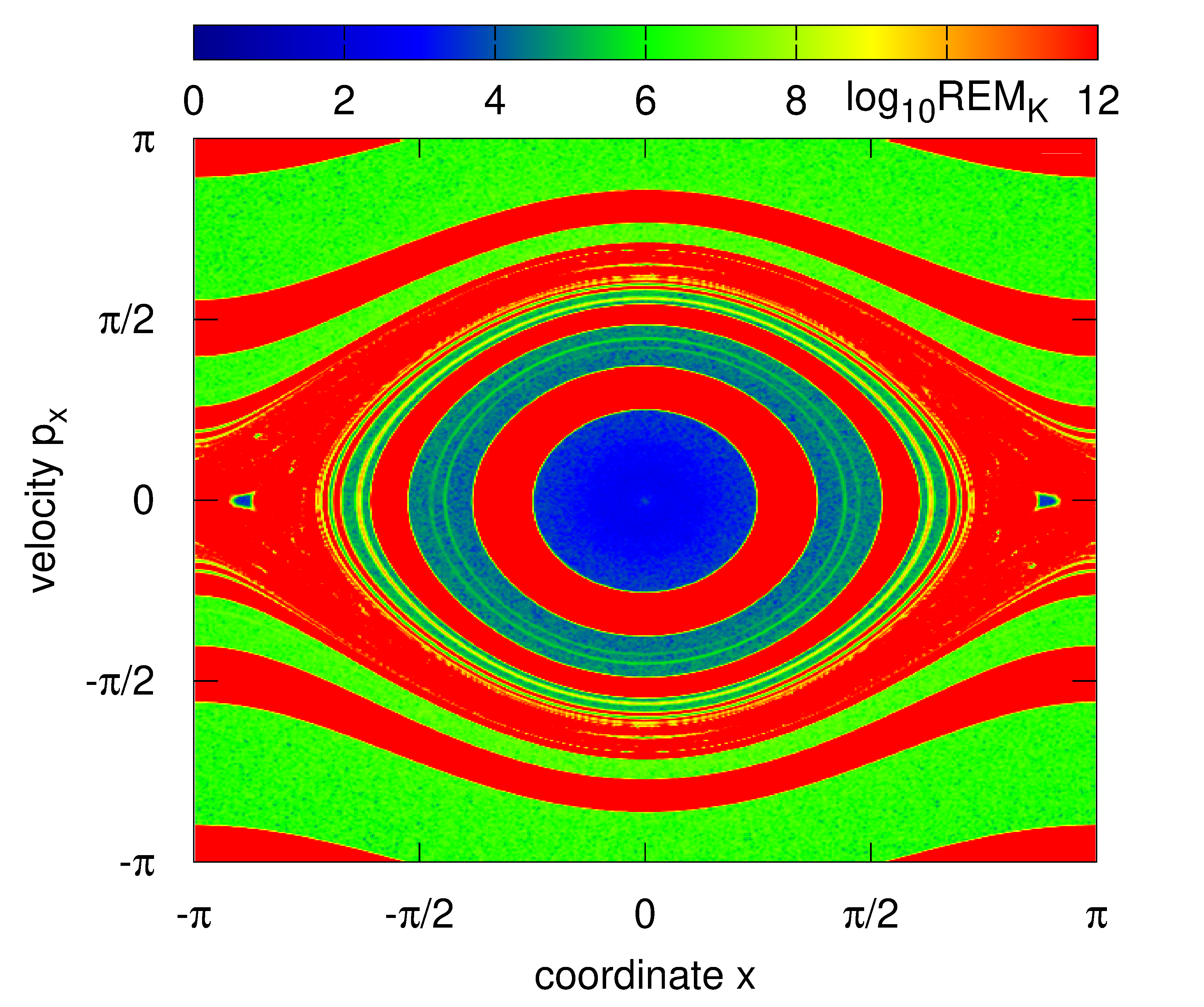}
}
\caption{
 First frame:  stroboscopic phase portrait 
 computed for $10^3$ dynamical periods, in the $(x ,p_x)$-plane for the  coupled pendulums
 with time dependence $\eps=0.01,~\Omega=1$,
   with initial   conditions in the same plane.
   The parameters are    $\omega_x^2=5,\,\omega_y^2=4,\,\mu=4$ the integration has been
   performed by the second order symplectic integrator described in  \cite{Yoshida1990}, with $\Delta t=2\pi/100$.   
  Second  frame: LE color plot at $t= 2\pi \times 50$.
  Third frame: RE color plot at $t= 2\pi \times 50$.
  Fourth frame: REM  color plot at at $t= 2\pi \times 50$.
 Fifth frame:  REM with kick
  at $t= 2\pi \times 50$ . }
\label{fig:fig_2} 
\end{figure}

The next step is the  introduction of the time dependence by choosing $\eps\not=0$
in order to create a chaotic layer,  around the separatrix
of the first pendulum in the $(x ,p_x)$-plane, whose  size grows with   $\epsilon$.
The first pendulum  is decoupled from the second one when $\mu=0$ or when
$\mu\not=0$ but we choose  $y=p_y=0$ as initial
conditions  since the $(x ,p_x)$ plane is invariant.
In this last case, the stroboscopic map gives
the phase space portrait in the $(x ,p_x)$ plane, see  the first frame of  Fig.~\ref{fig:fig_2}.
The next frames of  Fig.~\ref{fig:fig_2}, show the results for LE and RE 
when  $y=p_y=0$: thetime  dependence leads to a progressive merge of the unstable annular
regions present when $\eps=0$  destroying the  stable annular regions among them.
In this case the REM portrait exhibits only the chaotic region around  the separatrix
since $(x ,p_x)$-plane is invariant, but if a small kick is applied before reversing the motion,
then the same pattern provided by  RE   is recovered,
\\
If initially  $y\not =0$ or $p_y\not =0$, the Poincar\'e section $y=0$ with $p_y>0$,   leading to
the  $(x ,p_x)$-plane phase  portrait, cannot be computed since  $H$ is no longer a first integral
and we cannot choose the pointson $H=E$ manifold  to project the corresponding orbits
on the $(x,p_x)$-plane. On the contrary, the fast indicators $E_L$  and $E_{BF}$   provide the
stability portrait in the  $(x,p_x)$-plane for any initial choice of $y,p_y$.
When  no  phase  portrait is available, since the Poincar\`e section cannot be computed,
the  stability  portraits are provided by the  proposed fast indicators choosing
the initial points in any selected two--dimensional   plane of phase space.
The stability portrait of a three--dimensional  hyper-plane of phase space can be obtained
by  showing a sequence  of  two--dimensional phase portraits, with an animation tool. 
 
\begin{figure}[!ht]
\centerline{
\includegraphics[width=2.75 cm, height=2.75 cm]{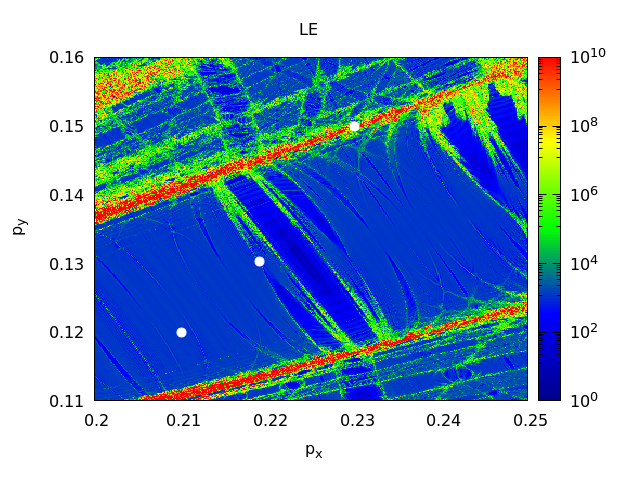}    
\includegraphics[width=2.75 cm, height=2.75 cm]{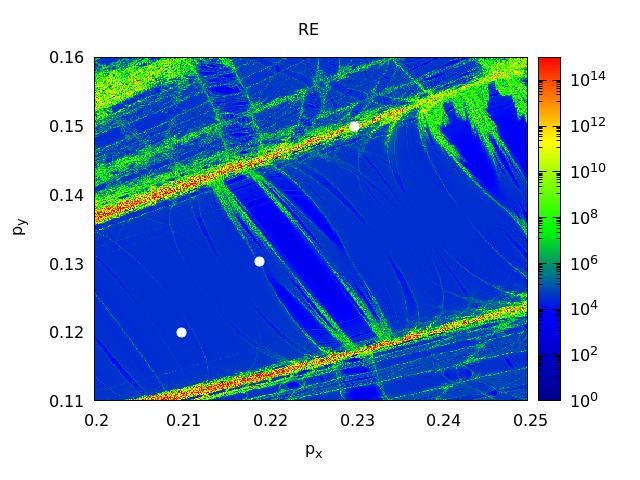}    
\includegraphics[width=2.75 cm, height=2.75 cm]{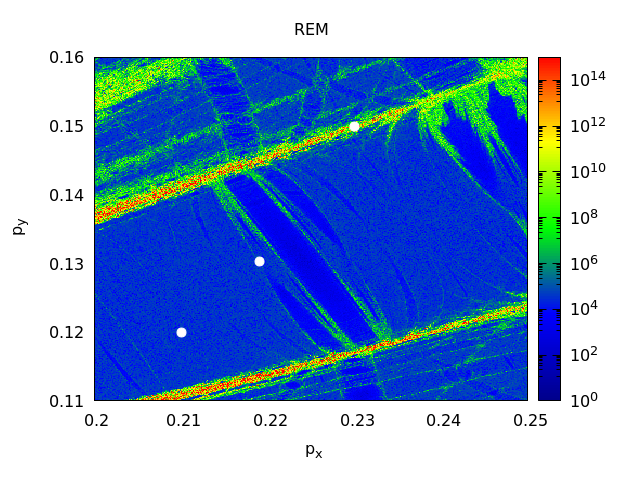}     }
\caption{ Plot of errors with a magnification  of the $p_x,p_y$ plane for $\mu=1$ and $N=1000$.
  From left to right LE, RE and REM.
  The white circles correspond to the initial conditions for which the errors variation with the iteration
number is shown in the next figure.}
\label{fig:figure_4} 
\end{figure}

\begin{figure}[!ht]
\centerline{
\includegraphics[width=2.75 cm, height=2.75 cm]{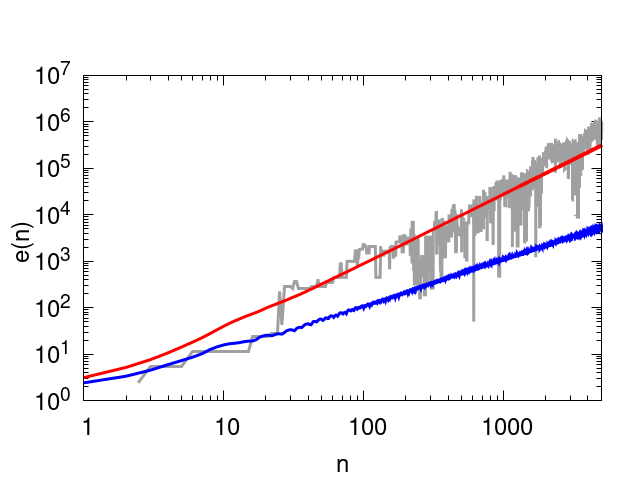}    
\includegraphics[width=2.75 cm, height=2.75 cm]{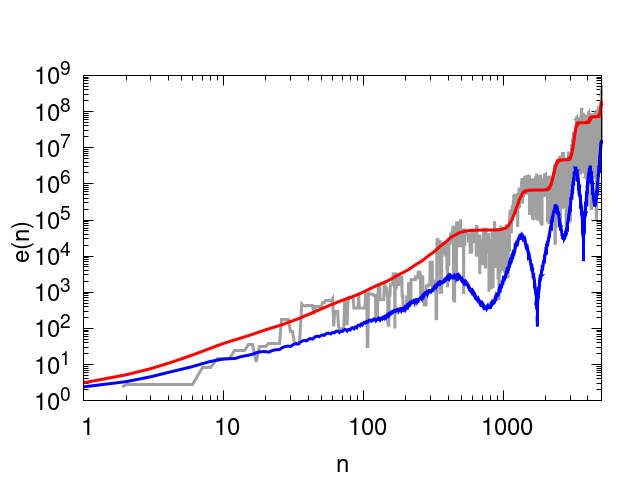}   
\includegraphics[width=2.75 cm, height=2.75 cm]{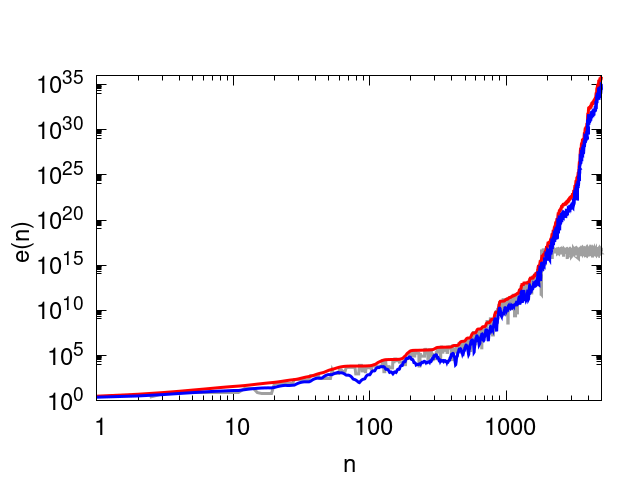}     }
\centerline{
\includegraphics[width=2.75 cm, height=2.75 cm]{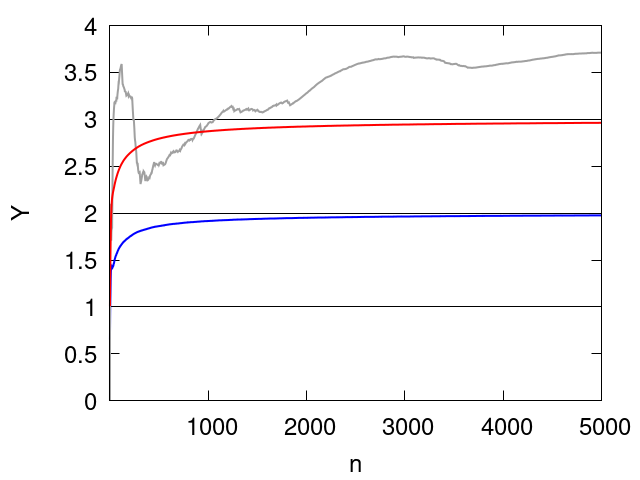}    
\includegraphics[width=2.75 cm, height=2.75 cm]{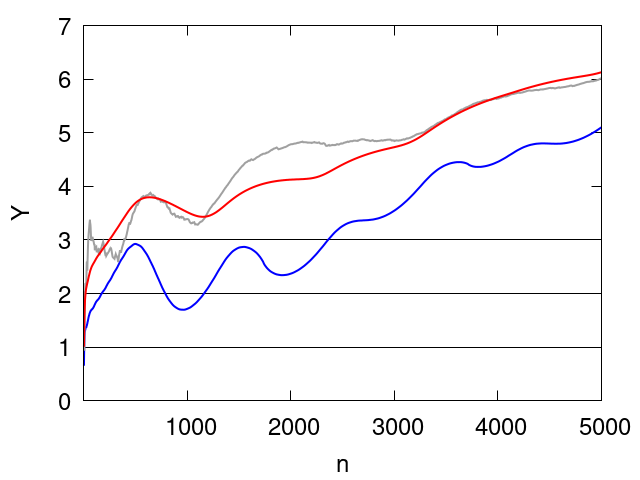}   
\includegraphics[width=2.75 cm, height=2.75 cm]{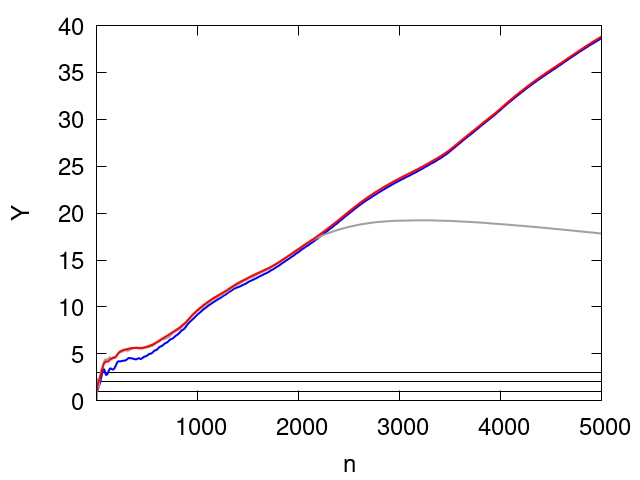}     }

\caption{ Plot of errors for the symplectic map with $\mu=1$ as a function of the iteration number $n$.
  Upper frames: plot of  errors LE, blue curve, RE,  red  curve, REM, grey curve
 for a single orbit with initial condition  $x=0,y=0$ and different
 values of $p_x,p_y$ corresponding to  regular, weakly chaotic and stronhly chaotic orbits.
 First  frame $p_x=0.21,\,p_y=0.12$,  second  frame
 $p_x=0.219,\,p_y=0.1303$, third  frame (right)  $p_x=0.23,\,p_y=0.15$. The  orbits have the initial points
 shown by the white circles in the previous figure.
Lower frames:  plot of the MEGNO filter applied to the previous orbits, in the same order. }
\label{fig:fig_5} 
\end{figure}

\spa
 We have analyzed  a symplectic map, previously proposed to explore the Arnold web of resonances.
 This map  is the   first order symplectic scheme for the following Hamiltonian 
\begin{equation}
H= {p_x^2 + p_y^2\over 2} + {1\over (2\pi)^2}\, {\mu\over  4 + \cos(2\pi x)+ \cos(2\pi y)}~,   
\label{eq_7_4}    
\end{equation}
computed on the $\Toro^4$ torus.  
See appendix X for the expression of the map $M(\xbf)$ and the tangent map $DM(\xbf)$.
For $0 \leq \mu \ll 1$  the map is quasi-integrable and  the flow of $H$
interpolates the orbits of the map. For $\mu \sim 1 $ the map is manifestly non-integrable
and exhibits the web of resonances.
This is the Chirikov regime since   the single resonances appear as a series of
strips of regular orbits with a rather thick boundary of chaotic orbits in the actions  $(p_x,p_y)$-plane.
We evaluate the errors for initial conditions in this  plane  with  $x=y=0$.
In Figure \ref{fig:fig_3}, we show  the error plots  for $\mu=1$. The symmetry for inversion of the actions
in the $(p_x,p_y)$-plane allows us to consider only the square $[0,1/2]^2$ where the web of resonances appears.
We have chosen the scales $[0,10^{10}]$ for $E_L$ and  $[0,10^{15}]$  for $E_{BF}$  respectively, in order to take
into account the law of growth for regular orbits which is  $n$ for $E_L$
and $n^{3/2}$ for $E_{BF}$ , which  roughly holds   also for REM. With this choice, the plot of errors are extremely
similar as it was previously observed in other models. 
\begin{figure}[!ht]
\centerline{
\includegraphics[width=2.75 cm, height=2.75 cm]{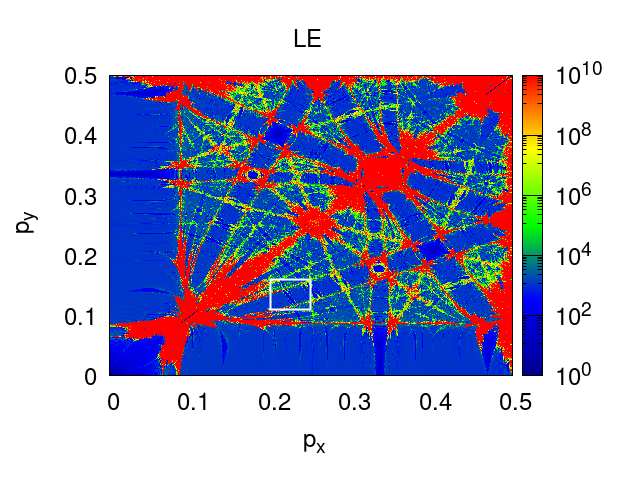}        
\includegraphics[width=2.75 cm, height=2.75 cm]{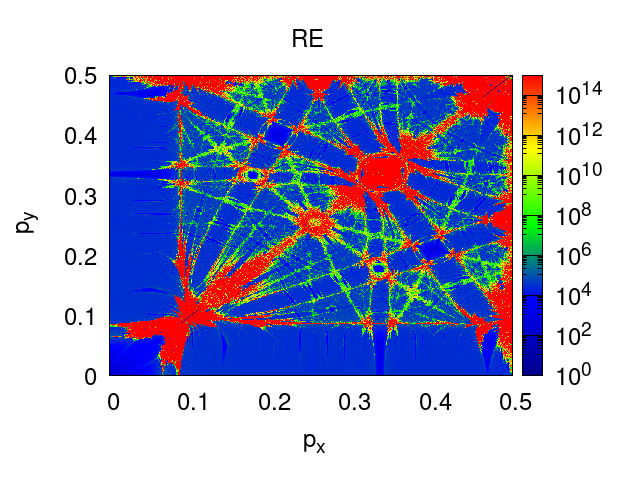}   
\includegraphics[width=2.75 cm, height=2.75 cm]{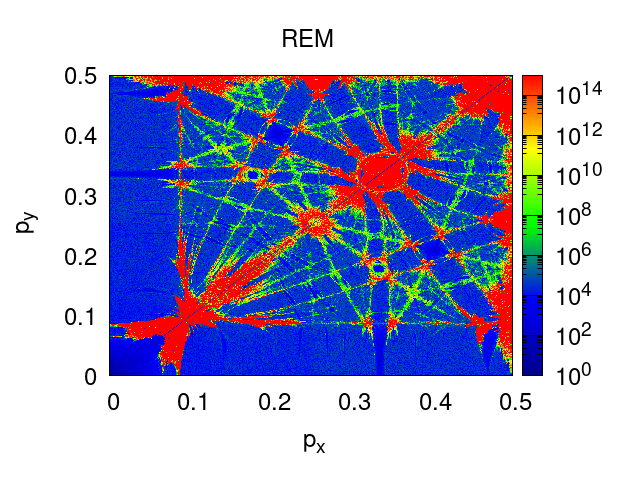}     }
\caption{ Plot of errors for the symplectic map with $\mu=1$  and $N=1000$ iterations
  for initial conditions in the domain   $[0,1/2]^2$   of the   $p_x,p_y$  torus $\Toro^2=[-1/2,1/2]^2$  with 
  $x=y=0$.
From left to right, we plot LE, RE and REM, respectively. 
In the plot of LE the white square corresponds to the region which is magnified in the next figure.}
\label{fig:fig_3} 
\end{figure}
%
%
In  Figure  \ref{fig:figure_4},  we  show  the error plots in a small square region of the
$(p_x,p_y)$-plane choosing  three  points  for which  the evolution  with the iteration number
is shown in Figure \ref{fig:fig_5}. In this last figure, we show also  the results obtained by
applying the MEGNO  filter. The first   orbit is  regular, the second one is  weakly chaotic,
the third one is strongly   chaotic.
%
We notice that  $E_L$   exhibits a power law behaviour with oscillations,   $E_{BF}$ 
exhibits a power law behaviour without oscillations, and REM is close to  $E_{BF}$  but exhibits large fluctuations.
For large values of $n$,  $E_L$ and $E_{BF}$  grow as $n^\alpha$  with  exponents $1$ and $3/2$, corresponding
to the asymptotic values of 2 and 3 for the MEGNO filter, whereas  the exponent of REM 
varies significantly with respect to $3/2$ due to its fluctuating growth.  Finally, we notice that  REM saturates at
$10^{17}$ for chaotic orbits since it is defined as the distance of the reversed orbit from the initial point divided
by the machine accuracy. This is the reason why we have chosen $10^{15}$ as the scale for REM and RE 
which implies a scale $10^{10}$ for $E_L$.
The comparison of REM with  other indicators such as FLI and SALI was performed in \cite{Mestre2011}.
In this paper, we have extended the comparison to $E_{BF}$, which has a solid theoretical justification,
and is simply related to $E_L$.
%
%

\section{Conclusions}
 \label{S:conclusions}
 
 In this paper, we have reviewed and generalized the theory of two fast dynamical indicators, namely, the Lyapunov Error (LE) and Reversibility Error (RE).

 With this generalization, the full set of  Lyapunov, Forward and Reversibility Errors are now invariant under the choice of initial conditions and are defined
 as the square roots of the corresponding invariants. The first LE and RE were previously  proposed, jointly with the  round-off induced fast dynamical indicator REM. The set of LE differ from FLI and GALI since they do not  depend on the initial deviations.
 The key point is that we consider the linear response to a  small random initial displacement
 and the  invariants of the corresponding covariance matrix.
 We consider also  the linear response to a small noise along the orbit up to time $t$
 and  the linear response to  a \textit{noisy} BF process in which
 the forward evolution (F), up to time $t$, is followed by the backward evolution (B) from $t$ to $2t$.
 The  fundamental matrix of the tangent flow  is the first step to build LE, F and RE
 and RE error covariance matrices. 
 The  original  motivation  for considering  RE was that,  for a map,  it can be
 compared with REM, where the small random displacement is replaced by
 the round-off. As a consequence, REM allows to estimate  the sensitivity  to the round-off
 in numerical computations while RE, which proves to be comparable,
 provides a theoretical foundation for its use.
 The key differences between RE and REM are the following: 
 \begin{itemize}
  \item REM provides just  the first invariant whereas for RE the full set can be computed.
  \item REM corresponds to a single  realization of  a pseudo-random process such as the round-off.
  \item Finally REM  corresponds to a linear approximation since the noise amplitude is finite even though very small.
\end{itemize}

The asymptotic behaviour of the LE invariants $I_L^{(j)}$, is exponential with a coefficient 
 given by twice the sum of the first $j$ Lyapunov  exponents (ordered in a non-increasing sequence), and the following properties has been proved:
 \begin{itemize}
  \item For integrable  Hamiltonian systems the invariants   $I_L^{(j)}$  grow according to a power law.
  \item For  non-integrable  Hamiltonian systems the regions where the Lyapunov errors $I_L^{(j)}$
 exhibit a power-law growth, correspond to regular  motions.
  \item For  non-integrable  Hamiltonian systems the regions where the Lyapunov errors $I_L^{(j)}$
 exhibit an exponential growth, correspond to chaotic  motions.
\end{itemize}

 The F error invariants $I_F^{(j)}$  have an asymptotic  power-law growth with a coefficient which is twice
 the sum of the positive Lyapunov exponents among the first $j$.
 The RE invariants $I_{R}^{(j)}$  have an asymptotic  power-law growth with a coefficient
 which is twice the sum of the absolute value of the negative Lyapunov exponents among the last $j$.
 The return at time $2t$ to the initial condition  in absence of noise implies the absence of a drift
 in the stochastic process corresponding to the noisy BF orbit at time $2t$.
 This  simplifies  the transition probability  for the BF process   since for linear systems it is  
 invariant  by  translation in phase space.

 If we change $t$ into $-t$ the  F  and  B evolution interchange.  A drastic difference 
 is observed  in the asymptotic behaviour since   attractors  and  repellers interchange.
 The change of sign in time is asymptotically equivalent to a sign  change  of the Lyapunov exponents.
 Autonomous Hamiltonian systems are time-reversal invariant if $H(\xbf)=H(I_R\xbf)$
 where  $\Iop_R$ is the linear  operator which changes the sign of the momentum $\pbf$. Indeed we have 
 $S_{-t}(\xbf)= \Iop_R S_t(\Iop_R \xbf)$ and $\Lop(\xbf,-t)=\Iop_R \,\Lop(\Iop_R \xbf,t)\Iop_R$.
 As a consequence the B error or the  BF Reversibility  error invariants are the same as the
 F  or  BF invariants provided that we change $\pbf$ into $-\pbf$.
 For any other system the
 behaviour of the F and B error invariants or  BF and FB Reversibility error invariants
 is  drastically  different  and reflects    the violation of the time reversal invariance.
 
 As a consequence, the limit of the Gibbs entropy divided by $t$  is just the sum of all the positive
 Lyapunov exponents just as for the Kolmogorov-Sinai entropy. 
 We have introduced the local  fidelity for BF process, for which, differently from the classical random  fidelity,  an analytic expression can be obtained. The basic idea is to consider the autocorrelation for the
 BF process for observables on a torus of side $\ell$ taking the  limit of the noise amplitude
 $\eps$ and of the torus size  $\ell$ to zero, while keeping their ratio $\rho=\eps/\ell$ constant.
 The asymptotic limit  for $t\to \infty$ can be easily worked out since the local BF fidelity
 depends on the BF Reversibility error correlation matrix.
\\
A simple procedure to numerically analysis the dynamics of a given dynamical  system, are:
 \begin{enumerate}
  \item Start from  a map $M(\xbf)$, which in the case of a flow is a suitably  chosen integrator.
  \item Compute the tangent map $L_n(\xbf)$ to obtain  the LE, F and  RE covariance matrices.
  \item Compute the corresponding invariants.
\end{enumerate}

The computation of the F and BF invariants requires the inversion  of the Lyapunov
matrix which could be ill-conditioned when  $(\lambda_1-\lambda_d)t$ is large.
The evaluation of the tangent map  may be avoided by replacing it  with finite differences.
Concerning RE, we can avoid the  evaluation of the tangent map by
computing REM,  which requires only the forward
and backward iterations of the map, but unlikely RE is affected by significant fluctuations.
\\
We recommend the use of REM for a first screening of any dynamical system, since the implementation of REM is straightforward and  its computational cost is just two times the cost of the computation
of the orbit for any initial point. The fluctuations can be  filtered out by a running
average in the phase space.
\\
In the literature and in this paper, the systems considered up to now were mainly Hamiltonian flows and symplectic maps  in  $\Reali^{2d}$ or $\Toro^{2d}$ with $d=1,2$.   For $d=2$,  the  second  invariant provides an  additional information since its  asymptotic behaviour depends on the first two Lyapunov exponents which are non negative.  For  generic  systems in $\Reali^d$  with $d\ge 3$  all the invariants should be considered.
 If the volumes contraction rate is uniform  $\lambda_1+\ldots+\lambda_d$  is constant
  only the first $d-1$ invariants  need   to be examined.
\\ 
The  models  considered in this paper, confirm the validity of the proposed indicators.
For Hamiltonian systems the invariants allow to discriminate the regions of regular and chaotic motion
choosing the initial conditions in a selected  plane of the phase space. In the case of a  slow   
modulation, where  the frequency map analysis  fails  due to the large  separation of frequencies, 
the proposed indicators are still applicable. 
\\
Finally, we observe that the ratio of the  weights  of a stochastic trajectory for the F and B processes is related to  the Lyapunov exponents, whereas the ratio of weights of the trajectories where
the initial and final points are interchanged, as in the fluctuation theorems, it is given by  the Boltzmann weights.                 

%
%
\section{Overview}  
We resume the main properties of the LE, RE, REM fast indicators
and  higher order invariants  for LE and RE.
\spa
{\bf Lyapunov error LE}
\spa
Given the flow $S_t(\xbf)$ of a dynamical system generated by a 
a  smooth vector field $\Phibf(\xbf)$ in  $\Reali^d$ the linear
response  $\Xibf(\xbf,t) $ to an initial random deviation of amplitude
$\eps$  and its covariance matrix $\Sigma^2(\xbf,t)$ are defined by 
\begin{equation}
 \begin{split} 
   \Xibf(\xbf,t)& = \lim_{\eps \to 0}{S_t(\xbf+\eps \xibf)- S_t(\xbf)\over \eps}= \Lop(\xbf,t)\xibf \\ \\
   \Sigma^2(\xbf,t)&=\mean{\Xibf(\xbf,t)\Xibf^T(\xbf,t)} =\Lop(\xbf,t)\,\Lop^T(\xbf,t)\quad \,\,
\end{split}
\label{eq_UG1}
\end{equation} 
where $\Lop(\xbf,t)= D S_t(\xbf)$  satisfies a linear equation and $\xibf$ is a random vector
with unit mean and unit covariance matrix $\mean{\xibf\xibf^T}=\Iop$.
The matrix $\Lop\Lop^T$ has the same invariants  $I_L^{(k)}$ for $1\le k \le d$
as the Lyapunov matrix $\Lop^T\Lop$. The following notation is used for the errors 
\begin{equation}
 \begin{split} 
   \hbox{LE} \phantom{ of order \quad j\ge 2}  &  \qquad
     E_L(\xbf,t) = \Bigl( I_L^{(1)}(\xbf,t) \,\Bigr)^{1/2}  \\ 
    \hbox{LE of order} \quad k\ge 2   & \qquad E_L^{(k) }(\xbf,t) = \Bigl( I_L^{(k)}(\xbf,t) \,\Bigr)^{1/2}
\end{split}
 \label{eq_UG2}
\end{equation} 
Notice that letting  $\ebf_i$ be the base vectors $(\ebf_i)_j= \delta_{ij}$  and
$\ebf_i(t)= \Lop(\xbf,t) \ebf_i$ their transforms  we have 
\begin{equation}
 \begin{split} 
      E_L(\xbf,t) = \Bigl( \Tr(\Sigma^2_L(\xbf,t))\,\Bigr)^{1/2} =
      \Bigl(  \sum_{j=1}^d \Vert \ebf_j(\xbf,t)\Vert^2 \,\Bigr)^{1/2}
\end{split}
 \label{eq_UG3}
\end{equation} 
Replacing the base $\ebf_j$ by another ortonormal base $\ebf'_j= \Rop \ebf_j$ the result
does not change. The error $E_L^{(k)}(\xbf,t)$ is the square root of the invariant
given by the sum of the volumes squared of the ${d\choose k}$ parallelotops whose
edges  are  the distinct  choices of $k$ among the $d$ vectors $\ebf_1(\xbf,t),\ldots,\ebf_d(\xbf,t)$.
The volumes squared are the Grahm matrices of these vectors. The   QR decomposition can
be used to compute each volume. A detailed discussion can be found in Appendix 1.
\\
The Lyapunov matrix  can be written as
\begin{equation}
 \begin{split} 
   & \Lop^T(\xbf,t)\,\Lop(\xbf,t) = \Wop(\xbf,t) \, e^{2t\,\Lambda(\xbf,t)\, }\, \Wop(\xbf,t) \\
   & \Lambda(\xbf,t)= \diag\bigl(\lambda_1(\xbf,t),\ldots,\lambda_d(\xbf,t)\bigr)
\end{split}
 \label{eq_UG4}
\end{equation} 
where $\lambda_j(\xbf,t)$ are ordered in a non decreasing sequence. The existence
of a limit of $\Wop(\xbf,t)$ and $\Lambda(\xbf,t)$ for
$t \to +\infty$ is insured by the  Oseledet theorem. Denoting with
$\Lambda(\xbf)=\diag\bigl(\lambda_1(\xbf,\ldots,\lambda_d(\xbf)\bigr)$
the limit of $\Lambda(\xbf,t)$ whose entries are the Lyapunov exponents we have
\begin{equation}
 \begin{split} 
   \lim_{t\to +\infty} \,\,{1\over t} \log E_L^{(k)}(\xbf,t)= \lambda_1(\xbf)+\ldots+\lambda_k(\xbf)
\end{split}
 \label{eq_UG5}
\end{equation} 
The added value of LE is the independence from the initial deviations and consequently
form the choice of the orthogonal base in phase space.
For a linear system $\Lop(t)= e^{\Aop t}$  where $\Aop$ is a real matrix with eigenvalues $\lambda_j$
we have $t^{-1} \,\log E^{(k)}_L(t)= \lambda_1+\ldots+\lambda_k + O(t^{-1})$
and $d\log E^{(k)}_L(t)/dt = \lambda_1+\ldots+\lambda_k + O(\eta)$ where  
$\eta = \exp(-t\,\min_{i\not = j}\, |\lambda_i-\lambda_j|)$ for $i\not = j$. A proof is given in Appendix 4
but  computer assisted proofs could  be extended to $d >2$.
The convergence with $t^{-1}$ of $t^{-1}\log E_L(t)$ and the exponential convergence
rate of $d\log E_L(t)/dt$ to a negative Lyapunov exponent has been proved for $d=1$ in Appendix 5.
The asymptotic behaviour of $E(\xbf,t)$ has been worked out for nonlinear Hamiltonian systems
in $\Reali^2$  using normal the forms in section 5.
\spa
  {\sl The map case}
\spa
 Letting $M(\xbf)$ be a map in $\Reali^d$ and denoting with $M_n(\xbf)=M(M_{n-1}(\xbf))$ its iterates with
$M_0(\xbf)(=\xbf$ and  $\Lop_n(\xbf) = DM_n(\xbf)$  the tangent map,  the previous definitions
hold after replacing $\Lop(\xbf,t)$ with $\Lop_n(\xbf)$ and the integrals with sums.
Denoting with $I^{(k)}_{L\,\,n}(\xbf)$ the  invariants of the covariance matrix $\Lop_n\Lop^T_n$
or the Lyapunov matrix  $\Lop_n\Lop^T_n$,   the Lyapunov error of any order  is  denoted by
\begin{equation}
 \begin{split} 
   \hbox{LE} \phantom{ of order \quad j\ge 2}  &  \qquad
     E_{L\,\,n}(\xbf) = \Bigl( I_{L\,\,n}^{(1)}(\xbf) \,\Bigr)^{1/2}  \\ 
    \hbox{LE of order} \quad k\ge 2   & \qquad E_{L\,\,n}^{(k) }(\xbf) = \Bigl( I_{L\,\,n}^{(k)}(\xbf) \,\Bigr)^{1/2}
\end{split}
 \label{eq_UG6}
\end{equation} 
\spa
{\bf Forward  error FE}
\spa
It is obtained  from the linear response to an additive noise during the
evolution up to time $t$. Letting $S_t(\xbf;\,\eps)$  be noisy flow
due to an additive random perturbation $\eps \,\xibf(t)$  where $\xibf(t)$ is
a white noise of unit covariance matrix $\mean{\xibf(t)\,\xibf^T(t')}= \Iop
\delta (t'-t)$, the linear response is a random vector given  by
\begin{equation}
 \begin{split} 
   \Xibf_{F}(\xbf,t)& = \lim_{\eps \to 0}{S_t(\xbf,\, \eps)- S_t(\xbf) \over \eps}= \\ 
   &\qquad \qquad  = \Lop(\xbf,t)\, \int_0^t \, \Lop^{-1}(\xbf,t')\,\xibf(t')\,dt'
\end{split}
\label{eq_UG7}
\end{equation}  
The mean value of $\Xibf_{F}(\xbf,t)$ is  vanishes  and its covariance matrix is
\begin{equation}
 \begin{split} 
   & \Sigma^2_{F}(\xbf,t)= 
 \Lop(\xbf,t) \int_0^t\, (\Lop^T(\xbf,t') \, \Lop(\xbf,t')\,)^{-1}\, dt'\,\,\,\,\Lop^T(\xbf,t) 
\end{split}
\label{eq_UG8}
\end{equation}  
The invariants of this matrix are denoted by $I_F^{(k)}(\xbf,t)$ and their square roots are
the forward errors $E_F^{(k)}(\xbf,t)$.  If the flow is replaced by a map similar expressions are
obtained.
\spa
{\bf Reversibility error RE}
\spa
We consider a Forward noisy evolution up to time $t$ followed by the Backward evolution, generated by $-\Phibf(\xbf)$ from time $t$ us to $2t$. The linear response to this BF process is given by the random vector
\begin{equation}
 \begin{split} 
 \hskip -.5 truecm   \Xibf_{BF}(\xbf,2t)& = \lim_{\eps \to 0}{S_{-t}\bigl(S_t(\xbf,\,\eps)\bigr) - \xbf \over \eps}= \\
   & = \int_0^t \, \Lop^{-1}(\xbf,t')\,\xibf(t')\,dt'
\end{split}
\label{eq_UG9}
\end{equation}          
The mean value  vanishes  and the covariance matrix is
\begin{equation}
 \begin{split} 
   \Sigma^2_{BF}(\xbf,t)&=   \mean{\Xibf_{BF}(\xbf,2t)\Xibf_{BF}^T(\xbf,2t)}  =  \\ & = 
      \int_0^t\, (\Lop^T(\xbf,t') \, \Lop^T(\xbf,t')\,)^{-1}\, dt' 
\end{split}
\label{eq_UG10}
\end{equation}
We denote the invariants of this matrix $I_{BF}^{(k)}(\xbf,t)$. The BF Reversibility
errors  are given by their square roots and we denote them
with $E_{BF}^{(k)}(\xbf,t)$. If The noise affects also
the Backward evolution then the linear response if given by 
\begin{equation}
 \begin{split} 
   \Xibf_{BF}(\xbf,2t)  = 
    \int_0^t \, \Lop^{-1}(\xbf,t')\,\bigl( \xibf(t')+\xibf(2t-t')\,\bigr )\,dt'
\end{split}
\label{eq_UG11}
\end{equation}          
but the covariance matrix is the same  as (\ref{eq_UG10}) provided we insert a factor $1/2$ in
its definition since the noise in $[0,t]$ is uncorrelated from the noise in $[t,2t]$. 
\spa
If we reverse the process by considering first the B process follows by the F process  the
FB Reversibility error covariance matrix is given  by 
\begin{equation}
 \begin{split} 
   \Sigma^2_{FB}(\xbf,t)&= 
      \int_{-t}^0\, (\Lop^T(\xbf,t') \, \Lop(\xbf,t')\,)^{-1}\, dt' 
\end{split}
\label{eq_UG12}
\end{equation}
\spa
    {\sl The map case}
\spa
If we have and invertible map $M(\xbf)$ and $\Lop_n(\xbf)$ is the tangent map, the BF
Reversibility error covariance matrix for  $n$  iterations of the noisy map
followed by $n$ iterations of the inverse map is given by
\begin{equation}
 \begin{split} 
   \Sigma^2_{FB\,\,n}(\xbf)&= \sum_{n'=1}^n   (\Lop^T_{n'}(\xbf) \, \Lop_{n'}(\xbf)\,)^{-1} 
\end{split}
\label{eq_UG13}
\end{equation}
The BF Reversibility errors given by the square root of the invariants are denoted with
$E_{BF\,\,n}(\xbf)$.  If the noise is present also  when 
the inverse map is applied, the covariance matrix defined by inserting a factor $1/2$ as in  the continuous case is given by 
\begin{equation}
 \begin{split} 
\Sigma^2_{FB\,\,n}(\xbf)& = {1\over 2}\,\Iop \,+\,\sum_{n'=1}^{n-1}   (\Lop^T_{n'}(\xbf) \, \Lop_{n'}(\xbf)\,)^{-1} + \\
& + {1\over 2}(\Lop^T_{n}(\xbf) \, \Lop_{n'}(\xbf)\,)^{-1}
 \end{split}
\label{eq_UG14}
\end{equation}
The difference between (\ref{eq_UG13}) and (\ref{eq_UG14}) is negligible for large values of  $n$.
If  $n$  iteration with the inverse noisy map are followed by $n$ iterations of the map the
BF Reversibility error covariance matrix is given by
\begin{equation}
 \begin{split} 
  \Sigma^2_{FB\,\,n}(\xbf)&= \,\sum_{n'=1}^{n}   \Bigl  (\Lop^T_{-n'}(\xbf) \, \Lop_{-n'}(\xbf)\,\Bigr)^{-1} 
 \end{split}
\label{eq_UG15}
\end{equation}
where  $\Lop_{-n}(\xbf)= D M_{-n}(\xbf)$ denote the tangent inverse map having defined $M_{-n}(\xbf)=
M^{-1}\bigl(\,M_{-n+1}(\xbf)\,\bigr)$.  If the noise affects  the B and F iterations then 
$\Sigma^2_{FB\,\,n}$ is given by (\ref{eq_UG14}) where $\Lop_{n'}$ are replaced by
$\Lop_{-n'}$ for $1\le n'\le n$.  
\spazio
{\bf Reversibility error method REM}
\spa
The label REM was chosen for the round off induced Reversibility error   on an
invertible map $M(\xbf)$, eventually
given by  the integrator of a flow with a time step $\Delta t$.
We denote with $\eps_*$ the round off amplitude, with  $M_{\eps_*}(\xbf)$ the map
evaluated with round-off, and  with  $M^{-1}_{\eps_*}(\xbf)$ the inverse map computed
with round-off where $ M^{-1}_{\eps_*}\bigl (M_{\eps_*}(\xbf)\bigr)=\xbf + O(\eps_*)$.
Notice that REM is defined by  
\begin{equation}
 \begin{split} 
   REM_n(\xbf)= {\Vert \,M_{\eps_*}^{-n}\bigl( M_{\eps_*}^n (\xbf)\,\bigr) -\xbf \Vert
   \over \sqrt{2} \,\,\eps_*}
 \end{split}
\label{eq_UG16}
\end{equation} 
where $\eps_*$ has a fixed non-vanishing value. Notice 
REM can be computed also for the BF process, where we first iterate $n$ times
the inverse  map $M^{-1}_{\eps_*}$  and  then  $n$  times  the map $M_{\eps_*}$.
The factor $\sqrt{2}$ is inserted to have a correspondence
with $RE$ when the noise affects both the map $M$ and its inverse.   We may compare
REM with RE taking into account that there are two major differences.
The round off can be considered  a  multiplicative noise  defined by
\begin{equation}
 \begin{split} 
   M_{\eps_*}(\xbf)=   (\Iop +\eps_* \Xop_*) \,M(\xbf)   \qquad
   \Xop_*=\diag(\xi_{1\,*},\ldots,\xi_{d\,*})
 \end{split}
\label{eq_UG17}
\end{equation}
where  $\xi_{j\,*}$   are random variables  in $[-{1\over 2}, \,{1\over 2} ]$ with zero mean,
and a quasi uniform distribution.
Letting ${\bf 1}$ the vector whose entries are equal to 1 and $\xibf_*=(\xi_{1\,*}, \ldots,
\xi_{d\,*})$ we  have
\begin{equation}
 \begin{split} 
   & \xibf_*= \Xop_*\,{\bf 1}  \\ \\
   & \Xop_* \,M(\xbf) = \Mop(\xbf) \,\xibf_*   \\ \\
   & \xibf_*=(\xi_{1\,*}, \ldots, \xi_{d\,*})^T \\ \\
   & \Mop(\xbf)=\diag\bigl(M_1(\xbf), \ldots, M_d(\xbf)\,\bigr)
 \end{split}
\label{eq_UG18}
\end{equation}
Since the additive noise is $ \xibf_*$ and the multiplicative noise is $\Mop\,\xibf_*$, the
covariance matrix is $\Iop$ for  the additive noise,   $\Mop^2(\xbf)$. for the multiplicative noise,
As a consequence the  covariance matrix for $\Xibf_{BF\,\,n}$ is given by
\begin{equation}
 \begin{split} 
   \Sigma^2_{BF\,\,n}(\xbf)& = {1\over 2}\,\Iop + \sum_{n'=1}^{n-1}   \, \Lop_{n'}^{-1}(\xbf)\,\Mop^2(\xbf_{n'-1})
   \bigl(\Lop_{n-1}^{-1}(\xbf)\,\bigr)^T + \\
        & + {1\over 2}\,    \Lop_{n}^{-1}(\xbf)\,\Mop^2(\xbf_{n-1})    \bigl(\Lop_{n}^{-1}(\xbf)\,\bigr)^T
 \end{split}
\label{eq_UG19}
\end{equation}
The result is obtained  with 
the procedure  of  section 4  from  equation (\ref{eq_4_35})  to  equation 
(\ref{eq_4_41}),  reported here as  (\ref{eq_UG14}),  by replacing $
\xibf_{n'}$ with $ \Mop(\xbf_{n'-1})\,\xibf_{n'}$.
When the noise is additive  the matrix $\Mop(\xbf)$ is replaced by $\Iop$ in (\ref{eq_UG19})
and  the standard  result given by  (\ref{eq_UG14}) is recovered.
We notice that the  (first)  BF Reversibility error is given by  
\begin{equation}
 \begin{split} 
   E_{BF\,\,n}(\xbf)& = \Bigl( \Tr \bigl(   \Sigma^2_{BF\,\,n}(\xbf)\,\bigr) \Bigr ) ^{1/2}=  \\ \\
   &=  \lim_{\eps\to 0}  \parton { \left\langle \,\,\Vert \,M_{\eps}^{-n}
     \bigl( M_{\eps}^n (\xbf)\,\bigr) -\xbf \Vert^2 \,\,\right \rangle  \over 2 \,\eps^2 }^{1/2}  
 \end{split}
\label{eq_UG20}
\end{equation}
The reversibility error   RE defined by (\ref{eq_UG20})
but computed  without taking the  average on the noise, assumed to be multiplicative, 
and  evaluated   for $\eps=\eps_*$   rather than taking the $\eps\to 0$ limit,
agrees with REM defined by (\ref{eq_UG16}).  As a consequence REM  agrees with  RE
for a specific realization on noise $\xibf_*$  which
corresponds to the round off error.  Such a realization certainly exists even though
is can be determined only a posteriori.  In addition to the  missing average,  the  condition
for the  correspondence  between RE and REM is that  if we evaluate RE by choosing $\eps$  
equal to  $\eps_*$   or by taking the $\eps\to 0$  limit which leads to  (\ref{eq_UG18})
the  difference should be negligible.  This is true if the linear approximation
is valid for $\eps\le \eps_*$.
Replacing the additive noise $\xibf_n$  with the multiplicative noise  $\Mop(\xbf_{n-1})\xibf_n$
is equivalent to replace a noise with unit covariance matrix  with a noise whose
covariance matrix is $M^2(\xbf_{n-1})$. As a consequence 
if the map in defined on a compacte set like torus  $Toro^d([0,1])$ then the entries
of the covariance matrix  belong to  $[0,1]$ rather than being equal to 1.
\\
\\
To conclude a strict comparison  with REM  can  be performed  by computing RE with a unique realization
of the  a multiplicative noise with amplitude $\eps_*$.
The smoothing of the fluctuations occurring when a single realizaztion of the noise is considered
can  be achieved  with a running average or a phase space average on a small
ball and the same procedure can be used to smooth the fluctuations of REM.
The Kolmogorov-Smirnov test might be applied to examine the statistical  properties of the round-off.
If our goal is not to find a correspondence  with REM, the computation of RE with an additive noise s
should be performed.  The definition we have chosen for $RE$ is for an additive noise since
in this case we have a  very simple realtion  beteen  RE and RE in the case of Hamiltonian
systems.
\spazio
{\bf Hamiltonian systems}
\spa
If a system is Hamiltonian the fundamental  matrix  of the tagent flow  $\Lop(\xbf,t)$
is symplectic. The  Lyapunov matrix $ \Lop^T\Lop$   and
the covariance matrix $\Lop\,\Lop^T$ are also symplectic. In this case the  BF Reversibility error
matrix can be written as
\begin{equation}
 \begin{split} 
  \Sigma^2_{BF}(\xbf,t)=  \Jop  \int_0^t\,\Lop^T(\xbf,t')\,\Lop(\xbf,t')\,dt' \,\Jop^{-1}
 \end{split}
\label{eq_UG21}
\end{equation}  
This similarity transformation implies that the  the matrix defined by the  integral in $[0,t]$
of  $\Lop^T\Lop$  has  the same  invariants  as $\Sigma^2_{BF}$. The  BF Reversibility error can be written as 
\begin{equation}
 \begin{split} 
  E_{BF}(\xbf,t) & =   \parton{\int_0^t\,E_L^2(\xbf,t')\,dt' \,}^{1/2} = \\
   & = t^{1/2}\,\, \Bigl \langle E^2_L(\xbf,t)\Bigr \rangle ^{1/2}
 \end{split}
\label{eq_UG22}
\end{equation}  
where $\mean{f(t)}= t^{-1}\,\int_0^t \,f(t')\,dt' $ denotes here the time average.   
For any initial point in a integrable region $E_{BF} $ has a linear growth
$(2+\alpha^2 t^2)^{1/2} \,A(t)$ where $A=1$ if the coordinates are normal
and  and $A(t)$ is periodic function for generic coordinates. Then 
$E_{BF}(t)=(2t+\alpha^2\,t^3/3)^{1/2}\,  B(t)$ where  $B=1$  for normal coordinates
and reaches a constant value with damped oscillations of amplitude $1/t$ for generic coordinates.  T
The   time average of the logarithmic derivative $d \log E_L^2(t)/d\log t$  oscillates,
the double time average known as MEGNO reaches a constant value  with damped oscillations.
The  time average of the logarithmic derivative   $d \log E_{BF}^2(t)/d\log t$
reaches a constant value with damped oscillations of amplitude $t^{-1}$.
As a consequence unlike $E_L(t)$ the reversibility error $E_{BF}(t)$
does not need to be averaged   to filter the oscillations.   The   logarithmic derivative
of $E^2_{BF}(t)$  oscillates   but the oscillations are damped  after a single time average.

{\bf Asymptotic behaviour of $E_L$ and $E_{BF}$}
\spazio
In a integrable region e have   $E_L^2(t)= (1+\alpha^2 t^2)\, A(t)$ where  $A(t)=e^{f(t)}$
where $A(t)$  is a  periodic function  of  period $T=2\pi/\omega$. For simplicity 
supposing that $A(t)= 1+ \eta  \cos(\omega t)$  we obtain 
\begin{equation}
 \begin{split} 
   E_{BF}^2(t)  =  \alpha^2 {t^3\over 3 }   \parton{ 1 + 3 \eta \,{\sin \omega t \over \omega t}
   + O\parton{1\over t^2}} 
 \end{split}
\label{eq_UG23}
\end{equation}  
the same result is obtained if $A(t)$ has a Fourier expansion with a time average 1  in $[0,T]$.
\\
\\
We recall that MEGNO is defined as the double time average of $d\log E_L^2/d\log t$.
The first time average is 2 times plus an oscillating term. The second time average
converges to 2 since   the oscillations are asymptotically damped with an amplitude
decaying as $1/t$.
Choosing $E_L^2(t)= (1+\alpha^2 t^2)\, A(t)$ with $A(t)= e^{\eta \,\cos(\omega t)}$ we have
\begin{equation}
 \begin{split} 
   \hskip -.3 truecm  \parmean{ d \log E^2_L(t)\over d\log t} =
   2- 2 {\hbox{atan}(\alpha t/\sqrt{2})\over 
\alpha t/\sqrt{2} }  + \eta \cos(\omega t)  - \eta {\sin(\omega t)\over \omega t}
 \end{split}
\label{eq_UG24}
\end{equation} 
After the second time mean the oscillations are reduced and the limit 2 is asymptotically reached
\begin{equation}
 \begin{split} 
   & \parmean{\parmean{ d \log E^2_L(t)\over d\log t}} = 2 + \eta  \,{\sin(\omega t) \over \omega t } 
  + O\parton{\log t\over  t} 
 \end{split}
\label{eq_UG25}
\end{equation}  
\\
\\
If we  compute $d\log E^2_{BF}(t)/d\log t$, we find it is equal to 3  with a remainder
which oscillates indefinitely.  The first time  average has a finite limit for $t\to +\infty$

\begin{equation}
 \begin{split} 
   & \parmean{ d \log E^2_{BF}(t)\over d\log t} = 3 + 3\eta  \,{\sin(\omega t) \over \omega t } +
  + O\parton{\log ( t)\over  t }
 \end{split}
\label{eq_UG26}
\end{equation}  
A similar result is obtained if we have $E_L(t)= t^2 e^{\eta\,\cos(\omega t)}$.
We have taken into account that 
\begin{equation}
 \begin{split} 
   {1\over t}\,\parbar{ \int_0^t {\sin (\omega t') \over \omega t'} \, dt'} \le 1+
   {1\over \omega t}\,\,\int_1^{\omega t} \, {d\tau\over \tau} = 1 +
   {\log (\omega t) \over \omega t}
 \end{split}
\label{eq_UG27}
\end{equation}  
To conclude if $E_L^2(t)= t^2 \,A(t)$  where $A(t)$ oscillates indefinitely then $E_{BF}^2(t)=t^3 B(t)$ where
$B(t)$ reaches a constant limit with damped oscillations.  The time average of the the first
logarithmic derivative of $E_L^2(t)$  oscillates whereas the double time average reaches a constant value
with damped oscillations. The  time average of the logarithmic derivative of $E_{BF}^2(t)$ reaches
a constant value with damped oscillations.
\\
\\
To conclude we notice that when  MEGNO is defined \cite{Cincotta2000, Maffione2011b}
it is assumed assume that $E_L^2(t)= (1+\alpha t)^2\, e^{f(t)} $ for regular orbits 
where $f(t)$ is periodic with zero mean.  Assuming $f(t)= \eta \cos(\omega t)$  we obtain
\begin{equation}
 \begin{split} 
    \parmean{ d \log E^2_L(t)\over d\log t} =
   2-  2  {\log (1+\eta t) \over \eta t}  + + \eta \cos(\omega t)  - \eta {\sin(\omega t)\over \omega t}
 \end{split}
\label{eq_UG28}
\end{equation} 
to be compared with (\ref{eq_UG24}). For a generic periodic function $f(t)$ with zero mean
\begin{equation}
 \begin{split} 
    \parmean{ d \log E^2_L(t)\over d\log t} =
   2-  2  {\log (1+\eta t) \over \eta t}  + f(t) - {1\over t}\,\int_0^t f(t') dt'
 \end{split}
\label{eq_UG29}
\end{equation} 
This result shows that  a second time  average is needed to damp the oscillations.
Indeed  $ft)$ has  zero  average and  the last term average vanishes
as $ t^{-1}\log t$.

\bibliography{Fast_invariant_indicators}{}
\bibliographystyle{plain}

%
\section{Appendix I  Numerical procedures}
\def\diag{\hbox{diag}}
\def\Vol{\hbox{Vol}}
The numerical computation of the invariants of the Lyapunov matrix $\Lop_n^T\Lop_n$
using the Faddeev-Leverrier  recurrence  faces the same numerical difficulties as
the computation of its eigenvalues. For a  positive  matrix the Jacobi method is
very efficient being limited only by machine accuracy.
\\
We consider  a map $M$ in $\Reali^d$  since any nonlinear   flow  is
computed  numerically  by using an integrating map with a suitable time step.
 Letting $\Lop_n(\xbf)= DM^{\circ \,n}(\xbf)$ be the tangent map obtained from the recurrence 
 $\Lop_n(\xbf)= DM(\xbf_{n-1}) \,\Lop_{n-1}(\xbf)$ with $\Lop_0=\Iop$,   the Lyapunov
 and BF reversibility error  covariance matrices $\Sigma^2_L$ and $\Sigma^2_{BF}$
 are given by
\begin{equation}
\begin{split}
 &  \Sigma^2_{L\,\,n}  = \Lop_n\Lop_n^T \\
 &  \Sigma_{BF\,\,n}^2 = {1\over 2}\Iop + {1\over 2}(\Lop_n^T\Lop_n)^{-1} + \,\sum_{n'=1}^{n-1} \,
  \bigl(\Lop_{n'}^T \Lop_{n'}\,\bigr)^{-1}
\label{eq_AI_1}
\end{split}
\end{equation}
where we have omitted to write explicitly the dependence on the initial condition $\xbf$.
For the reversibility error covariance matrix if  the noise is present only in the F process,
then  in the r.h.s. of (\ref{eq_AI_1}) the first 2 terms are absent and the sum runs form 1 do $n$.
The matrix  $\Lop_n\,\Lop_n^T$  and the Lyapunov matrix  $\Lop_n\Lop_n^T$  have  the same
invariants.  The real matrix $\Lop_n$  and the symmetric matrices $\Lop_n^T\Lop_n$,
$\Lop_n\Lop_n^T$ can be written as 
\begin{equation}
\begin{split}
  & \Lop_n = \Rop_n \,\Dop_n \Wop^T_n  \qquad \qquad \Dop_n= e^{n\,\Lambda(n)} \\ \\
  & \Lop_n^T\Lop_n= \Wop_n\,\Dop_n^2\,\Wop_n^T \qquad \quad  
   \Lop_n\Lop_n^T= \Rop_n, \,\Dop_n^2\,\Rop_n^T 
\label{eq_AI_2}
\end{split}
\end{equation}
were $\Rop_n$ and $\Wop_n$ are orthogonal and $\Lambda(n)= \diag(\lambda_1(n), \ldots, \lambda_d(n)\,) $
is a  diagonal matrix whose entries for any $n$  are ordered in a non increasing sequence 
$\lambda_1(n)\ge\lambda_2(n) \ge \ldots\ge \lambda_d(n)$. 
In the more general case of a rectangular matrix the expression written above for $\Lop_n$
is known s singular value decomposition.   The columns of  matrix $\Lop_n$ are
the vectors $\ebf_i(n)= \Lop(n)\ebf_i$ where $\ebf_i$ are the  orthonormal base vectors
defined by  $(\ebf_i)_j=\delta_{ij}$.  As a consequence taking (\ref{eq_AI_2}) into account
and letting $\Wop_n=(\wbf_1(n),\ldots,\wbf_d(n))$, where $\wbf_i(n)$ are aorthonormal
vectors can write  
\begin{equation}
\begin{split}
  \Lop_n &= (\ebf_1(n),\ldots,\ebf_d(n))  \\ \\ 
   \ebf_i(n)&= \Rop_n \,\,\begin{pmatrix} e^{n\,\lambda_1(n)} \,\, \ebf_i\cdot \wbf_1(t) \cr   \vdots \cr
         e^{n\,\lambda_d(t)} \,\, \ebf_i\cdot \wbf_d(n)      \end{pmatrix}
\label{eq_AI_3}
\end{split}
\end{equation} 
Let $\mu_j(n)=e^{2n\,\lambda_j(n)}$ be the eigenvalues of the Lyapunov matrix, 
where  $\lambda_j(n)\to \lambda_j$ for $n\to \infty$. The Jacobi method can be used to 
compute the eigenvalues and  the eigenvectors os the Lyapunov matrix..
Supposing that $\lambda_1>0$ the computation of the largest eigenvalue is limited
by the overflow condition. The computation of the second eigenvalue $\mu_2(n)$  is limited by
machine accuracy $\eps_*$. Indeed as the ratio $\mu_2(n)/\mu_1(n)$ decreases the accuracy
of $\mu_2(n)$ decreases  too  and  the last significan digit  is lost  for
$n\ge n_*=  (\log \eps_*)/(2(\lambda_1-\lambda_2))$.
iterations.
For the typical value $\eps_*=10^{-16}$ we have   $n_* \sim 18.4/(\lambda_1-\lambda_2)$.
The  condition to obtain  the  next eigenvalues is obvious. 
The accuracy for an eigenvector is comparable with the accuracy of the corresponding eigenvalue.
\\ \\
The orthogonal matrix $\Rop_n$ is obtained by diagonalizing $\Lop_n\Lop_n^T$.
It may not have a limit as $n\to \infty$. An example was provided 
for a nonlinear   integrable system written in normal coordinates, 
see section V (\ref{eq_5_18}) where the corresponding matrix $\Rop_L(t)$
was explicitly computed.
\\
When the  eigenvalues of the Lyapunov matrix  follow  a power law, as for Hamiltonian
systems in the integrable regions, the loss of accuracy is much slower.
\\ \\
The cancellations of the leading terms when the Faddeev-Leverrier  formula  is used to compute
the invariants $I^{(j)}_n$ with $n\ge 2$   occurs 
also when  Lyapunov matrix is in diagonal.
Indeed supposing $\Lop^T_n\Lop_n= \diag(\mu_1(n),\,\mu_2(n))$.
we have for $d=2$  
\begin{equation}
\begin{split}
  I^{(2)}_n= {1\over 2} \,\Bigl( \bigl(\mu_1(n)+\mu_2(n)\bigr )^2 -\mu^2_1(n)-\mu^2_2(n)  \Bigr) =
  \mu_1(n)\mu_2(n)
\label{eq_AI_4}
\end{split}
\end{equation}
\\
The result is due to the cancellaton   of the  leading  $\mu_1^2(n)$.
 For a generic map in $\Reali^2$ 
 the Lyapunov matrix is the Grahm matrix $\Gop(n)$ of the vectors  $\ebf_1(n)$ and $\ebf_2(n)$
 where $\Gop_{i\,j}(n)=\ebf_i(n)\cdot \ebf_j(n)$.
The Faddev-Leverrier formula for the second invariant is given by  
\begin{equation}
\begin{split}
  & I^{(2)}_n  = {1\over 2}\Bigl(\,  \Tr^2(\Gop(n))-\Tr(\Gop^2(n))\Bigr) =  \det (\Gop(n)) = \\ \\
  & =  (\ebf_1(n)\cdot\ebf_1(n))\, (\ebf_2(n)\cdot\ebf_2(n))\,-\,(\ebf_1(n)\cdot\ebf_2(n))^2 = \\ \\
  & =e^{2n(\lambda_1(n)+\lambda_2(n))} \,\det \begin{pmatrix} \wbf_1(n) \cdot \ebf_1 & \wbf_2(n)\cdot \ebf_1 \\ \\
  \wbf_1(n) \cdot \ebf_2 & \wbf_2(n)\ebf_2 \end{pmatrix} 
\label{eq_AI_5}
\end{split}
\end{equation}
Notice that  the last tem is equal to $\det(\Wop(n))=1$  because  $\Wop(n) =(\wbf_1(n),\wbf_2(n))$
is an orthogonal matrix.
The exact result $e^{2n(\lambda_1(n)+\lambda_2(n))}$ is due
to the cancellation of the leading terms $e^{4n\,\lambda_1(n)}$.
In this case the ratio betwen the exact result and the leading terms which cancel is
$e^{2n(\lambda_1(n)+\lambda_2(n))}/e^{4n\lambda_1(n)}= e^{-2n(\lambda_1(n)-\lambda_2(n))}$
and therefore for a given numerical accuracy  the value of $n$ for which the
last significant digit of the result  is lost is the same as for Jacobi method.
\\
\\
We consider now the second invariant for   the Laypunov matrix in the case of
a map in $\Reali^d$. We recall that $\Lop_n^T\Lop_n=\Gop(n)$
where $\Gop(n)$ denotes the Grahm matrix of the vectors $\ebf_1(n),\ldots,\ebf_d(n)$
In this case the second invariant computed with  the Faddeev-Leverrier formula  is given by 
\begin{equation}
\begin{split}
 &  I^{(2)}(n)= {1\over 2} \Bigl(  \Tr^2\bigl(\Gop(n)\bigr)-\Tr\bigl(\Gop^2(n)\bigr)^2\Bigr) = \\ 
&\; = {1\over 2}\Bigl( \sum_{i=1}^d\,\sum_{j=1}^d \bigl( G_{ii}(n)\,G_{jj}(n)-G_{ij}^2(n)\,\bigr) \Bigr)\equiv\\ 
  & \hskip -.25 truecm  \equiv  {1\over 2}\Bigl( \sum_{i=1}^d\,\sum_{j=1}^d \bigl ( \ebf_i(n)\cdot\ebf_i(n)\,
  \ebf_j(n)\cdot\ebf_j(n) -(\ebf_i(n)\cdot\ebf_j(n))^2 \Bigr ) =\\ 
  &\;  = \sum_{1\le i<j}^d  \bigl ( \ebf_i(n)\cdot\ebf_i(n)\,
  \ebf_j(n)\cdot\ebf_j(n) -(\ebf_i(n)\cdot\ebf_j(n))^2 \Bigr ) \equiv \\ 
  &\;  \equiv \sum_{1\le i<j}^d \,\det\bigl( \Gop^{(i,j)}(n) \bigr) = \sum_{1\le i<j}^d \,\Vol^2\bigl(\Pcal^{(i j)}(n)\bigr)
\label{eq_AI_6}
\end{split}
\end{equation}
where we denoted with $\Gop^{(i,j)}(n)$ the $2\times 2$ Grahm matrix of the vectors
$\ebf_i(n)$ and $\ebf_j(n)$ and $\Pcal^{(ij)}(n)$ is the parallelogram whose sides are
$\ebf_i(n)$ and $\ebf_j(n)$.
It is evident that the Faddeev Leverrier formula given by the first
three  lines in (\ref{eq_AI_6})  involves the calcellation of the leading terms
and cannoto be used in numerical computations.  In the last two lines the
cancellation has been carried out and these expressions are suitable for
numerical computations. As a consequence the second invariant must be computed
as the sum of the squares of the volumes $\Pcal^{(i j)}(n)$ which are given by the
determinants of the $2\times 2$ Grahm matrices $\Gop^{(i,j)}$, which do not
involve cancellations.
\\
If the Faddeev-Leverrier recurrence is used to compute the higher order invariants $I^{(k)}(n)$ for
$k\ge 3$  the cancellation of the leading terms occurs as well. As a consequence this procedure
cannot be used for numerical computations. However  just as for $k=2$ the  invariant of order
$k\ge 3$ is given by the sum of the squares of all the volumes $\Pcal^{(j_1\,j_2\,\ldots\, j_k)}(n)$,
which are equal to  the determinants of the  Grahm matrices $\Gop^{(j_1\,j_2\,\ldots\, j_k)}(n)$,
whese computation involves no cancellations.
This can be proved using esternal algebra. Indeed   $\Vol(\Pcal^{(j_1\ldots j_k)}(n))$ is the norm of
the wedge product $\ebf_{j_1}(n)\wedge\ldots\wedge\ebf_{j_k}(n)$ and that the $k$-th invariant
of $\Gop(n)$ is the sum  of the squares of the volumes of the ${d\choose k}$ parallelotpes
$\Pcal^{(j_1\,j_2\,\ldots\, j_k)}(n)$ whose sides are $(\ebf_{j_1}(n),\ebf_{j_1}(n),\ldots,\ebf_{j_k}(n))$
\begin{equation}
\begin{split}
  I^{(k)}(n) & = \sum_{1\le j_1<j_2<\ldots<j_k}^d \Vol^2\bigl(\Pcal^{(j_1\,j_2\,\ldots\, j_k)}(n) \bigr)\equiv \\ 
  & =\sum_{1\le j_1<j_2<\ldots<j_k}^d \, \det\,\bigl(\,\Gop^{(j_1\,j_2\,\ldots\, j_k)}(n) \,\bigr)
\label{eq_AI_7}
\end{split}
\end{equation}
This is the expression obtained from the Faddeev-Leverrier recurrence once
the cancellations  of the leading terms have been carried out and it is suitable
for numerical  computations.
The direct use of the Faddeev-Leverrier recurrence  must be avoided in numerical
computations since the cancellations rise the same accuracy problems as
the direct computation of the eigenvalues of the Lyapunov matrix $\Lop^T(n)\Lop(n)$
using the Jacobi method.
\\
The   computation of the determinant of $\Gop^{(j_1\,j_2\,\ldots\, j_k)}(n)$
may face overflow (underflow)  problems when $n$ grows.  The problem  is  be solved with the QR
decomposition and a renormalization procedure.
\subsection{The QR decomposition}
%
%
%
The volume of parallelotope $\Pcal^{(1,2,\ldots,k)}(n)$   is provided by
the $QR$ decomposition of  the rectangular $d\times k$ matrix $(\ebf_1(n),\ebf_2(n),\ldots,\ebf_k(n))$
which corresponds to the  Grahm-Schmidt ortonormalization of the column vectors of this matrix .
The matrix  can be written as 
$ \Qop^{(k)}(n)\,\Top^{(k)}(n)$ where $\Qop^{(k)}(n)=(\epsilonbf_1(n),\epsilonbf_2(n),\ldots,\epsilonbf_k(n))$
is a rectangular $d\times k$  matrix whose $k$ columns  are orthonormal vectors $\epsilonbf_j(n)$
and $\Top_n^{(k)}$ an upper triangular $k\times k$ matrix.
The explicit form is 
\def\epsbf{\epsilonbf}
\begin{equation}
\begin{split}
  & \bigl(\ebf_1(n),\ldots,\ebf_k(n)\,\bigr)=   \bigl(\epsbf_1(n),\epsbf_2(n), \ldots,\epsbf_k(n)\,\bigr)\,
  \times  \\ \\
  &  \times  \begin{pmatrix} \epsbf_1(n)\cdot \ebf_1(n) & \epsbf_1(n)\cdot \ebf_2(n) &
     \ldots  &  \epsbf_1(n)\cdot \ebf_k(n)\\ \\
    0  & \epsbf_2(n)\cdot \ebf_2(n) & \ldots & \epsbf_2(n)\cdot \ebf_k(n)  \\
    \vdots   &   \vdots & \ldots & \vdots \\
    0 & 0 &  \ldots &  \epsbf_k(n)\cdot \ebf_k(n) \end{pmatrix}
\label{eq_AI_8}
\end{split}
\end{equation}
where the orthonormal vectors $\epsbf_k(n)$ are determined by the
recurrence $\ubf_j(n)= \ebf_j(n)-\sum_{\ell=1}^{j-1}\,  \epsbf_\ell(n) \,\,\epsbf_\ell(n) \cdot \ebf_j(n)$
and  $\epsbf_j(n)= \ubf_j(n)/\Vert \ubf_j(n)\Vert$.  After computing  all the orthonormal
vectors $\epsbf_j(n)$ we notice that we can write
$\ubf_j(n)= \epsbf_j(n)\,\,\, \epsbf_j(n)\cdot  \ebf_j(n)$ so that finally
\begin{equation}
\begin{split}
  \ebf_j(n) = \sum_{\ell=1}^j \, \epsbf_\ell(n)\,\,\,\epsbf_\ell(n)\cdot \ebf_j(n)   \qquad j=1,\ldots,k
\label{eq_AI_9}
\end{split}
\end{equation}
The volume of the $\Pcal^{(1,2,\ldots,k)}$ is given the product of the projections of the vectors
$\ebf_j(n)$ on the correspnding orthonormal  vectors $\epsilonbf_j(n)$   
\begin{equation}
\begin{split}
  \Vol\bigl(\Pcal^{(1,2,\ldots,k)} \bigr)= \prod_{j=1}^k \epsbf_j(n)\cdot\,\ebf_j(n) = \det\bigl(\,\Top^{(k)}(n)\,\bigr)
\label{eq_AI_10}
\end{split}
\end{equation}  
We can easily check the result just obtained by observing that $\bigl(\Top^{(k)}\,
\bigr)^T\Top^{(k)}$ is equal to the Grahm matrix $\Gop^{(k)}(n)$ of the vectors
$(\ebf_1(n),\ldots,\ebf_k(n))$. As a consequence
since the determinant of the Grahm matrix $\Gop^{(k)}(n)$  is the squared volume of $\Pcal^{(1,2,\ldots,k)}(n)$
we have 
\begin{equation}
\begin{split}
  & \Vol^2(\Pcal^{(1,2,\ldots,k)}(n)) = \det(\Gop^{(k)}(n))= \\ \\
  & \quad =\det\Bigl( \bigl(\Top^{(k)}(n)\,
\bigr)^T\Top^{(k)}(n)\,\Bigr)= \det^2\bigl(\Top^{(k)}(n)\,\bigr)
\label{eq_AI_11}
\end{split}
\end{equation} 
 We finally remark that the QR decomposition to compute $\Vol\bigl(\Pcal^{(j_1\,j_2\,\ldots\, j_k)}(n)\bigr)$
 for any choice of  $k$ indices involves  $d(k-1)k/2$ sums and multiplications plus $k$ 
 multiplications and normalizations  of a vector whereas   the Gauss method to computes the determinant of the $k\times k$  Grahm matrix and involves  $d(k-1)k/2$ sums and multiplications plus $k$ multiplications. As a consequence, the computational complexity is comparable.
 To compute the invariant $I^{(k)}(n)$ these operations must be repeated ${d\choose k}$ times.
\subsection{Avoiding overflows and underflows}
For $n$ large the volumes can increase or decrease beyond  the machine limits
causing overflows or underflows. To this end we consider a sequence of QR transformation
one for  each iteration of the map (this might be replaced with  the iterations of $M^{\circ m}$
 but for simplicity, we   consider just $m=1$). We first recall that
\begin{equation}
\begin{split}
 & \ebf_j(n)=  DM(\xbf_{n-1})\,\ebf_j(n-1)  \\
 & \Lop_{n}= DM(\xbf_{n-1})\Lop_{n-1}  
\label{eq_AI_12}
\end{split}
\end{equation}
Consider then  the following sequence of QR transformations applied to the first $k$ vectors
$(\ebf_1,\ldots,\ebf_k)$ which are evolved by applying subsequently $DM(\xbf), DM(\xbf_1), \ldots $
\begin{equation}
\begin{split}
  & (\ebf_1(1),\ldots,\ebf_k(1)) =  \Bigl(\epsbf_1(1),\epsbf_2(1), \ldots,\epsbf_k(1)\Bigr)\, \Top^{(k)}(1;0) \\ \\ 
  & \hskip-.5 truecm  \Top^{(k)}(1;0)  =  \\ 
  & = \begin{pmatrix} \epsbf_1(1)\cdot \ebf_1(1) & \epsbf_1(1)\cdot \ebf_2(1) &
     \ldots  &  \epsbf_1(1)\cdot \ebf_k(1)\\ \\
    0  & \epsbf_2(1)\cdot \ebf_2(1) & \ldots & \epsbf_2(1)\cdot \ebf_k(1)  \\
    \vdots   &   \vdots & \ldots & \vdots \\
    0 & 0 &  \ldots &  \epsbf_k(1)\cdot \ebf_k(1) \end{pmatrix}
\label{eq_AI_13}
\end{split}
\end{equation}
To obtain  $(\ebf_1(2),\ldots,\ebf_k(2))$, we apply  $DM(\xbf_1)$ to $(\ebf_1(1),\ldots,\ebf_k(1))$
and denoting $(\epsbf_1(2),\ldots,\epsbf_k(2))$ the ortho-normalized vectors obtained from the set 
$DM(\xbf_1)(\epsbf_1(1),\ldots,\epsbf_k(1))$ rather than from  the set $(\ebf_1(2),\ldots,\ebf_k(2))$)
we can write
\begin{equation}
\begin{split}
  &(\ebf_1(2),\ldots,\ebf_k(2)) =  DM(\xbf_1) \Bigl(\epsbf_1(1),\epsbf_2(1), \ldots,\epsbf_d(1)\Bigr)\,\times \\ 
          &\qquad  \times \Top^{(k)}(1;0)=  (\epsbf_1(2),\ldots,\epsbf_k(2)) \, \Top^{(k)}(2;1) \, \Top^{(k)}(1;0)\\ \\
  &   \Top^{(k)}(2;1)  = \\
  & \begin{pmatrix} \epsbf_1(2)\cdot \etabf_1(2) & \epsbf_1(2)\cdot \etabf_2(2) &
     \ldots  &  \epsbf_1(2)\cdot \etabf_k(2)\\ \\
    0  & \epsbf_2(2)\cdot \etabf_2(2) & \ldots & \epsbf_2(2)\cdot \etabf_k(2)  \\
    \vdots   &   \vdots & \ldots & \vdots \\
    0 & 0 &  \ldots &  \epsbf_k(2)\cdot  \etabf_k(2) \end{pmatrix}  \\ \\
   & \qquad  \qquad \etabf_j(2)= DM(\xbf_1)\,\epsbf_j(1) \qquad j=1,\ldots,k 
\label{eq_AI_14}
\end{split}
\end{equation}
Iterating the process we find that
\begin{equation}
\begin{split}
  &(\ebf_1(n),\ldots,\ebf_k(n)) = \Bigl(\epsbf_1(n),\epsbf_2(n), \ldots,\epsbf_k(n)\Bigr)\, \Top^{(k)}(n)\\ \\ 
  & \Top^{(k)}(n)= \Top^{(k)}(n,n-1)\,  \Top^{(k)}(n-1,n-2)\cdots \Top^{(k)}(1,0) \\ \\ 
  & \hskip -.4 truecm   \Top^{(k)}(n;n-1)  =
  \begin{pmatrix} \epsbf_1(n)\cdot \etabf_1(n) & \epsbf_1(n)\cdot \etabf_2(n) &
     \ldots  &  \epsbf_1(n)\cdot \etabf_k(n)\\ \\
    0  & \epsbf_2(n)\cdot \etabf_2(n) & \ldots & \epsbf_2(n)\cdot \etabf_k(n)  \\
    \vdots   &   \vdots & \ldots & \vdots \\
    0 & 0 &  \ldots &  \epsbf_k(n)\cdot  \etabf_k(n) \end{pmatrix}  \\ \\
   & \qquad  \qquad \etabf_j(n)= DM(\xbf_{n-1})\,\epsbf_j(n-1) \qquad j=1,\ldots,k 
\label{eq_AI_15}
\end{split}
\end{equation}
Finally, the  volume of the parallelotope $\Pcal^{(1\ldots k)}(n)$ is given by the product
of the $k$ diagonal elements
of the matrices $\Top^{(k)}(m,m-1)$ for $m=1,2,\ldots,n$. The log of the volume  divided by $n$ is given by
\begin{equation}
\begin{split}
 & {1\over n}\,\log \Vol\bigl(\Pcal^{(1,2,\ldots,n)} \bigr)=   {1\over n}\sum_{n'=1}^n \,\,\sum_{j=1}^d \, \log\Bigl( \eps_j(n')
  \cdot \etabf_j(n')\Bigr ) \\ \\
  & \qquad \qquad  \etabf_j(n')= DM(\xbf_{n'-1}) \eps_j(n'-1) \qquad \eps_j(0)\equiv \ebf_j
\label{eq_AI_16}
\end{split}
\end{equation}
The invariant $I^{(k)}(n)$  i given by the sum over the squares of the $(d\choose k)$
parallelotopes $\Pcal^{(j_1,j_2,\ldots,j_k)}$ which are obtained from the previous algorithm
by replacing the initial  set of unit vectors $(\ebf_1,\ldots,\ebf_k)$ with
$(\ebf_{j_1},\ldots,\ebf_{j_k})$. Denoting with 
$\Pcal^{(m_1,\ldots,m_k)}(n)$ the parallelotope  with the largest volume   we can write the $\log$  the
of the invariant $k$ as 
\begin{equation}
\begin{split}
  & \log I^{(k)}(n)   = \log\parton{ \sum_{1\le j_1<\ldots<j_n} \Vol^2(\Pcal^{(j_1,\ldots,j_k)}(n)) } = \\ \\ 
  & \qquad =  \;\log \Bigl(\Vol^2(\Pcal^{(m_1,\ldots,m_k)}(n))\Bigr) \;+  \\ \\
  & \qquad \quad + \log \biggl( 1 +  \sum_{ \begin{matrix}
      \scriptstyle    1\le j_1<\ldots<j_k \\ \scriptstyle  j_i\not= m_i \end{matrix} }  
   {\Vol^2\bigl( \Pcal^{(j_1,\ldots,j_k)}(n)\bigr) \over \Vol^2\bigl( \Pcal^{(m_1,\ldots,m_k)}(n)\bigr)} \biggr)
\label{eq_AI_17}
\end{split}
\end{equation}
The second term   ranges between zero and $\log\Bigl( 1+ {d\choose k} \Bigr)$  and asymptotically
can be negelcted with respect to the first one.
Notice that the indices $m_1,\ldots,m_k$ can vary with $n$  even thought we did
not indicate explicitly this possible dependence. 
\subsection{Reversibility error invariants}
The  BF  Reversibility covariance matrix is the sum of $(\Lop^T_{n'}\Lop^T_{n'})^{-1}$
from 1 to $n$ and the computation of the inverses  faces problems when the matrix becomes ill conditioned.
For Hamiltonian systems  the matrix obtained by replacing $(\Lop^T_{n'}\Lop^T_{n'})^{-1}$ with 
$\Lop^T_{n'}\Lop^T_{n'}$ in the sum, has the same invariants.
The computation of the invariants must be carried out using the algorithms which avoid  the
cancellation of the leading terms. For instance the second invariant of the matrix
must be computed according to
\begin{equation}
\begin{split}
  I_{R}^{(2)}(n)= \sum_{1\le i<j}^d \Bigl( (\Sigma^2_{R}(n))_{ii}(\Sigma^2_{R}(n))_{jj}
  -(\Sigma^2_{R}(n))^2_{ij}\, \Bigr)
\label{eq_AI_18}
\end{split}
\end{equation}
For the invariant $k$ the expression to be used is  \textcolor{red}{\bf to be checked}
\begin{equation}
\begin{split}
  I_{R}^{(k)}(n)= \sum_{1\le j_1 < j_2<\ldots<j_k}^d  \,\,\det (\Sop^{(j_1,j_1,\ldots,j_k)}(n)) \\ \\
  \Bigl(\Sop^{(j_1,j_1,\ldots,j_k)}(n) \Bigr)_{i\,\ell} =(\Sigma^2_{R}(n))_{j_i,\,j_\ell}(n)
\label{eq_AI_19}
\end{split}
\end{equation}
which is tha analogue of the sum of the squared volumes of the ${d\choose k}$ parallelotopes
for the Lyapunov invariants
see (\ref{eq_AI_7}).   Since $\Sop^{(j_1,j_2,\ldots,j_k)}(n)$ are not Grahm matrices
but rather sums of Grahm matrices the QR decompositin method cannot be used.




%

\section{ Appendix II.  Lyapunov error and  linear approximation}
\spa
%
%
The   linear response   allows to  determine the  sensitivity  of a dynamical
system  to   deviations  of amplitude $\eps$   from the  initial condition 
 in the limit $\eps\to 0$. This limit allows
to  obtain a response at  any time and to investigate  its 
asymptotic  behaviour when $t\to \infty$.  If we consider the linear
approximation for a finite amplitude $\eps$  its validity limits
can be established  if we are able to   bound the remainder
determining the conditions under wich it  can be neglected with respect
to the  linear term.
%
%
%
%
%
%
%
\def\epsbar{\overline{\eps}}
We consider first an initial deviation 
$\eps \,\etabf$  with respect to the initial condition with
$\Vert\etabf\Vert=1$. The solution at time $t$ can be written as 
\begin{equation}
  \begin{split}
   &  \ybf(t)= S_t(\xbf+\eps\etabf)= S_t(\xbf) + \eps \,\Lop(\xbf,t)\,\etabf  + \eps^2    \Rbf(\xbf,t;\eps\etabf)   \\ \\
    &   \Rbf(\xbf,t;\eps\etabf)= {1\over 2}\,{\partial^2\over \partial x_i\partial x_j}
    \,S_t(\xbf+ \epsbar\,\etabf)\,\eta_i\,\eta_j
    \qquad 0<|\epsbar|<|\eps|
  \end{split}
 \label{eq_AII_1}
 \end{equation}
The linear approximation is valid  if the ratio $r$ between the norm of the 
remainder and the norm of linear term is small  $r\ll 1$
\begin{equation}
  \begin{split}
   &  r(\xbf,\etabf,t;\eps)= {\eps^2\Vert \Rbf(\xbf,\etabf,t;\eps)\Vert\over |\eps| \,\Vert \Lop(\xbf,t)\,\etabf\Vert }
     \end{split}
 \label{eq_AII_2}
 \end{equation}
Since this ratio depends on $\eta$ one might take either the average with respect to $\etabf$
or the maximum of $\etabf$ on the unit sphere.
The ratio $r$  can be related to the ratio  $|\eps|/|\eps_c|$   where $\eps_c$
is the singularity  closest
to the origin in the complex $\eps$ plane.  Usually $S_t(\xbf+\eps\etabf)$  has movable singulatities
which depend on $\xbf,\etabf$ and $t$ and the search of the closest singularity  is not an easy task.
\\
\\
To this end we consider the one dimensional flows with    $x\in \Reali$ and  $\eta=1$. The solution
for real dispacements $\eps\in \Reali$ is extended to the complex $\eps$ plane still keeping $x,\,t$ real
and we denote  $\eps_c=\eps_c(x,t)$ the singularity closest to the origin.
\\
\\
Given and analytic function whose singulary closest to the origin is $z_c$
from the Cauchy theorem we get the 
first  order  Taylor expansion with an explicit expression for the  remainder
\def\epsbar{\overline{\eps}}
\begin{equation}
  \begin{split}
 f(z)& = {1\over 2 \pi\,i}\oint {f(\zeta)\over \zeta-z}\, d\zeta = f_0\,+\,z\, f_1\, \,+ z^2\, R(z) \\ \\
 R(z)& =  {1\over 2\pi\,i } \, \oint {f(\zeta)\over \zeta^2(\zeta-z)}\, d\zeta
  \end{split}
 \label{eq_AII_3}
 \end{equation}
The integral is on a circle $|\zeta|=\rho <|z_c|$ where $|f(\zeta)|\le M(\rho)$  and  $M(\rho)$ may 
diverge as $\rho$ approaches $|z_c|$.  We have used the identity
$1/(\zeta-z)=1/\zeta+z/\zeta+ z^2/(\zeta^2(\zeta-z))$.  Since $\zeta$
varies on a circle ad $z$ is in its interior
we have $|\zeta-z|\ge |\zeta|-|z|= \rho -|z|$ and we have the estimate on $R(z)$ setting $\zeta= \rho e^{i\theta}$
\def\epsbar{\overline{\eps}}
\begin{equation}
  \begin{split}
 |R(z)|  \le  {1\over 2\pi} \,\oint {|f(\zeta)|\over |\zeta|^2(|\zeta|-|z|)} \,|d\zeta| =
   {M(\rho)\over \rho(\rho-|z|)
}  \end{split}
 \label{eq_AII_4}
 \end{equation}
where $M(\rho)$ is the average of $|f(\rho\, e^{i\theta})|$  on the circle of radius $\rho$.
The absolute value of the ratio $r$ between the remainder $z^2R(z)$
and the first order term  $zf_1$  is bounded by
\def\epsbar{\overline{\eps}}
\begin{equation}
  \begin{split}
 |r(z)| = \parton {|z^2 R(z)|\over |z\,f_1|} \le {M(\rho)\over |f_1|\, \rho } \,\, {|z|\over \rho-|z|}
  \end{split}
 \label{eq_AII_5}
 \end{equation}
For $|f_1$ we have the following upper bound
\def\epsbar{\overline{\eps}}
\begin{equation}
  \begin{split}
 |f_1| \le {1\over 2\pi} \,\oint { |f(\zeta)|\over |\zeta^2|}\,|d\zeta| ={1\over \rho} \,{1\over 2\pi}\,\oint
 |f(\rho\,e^{i\theta})|\,d\theta = {M(\rho)\over \rho}
  \end{split}
 \label{eq_AII_6}
 \end{equation}
If we might replace $|f_1|$  with  this  upper bound in (\ref{eq_AII_5}) then we would obtain
$|r(z)| \le |z|/(\rho-|z|)$   for $\rho\le |z_c|$. 
\\
\\
We can provide an accurate estimate to  the remainder for one dimensional flows.
To this end we consider the logistic equation
$dx(t)/dt= x(t)(1-x(t))$ whose flow $x(t)\equiv S_t(x)$  explicitly reads 
\def\epsbar{\overline{\eps}}
\begin{equation}
  \begin{split}
  S_t(x) = {x\over x+(1-x)e^{-t}}
   \end{split}
 \label{eq_AII_7}
 \end{equation}
The singularity in $x$  for $t>0$   is a pole  on the negative real axis  at $x_c(t)$
\begin{equation}
  \begin{split}
  x_c(t)= -{e^{-t} \over 1-e^{-t}} 
   \end{split}
 \label{eq_AII_8}
 \end{equation}
The  partial derivatives of $S_t(x)$  with respect do $x$ are given by 
\def\epsbar{\overline{\eps}}
\begin{equation}
  \begin{split}
 &  \Lop(x,t)\equiv \derp{}{x}\,S_t(x) = {e^{-t}\over \bigl(x+(1-x)e^{-t}\bigr)^2} \\
 &  {\partial^2\over \partial x^2}\,S_t(x) = -2 \,\,{e^{-t}(1-e^{-t})\over  \bigl(x+(1-x)e^{-t} \bigr)^3}  \\ 
 &  {\partial^n\over \partial x^n}\,S_t(x) =(-1)^{n+1}\,n!\, {e^{-t}(1-e^{-t})^{n-1}\over
   \bigl(x+(1-x)e^{-t} \bigr)^{n+1}}
    \end{split}
 \label{eq_AII_9}
 \end{equation}
The remainder is easily obtained by summing a geometric series
\begin{equation}
  \begin{split}
  & \eps^2 R(x,t;\eps)= \sum_{n=2}^\infty {\eps^{n}\over  n!} {\partial^n\over \partial x^n}\,S_t(x) = \\ \\
  & = -\eps^2\,{e^{-t}\over \bigl(x+(1-x)e^{-t}\bigr)^2}\,\,{1-e^{-t}  \over x+(1-x)e^{-t} + \eps(1-e^{-t})}= \\ \\
  & = -\eps^2\,\,\Lop(x,t)\,\, {1-e^{-t}  \over x+(1-x)e^{-t} + \eps(1-e^{-t})}=-{\eps^2 \Lop \over  \eps-\eps_c }
    \end{split}
 \label{eq_AII_10}
 \end{equation}
where $\eps_c$ is the pole in in $\eps$ of $S_t(x+\eps)$ so that $\eps_c=x_c-x$.
As a consequence the ratio between the remainder and the first order term can be written as  
\begin{equation}
  \begin{split}
     {\eps^2 R\over \eps \Lop }  &   =  -{\eps\over \eps-\eps_c(x,t)} =
    -\eps {S_t(x+\eps)\over x+\eps}\,(1-e^{-t})  \\ \\
    \eps_c  &= -x +x_c = -x -{e^{-t}\over 1-e^{-t}}
    \end{split}
 \label{eq_AII_11}
 \end{equation}
If  $x$ is on the positive real axis  the solution  $S_t(x)$ is bounded for any $t>0$.
As a consequence we choose   $x+\eps>0$.
In order to obtain an estimate uniform in $t$ we move at some
distance $\ell\ll 1$ from the origin choosing  $x+eps \ge \ell$
and $|\eps|\ll \ell$.
We notice that for $x>1$ the solution decreases with $t$ so that  $S_t(x)\le x$
whereas for $0<x<1$ it  increases so that  $S_t(x)\le 1$.
The following estimate holds 
\def\dis{\displaystyle}
\begin{equation}
  \begin{split}
 {S_t(x+\eps)\over x+\eps}\le \begin{cases}
 {\dis1\over \dis x+\eps}\le {\dis 1\over \dis  \ell}  & \qquad  \ell\le x +\eps \le 1  \\ \\
  1  & \qquad x +\eps > 1
  \end{cases}
    \end{split}
 \label{eq_AII_12}
 \end{equation}
Choosing $|\eps|\ll \ell   \ll   1 $  from the the last expression of $r$
in (\ref{eq_AII_11}) and (\ref{eq_AII_12})  we obtain  the following  bound on  $|r|$ 
valid for any $t\ge 0$  
\def\dis{\displaystyle}
\begin{equation}
  \begin{split}
 r(x,t;\eps) \le \begin{cases}  {\dis |\eps|\over\dis  \ell}  & \quad \ell\le x+\eps < 1 \\  \\
 |\eps|    & \quad  x +\eps > 1
 \end{cases}
    \end{split}
 \label{eq_AII_13}
 \end{equation}
Since  $|\eps|\ll \ell   \ll   1 $ it  is   evident that we can approximate the bound
to $r$ with 
\def\dis{\displaystyle}
\begin{equation}
  \begin{split}
 r(x,t;\eps) \le \begin{cases}  {\dis |\eps|\over\dis  \ell}  & \quad \ell\le x < 1 \\  \\
 |\eps|    & \quad  x > 1
 \end{cases}
    \end{split}
 \label{eq_AII_14}
 \end{equation}
Provided the the initial condition $x$ is at a finite distance not less than $\ell$
from the unstable point $x=0$ the linear approximation is valid if  $|\eps|/\ell\ll 1$.  
If we do not  choose  the initial condition  at a finite distance from the origin
we find that as $x+\eps$  approaches 0  with $|\eps|< x$  when  $\eps$ is negative 
the estimate on the  absolute value of the remainder becomes 
\def\dis{\displaystyle}
\begin{equation}
  \begin{split}
 |r(x,t)| \le |\eps| \,e^t \ll  1  \qquad \longrightarrow \qquad t\ll \log{1\over |\eps|}
    \end{split}
 \label{eq_AII_15}
 \end{equation}
As a consequence if we do not impose the condition $x>\ell$ the linear approximation holds
only for finite times wich grow  with the log of $1/|\eps|$ as $|\eps|$  decreases.
The singularity in $\eps$ of $S_t(x+\eps)$ is  $\eps_c$ given by $x+\eps_c=x_c$  according to
the (\ref{eq_AII_11}).
From the last equation (\ref{eq_AII_10})   the following expresion  is obtained   
\def\dis{\displaystyle}
\begin{equation}
  \begin{split}
  |r(x,t;\eps)| = {|\eps|\over |\eps_c(x,t)-\eps|}\le {|\eps|\over |\eps_c(x,t)|-|\eps|}
     \end{split}
 \label{eq_AII_16}
 \end{equation}
for $|\eps|<|\eps_c|$.
We notice that  from (\ref{eq_AII_16}) the the bound (\ref{eq_AII_5}),
where we change $z$ into $\eps$,    immediately follows since
$|\eps|<\rho<|\eps_c|$  
\def\dis{\displaystyle}
\begin{equation}
  \begin{split}
  r(x,t;\eps) \le {|\eps|\over |\eps_c(x,t)|-|\eps|}\le {|\eps|\over \rho-|\eps|}
     \end{split}
 \label{eq_AII_17}
 \end{equation}
\subsection{Random initial deviations}
When the initial condition has a radom deviation $\eps \xi$ where
$\xi$ is a random variable with zero mean and unit variance having a Gaussian distributions,
we have to
restrict it to a finite interval $[-m,m]$  with $m\ge 3$
to avoid  the singularity for large values of $|\xi|$. We recall that
the probabiity that the probability that $\xi \in [-m,m]$ is
$0.997$ for $m=3$ and converges rapidly to 1 when $m$ increases.
In this case we consider the ratio  $|r|$ between the average
absolute value  of the  remainder and the absolute value  of the of the linear term,  where the mean
with respect to $\xi$ is taken over $[-m,m]$ with $m\ge 3$.  
\def\dis{\displaystyle}
\begin{equation}
  \begin{split}
    &  S_t(x+\eps\xi)= S_t(x) + \eps\xi \Lop(x,t) + \eps^2 \xi^2 R(x,t:\eps,\xi)   \\ \\
    &   |r|(x,t;\eps)  \equiv { \mean {|\eps^2 \xi^2 \,R(x,t;\eps\xi)|} \over \mean {|\eps\,\xi\,\Lop(x,t)|} } 
     = {|\eps|\over \mean{|\xi|} } \,\parmean {   \xi^2\over |\xi \eps-\eps_c(x,t)|} =  \\ \\ 
     & \;\;   = {|\eps| \over \mean{|\xi|} \,|\eps_c }\parmean {   \xi^2\over  1- |\xi|\,|\eps|/|\eps_c|} \le
     2\,{\mean{\xi^2}\over \mean{|\xi|}} \, {|\eps|\over |\eps_c}\ \simeq 
     \sqrt{2\pi}\,  {|\eps| \over |\eps_c(x,t)}  \\ \\
     & \quad \hbox{for}\qquad  {|\xi|\,|\eps|\over \eps_c(x,t)|}\le
    m\,{|\eps|\over |\eps_c(x,t)|} <{1\over 2}  \quad \longrightarrow \quad  |\eps|<{|\eps_c|\over 2m}
     \end{split}
 \label{eq_AII_18}
 \end{equation}
Where we took into account that $\mean{\xi^2}\simeq 1$ and $\mean{|\xi|}\simeq 2/\sqrt{2\pi}$  when the averages are
computed in $[-m,m]$ with $m\ge 3$.
\\
\\
This result can be extended to other  systems with  $d=1$  defined by $dx(t)/dt=\Phi(x(t))$
when  $\Phi(x)$ is a polynomial. For instance the flow of the Duffing equation
defined by $\Phi(x)=x(1-x^2)$ is given by 
\begin{equation}
  \begin{split}
    S_t(x)  & ={x\over \sqrt{x^2+(1-x^2) e^{-2t}}}  \qquad x\in \Reali  \\ \\
   L(x,t) &= \derp{S_t(x)}{x}= {e^{-2t}\over (x^2+(1-x^2) e^{-2t})^{3/2}}
  \end{split}
 \label{eq_AII_19}
 \end{equation}
The singularities in $\eps$ of $S_t(x+\eps)$ is given by  two
cuts on a straight line parallel to the imaginary axis whose branch points are
$\eps_s= -x\pm i\,e^{-t}/\sqrt{1-e^{-2t}}$.   The validity condition for
the linear approximation of  $S_t(x+\eps)$   is given
by $|\eps|/|\eps_s|\ll 1$ and for any $t\ge 0$  becomes $|\eps|/\ell\ll 1$  where
$|x|\le \ell$ is the the excluded neighborhood of the unstable fixed point.
For any $x\in \Reali$  the  linear approximation holds for $t\ll \log(1/|\eps|)$
being determined by the neigborhood of the  unstable point.
For a generic $d=1$ system the stable and unstable fixed points alternate on the real axis.
Let $a<c$ be two consecutive unstable fixed  points and $b\in [a,c]$  be the stable
fixed point. In the  interval $[a+\ell,c-\ell]$ which excludes a neigborhood of
the untable fixed points $a,\,c$, the linear approximation to $S_t(x+\eps)$ holds for any
$t>0$ provided that  $\eps|/\ell \ll 1$. On the whole real axis the approximation
holds only for finite times $t\ll \\lambda^{-1}\log(1/|\eps|)$  where $lambda$
is the largest Lyapunov exponnet at the unstable oints, and is of no practical interest. 
If the displacemet is  stochastic $\eps \xi$ and has a Gaussian distribution with unit
variance one  chooses $\xi$ in $[-m,\,m]$ as in the case of the logistic equation. 
The extension to $d\ge 2$  is difficult, but it is plausible that the the following statement is
correct
\spa
    {\sl For any autonomous system with a real  holomorphic vector field
    $\Phibf(\xbf)$ in $\Reali^d$  within the  basin $\Bcal$ of any attractive fixed point $\xbf_c$
    of the flow $S_t(\xbf)$ the linear approximation 
    $S_t(\xbf)+\eps\,\Lop(\xbf,t)\,\etabf$   to $S_t(\xbf+\eps\,\etabf)$  holds  for any $t\ge 0$
    and  any initial  condition   $\xbf\in \Bcal$ whose distance  form the boundary is 
    $\ge \ell>0$  and for any    $\etabf$ on the unit sphere,    provided
    that $|\eps|/\ell \ll 1$.
    The linear approximation holds  for any $\xbf \in \Bcal$  only for short times
    $t\ll \lambda^{-1} \log( 1/|\eps|)$ where $\lambda>0$  largest Lyapunov
    exponent  on the boundary of  $\partial \Bcal$. The result can be extended   to an  initial random 
    displacement as for a flow in $\Reali$.}
\spa
The reason why the statement is plausible is that  the flow is linearizable near
a  stable fixed point according to the Poincar\'e Dulac theorem.
Choosing the origin at the stable point   the  analytic transformation
$\Xbf=\psi(\xbf)$ tangent to the identity,  whose inverse we denote $\xbf=T(\Xbf)$,
defines  a  conjugation  $S_t(\xbf)=T(e^{\Lambda t}\psi(\xbf))$ analytic on a  polydisc. The map
$\psi(\xbf)$ can be analytically continued up to the  basin boundary $\Bcal$ where
repulsive or saddle type fixed points are  present  and where  $\psi(\xbf)$ is singular.
By approaching these points  the flow is no longer linearizable and   the linear
approximation of $S_t(\xbf+\eps\etabf)$  fails to hold for $t$ large.
\spa
\subsection{ Hamiltonian systems}
\spa
The conditions which justify the linear approximation
can be established for an autonomous  Hamiltonian 
system $H(q,p)=p^2/2+ V(q)$ whose orbits are bounded. In this case the phase space splits
into several regions, each delimited  by a separatrix issued from an a hyperbolic point, and having an
elliptic point in its interior. In each of these regions  action angle variables $\xbf=(\theta,J)^T$
can be introduced so that $H(q,p)= \hat H(J)$. This function and its derivative $\Omega(J)=d\hat H/dJ$ 
have a logaritmic singularity at the separatrix  whose inner  action we denote by $J_s$.
We refer to the Duffing
oscillator whose Hamiltonian is $H(q,p)={1\over 2}(p^2-q^2)+{1\over 4}q^4$ and denote with $J_s$ the action
the $q\ge 0$ halph plane which is the area or the orbit with $H(q,p)=E<0$ with $q>0$ in the limit $E\to 0$.
We set also $E(J)\equiv \hat H(J)$ the energy as a function of the action which can be expliclty
evaluated  with elliptic functions.  Notice that within the separatrix  $0\le J\le J_s$ we have
$-1/4\le E\le 0$ and the period approaching the separatrix becomes $T(E)\simeq \log(1/|E|)$ so that 
frequency is approximated by   $\Omega(E) \sim 2\pi/\log(1/|E|)$.
Close to the separatrix we  can evaluate $J$ as a function of $|E|=-E$ recalling that $-1/4\le E\le 0$
and integrating $d |E|/(d(J_s-J)= \Omega(J)= -\pi/\log(|E|)$
\def\Jbar{\overline{J}}
\def\epsbar{\overline{\eps}}
\begin{equation}
  \begin{split}
  \pi\,(J_s-J) = f(|E|)\equiv |E|-|E| \log (|E|)
  \end{split}
 \label{eq_AII_20}
 \end{equation}
where $y= f(x)= x-x\log x $ is positive with a maximum at $x=1$ and vanishes at $x=e$. Even though this
result is accurate for $|E|$ close to  zero we choose $|E|\ge \ell>0$ where $\ell\ll 1 $. Since  in the
interval $[0,1/4]$ where $|E|$ is defined the function is monotonic increasing and from $|E|\ge \ell$
follows $f(|E|)\ge f(\ell) $  and  $J_s-J=f(|E|)\ge f(\ell)$. As a consequence we limit our
analysis to $0\le J\le J_s-f(\ell)$ that is we exclude a small layer near the separatrix.
The solution in action angle coordinates at $\xbf+\eps \etabf$  for $\eps>0$ and $\etabf=(\cos \phi, \sin\phi)^T$
with $0\le \phi\le \pi$ so that   $\eta_2\ge 0$  and $J+\eps\eta_2>0$.  From
$S_t(\xbf)= (\phi+\Omega(J)t, J)^T$  we compute the $\eps$ expansion  for  the  displacement $\eps \etabf=
\eps(\eta_1,\eta_2)^T $ is   
\begin{equation}
  \begin{split}
   & S_t(\xbf+\eps\etabf)=S_t(\xbf)+ \eps \Lop(\xbf,t)\etabf + \eps^2 \Rop(\xbf,\etabf,t;\eps) \\ \\
   & \Lop \etabf=  \begin{pmatrix} \Omega'(J) t \, \eta_2 + \eta_1  \\ \\ \eta_2 \end{pmatrix}
   \qquad \Rop =   \begin{pmatrix} {1\over 2}\Omega''(\Jbar) t \eta_2 ^2 \\ \\  0 \end{pmatrix}
  \end{split}
 \label{eq_AII_21}
 \end{equation}
The remainder of the first order Taylor expansion in $\eps$ has been written
defining $\Jbar= J+\epsbar \eta_2$
where $0<\epsbar<\eps$ so that $J<\Jbar<J+\eps$.
The  the first and second derivatives of
$\Omega(J)$ with respect to $J$  can be easily computed for  $J$ close to $J_s$
and $E$ negative close to zero  starting from $\Omega(J)\simeq \pi/\log(1/|E(J)|)$   and $dE/dJ= \Omega(J)$
\def\Jbar{\overline{J}}
\def\epsbar{\overline{\eps}}
\begin{equation}
  \begin{split}
    \Omega'(J)&= \Omega {d\Omega\over dE}= -{(2\pi)^2\over |E| \log^3{\dis 1\over \dis|E|}} \\ \\
    \Omega''(J)& = \Omega {d\Omega'\over dE} =-{  (2\pi)^2 \over |E|^2\log^6{\dis 1\over \dis|E|} }
      \Bigg(  \log^3{\dis 1\over \dis|E|} - \\
      & - 3|E| \log^2{\dis 1\over \dis|E|} \Bigg) \simeq \\ \\
      &  \simeq -{ \Omega(J) (2\pi)^2 \over |E|^2\log^3{\dis 1\over \dis|E|}  }= {\Omega(J) \Omega'(J)\over |E|}
  \end{split}
 \label{eq_AII_22}
 \end{equation}
Denoting with $\mean{\phantom{x}}$  the average on $\phi\in [0,\pi]$
for  the components of  $\etabf=(\cos\phi,\sin\phi)^T$  we have  $\mean{\eta_i\eta_j}={1\over 2}\delta_{ij}$.
The relative error $|r|$    is  given by    

\begin{equation}
\begin{split}
  & |r|= \eps {   \mean{ \Vert \Rbf\Vert } \over \parmean {\Vert \Lop \etabf\Vert^2 }^{1/2}  }  \qquad \qquad 
   \mean{ \Vert \Rbf\Vert }\le  {1\over 4} \,|\Omega''(J+\eps))| t   \\ \\
  & \parmean {\Vert \Lop \etabf\Vert^2 }= 1 +{1\over 2}\,{\Omega'}^2(J)\,t^2 
  \end{split}
 \label{eq_AII_23}
 \end{equation}  
Since $\Jbar= J+\epsbar \eta_2$  where $0<\epsbar<\eps$ and $0\le \eta_2\le 1$ using the mean value theorem
we find that $\mean{|\Omega_2(\Jbar) \eta_2^2}= |\Omega_2(\overline{\Jbar})|\,  \mean{\eta_2^2}$
where $ 0 <\overline{\Jbar}<J+\eps$. taking into account that $|\Omega''(J)|$ is an increasing function of
$J$ the above estimate follows.  Finally  using $\Omega''(J)\simeq \Omega'(J)\Omega(J)/E(J) $ 
\begin{equation}
\begin{split}
  & r\le {\eps\over 4}\,{\Omega(J+\eps)\over |E(J+\eps)|}\,{t \, |\Omega'(J+\eps)|\over \parton{1+{1\over 2}(\Omega'(J)t)^2 }^{1/2}} \le \\ \\
  & \le {\eps\over 2\sqrt{2}}\,{\Omega(J+\eps)\over |E(J+\eps)|}{|\Omega'(J+\eps)|\over |\Omega'(J)|}
  \end{split}
 \label{eq_AII_24}
\end{equation}
The final estimate is easily obtained by choosing $J$ such that $|E(J+\eps)|\le \ell $  and
and $\eps$ such  that $\eps/\ell \ll 1$.
We choose  $\eps$ so small such that $|\Omega'(J+\eps)|/|\Omega'(J)|\le 2\sqrt 2$.
As a consequence we obtain
\begin{equation}
\begin{split}
  & r\le \eps\,{ 2\pi \over |E(J+\eps)| \log(1/|E(J+\eps)|)}\le \,{\eps\over \ell}\,\,\,{2\pi\over \log(1/\ell)}
  \end{split}
 \label{eq_AII_25}
\end{equation}
where we have taken into account that $E(J+\eps)\ge \ell$
and that $f(x)=x\,\log(1/x)$ is increasing for $0<x<e^{-1}$.
As a consequence $|r|\ll 1$ for $\eps \ll \ell$.  If we 
choose the initial condition far enough from the  the separatrix so that   $J+\eps$
is sufficiently smaller than $J_s$ then  $E(J+\eps)>\ell$ is satified for a given $\ell$.
Since   $f(x)$ is increasing for $0\le x<e^{-1} $.
the linear approximation holds if $\eps/\ell \ll 1$ where   $E=E(J+\eps)\ge \ell$
implies 
\begin{equation}
\begin{split}
   J_s-J &= \pi^{-1}\,f(|E|) \qquad \qquad f(x)= x-x\log x    \\ \\
   J+\eps& =J_s-\pi^{-1} f(|E(J+\eps)|) \le  J_s -f(\ell)     
  \end{split}
 \label{eq_AII_26}
\end{equation}
and finally $ J \le J_s-\pi^{-1}f(\ell)-\eps$.
Finally the  linear approximation is valid if the the difference $J_s-J$ is larger than $\pi^{-1}f(\ell)+\eps$.
Far enough  from the sepatrix and for $\eps/\ell\ll 1$
the linear approximation holds for any time.  
By approaching further the separatrix the linear approximation holds for smaller and smaller values of $\eps$
and on the sepatrix itself the linear approximation holds only for times shorter than $\lambda^{-1} \,\log(1/\eps)$
where $\lambda$ is the Lyapunov exponent.  The extension to higher dimensions is difficult.
However it is pausible that
in the regions of regular motion the linear approximation holds at a finite distance from  the regions of
chaotic motion provided that $\eps$ is small enough.

\section{Appendix III.  Forward, reversibility error and linear approximation}
To justify the linear approximation  to  flows with a weak noise is  difficult.
We consider first  the case in which we add a constant random vector $\eps\xibf$ to
the vector field. We can state the  conditions for the validity of the linear
approximation for  one dimensional systems and  plane autonomous Hamiltonian systems with
bounded orbits. The result  are  similar to the ones obtained for an initial  random deviation. 
For a one dimensional  system the linear approximation 
is valid  for any point  in the basin of attraction of a stable fixed point
having a distance $\ge \ell$ from the unstable fixed point(s), 
provided that the  gaussian random variable is limited to  $|\xi|\le m$ where $m\ge 3$
and $|\eps|/\ell\ll 1$. Relaxing this condition the approximation holds for $t\ll \log(1/|\eps|)$.
For a plane Hamiltonian system the linear approximation holds provided the distance
from the separatrix is larger than  $ \ell$ and $|\eps|/\ell \ll 1$. If we replace the constant
random vector with a white noise the approximation holds for  initial conditions at a  distance
not less than $\ell$ from the separatrix but only for  $t\ll(\ell/\eps)^2$.
\spa
For the logistic equation  the solution $y(t)$ corresponding to the perturbed vector field
$\Phi(x)=x(1-x)+\eps $ is given by 
\begin{equation}
  \begin{split}
& y(t) = {  x_+(x-x_-) -x_-(x-x_+) e^{-\lambda t}\over x-x_- - (x-x_+) e^{-\lambda t}  } \qquad 
\lambda= \sqrt{1+4\eps\,} \\ \\
& x_- = {1\over 2}\Bigl((1-\sqrt{1+4\eps\,}\Bigr) \qquad
\qquad x_+={1\over 2}\Bigl (1+\sqrt{1+4\eps\,} \Bigr)
\end{split}
  \label{eq_AIII_1}     
\end{equation}  
The solution exhibits a fixed singularity in  $\eps$ at
$\eps_s=-1/4$. To analize the movable singularities
which depend on $x,t$ is is convenient to  set  $x_-=z  $ so that 
$x_+=1-z$ and $\lambda=1-2z$. The solution can be written as $y(t)=P(z,t)/Q(z,t)$
and  the movable
singularities are the  poles of $Q(z,t)$. Notice that $z=1/2$ is not a pole since
both $P$ and $Q$ vanish but corresponds to the branch point in $\eps$ where the square root
vanishes. The poles   $Q(z,t)=0$  are the solutions of 
\begin{equation}
  \begin{split}    
e^{-(1-2z)t}= {x-z\over x-1+z} 
  \end{split}
\label{eq_AIII_2}     
\end{equation}   
Notice that the exponential   on the l.h.s. of   (\ref{eq_AIII_2})  is an increasing
function of $z$ since $t\ge 0$  and its value is 1 for $z=1/2$. 
The rational function  on the r.h.s. of (\ref{eq_AIII_2})  is an increasing
function of $z$ for $x<1/2$, decreasing for $x>1/2$. Indeed, it has a zero at $z=x$,
a pole at $z=1-x$,  its derivative is $(1-2x)/(x-1+z)^2$ and its value is $1$ at $z=1/2$. 
As a consequence equation  (\ref{eq_AIII_2}) has only one additional solution $z=z_s(x,t)$ 
beyond $z=1/2$  if $x<1/2$, no additional solution if $x>1/2$.  
The solution $z_s(x,t)$ for a fixed  initial condition  $0<x<1/2$  starts  with
$z_s=1/2$ for $t\ge t_*(x)$,  reaches $x$ for $t\to \infty$ and $t_*(x)$
is an increasing function of $x$.  The movable singularity $\eps_s(x,t)$
is given by $\eps_s(x,t)= z_s(x,t)(z_s(x,t)-1)$.   Letting $\eps^2 R(x,t)$ be remainder
of the linear approximation $x(t)+\eps y_1(x,t)$ to $y(t)$  where
$y_1=L(x,t)\int_0^t \,L^{-1}(x,t') \,dt'$,  we compute the
relative error  choosing $|\eps|\le {1\over 2}\,|\eps_s|$.   The following inequality holds 
\begin{equation}
  \begin{split}    
  &  r  = |\eps| { |R|\over |y_1|}\le  {|\eps|\over |\eps_s|}= \\
   & = {|\eps|\over z_s(1-z_s)} \le \begin{cases} {\dis |\eps|\over \dis x(1-x)}
      \;\; 0\le x\le {1\over 2} \\ \\
     \;\; 4\,|\eps| \qquad  \qquad  x\ge{1\over 2} \end{cases}
 \end{split}
\label{eq_AIII_3}    
\end{equation}   
where the last inequality follows from $z_s(x,t)\ge x $   which implies $z_s(1-z_s)\ge x(1-x)$.
To justify $|\eps_s|\,|R|\,/|y_1|\le 1$ we notice that the Taylor coefficient $y_n(x,t)$ of the Taylor expansion
of $(x,t;\eps)$ defined by (\ref{eq_AIII_1})  are bounded by $c/|\eps_s|^n$ so that for the remainder we have
\begin{equation}
  \begin{split}
     |R|\le {c\over \eps_s^2}\,\parton{1-{|\eps|\over|\eps_s| }}^{-1}\le \,{c\over \eps_s^2}\,\parton{1+2{|\eps|\over |\eps_s|}  }  
  \end{split}
\label{eq_AIII_4}     
\end{equation}   
where the last inequality is valid for $|\eps|/|\eps_s|<1/2$. This bound should be
compared with the bound $c/\eps_s^2$
to the second order term $y_2$. As a consequence it
safe to assume that $|\eps_s|\,\,|R|/|y_1|$ has the same bound as  $|\eps_s|\,\,\|y_2|/|y_1|$
for $|\eps/|\eps_s|$ sufficiently small.
We have verified numerically within  machine accuracy that $| \eps_s\, y_2/y_1|\le 1$ and that
$| \eps_s\,R/y_1|\le 1$ to a high accuracy for $|\eps|/|\eps_s|$ sufficiently small.
As a consequence we consider the proposed bound to $r$ as a computed assisted proof.
\\
\\
If the constant forcing  $\eps$  is replaced with a constant random
forcing $\eps \xi$ where $\xi$ is a gaussian variable with cut-off $|\xi|\le 3$,  then to compute the relative
error we  replace $|R|$ with of $\mean{\xi^2\,|R|}$ and $|y_1|$  with $\mean{|\xi|} |y_1|$.
The remained $|R|$ can be replaced by $|y_2|$ provided that
$|\eps \xi|/|\eps_s| \le 3|\eps|/|\eps_s|\ll 1$. In this case we find
\begin{equation}
  \begin{split}
    r&= |\eps| {\mean {|R|\,\xi^2 }\over  |y_1|\, \mean{|\xi|} } \le {|\eps|\over |\eps_s|}\, \parton{1+6{|\eps|\over |\eps_s|}}\,
    {|\eps_s\, y_2|\over |y_1|} {\mean{\xi^2}\over \mean{|\xi|} } \le  \\ \\
   & \le \,\sqrt{\pi\over 2}\,{|\eps|\over |\eps_s|}  \parton{1+6{|\eps|\over |\eps_s|}}\le 2{|\eps|\over |\eps_s|} \qquad \qquad 
    {|\eps_s|\over |\eps_s|}\le {1\over 10}
  \end{split}
\label{eq_AIII_5}    
\end{equation} 
To conclude for a constant forcing  or time independendt random forcing,
the linear approsimation holds for any
$x>0$ at distance not less tha $\ell$ from the unstable point
provided that $|\eps|/\ell \ll 1$
and we believe that this result extends to higher dimensions.
\spazio
  \subsection{Hamiltonian systems} 
\spazio
  We consider an autonomous Hamiltonian system in $\Reali^2$ with bounded orbits
  and a domain  whose boundary is a sepratrix. In its interior there is an
  elliptic fixed point. 
  In action angle variables $(\theta,J)$    
  in this region the Hamiltonian is $H=H(J)$ and has a logarithmic singularity for $J_s$ where
  $J_s$ is the action of the separatrix. In a region $\ell\le J\le J_s-\ell$  whose points have a final
  distance from the elliptic point and the separatrix $H(J)$ is regular and its derivatives $\Omega=H'J)$
  and $\Omega'(J), \,\Omega''(J)$ are  bounded.   Introducing an
  additive white noise in $[0,t]$  the equations of
  motion are $\dot \theta(t')= \Omega(J(t')) +\eps \xi_1(t')$ and $\dot J(t')= \eps \xi_2(t')$.
  The first order Tayor expansion  of the solution at time $t$ reads  
\begin{equation}
  \begin{split}
    \theta(t)&= \theta + \Omega(J) t + \eps \,\parton{w_1(t)+\Omega'(J) \int_0^t \,w_2(\tau)\,d\tau  } + \\ \\
    & + \eps^2 R(t)   \qquad \qquad  \quad      J(t)= J+ \eps w_2(t) \\   \\ 
R   &={w_2^2(t) \over 2}\,\int_0^t\,\Omega''\bigl(J+\eps\,\lambda(\tau)
    \,w_2(\tau)\bigr)\,d\tau
  \end{split}
\label{eq_AIII_6}      
\end{equation}  
where $w_1(t),\,w_2(t)$ denote the Wiener noise corresponding to $\xi_1(t),\,\xi_2(t)$.
The first order terms evaluated at time $t$  define the linear response that can be obtained from
the fundamental matrix $\Lop(J,t)= \begin{pmatrix}1  & \Omega'(J) \,t  \\ 0 & 1 \end{pmatrix}$  according to 
\def\Jbar{\overline{J}}
\begin{equation}
  \begin{split}
   &  \begin{pmatrix} \theta_1(t) \\ \\J_1(t)\end{pmatrix}= \Lop(J,t)\,\int_0^t \,\Lop^{-1}(J,\tau)\,
    \begin{pmatrix} \xi_1(\tau) \\ \\\xi_2(\tau)\end{pmatrix}\,d\tau= \\ \\
    & =  \begin{pmatrix}  w_1(t) + \Omega'(t)\,\Bigl(tw_2(t)-{\int_0^t}  \tau dw_2(\tau) \Bigr)
      \\ \\ w_2(t)\end{pmatrix}
  \end{split}
\label{eq_AIII_7}     
\end{equation}  
from which we recover the previous results after an integration by parts  of the integral 
within the brackets.
Recalling that $w_2(t)$ has a Gaussian distribution and that its variance is $\mean{w_2^2(t)}=t$,
the remainder $R$ remains bounded with probability close to 1  as long as 
$3 \eps \,\sqrt{t}\le \ell$ for some $\ell\ll 1$ and $\eps \ll \ell$ since $\Omega''$
remains bounded. Indeed  the  argument  of $\Omega''$ varies within $J-\ell, J+\ell$ and choosing
$2\ell\le J \le J_s-2\ell$ the singularity at $J_s$ is not met  and the argument is positive.
The estimate is limited to a finite time interval $\sqrt{t}<\ell/(3 \eps)$.
We  evaluate now the relative error  defined by
$r= |\eps|\mean{|R|}/\mean{(\theta_1^2+J_1^2)^{1/2}}$.
Using the inequality    $\mean{(\theta_1^2+J_1^2)^{1/2}}\ge(2/\pi)^{1/2}\mean{\theta_1^2+J_1^2}^{1/2}$
we obtain for $\sqrt{t}<\ell/(3 \eps)$
\begin{equation}
  \begin{split}
   & r = \sqrt{\pi\over 8}\, {|\eps|\over 2} \, \, \cdot \\
   & \cdot {\mean{w_2^2(t)}\,t\,|\Omega_2|\over
      \Bigl <  \Bigl(w_1(t)+ \Omega'(J) \,\int_0^t \,w_2(\tau)\,d\tau\,\Bigr)^2+w^2_2(\tau) \Bigr>^{1/2}} \le
    \\   \\
    & \le |\eps| { |\Omega_2|\,t \over \Bigl(2t +{\Omega'}^2(J) \,t^3/3\Bigr)^{1/2} } \le
        |\eps| \,\sqrt{t}\,C \le C{\ell\over 3}
  \end{split}
\label{eq_III_8}    
\end{equation}  
where $|\Omega_2|$ denotes the maximum of $|\Omega''(J)|$  for $ \ell\le J \le J_s-\ell$
and  $C= |\Omega_2|/\sqrt{2}$.  We have  taken into account
$\mean{\bigl(\int_0^t w_2^2(\tau)\,d\tau\bigr )^2}=t^3/3$.
Finally the linear approximation is valid only for finite times  which grow as $(\ell/\eps)^2$..
Indeed unlikely the case of the initial deviation in which a sufficient initial distance from the
separatrix insures that the separatrix  is never  reached,  so that  the linear approximation  holds  for  $t>0$,
in this case the action diffusion (or the linear time variation if the system has constant forcing)
causes as $\eps \,\sqrt{t}$  so that imposing that we remain at a finite distance from the separatrix for
finite times, and this limits the validity  of  the linear approximation 
%
%
%
%
\spa
We  show  now  that the linear approximation for the  BF reversibility process holds under the
same  condition. Indedd the evolution in $[t,2t]$ determined by $\dot \theta(t')= -\Omega(J(2t-t'))$
and  $\dot J(t')=0$ so that for $t=2t$ we find 
\begin{equation}
  \begin{split}
    & \theta(2t) = \theta(t) - \Omega(J(t)) t = \theta +\Omega(J)t -\Omega(J+\eps w_2(t))\,t + \\ 
    & \qquad + \eps w_1(t) + \eps\Omega'(J) \,\int_0^t \,w_2(\tau)\,d\tau + \eps^2 R(t) \\
   &  J(2t)= J+ \eps w_2(t)   
  \end{split}
\label{eq_AIII_9}     
\end{equation}  
As a consequence the final result is $\theta(2t)= \theta+ \eps \theta_1(2t)+ \eps^2 R(2t)$ and
$J(2t)= J + \eps w_2(t)$   where the first order terms and the remainder  for $\theta(2t)$ are given by 
\begin{equation}
  \begin{split}
    \theta_1(2t) & = w_1(t)+\Omega'(J)\,\int_0^t \,w_2(\tau) \,d\tau  \\  
    R(2t)  &= {1\over 2}\, w_2^2(t)\parton{t\,\Omega''(\Jbar(t)) +\int_0^t\,\Omega''(J+\lambda(\tau)w_2(\tau))\,d\tau }
  \end{split}
\label{eq_AIII_10}    
\end{equation}  
\def\taubar{{\overline \tau}}
where $\Jbar(t)= J+\lambda(t) w_2(t)$. Using the mean theorem and restricting the action to the interval
$2\ell\le J\le J_s-2\ell$   and letting  $3 \eps ,\sqrt{t}\le \ell $ the relative error is  bounded by by 
%
\begin{equation}
  \begin{split}
    r  & \le \sqrt{\pi\over 2}|\,\eps| \, {\mean{|R(2t)|}  \over \mean {\theta_1^2(2t)+J_1^2(2t)) }^{1/2} }\le \\ \\
      & \le |\eps|{2 |\Omega_2| \,t\mean{w_2^2(t)}\over \Bigl(2t+ {\Omega'}^2(J)t^3/3)\Bigr)^{1/2} } \le  |\eps|\,C\,\sqrt{t}   \le C\,{\ell\over 3}
  \end{split}
\label{eq_AIII_11}      
\end{equation}  
To conclude  the linear approximation for the  F   and  BF evolution with noise holds under
the same conditions: $2\ell\le J\le J_s-2\ell$ for $t\le (\ell/3\eps)^2$ and for $\eps\ll\ell\ll 1$
the relative error is smaller that $\ell\,C/3$ for rather long times.
\\
\\
We close  this section by  quoting    the covariance matrix  $\Sigma^2_R(J,t)$  for  BF Reversibility 
noisy processes. Stating from the tangent map $\Lop(J,t)$
\begin{equation}
  \begin{split}
    \Lop(J,t)= \begin{pmatrix} 1 & \Omega'(J)\,t \\ \\ 0 & 1\end{pmatrix}
  \end{split}
\label{eq_AIII_12}    
\end{equation}  
We find the following  expression for $\Sigma^2_{BF}$
\begin{equation}
  \begin{split}
    \Sigma^2_{BF}= \int_0^t \,\bigl(\Lop^T(J,\tau)\Lop(J,\tau)\bigr)^{-1}\,d\tau =
    \begin{pmatrix} t+{1\over 3} {\Omega'}^2\,t^3 & -{1\over 2} \,\Omega' \,t^2 \\  \\
   - {1\over 2}\,\Omega' \,t^2   & t\end{pmatrix}
  \end{split}
\label{eq_AIII_13}    
\end{equation}  
In   (\ref{eq_5_34}) we have given the expression of $\Sigma^2_{BF}$ using normal coordinates, rather than 
action-angle coordinates. The invariants are the same as for (\ref{eq_AIII_13})
provided that we change $\Omega'$ into $2J\,\Omega'$.
\\
The espression of the covariance matrix  $\Sigma^2_F$
for the Forward noisy process is given by 
\begin{equation}
  \begin{split}
    \Sigma^2_{F}& = \Lop(J,t) \int_0^t \,\bigl(\Lop^T(J,\tau)\Lop(J,\tau)\bigr)^{-1}\,d\tau \,\Lop^T(J,t) = \\ 
  &=    \begin{pmatrix} t+{1\over 3} {\Omega'}^2\,t^3 & -{1\over 2} \,\Omega' \,t^2 \\  \\
   - {1\over 2}\,\Omega' \,t^2   & t\end{pmatrix}
  \end{split}
\label{eq_AIII_14}     
\end{equation}  
and in this case we have $\Sigma^2_F=\Sigma^2_{BF}$.
%
%

%
\def\diag{\hbox{diag}}
\section{Appendix IV. Asymptotic behaviour of Lyapunov  and reversibility
error invariants  for linear flows}
%
The linear response to an initial random displacement or to the noise along the orbit
provides the Lyapunov, Forward and   Reversibility  error  covariance matrices.
Their asymptotic behaviour  is determined by the Lyapunov exponents.
Analytical results have been obtained  for linear  Hamiltonian systems in $\Reali^2$.
For nonlinear Hamiltonian systems in $\Reali^2$ the results were  obtained
using normal coordinates  near a fixed point. 
Linear systems in $\Reali^d$ where considered only for symmetric matrices, a case 
in which the results are straightforward.
We consider here a generic linear flow    $\Lop(t)\,\xbf$  for $\xbf \in \Reali^d$
and $\Lop(t)=e^{\Aop t}$  where   $\Aop$ is a any real  matrix.   
The Lyapunov matrix can be written as  $\Lop^T(t)\Lop(t)=\Wop(t)e^{2t\,\Lambda(t)}\,\Wop^T(t)$
where the diagonal matrix $\Lambda(t)$ converges to the Lyapunov exponents $\Lambda$
and $\Wop(t)$ to the matrix of Lyapunov vectors $\Wop$. 
For $d=2$ we show that if the eigenvalues of $\Aop$ are real then $\Lambda(t)$
converges to these eigenvalues as $1/t$. If the eigenvalues  are complex conjugate then 
$\Lambda(t)$ has the same type of convergence to their real part.
For $d=2$ the asymptotic behaviour of the Forward and  BF Reversibility 
error is obtained.
For linear systems in $\Reali^d$ we  suggest  the  asymptotic behaviour of   L, F and BF
invariants  as a natural extension of the results obtained for $d=2$  and for $d>2$
when the matrix $\Aop$ is symmetric.
\\
\\
When the system is nonlinear  and the field $\Phibf(\xbf)$  vanishing at $\xbf=0$
is holomorphic with a non resonant linear part $\Lambda \,\xbf$,
then  according to the Poincar\'e-Dulac theorem there  is an an analytic
change of coordinates $\Xbf=\psi(\xbf)$, tangent to the identy,
which linearizes the field. If the fixed point of the flow $\xbf=0$ is attractive,
then the flow is linearizable in its basin of attraction.
If the fixed point is a saddle then according to Hadamard-Perron theorem  or Hartman-Grobmann
theorems there are two manifolds  $\Wcal_s$  and  $\Wcal_u$ tangent at $\xbf={\bf 0}$
to the eigenspaces $\Ecal_s$ and $\Ecal_u$ of the positive and negative eigenvalues.
If  $\xbf\in \Wcal_s$ then $S_t(\xbf)\to 0$,  if $\xbf\in \Wcal_u$ then $S_{-t}(\xbf)\to 0$
and in both cases the system is linearizable. The non linear system
inherits the properties of the linear system.
\subsection{A similarity transformation in \texorpdfstring{$\Reali^d$}{TEXT} }
\spa
We analyze the  invariants of Lyapunov and reversibility  error matrices for linear systems
in $\Reali^d$ with $\Phibf=\Aop\xbf$ when $\Aop$ has real distinct eigenvalues. The
matrix $\Aop$ is conjugated to $\Lambda=\diag(\lambda_1,\ldots, \lambda_2)$
with $\lambda_1>\lambda_2>\ldots>\lambda_d$ and the  diagonalizing transformation
$\Top$ is real though not orthogonal and $\det(\Top)=1$.
\begin{equation}
  \begin{split}    
     \Aop=\Top \Lambda \Top^{-1} \qquad \quad  \Lop(t)= e^{\Aop t} = \Top \,e^{t\,\Lambda}\,\Top^{-1} 
  \end{split}
  \label{eq_AIV_1}
\end{equation}   
A similar decomposition when the eigenvalues are complex or when there a unique real eigenvalue
of multiplicity 2 but  this case $\Lambda$ is no longer diagonal 
In order to analyze the  asymptotic behaviour of
the Lyapunov matrix $\Lop^T(t)\Lop(t)$ we introduce a matrix $\Xop(t)$  
obtained with  a similarity tramsformation
which does not alter its  invariants and the eigenvalues 
\begin{equation}
  \begin{split}    
    & \Lop^T(t)\Lop(t)=\Top \,\Xop(t)\,\Top^{-1}    \phantom{\int}\\
    & \Xop(t)= \Vop^{-1}\,e^{t\Lambda}\,\Vop\,e^{t\Lambda}  \qquad \Vop=\Top^T\Top   
  \end{split}
  \label{eq_AIV_2}
\end{equation} 
In order to analyze the  BF Reversibility error covariance matrix we itroduce the
new matrix   $\Yop(t)$ defined as the the integral od $\Xop(t')$ in $[0,t]$
\begin{equation}
  \begin{split}    
    & \Yop(t) =\int_0^t \Xop(t') \,dt' \equiv \int_0^t \,\Vop^{-1} \,e^{\Lambda^T\,t'}\, \Vop\,
       e^{\Lambda\,t'}\,dt'
  \end{split}
\label{eq_AIV_3}
\end{equation} 
The covariance matrix $\Sigma^2_{BF}$ is related by a similarity transformation to a matrix
$\Yop_{BF}(t)$ defined by 
\begin{equation}
  \begin{split}
   \Yop_{BF}(t) & =\int_0^t \,\Vop \,e^{-\Lambda\,t'}\, \Vop^{-1}\,
       e^{-\Lambda\,t'}\,dt' \\
    \Sigma^2_{BF} (t)&=\bigl(\Top^T\bigr)^{-1} \,\Yop_{BF}\,\Top^T 
  \end{split}
\label{eq_AIV_4}
\end{equation} 
We notice that  $\Yop_{BF}(t)$ is obtained from $\Yop(t)$ by  exchanging
$\Vop$ with $\Vop^{-1}$ and $\Lambda$ with  $-\Lambda$.
\spa
\subsection{Invariants for linear systems in \texorpdfstring{$\Reali^2$}{TEXT} }
\spa
For a plain system if the matrix $\Aop$ has real distinct eigenvalues the covariance matrices
$\Sigma^2_L$ and $\Sigma^2_{BF}$ are related by a similarity transformation to the matrices
$\Yop$ and $\Yop_{BF}$ defined above. 
If $\Aop$ has complex eigenvalues $\lambda\pm i\omega$ then it is conjugated by a real
transformation $\Top$ to the matrix $\Lambda= \lambda I +\omega \Jop$   where
$\Jop=\begin{pmatrix} 0 & 1 \\ -1\ & 0\end{pmatrix}$. In this case 
$e^{\Aop t}$ is conjugated  to $e^{\Lambda t}= e^{\lambda t} \Rop(\omega t)$ where
$\Rop(\omega t)= e^{\omega t \,\Jop}$ is a rotation matrix.
\\
If $\Aop$ has an eigenvalue $\lambda$  of multiplicity 2 then $\Aop$ is conjugated to
$\Lambda= \lambda \Iop+\alpha \Nop$ where $\Nop=\begin{pmatrix} 0 & 1 \\ 0 & 0\end{pmatrix}$
and in this case  $e^{\Aop t}$ is conjugated to
$e^{\Lambda t}=  e^{\lambda t}\,\parton{\Iop+\alpha t \Nop}$.
\\
The matrix  $\Yop_{BF}(t)$ is obtained from $\Yop(t)$ by  exchanging
$\Vop$ with $\Vop^{-1}$ and $\Lambda$ with  $-\Lambda$ .
\\
\\
When $\Lambda$ is not diagonal  the matrix $\Xop$ conjugated to $\Lop^T\Lop$ is defined by
 $\Xop(t)=\Vop^{-1}\,e^{t\Lambda^T}\,\Vop\,e^{t\Lambda} $ and $\Yop(t)$ is defined by
 its integral.  The matrix $\Sigma^2_{BF}(t)$ is conjugated to  $\Yop_{BF}(t)$
obtained from $\Yop(t)$ by exchanging $\Vop$ with $\Vop^{-1}$ and $\Lambda$ with
$-\Lambda^T$. In the case of complex eigenvalues this amounts to change $\lambda$ into
$-\lambda$ leaving $\omega$ unchanged.  
\spazio
\phantom{xxxxxxxxxxxx}
\spazio
{\bf Parametrization of the matrices $\Top$ and $\Vop$ }
\spa
We propose  the following  parametrization of the matrix $\Top$  and of $\Vop$
\begin{equation}
  \begin{split}    
    \Top& =\begin{pmatrix} 1+BC & B \\ \\ C & 1     \end {pmatrix}  \\ \\
    \Vop& = \begin{pmatrix} (1+BC)^2 +C^2 & \qquad  B(1+BC)+C  \\ \\ B(1+BC)+C & \qquad  1+B^2     \end {pmatrix}
  \end{split}
\label{eq_AIV_5}
\end{equation}
We shall often use the following notation for the  elements of $\Vop$
and its inverse   
\begin{equation}
  \begin{split}    
    \Vop=\begin{pmatrix} a & b \\ \\ b &  c     \end {pmatrix}  \qquad 
     \Vop^{-1}= \begin{pmatrix}  c & -b \\ \\ -b & a      \end {pmatrix}  \qquad  ac-b^2=1
  \end{split}
\label{eq_AIV_6}
\end{equation}   
\\
  {\bf  Real distinct eigenvalues  $\lambda_1,\,\lambda_2$ }
\\
\\
In this case  $\Lambda=\diag(\lambda_1,\lambda_2)$ and using the  notation
(\ref{eq_AIV_6}) for the matrix $\Vop$ the matrix $\Xop$  and explicitly reads 
\begin{equation}
  \begin{split}    
 &  \Xop(t)  = \\
& = \begin{pmatrix} ac\,e^{2\lambda_1 t}-b^2\,e^{(\lambda_1+\lambda_2)t} &  \quad bc\,\bigl( e^{(\lambda_1+\lambda_2)t}
    -e^{2\lambda_2\,t} \bigr ) \\ \\ ab\, \bigl( e^{(\lambda_1+\lambda_2)t}
   - e^{2\lambda_1\,t}\bigr )     &  \quad  ac\,e^{2\lambda_2 t}-b^2\,e^{(\lambda_1+\lambda_2)t}   \end {pmatrix} 
  \end{split}
\label{eq_AIV_7}
\end{equation}   
The invariants of $\Xop(t)$  are given by
\begin{equation}
  \begin{split}    
   & \Tr(\Xop(t))= ac\bigl( e^{2\lambda_1 t} +e^{2\lambda_2 t}) -2b^2\,e^{(\lambda_1+\lambda_2)t}  \\ \\
   &  \det(\Xop(t))= e^{2(\lambda_1+\lambda_2)\,t}
  \end{split}
\label{eq_AIV_8}
\end{equation}  
It is convenient to write the  trace in a different form by factorizing the
leading term $e^{2\lambda_1 t} $  
\begin{equation}
  \begin{split}
     {1\over 2}\Tr(\Xop(t)) & = {ac\over 2} e^{2t\lambda_1}\Bigl( 1- {2b^2\over ac} e^{-t(\lambda_1-\lambda_2)} + \\ 
    & e^{-2t(\lambda_1-\lambda_2)} \Bigl)
  \end{split}
\label{eq_AIV_9}
\end{equation} 
and to make the same factorization for  the square root of 
$\Delta = {1\over 4} \Tr^2(\Xop(t)) - \det(\Xop(t)) $ 
\begin{equation}
  \begin{split}    
 & \sqrt{\Delta}  =
{ac \over 2} e^{2t\lambda_1} \Bigl[ 1-  {4b^2\over  ac} e^{-t(\lambda_1-\lambda_2)}
  +2e^{-2t(\lambda_1-\lambda_2)}  + \\  
& + \parton{2b^2\over ac}^2 e^{-2t(\lambda_1-\lambda_2)} - \\
& \parton{2\over ac}^2 e^{-2t(\lambda_1-\lambda_2)} + O(\eta^3) \Bigl ]^{1/2}
  \end{split}
\label{eq_AIV_10} 
\end{equation}
where we have set $\eta=e^{-t(\lambda_1-\lambda_2)}$. By expanding  $(1+x)^{1/2}=1+x/2-x^2/8+\ldots$
we notice that the forth term in square bracket multiplied by $1/2$ cancels the square of the second
term multiplied by $-1/8$ finally we obtain the final form for $\sqrt{\Delta}$ 
and 
\begin{equation}
  \begin{split}    
    \sqrt{\Delta} & ={ac\over 2}\,e^{2t\lambda_1} \,\biggl[  1  -{2b^2\over ac} e^{-t(\lambda_1-\lambda_2)}
      +e^{-2t(\lambda_1-\lambda_2)}  -  \\
      & \hskip 2.5 truecm - {2\over (ac)^2} e^{-2t(\lambda_1-\lambda_2)} +O(\eta^3) \biggr] 
  \end{split}
\label{eq_AIV_11}
\end{equation}
The eigenvalues  $\mu_1^2,\mu_2^2$ of $\Xop$   given by   ${1\over 2} \Tr(\Xop)\pm \sqrt{\Delta} $ explicitly read
\begin{equation}
  \begin{split}    
    \mu_1^2(t)& = ac\,e^{2\lambda_1 t}\,\Bigl( 1 -{2b^2\over ac} e^{-t(\lambda_1-\lambda_2)} + O(\eta^2)  \Bigr) \\ \\
    \mu_2^2(t)&={ac\over 2}\,e^{2\lambda_1 t}\,\Bigl({2\over (ac)^2} \,e^{-2t(\lambda_1-\lambda_2)}+ O(\eta^3) \Bigr) =\\
     &  = {1\over ac}\,e^{2\lambda_2 t}\,\bigl( 1+O(\eta)\bigr) \qquad \eta=e^{-t(\lambda_1-\lambda_2)}
  \end{split}
\label{eq_AIV_12}
\end{equation}
In the expression of $\mu_1$ it is important to keep the term of order
$\eta$ when the eigenvectors are computed  
The first  eigenvector $\vbf_1$ of  $\Xop$ is obtained by  solving the equation
\begin{equation}
  \begin{split}    
     \begin{pmatrix} b^2 \,e^{t(\lambda_1+\lambda_2)} \,I_\eta  & \quad bc\,e^{t(\lambda_1+\lambda_2)} I_\eta
     \\ \\
  -ab e^{2\lambda_1 t} I_\eta  &   \quad -ac e^{2\lambda_1 t }\,I_\eta  \end {pmatrix} \,\vbf_1=0 
  \end{split}
\label{eq_AIV_13}
\end{equation}  
where $I_\eta=1+O(\eta)$.  The second eigenvector  $\vbf_2$  is obtained from 
\begin{equation}
  \begin{split}    
     \begin{pmatrix}  ac\,e^{2t\lambda_1} \,I_\eta \,  & bc\,e^{2t\lambda_1} \eta \,I_\eta
      \\ \\
  -ab e^{2\lambda_1 t} I_\eta  &   \quad -ac e^{2\lambda_1 t }\,\eta\,I_\eta  \end {pmatrix} \,\vbf_2=0 
  \end{split}
\label{eq_AIV_14}
\end{equation}  
As a consequence the eigenvectors matrix is given by
\begin{equation}
  \begin{split}    
    \Top_\Xop = (\vbf_1,\vbf_2) =  N_T \begin{pmatrix}  c \, + O(\eta)   &   O(\eta) \\ \\
     -b + O(\eta) \, &  1 +O(\eta)\end {pmatrix} \qquad
  \end{split}
\label{eq_AIV_15}
\end{equation}  
where $N_T= (c+O(\eta))^{-1}$ is chosen so that $\det(\Top_\Xop)=1$ 
We can now compute the orthogonal matrix $\Wop$ which diagonalizes $\Lop^T\Lop$. Using the parametrization
(\ref{eq_AIV_5})  for the matrix $\Top$ we find that $\Wop(t)=\Top\,\Top_\Xop$ is given by 
\begin{equation}
  \begin{split}    
  &  \hskip -.1 truecm \Wop(t)  = 
   \begin{pmatrix} 1+BC & B \\ \\ C & 1     \end {pmatrix}
 \begin{pmatrix} 1+B^2 +O(\eta)&  O(\eta)    \\ \\ -B-C-B^2C +O(\eta)  & 1 +O(\eta)  \end {pmatrix}  \\ \\
 & \hskip -.1 truecm \Wop(t) ={1\over \sqrt{1+B^2+O(\eta)}}\, \begin{pmatrix} 1 +O(\eta) & B+O(\eta) \\ \\ -B +O(\eta)& 1  +O(\eta)
 \end {pmatrix} 
    \label{eq_AIV_16}
    \end{split}
\end{equation}  
We have taken into account that $\det(\Top_\Xop)=1$ and $\det(\Top)=1$. As a consequence the factor
$(1+B^2+O(\eta))^{1/2}$ is such that  $\det(\Wop)=1$ and $\Wop$ is an orthogonal matrix.
Letting  $\cos \theta= 1/\sqrt{1+B^2}$ and  $\sin  \theta=B/\sqrt{1+B^2}$  we see that
$\Wop(t)= \Rop(\theta) ( 1+O(\eta))$. As a consequence  the asymptotic limit of $\Wop(t)$ for
$t\to \infty$ is $\Wop=\Rop(\theta)$ and convergence is exponentially fast.
The result is summarized by 

\begin{equation}
  \begin{split}      
    & \Lambda(t)  =  \parton{\Lambda+{1\over 2t}\Lambda_1\,} \,\bigl(1+O(\eta)\bigr)   \\ \\
    & \Lambda_1=\log (ac)\,\diag(1,-1)  \\ \\
    & \Wop(t) = \Rop(\theta) \,\bigl(1+O(\eta)\bigr) \qquad \quad \eta=e^{-t(\lambda_1-\lambda_2)}
    \label{eq_AIV_17}
    \end{split}
\end{equation}  
The convergence of $\Lambda(t)$ to its limit $\Lambda$ is slow as $1/t$
due to the  costant factors $ac$ and $(ac)^{-1}$ in the eigenvalues of $\Xop(t)$
or $\Lop^T(t)\Lop(t)$,  but the remainder is exponentially small  $O(\eta)$.
While computing the eigenvalues
of $\Lop^T\Lop$ the slow convergence as $1/t$ can be made exponentially fast
as $\eta$ by a linear intepolation in $1/t$.
The convergence of $\Wop(t)$ to its asymptotic limit is also exponentially fast
as confirmed by  computing numerically the eigenvectors of $\Lop^T\Lop$.
\\
We finally observe that for the first invariant, the trace,  $(2t)^{-1}\log I^{(1)}(t) $
converges to $\lambda_1$  as $1/t$ just as the first eigenvalue,
whereas for the second invariant the determinant we have $(2t)^{-1}\log I^{(2)}(t)=\lambda_1+\lambda_2$
for any $t$. 
\spa
{\bf Behaviour of  $\Yop$}
\\
\\
The matrix $\Yop(t)$ obtained by integrating $\Xop(t)$ is given by  
\begin{equation}
  \begin{split}    
     \Yop(t)&=\begin{pmatrix} ac\,I_{11}-b^2\,I_{12} &  \qquad   bc\,\bigl(I_{12}    -I_{22} \bigr ) \\ \\
     ab\, \bigl( I_{12}  - I_{11}\bigr )     & \qquad    ac\,I_{22}-b^2\,I_{12}   \end {pmatrix}  \\ \\
     & I_{ij}= {e^{(\lambda_i+\lambda_j)t}-1\over \lambda_i+\lambda_j}
  \end{split}
\label{eq_AIV_18}
\end{equation}
The the invariants, which are the trace and the determinant of $\Yop(t)$,  are given by
\begin{equation}
  \begin{split}    
    \Tr(\Yop) & = ac(I_{11}+I_{22})-2b^2 I_{12}  \\  \\
    \det(\Yop) &= ac\,I_{11}\,I_{22} -b^2 \,I_{12}^2    
  \end{split}
\label{eq_AIV_19}
\end{equation}  
We consider now the asymptotic behaviour of the invariants in  three cases 
\\
\\
{\bf Case I   $\lambda_1>\lambda_2>0$  : unstable node }
\\
\\
The integrals  have an exponential growth  $I_{ij}\simeq e^{(\lambda_i+\lambda_j)t}/(\lambda_i+\lambda_j)$
so that
\begin{equation}
  \begin{split}      
  &  I^{(1)}(t)=\Tr(\Yop)  = {ac \over 2 \lambda_1} \, e^{2\lambda_1 t}     \Bigl(  1+O(\eta)\Bigr) \\
    & \eta=e^{-(\lambda_1-\lambda_2)t}\\ \\  
 &   I^{(2)}(t)=\det(\Yop ) =  e^{2(\lambda_1+\lambda_2)t} \parton{ {ac\over 4\,\lambda_1\,\lambda_2}
      - { b^2\over (\lambda_1+\lambda_2)^2}  } = \\ \\
    & =  e^{2(\lambda_1+\lambda_2)t} \;\;\;\; {ac(\lambda_1-\lambda_2)^2 +4\lambda_1\lambda_2\over 4\lambda_1\lambda_2\,
      (\lambda_1+\lambda_2)^2}
 \label{eq_AIV_20}
    \end{split}
\end{equation}    
\\
The last expression shows explicitly that $  I^{(2)}(t)$ is positive.
The limit of $(2t)^{-1}\,\log I ^{(j)}t)$  for $t\to +\infty $ is $\lambda_1$ for
$j=1$ and $\lambda_1+\lambda_2$ for $j=2$.
The eigenvalues $\mu_j^2$  of $\Yop$ are given by
\\
\begin{equation}
  \begin{split}
&\mu^2_{1,2}= \Tr\parton{\Yop\over 2}\,\parqua{1 \pm\parton{ 1- {\det(\Yop)\over \Tr^2\parton{\Yop/ 2} }  }^{1/2} }\\ \\
&     { \det(\Yop) / \Tr^2\parton{\Yop/ 2}}= O(\eta^2)\qquad \quad \eta=e^{-(\lambda_1-\lambda_2)}
  \end{split}
  \label{eq_AIV_21}
\end{equation}   
\\
The asymptotic behabviour of the eigenvalues is 
\begin{equation}
  \begin{split}
    &\mu_1= \Tr(\Yop)\,\Bigl( 1+O(\eta^2)\Bigr)= {ac\over 2\lambda_1}\,e^{2\lambda_1 \,t}\Bigl(1+O(\eta) \Bigr)\\ \\
    & \mu_2 ={\det(\Yop)\over \Tr((\Yop)} \Bigl(1+O(\eta^2)\Bigr)= \\ 
    & = e^{2\lambda_2 t}\;\;{2\lambda_1\over ac}\;
    {ac(\lambda_1-\lambda_2)^2+4\lambda_1\lambda_2 \over 4\lambda_1\lambda_2 (\lambda_1+\lambda_2)^2}\times\Bigl(1+O(\eta) \Bigr)
  \end{split}
  \label{eq_AIV_22}     
\end{equation}
\\
\\
{\bf Case II    $\lambda_1>0>\lambda_2$ : saddle}
\\
\\
Asymptotically we have    $I_{11}\sim e^{2\lambda_1t} (2\lambda_1)^{-1}$  and 
$I_{22}\sim (2|\lambda_2|)^{-1}$.  The last term  behaves as   $I_{12}\sim e^{(\lambda_1-|\lambda_2|)t}
((\lambda_1-|\lambda_2|)^{-1}$ if $\lambda_1>|\lambda_2|$ and $I_{12}\sim (|\lambda_2 -\lambda_1))^{-1}$
if  $|\lambda_2|>\lambda_1$.  The leading term for $\Tr(\Yop)$ is $\c\,I_{11}$  and for $\det(\Yop)$ it is
$ac\,I_{11}I_{22}$. the explicit expression is  
\begin{equation}
  \begin{split}
   & \Tr(\Yop) =  {ac \over 2 \lambda_1}\,e^{2\lambda_1 t} \, \Bigl (1+O(\eta)  \Bigr) \\ \\
   &  \det\Yop) = {ac \over 4\,\lambda_1\,|\lambda_2|}\,e^{2\lambda_1t} \,\Bigl ((1+O(\eta) \Bigr) \\ \\ 
   &  \eta= e^{-(\lambda_1+\alpha) \,t}\qquad \qquad \alpha=\min(\lambda_1,\,\,|\lambda_2|)
    \label{eq_AIV_23}
    \end{split}
\end{equation}    
The eigenvalues are given by
\begin{equation}
  \begin{split}      
    & \mu_1^2(\Yop) = {ac \over 2\lambda_1}\,e^{2\lambda_1t} \Bigl(1+ O(\eta)  \Bigr) \\ \\
    & \mu_2^2(\Yop) = {\det(\Yop)\over \Tr(\Yop)}\Bigl(1+O(e^{-2\lambda_1 t})\Bigr) = 
    {1\over2|\lambda_2|}\,\Bigl( 1+O(\eta)\Bigr)
    \label{eq_AIV_24}
    \end{split}
\end{equation}   
\\
\\
\\
{\bf Case III    $0>\lambda_1>\lambda_2$ : stable  node}
\\
\\
Al the integrals reach a constant limit $I_{ij}\simeq (|\lambda_i|+|\lambda_j|)^{-1}$ so that the invariants
become 
\begin{equation}
  \begin{split}    
    &  \hskip -.1 truecm \Tr(\Yop(t)= \parton{   { ac(|\lambda_1|+|\lambda_2|) \over 2|\lambda_1|\,\,|\lambda_2|} 
    - {2b^2 \over |\lambda_1|+|\lambda_2| }  }\,\Bigl( 1+O(\eta)\Bigr)\\ \\
   &  \hskip -.1 truecm \det(\Yop(t) =\parton{   {ac\over 4 |\lambda_1 |\, |\lambda_2| }
      -{b^2\over(|\lambda_1|+ |\lambda_2|)^2 }  }     \Bigl(  1+ O(\eta )  \Bigr) 
    \label{eq_AIV_25}
    \end{split}
\end{equation} 
where $\eta=e^{-2|\lambda_1|t}$.
\\
\\
 From the previous asymptotic behaviour
it follows that   $(2t)^{-1}\log I^{(j)}(t)$
converges to $\lambda_1$ for $j=1$  and $\lambda_1+\lambda_2$ for $j=2$ if $\lambda_1>\lambda_2>0$.
It converges to   $\lambda_1$ for $j=1,2$  if $\lambda_1>0>\lambda_2$
and converges to 0  for $j=1,2$  if   $0>\lambda_1>\lambda_2$. 
In all cases the  convergence rate is $1/t$   but a  linear interpolation in $1/t$   renders the convergence
exponentially fast. 
\spazio
  {\bf Complex  eigenvalues $\lambda\pm i\,\omega$:  focus}
\\
\\
    \def\Xopbar{\Xop}
    \def\Yopbar{\Yop}
    The fixed point is a focus and  evolution matrix is  $\Lop(t)=e^{\lambda t} \Top \Rop(\omega t)\Top^{-1}$.
    The Lyapunov matrix is conjugated
    to $\Xop(t)= e^{\lambda t}\,\Vop^{-1}\Rop(-\omega t)    \Vop\,\Rop(\omega t)$.
    The matrix $\Yop(t)$  given by  the integral of $\Xop(t)$ is simply related to the  BF
    Reversibility error covariance matrix.
    Letting $C(t)= \cos(\omega t)$ and
    $S(t)=\cos(\omega t)$  we write the rotation matrix as  $\Rop(\omega t)=\Iop C(t)+\Jop\,S(t)$
    and  $\Xop(t)$  takes  the following form according to (\ref{eq_5_5})
\begin{equation}
 \begin{split} 
    & \Xop(t)= 
   e^{2\lambda t}\,\Bigl( \Iop + \Uop
   \bigl(-\Iop+C_2(t)\Iop +S_2(t) \Jop\bigr)\,\Bigr) \\  
   & \Uop = {1\over 2}\bigl(\Iop+\Vop^{-1}\Jop\Vop\Jop \bigr)
\end{split}
\label{eq_AIV_26}
\end{equation}  
where $C_2(t)=\cos(2\omega t)$ and $S_2(t)=\sin(2\omega t)$.
Using the expression for the trace and the determinant of $\Uop$
quoted in  (\ref{eq_5_5})  the trace and the determinant of $\Xop(t)$ are given by 
\begin{equation}
 \begin{split} 
   \Tr(\Xop(t)) = & 2 e^{2\lambda t} \Bigl[ \parton{a+c\over 2}^2 - \\
   & - \parton{\parton{a+c\over 2}^2 -1 \Bigl]
     \cos(2\omega t)} \\ 
    \det (\Xop(t))=  & e^{4\lambda t}
\end{split}
\label{eq_AIV_27}
\end{equation}  
Integrating $\Xopbar(t')$ from 0 to $t$  over $t'$ we obtain $\Yop(t)$.  
\begin{equation}
 \begin{split} 
   \Yop(t)
  & =  E_I(t) \biggl( \Iop  + \Uop\Bigl (-\Iop   +  {C_I(t)\over E_I(t)}
  \,\Iop +\,{S_I(t)\over E_I(t)} \,\Jop \Bigr) \,\bigg)
\end{split}
\label{eq_AIV_28}
\end{equation}  
where $E_I(t)$, $C_I(t)$ and $S_I(t)$ denote the integrals of  $e^{2\lambda t'}$, 
$e^{2\lambda t'}\, C_2(t')$ and $e^{2\lambda t'}\, S_2(t')$
from 0 to $t$.   The trace  of $\Yop(t)$ is  given by
\begin{equation}
 \begin{split} 
   &  \Tr(\Yop(t)) =
   2E_I\, \parqua{ \parton{a+c\over 2}^2 \,+\,{C_I\over E_I} \parton{1-\parton{a+c\over 2}^2 } }
\end{split}
\label{eq_AIV_29}
\end{equation}  
The  determinant  of $\Yop(t)$ is  obtained by using $\det(\Iop+\Aop)= 1+\Tr(\Aop) +\det(\Aop)$ and
$\det(-\Iop+\Aop)= 1-\Tr(\Aop) +\det(\Aop)$. The result is 
\begin{equation}
 \begin{split} 
   &  \det(\Yop(t)) =E_I^2(t) \biggl[ 1+ \Tr(\Uop) \parton{ {C_I\over E_I}-1} + \\ \\
   & \qquad +\det(\Uop) \parton{   1+ 
   \det\parton{  {C_I\over E_I}\Iop +  {S_I\over E_I}\Jop}     -2   {C_I\over E_I}  } \biggr]   \\  \\
 &\qquad  = E_I^2(t) \biggl[ 1+\det(\Uop) \parton{  {C_I^2+S_I^2\over E_I^2}-1 } \biggr]
   \end{split}
\label{eq_AIV_30}    
\end{equation}  
wehere we took into account $\Tr(\Uop)= 2 \det(\Uop)= 2\Bigl( 1-(a+c)^2/4\Bigr)$.
If the fixed point is an unstablefocus   $\lambda>0$ the invariants have an exponential
growth $I^{(j)}(t)\simeq^{2j\,\lambda t}$.
More explicitly we  have $\Tr(\Yop)= e^{2\lambda \,t}\,  (a+c)^2/(4\lambda) \bigl( 1+O(e{-2\lambda t}\bigr)$
and $\det(\Yop)= e^{4\lambda \,t}\,  (a+c)^2/(16\lambda^2) \bigl( 1+O(e{-2\lambda t}\bigr)$.
If the fixed point  is  a  center
$\lambda=0$ the invariants  have  a power law growth $I^{(j)}(t)\simeq t^{j}$.
If the fixed points  is a stable  attracting focus $\lambda<0$  then the invariants $I^{(j)}(t)$
reach a constant limit as $t\to +\infty$.
\subsection{Limits for Lyapunov,  RF and FB   error matrices in \texorpdfstring{$\Reali^2$}{TEXT} }   
  \def\lambdab{\overline {\lambda}}
  The asymptotic estimates  follow  from the estimates obtained for the matrix $\Yop$.
  We recall that the invariants  of $\Sigma^2_{R}(t)$ are the same as for $\Yop(t)$
  where we exchange $\Vop$ with $\Vop^{-1}$ and $\lambda_j$ with $-\lambda_j$ if
  the the eigenvalues of $\Aop$ are real and $\lambda $ with $-\lambda$ if
  they are complex conjugate ($\lambda\pm i\omega)$.
  Since  the invariants of $\Yop(t)$   depend only on $ac =\Vop_{11}\,\Vop_{11}^{-1}$
  the invariants of  $\Sigma^2_{R}(t)$  are the same as for  $\Yop$ after changing
  $\lambda_j$ into $-\lambda_j$ if the eigenvalues are real and  $\lambda$ into $-\lambda$
  if they are complex conjugate.
  \\
  We summarize the results  for the Lyapunov and Reversibility error  invariants
  for the different stability conditions of the fixed point.
  \spa
  {\bf  Unstable node   }
          \spa
          The eigenvalues of $\Aop$ are  real with $\lambda_1>\lambda_2>0$. The asymptotic 
          limits   are 
      \begin{equation}
  \begin{split}    
  &  \lim_{\to \infty}\,(2t)^{-1}\,\log I^{(j)}(t)= \\
    & = \left\{\begin{matrix}  L &  F  & BF   & \phantom{\int} \\ 
      \lambda_1    &        \lambda_1  &    0    &  \;j=1    \phantom{\biggr)}  \\
      \lambda_1+ \lambda_2   &  \lambda_1+\lambda_2   & 0  &  j=2 
    \end{matrix} \right .
    \label{eq_AIV_31}    
    \end{split}
\end{equation}  
      \spa
      {\bf Saddle  }
          \spa
          The eigenvalues of $\Aop$ are real $\lambda_1>0>\lambda_2 $ . The asymptotic limits   are 
      \begin{equation}
  \begin{split}    
   & \lim_{\to \infty}\,(2t)^{-1}\,\log I^{(j)}(t)= \\
    & = \left\{\begin{matrix}  L &  F  & \quad BF   & \phantom{\int} \\ 
      \lambda_1    &     \lambda_1    & \quad |\lambda_2|        &  \;j=1    \phantom{\biggr)}  \\
      \lambda_1 + \lambda_2  & \lambda_1    & \quad |\lambda_2|    &   j=2 
    \end{matrix} \right .
    \label{eq_AIV_32}
    \end{split}
\end{equation}  
      \spa
      {\bf Stable node}
      \spa
      The eigenvalues of $\Aop$ are  all negative $ 0>\lambda_1>\lambda_2$ and the asymptotic limits
      of $(2t)^{-1}\,\log I^{(j)}(t)$ for the Lyapunov, BD and FB reversibility error matrices  are  
      \begin{equation}
  \begin{split}    
   & \lim_{\to \infty}\,(2t)^{-1}\,\log I^{(j)}(t)= \\
   & = \left\{\begin{matrix}  L &  F  & \quad  BF  & \phantom{\int} \\ 
      \lambda_1    &   0 &  \quad |\lambda_2|  \;   &   \;j=1    \phantom{\biggr)}  \\
      \lambda_1+\lambda_2  &  0 & \quad |\lambda_2| +|\lambda_1|   &  j=2 
    \end{matrix} \right .
    \label{eq_AIV_33}
    \end{split}
      \end{equation}
\\
{\bf  Unstable focus   }
          \spa
          The eigenvalues of $\Aop$ are  complex  $\lambda\pm i\omega $  with $\lambda >0$. The asymptotic
          limits   are 
      \begin{equation}
  \begin{split}    
    \lim_{\to \infty}\,(2t)^{-1}\,\log I^{(j)}(t)= \left\{\begin{matrix}  L & \quad  F  & \quad  BF  & \phantom{\int} \\ 
      \lambda    &   \quad   \lambda  &\quad  0   &  \;j=1    \phantom{\biggr)}  \\
      2\lambda   &   \quad  2\lambda  & \quad  0   &  j=2 
    \end{matrix} \right .
    \label{eq_AIV_34}
    \end{split}
\end{equation}  
      \spazio
          {\bf Stable focus }
          \spa
          The eigenvalues of $\Aop$ are   $ \lambda\pm i\omega $ with $\lambda<0$. The asymptotic limits   are 
      \begin{equation}
  \begin{split}    
    \lim_{\to \infty}\,(2t)^{-1}\,\log I^{(j)}(t)= \left\{\begin{matrix}  L & \quad  F  &\quad  BF   & \phantom{\int} \\ 
      \lambda    & \quad   0 & \quad |\lambda| \;      &  \;j=1    \phantom{\biggr)}  \\
      2\lambda   &  \quad 0 &  \quad 2|\lambda|         &  j=2 
    \end{matrix} \right .
    \label{eq_AIV_35}
    \end{split}
\end{equation}  
      
  %
  %
  %
\subsection{Invariants  for linear systems in \texorpdfstring{$\Reali^d$}{TEXT}}     
We compute now the asymptotic behaviour of the  Lyapunov error invariants
for arbitrary dimension $d$ with real and simple eigenvalues  given by
$\Lambda=\diag(\lambda_1,\ldots,\lambda_d)$.
Since  $\Lop=\Top e^{\Lambda t} \Top^{-1}$
and $\Lop^T\Lop= \Top \,\Xop\,\Top^{-1}$ where $\Vop=\Top^T\Top$
the invariants of $\Lop^T\Lop$ and $\Xop$ are the same.
We consider   matrix $\Yop$ defined by the  integral
$\Yop(t)=\int_0^t \,\Xop(t')\,dt'$ to which the Reversibility error covariance
matrix is  conjugated with simple changes previously described. 
\\
The entries $\lambda_i$  of the diagonal matrix 
$\Lambda$ are ordered $\lambda_1>\lambda_2>\ldots>\lambda_d$  and the matrices $\Xop(t)$ and
$\Yop(t)$ can be written as 
\def\Vopm{\Vop^{-1}}
\begin{equation}
  \begin{split}
    \Xop_{ij}(t)&= \sum_{k=1}^d \,\Vopm_{ik}\,\Vop_{kj}\,I_{kj}(t)  \quad \qquad  I_{kj}(t) =e^{(\lambda_k+\lambda_j)t} \\ \\
    \Yop_{ij}(t)&=\sum_{k=1}^d \,\Vopm_{ik}\,\Vop_{kj}\,I_{kj}(t)\qquad \quad
    I_{kj} (t)={e^{(\lambda_k+\lambda_j)t} -1\over \lambda_k+\lambda_j}
  \end{split}
    \label{eq_AIV_36}
\end{equation}   
The matrices $\Xop$ and $\Yop$ have the same expression, only the definition of $I_{kj}$ changes.
Notice the $I_{kj}(t)$ for $\Xop$ diverge exponentially if $\lambda_i+\lambda_j>0$, converge 
exponentially fast to 0 if  $\lambda_i+\lambda_j<0$
and  are equal to $t$  in the limit  $\lambda_i+\lambda_j\to0$.
The   $I_{kj}(t)$ defined  for $\Yop$ diverge exponentially  for $\lambda_i+\lambda_j>0$, converge
exponentially fast to   $1/(|\lambda_i+\lambda_j|)$ if $\lambda_i+\lambda_j<0$ and diverge as $t$ if
the limit $\lambda_i+\lambda_j\to 0$ is taken.
\\
\\
For comparison with the previous expression for $\Xop$ we write it for $d=2$
\begin{equation}
  \begin{split}    
 & \Xop= \\ 
 & = \begin{pmatrix}  \Vopm_{11}\Vop_{11}\,I_{11}+ \Vopm_{12}\Vop_{21}\,I_{21}  & \quad
    \Vopm_{11}\Vop_{12}\,I_{12}+ \Vopm_{12}\Vop_{22}\,I_{22} \\ \\
     \Vopm_{21}\Vop_{11}\,I_{11}+ \Vopm_{22}\Vop_{21}\,I_{21}  & \quad
    \Vopm_{21}\Vop_{12}\,I_{12}+ \Vopm_{22}\Vop_{22}\,I_{22} 
       \end {pmatrix} 
  \end{split}
\label{eq_AIV_37}
\end{equation}   
Notice that $I_{21}=I_{12}$ and that $\Vop,\,\Vop^{-1}$ are symmetric.
By comparing with (\ref{eq_AIV_5}) we have  $\Vop_{11}=a,\;\Vop_{12}=\Vop_{21}=b,\;\Vop_{22}=c$
and  $\Vopm_{11}=c,\;\Vopm_{12}=\Vopm_{21}=-b,\;\Vopm_{22}=a$. and  (\ref{eq_AIV_7}) corresponds to (\ref{eq_AIV_37}).
\\
\\
  {\bf First invariant   of   $\Xop(t)$  and  $\Yop(t)$} 
\\
\\
The asymptotic expression of   the first invariant  $I^{(1)}(\Xop)= \Tr(\Xop)$ for a linear flow in $\Reali^d$
is immediately  obtained. 
\begin{equation}
  \begin{split}
    I^{(1)}(\Xop) &= \Vopm_{11}\Vop_{11} \,e^{2\lambda_1 t} \,\,\bigl(1+O(\eta_{})\bigr) \qquad \eta_{}= e^{-t\,(\lambda_1-\lambda_2)} 
  \end{split}
 \label{eq_AIV_38}
\end{equation}    
In a similar way for  the first invariant of $\Yop(t)$ we obtain 
\def\dis{\displaystyle }
\begin{equation}
  \begin{split}
    \hskip -.4 truecm  I^{(1)}(\Yop) =\begin{cases} {\dis \Vopm_{11}\,\Vop_{11}\over \dis 2\lambda_1}
    \,e^{2\lambda_1 t} \,\,\bigl(1+O(\eta_{})\bigr)
    \quad \hbox{if}\quad \lambda_1>\lambda_2>0   \phantom{{1\over \dis\Bigr)}}\\  
 {\dis \Vopm_{11}\,\Vop_{11}\over \dis 2\lambda_1  } \,e^{ 2\lambda_1 t} \,\,\bigl(1+ O(e^{-(\alpha+\lambda_1) t})\bigr)
         \phantom{\biggr)} \;\;   \hbox{if}\;\; \lambda_1>0>\lambda_2 \\  
      \sum_{i,k=1}^d \, {\dis \Vopm_{ik}\Vop_{ki}\over \dis  |\lambda_i|+|\lambda_k|}  \,\,
      \bigl(1+O(e^{-2t|\lambda_1|})\bigr) \quad \hbox{if}\quad   0>\lambda_1
    \end{cases}
    \end{split}
   \label{eq_AIV_39}    
\end{equation}
where $\alpha=\min(\lambda_1,\,|\lambda_2|)$.
%
\\
\\
  {\bf Second invariant   of   $\Xop(t)$ and $\Yop(t)$}
\\
\\
We recall that the second invariant of $\Xop$ is defined by
\begin{equation}
  \begin{split}
    I^{(2)}(\Xop) ={1\over 2}\,\bigl( \Tr^2(\Xop)-\Tr(\Xop^2)\bigr )= \sum_{1\le i<j}^d \,\,\bigl(\Xop_{ii}\,\Xop_{jj}-
    \Xop_{ij}\,\Xop_{ji}       \bigr)
    \end{split}
   \label{eq_AIV_40}    
\end{equation}    
The leading term in the sum is $\Xop_{11}\,\Xop_{22}-    \Xop_{12}\,\Xop_{21}$ and we show it is proportional
to  $I_{12}^2$.  To this end we compute first the four factors 
\begin{equation}
  \begin{split}
    \Xop_{11} & =  \Vopm_{11}\Vop_{11} \,I_{11}  +  \Vopm_{12}\Vop_{12} \,I_{12} +  O(I_{13}) \\ 
    \Xop_{12} & =   \Vopm_{11}\Vop_{12} \,I_{12}  +  \Vopm_{12}\Vop_{22} \,I_{22}   +O(I_{23}  )         \\ 
    \Xop_{21} & =    \Vopm_{21}\Vop_{11} \,I_{11}  +  \Vopm_{22}\Vop_{21} \,I_{21}  +  O(I_{13})      \\ 
    \Xop_{22} & =    \Vopm_{21}\Vop_{12} \,I_{12}  +  \Vopm_{22}\Vop_{22} \,I_{22}  +  O(I_{23}  )           \\ 
    \end{split}
   \label{eq_AIV_41}    
\end{equation}   
The evaluation of the leading term leads to 
\begin{equation}
  \begin{split}
    &  \Xop_{11}\,\Xop_{22}-    \Xop_{12}\,\Xop_{21}  =
    c_{12}\,I_{12}^2   +O(I_{12}\,I_{13}) + O(I_{12}I_{22}) \\ \\ 
   &  c_{12}= \bigl( \Vopm_{11} \Vopm_{22}  - (\Vopm_{12})^2\bigr)
    \,\bigl( \Vop_{11} \Vop_{22}\,-
    \Vop_{12}^2\bigr)
    \end{split}
   \label{eq_AIV_42}    
\end{equation}   
where we have taken into account $\Vop$ and its inverse are symmetric, 
that the largest  terms  $I_{11}I_{12}$
cancel and that  $ I_{11}I_{22}= I_{12}^2$. In addition
we have used the identity  $ I_{11}I_{23}= I_{12}I_{13}$.  The result can be written by
expressing the remainder in a different form
\begin{equation}
  \begin{split}
    & \Xop_{11}\,\Xop_{22}-    \Xop_{12}\,\Xop_{21}  = c_{12}\,\,e^{2t(\lambda_1+\lambda_2)}\, 
    \Bigl(1+ O(\eta_{12}) + O(\eta_{23})\Bigr ) \\ \\
    & \eta_{ij}= e^{-(\lambda_i-\lambda_j)t} 
    \end{split}
   \label{eq_AIV_43}    
\end{equation}    
Next, we evaluate $ \Xop_{11}\,\Xop_{jj}-\Xop_{1j}\,\Xop_{j1}$ for $j\ge 3$
\begin{equation}
  \begin{split}
    & \Xop_{11}=\Vopm_{11} \Vop_{11}\,I_{11}  +O(I_{12})  \\
    & \Xop_{jj}=  \Vopm_{j1} \Vop_{1j}\,I_{1j}\,+O(I_{2j}) \\
    &  \Xop_{1j}=  \Vopm_{11} \Vop_{1j}\,I_{1j}\,+O(I_{2j}) \\
    & \Xop_{j1}=  \Vopm_{j1} \Vop_{11}\,I_{11}\,+O(I_{12}) 
    \qquad  
    \end{split}
   \label{eq_AIV_44}    
\end{equation}  
Since $j\ge 3$ we find that 
\begin{equation}
  \begin{split}
    & \Xop_{11}\,\Xop_{jj}-    \Xop_{1j}\,\Xop_{j1}  =  O(I_{12}\,I_{1j}) 
    \end{split}
   \label{eq_AIV_45}    
\end{equation}   
where we took into account $I_{11}\,I_{2j}= I_{12}\,I_{1j}$.  Furthermore since $I_{1j}\le I_{13}$, we  can replace $ O(I_{12}\,I_{1j})$ with   $ O(I_{12}\,I_{13})$ for any $j\ge 3$.
\\
Finally we compute $\Xop_{ii}\,\Xop_{jj}-    \Xop_{ij}\,\Xop_{ji}$ for $2\le i<j$. 
Since $j\ge 3$ we find that 
\begin{equation}
  \begin{split}
   & \Xop_{ii}=  \Vopm_{i1} \Vop_{1i}\,I_{1i}\,+O(I_{2i}) \\
   & \Xop_{jj}=  \Vopm_{j1} \Vop_{1j}\,I_{1j}\,+O(I_{2j})  \\  
   &  \Xop_{ij}=  \Vopm_{i1} \Vop_{1j}\,I_{1j}\,+O(I_{2j})  \\
   & \Xop_{ji}=  \Vopm_{j1} \Vop_{1i}\,I_{1i}\,+O(I_{2i})
    \end{split}
   \label{eq_AIV_46}    
\end{equation}     
and  the result is 
\begin{equation}
  \begin{split}
  \Xop_{ii}\,\Xop_{jj}-    \Xop_{ij}\,\Xop_{ji}  = O(I_{12}\,I_{ij})  \qquad 2\le i<j
    \end{split}
   \label{eq_AIV_47}    
\end{equation}    
where we took into account $ I_{1i}\,I_{2j} = I_{1j}I_{2i}= I_{12}I_{ij}$.
Furthermore since $I_{1j}\le I_{13}$ we can replace  $O(I_{1i}\,I_{2j})$ with $ O(I_{12}I_{23})$. 
 for any $j\ge 3$. 
\\
\\
  {\bf Asymptotic behaviour of  $I^{(2)}(\Xop)$ }
  \\
  \\
  The  second invariant   has a  the following expression where   rate at which the
  asymptotic limit is reached is is explicitly written 
\begin{equation}
  \begin{split}
    & I^{(2)}(\Xop)  =  c_{12}\,I_{12}^2 + O(I_{12}\,I_{13}) + O(I_{12}I_{22}) = \\  \\
   & \;= c_{12}\,\,e^{2t(\lambda_1+\lambda_2)}\,\,
   \bigl(1+ O(\eta_{12}) + O(\eta_{23}) \bigr) 
    \end{split}
   \label{eq_AIV_48}    
\end{equation}   
As a consequence if $\lambda_1+\lambda_2$ is postive  or negative the invariant has an exponential growth
or decrease.
For $d=2$  we  find that  $c_{12}=  \det(\Vop^{-1})\,\det(\Vop)=1$. However for $d>2$
we have $c_{12}\not =1 $  so that $(2t)^{-1}\,\log(I^{(2)}(t)) = \lambda_1+\lambda_2 +O(1/t)$.
\\
\\
For any $d$ we have   $I^{(d)}(t)=  \det(\Xop(t))= e^{2(\lambda_1+\lambda_2+\ldots+\lambda_d)}(t)$.
This result follows  immediately  from the definition  $\Xop=\Vop^{-1}e^{\Lambda t}\Vop e^{\Lambda t}$
\\
\\
The  asymptotic behaviour of the higher order invariants $I^{(j)}(t)$ of $\Xop(t)$
and $\Lop^T(t)\Lop(t)$ for $j\ge 3$ is  expected to be 
\begin{equation}
  \begin{split}
     &  I^{(j)}(t)  = c_j \,e^{(\lambda_1+\ldots+\lambda_j)\,t}\,\Bigl(1+O(\eta) \Bigr) \\
  &  \eta =\exp\Bigl(- t\,\min_{1\le i<j} (\lambda_i-\lambda_j)\Bigr) 
    \end{split}
   \label{eq_AIV_49}    
\end{equation}   
where the constant $c_j$  has been  explicitly  determined for $j=1,2$.
The result might be confirmed numerically by generating randomly the coefficients
of the positive matrix $\Vop$.
The presence of the positive   constant $c_j$ 
renders  the convergence of $(2t)^{-1}\log I^{(j)}(t)$ to $\lambda_1+\ldots+\lambda_j$  linear in
$1/t$   but the remainder converges exponentially fast to 0. This observation  should be taken
into account also in the computation of the Lyapunov exponents.
\\
\\
  {\bf Asymptotic behaviour of  forward and reversibility error invariants } 
  \\
  \\
  A detailed analysis has been developed for the the $d=2$ case, for which  the asymptotic behaviour of
  $\Yop(t)$ has been obtained.
  The Forward error reversibility covariance  matrix $\Sigma^2_F(t)$ is conjugated to the the matrix $\Yop(t)$
  defined as the integral of $\Xop(t')=\Vop^{-1}e^{\Lambda t'}\Vop e^{\Lambda t'} $ in $[0,t]$. 
  The BF   Reversibility error  covariance matrix $\Sigma^2_{BF}(t)$  is  conjugated to  the
   matrix obtained from $\Yop$ by interchanging $\Vop$ and $\Vop^{-1}$ and changing the diagonal matrix 
   $\Lambda$   into $-\Lambda$.   We denote as usual with $n_+$ the first
  positive  entries and $n_-=d-n_+$ the last negative entries  of $\Lambda$.
  The forward error invariants   $I^{(k)}_F(t)$  of $\Sigma^2_F(t)$ are the same as the
  invariants of $\Yop(t)$  and are expected to be given by
\begin{equation}
   \begin{split}
     &I^{(k)}_F(t)  = c_k \,e^{2(\lambda_1+\ldots+\lambda_k) t} (1+O(\eta))  \qquad   \hbox{if}\quad k\le n_+ \\ \\
     & I^{(k)}_F(t) = c_k \,e^{2(\lambda_1+\ldots+\lambda_{n_+}) t} (1+O(\eta))
     \quad  \hbox{if}\quad n_++1\le k\le d \\ \\
     & I^{(k)}_F(t)  = c_k \,(1+O(\eta))   \qquad \qquad   \hbox{if}\quad  n_+=0 \quad  
  \end{split}
   \label{eq_AIV_50}    
\end{equation}  
where $c_k>0$ are positive constants and  $\eta(t)=e^{-\alpha t}$  where $\alpha=\min_{i<j} (\lambda_i-\lambda_j)$.
Even in this case the result quoted above can be supported  by numerical computations.
\\
\\
The asymptotic behaviour of the BF  Reversibility error invariants of the covariance matrix $\Sigma^2_F$      
is immediately obtained by interchanging $\Vop$ and $\Vop^{-1}$,  which
changes just the value of the  constants $c_k$, and $\Lambda$ into $-\Lambda$,
that is the sign of all the exponents.  As a consequence the asymptotic behaviour of the
invariants $I^{(k)}_{BF}(t)$ of $\Sigma^2_{BF}(t)$  is  expected to be given by 
\begin{equation}
   \begin{split}
     & I^{(k)}_{BF}(t)  = c_k \,e^{2(|\lambda_d|+\ldots+|\lambda_{d-k+1}|) t} (1+O(\eta))  \qquad   \hbox{if}\quad
      1\le  k\le n_- \\ \\    
     & I^{(k)}_{BF}(t) = c_k \,e^{2(\lambda_d+\ldots+\lambda_{n_++1}) t} (1+O(\eta)) \quad  \hbox{if}\quad n_-+1\le k\le d \\ \\ 
     & I^{(k)}_{BF}(t)  = c_k \,(1+O(\eta))   \qquad 1\le k\le d \qquad   \hbox{if}\quad  n_-=0  
  \end{split}
   \label{eq_AIV_51}    
\end{equation}  
The FB Reversibility error invariants are expected to have  the same  asymptotic behaviour
as the Forward error invariants.
\\
\\
In the nolinear case it is much more difficult to determine the asymptotic behaviour of the
Lyapunov,  Forward and BF Reversibility error invariants. Our analysis is confined to 
one dimensional systems such as the logistic equation, see Appendix V and to integrable Hamiltonian
systems on $\Reali^2$, see appendix VI. 
\clearpage 
\newpage

\section {Appendix V.  The logistic equation example}
\spa
We  consider    a specific one dimensional example, the logistic equation,
for which the Lyapunov and Reversibility error can be analytically computed.
As a consequence their asymptotic behaviour  is obtained  and 
the validity conditions  of the linear response is specified.
The logistic equation and its solution read 
\begin{equation}
  \begin{split}
& {d\over dt}x(t)= x(t)(1-x(t)) \\
& x(t)\equiv S_t(x)= {x \over x+(1-x)e^{-t}}
\end{split}
 \label{eq_AV_1}
\end{equation}
\spa
The solution for $x\ge 0$  is bounded  for any  $t\ge 0$  and for $x>0$
asymptotically
converges to the stable fixed point $x=1$. For $x<0$ the solution diverges
at a    finite  time $t_*= \log(1+|x|)$.  
We compute  first the Lyapunov, Forward and BF Reversibility errors in order to analyze
their asymptotic behaviour for $t\to +\infty$.  The Lyapunov error is defined by 
\begin{equation}
  \begin{split}
    E_L(x,t)= |L(x,t)|=\parbar{\derp{}{x}\,S_t(x) }= {e^{-t} \over (x+(1-x)e^{-t})^2}
\end{split}
 \label{eq_AV_2}
\end{equation}
The Lyapunov error  can be written as  $E_L(x,t)= e^{t\,\lambda(x,t)}$ where   the function $\lambda(x,t)$
is asymptotically for $t\to+\infty$   equal to the Lyapunov exponent $\lambda(x)$.
For a generic one dimensional
system we can write 
\begin{equation}
  \begin{split}
   \log L(x,t) \equiv   t\lambda(x,t)=  t \lambda(x)+  \lambda_1(x) + \eta(x,t)
\end{split}
 \label{eq_AV_3}
\end{equation}
where $\eta(x,t)$ converges to zero exponentially fast as $e^{-t}$. As a consequence, we have 
\begin{equation}
  \begin{split}
   &  {1\over t}\log L(x,t) \equiv   \lambda(x) + O(t^{-1}) \\  \\
   &  \derp{}{t} \log L(x,t) = {1\over L(x,t)}\,\derp{}{t}\log(x,t)= \lambda(x) + O(\eta)
\end{split}
 \label{eq_AV_4}
\end{equation}
In the specific case of the logistic equation, we have 
\begin{equation}
  \begin{split}
 &   \log L(x,t)  = -t -2\log x -2 \log\parton{1+{1-x\over x}e^{-t}} \simeq   \\ \\
   & \quad  \simeq -t -2\log x -2 {1-x\over x} e^{-t} \qquad \hbox{for} \quad t\gg \log{1\over x}
\end{split}
 \label{eq_AV_5}       
\end{equation}
The Lyapunov exponent $\lambda(x)$ is a discontinuous function equal to $-1$ for any $x>0$
and equal to 1 for $x=0$.  The function  $t^{-1}\,\log L(x,t)$  converges to $\lambda(x)$
as $t^{-1}$ whereas  the  time derivative of  $\log L(x,t)$ converges to $\lambda(x)$
as $e^{-t}$. 
\begin{figure}[!ht]
\centerline{
\includegraphics[width=3.5 cm, height=3.5 cm]{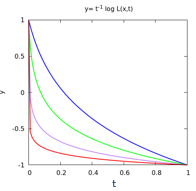}        
\includegraphics[width=4 cm, height=3.5 cm]{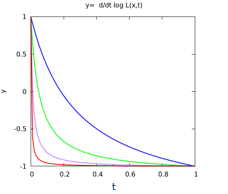}    }
\caption{Left side: plot of $t^{-1}\,\log L(x,t)$ for the logistic equation at different times $t=2.5$ blue line,
  $t=5$ green  line,   $t=10$ purple line, $t=20$ red line.
Right  side: plot of $d/dt \,\log L(x,t)$ for the logistic equation at different times $t=1.5$ blue line,
$t=3$ green  line,   $t=4.5$ purple line, $t=6$ red line.}
\label{fig:fig_3} 
\end{figure}

  \subsection   { Remainder of  the linear approximation to \texorpdfstring{$S_t(x+\eps)$}{TEXT} }
\spa
We write the  series expansion of $S_(x+\eps)$ as the linear term plus a remainder 
\begin{equation}
  \begin{split}
    S_t(x+\eps)&= S_t(x) + \eps \,L_1(x,t) + \eps^2 R(x,t;\eps) \\
    R(x,t;\eps)&= \sum_{n=2}^\infty\,{\eps^{n-2}\over n!}\,L_n(x,t)  \\ 
L_n(x,t)& = {\partial^n\over \partial x^n}S(x,t)= (-1)^{n+1}\, n! {e^{-t}\,(1-e^{-t})^{n-1} \over (x+(1-x)e^{-t})^{n+1}}
\end{split}
 \label{eq_AV_6}
\end{equation}
Notice that $\eps$ can be positive or negative.
A straightforward calculation shows that the remainder $R$ is given by 
\begin{equation}
  \begin{split}
  R(x,t;\eps= -L(x,t)\, {\alpha\over 1+\alpha \eps} \qquad \qquad \alpha= {1-e^{-t}\over x+(1-x)e^{-t}}
\end{split}
 \label{eq_AV_7}
\end{equation}
Observing thet $\alpha>0$ and $x,\,t,$   are positive  the ratio $\rho$ between the  remainder
and the linear approximation can be written as
\begin{equation}
  \begin{split}
    {\eps^2 R(x,t;\eps)\over  \eps L(x,t) }  = -{ \eps\alpha \over 1+\eps\alpha}=
    -{\eps\over \eps-\eps_c}
\end{split}
 \label{eq_AV_8}
\end{equation}
taking into account (\ref{eq_AII_11}). Here   we  have set $L=L_1$.
Since $ \alpha $ is strictly postive on $\Reali_+$ we have
\begin{equation}
  \begin{split}
    r(x,t)=|eps| {|R(x,t;\eps)|\over  |L(x,t)| }  \le {|\eps| \alpha \over 1-|\eps|\,\alpha} 
    \end{split}
 \label{eq_AV_9}
\end{equation}  
The estimate  valid  for any $t>0$  with  $x\ge \ell$ and $|\eps|\ll \ell\ll1$
has been previously obtained  in (\ref{eq_AII_11}). 
When  $x>1$ we have $S_t(x)\le x$ and therefore $r \le \eps$.
When $0\le x\le 1$  we have $S_t(x)\le 1$ so that $|r| \le \eps/x$
and with the restriction $x\ge \ell$
we can write $r\le \eps/\ell $ and the linear approximation holds for $\eps\ll \ell$.
Without this restriction  $x\ge \ell$  no  $t$ independent bound can be obtained. 
Indeed for $0\le x\le 1$ we have   $S_t(x)\le x e^t$ and finally we have  
\begin{equation}  
    |r(x,t)| \le  \begin{cases}  |\eps|    &  \qquad x   > 1 \\ \\
     { \dis |\eps|\over \dis \ell }  &  \qquad \ell \le x \le 1  \\ \\
              |\eps|  e^t   & \qquad 0\le x\le 1 \end{cases}
 \label{eq_AV_10}
\end{equation}
\spa
    \subsection {Random initial deviation} 
\spa
If the initial    deviation is stochastic  we  replace $\eps$ with  $\eps \xi$ where
now $\eps>0$   and $\xi$  is uniformly distributed in $[-\sqrt{3},\sqrt{3}]$
in order to have  zero mean and unit variance.
In this case   $r$ is obtained  by replacing $\eps$ with $\eps\xi$ and the 
estimate on $r$  is given by
\begin{equation}
  \begin{split}
    |r|  = & \eps |\xi|\,{\alpha\over |1+\eps\xi\alpha|}\le \sqrt{3} \eps \,{\alpha\over
      1-\sqrt{3}\,\eps \alpha}  \le  2 \sqrt{3}\,\eps \alpha    \\ \\
    &  \hbox{for}   \qquad \quad \eps \alpha \le 1/(2\sqrt{3})
\end{split}
 \label{eqF_AV_11}
\end{equation}
We  recall  that $\alpha=S_t(x)/x  <1 $ for $x>1$  and $\alpha\le 1/\ell$ for $\ell<x<1$.
As a consequence for $\eps\ll \ell\ll 1 $ the estimate on $|r|$ is the same as (\ref{eq_AV_10}) up to the multiplicative factor   $2 \sqrt{3}$.


%
\section{Appendix VI.  Lyapunov error  and separatrix}
\begin{figure}[!ht]
\centerline{
\includegraphics[width=4 cm, height=3.5 cm]{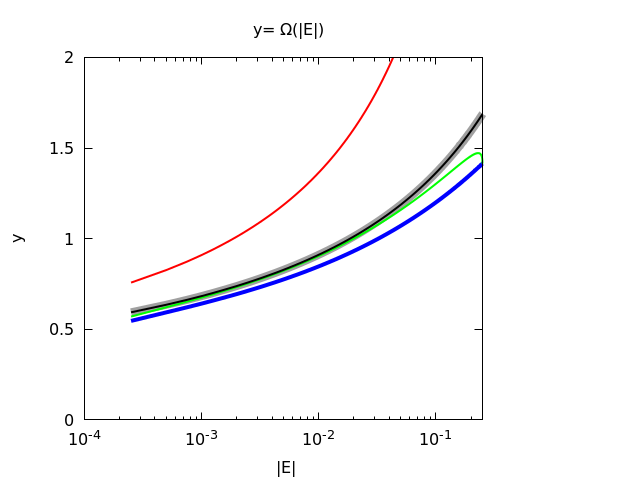}    
\includegraphics[width=4 cm, height=3.5 cm]{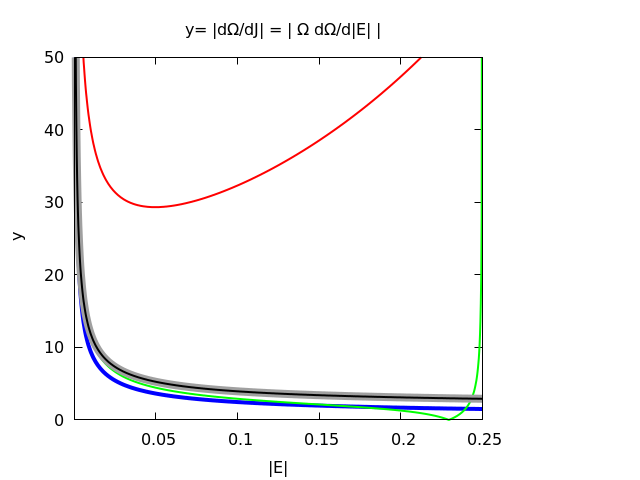}     }
%
%
\caption{Left side: plot of the  frequency $\Omega(|E|)$  for $E<0$ in the energy interval $0\le |E|\le 1/4$.
  Exact result (blue line) computed  accurately by the  trapezoidal rule with $\Delta E=0.25 1,10^{-3}$,
  approximation where in the integrand $4\sin^2(\phi/2)$ is replaced
  with $\phi^2$ (green line).  We show the result where the endpoints $x^2\pm$ are
  replaced with $|E|$ and $2-|E|$ respectively and  $4\sin^2(\phi/2)$ is replaced
  with $\phi^2$ (gray line). Finally the
  asymptotic  approximation   $\Omega= (2\pi)/\log(1/|E|))$   for $E\to 0$  (red line)
  is shown. Right side: the same plots for $|d\Omega/dJ|=
  |\Omega d\Omega/dE|$ where the different colors correspond to the same approximations as for the left side.  }
\label{fig:fig_sep} 
\end{figure}

We analyze the behaviour of the Lyapunov and reversibility error invariants
for an integrable system. As a model we choose the Duffing oscilator
whose Hamiltonian is given by
\begin{equation}
  \begin{split}
  H(x,p)= {p^2\over 2}-{x^2\over 2} +{x^4\over 4}
\end{split}
 \label{eq_AVI_1}
\end{equation}   
We first analyze the behavior of the  frequency $\Omega(E)$ as a function of the energy $E$
when $E\to 0$ that is approacing the separatrix in the region corresponding to the closed orbits
around the elliptic point $(1,0)$.
Each orbit in this region where $-1/4\le E<0$ has two inversion
points  $x_\mp$ which are the positive solutions  of the equation $x^4-2x^2-4E=0$
\begin{equation}
  \begin{split}
    x^2_- &= 1-(1+4E)^{1/2} =  2|E| +O(|E|^3) \\ \\
    x^2_+ &= 1+(1+4E)^{1/2} = 2- 2|E| +O(|E|^3)
\end{split}
 \label{eq_AVI_2}
\end{equation} 
From  $H(x,p)=E$ we obtain the equation for the orbit given by two symmetric arcs $p=\pm p(x)$ where 
\begin{equation}
  \begin{split}
  p(x) = {1  \over \sqrt{2}} \bigl( (x^2-x_-^2)(x_+^2-x^2) \bigr)^{1/2}\end{split}
 \label{eq_A_VI_3}
\end{equation} 
so that the period is given by
\begin{equation}
  \begin{split}
    & T= 2\int_{x_-}^{x_+}\,\,{dx\over p(x)} = 2 \sqrt{2} \, \int_{x_-}^{x_+}\, {1\over \sqrt{(x^2-x_-^2)(x_+^2-x^2) } }\,dx = \\ \\
     & = \sqrt{2} \,\,\int_0^\pi \,{d\phi\over x(\phi)}=
    \sqrt{2}\,\,\int_0^\pi \,\, {d\phi\over \sqrt{x_-^2+(x_+^2-x_-^2) \sin^2(\phi/2)} \,}= \\ \\
    & = 2\parton{x_+^2-x_-^2\over 2}^{-1/2}\,\, \int_0^\pi\,\,
    \parton{{4x_-^2\over x_+^2-x_-^2} +      4\sin^2(\phi/2) }^{-1/2}\,d\phi
  \end{split}
 \label{eq_AVI_4}
\end{equation} 
where we have changed the integration variable  by  parameterizing the orbit according to
$x(\phi)= x_-^2+(x_+^2-x_-^2) \sin^2(\phi/2)$. The integral can be evaluated
by observing that the integrand can
be written as $(a^2+4\sin^2(\phi/2)$ where $a^2= 4|E| +O(E^2)$
and that for $\eps\to 0$ the leading contribution comes from the integration in the region
$0\le \phi\le 2\sqrt{E}$ where we can  approximate $2\sin(\phi/2)$ with $\phi$. Making this
approximation in the  whole integration  region we obtain 
\begin{equation}
  \begin{split}
    T& \simeq  2\int_0^\pi \,\,{d\phi\over \sqrt{4|E|+\phi^2}} 
    = 2 \,\hbox{arcsh}\parton{\pi \over 2 \sqrt{|E|}}  =  \\ \\
   &  =\log{1\over |E|}+ O(1)
  \end{split}
 \label{eq_AVI_5}
\end{equation} 
We have used the asymptotic expansion of $\hbox{arcsh}(x)$ valid for large  $x>0$
\begin{equation}
  \begin{split}
    \hbox{arcsh}(x)= \log\Bigl(x+\sqrt{1+x^2} \Bigr)= \log(2x)+ {1\over 4x^2}+O(x^{-4})
  \end{split}
 \label{eq_AVI_6}
\end{equation} 
As a consequence for $x\gg 1$ we have   $\hbox{arcsh}(x)= \log x + O(1)$.
We  notice that the approximation obtained by replacing $4\sin^2(\phi/2)$ with $\phi^2$ in
(\ref{eq_AVI_4}) is rather accurate except close to $E=-1/4$. If we replace $x^2_\pm$
with their  first order expansion in $E$  in  (\ref{eq_AVI_4}) 
the result is almost as accurate  in the whole  energy interval $-1/4<E<0$.
The expression  $2 \,\hbox{arcsh}\bigl(\pi/ (2 \sqrt{|E| }\bigr)$  is
a good approximation to the period.
On the contrary the further approximation $\log(1/|E|)$ is valid only for $E$ very close to zero.
In general if we have a separarix whose energy is $E_{\hbix{sep}}$ the period close to 
the separatrix is given by $T\simeq \log(1/(E_{\hbix{sep}}-E))$ and the   frequency 
is given by  
\def\Esep{ E_{\hbix{sep}}}
\def\Jsep{ J_{\hbix{sep}}}
\begin{equation}
  \begin{split}
    \Omega(E)\simeq  {2\pi\over \log \bigl(\,( \Esep-E)^{-1})\,\bigr) }
  \end{split}
 \label{eq_AVI_7}
\end{equation} 
For the Duffing oscillator $\Esep=0$.
In order to evaluate the Lyapunov error when the separatrix is approached
we need the derivative of the frequency with respect to the action. As a consequence we have 
\begin{equation}
  \begin{split}
    { d\Omega\over dJ}= { d\Omega\over dE}\,{dE\over dJ} =\Omega(E) {d\Omega\over dE}
  \end{split}
 \label{eq_AVI_8}
\end{equation} 
In the case of the Duffing oscillator we obtain
\begin{equation}
  \begin{split}
   & { d\Omega\over dJ}= -{ \Omega^3(E) \over 2|E|\,\sqrt{4|E|+\pi^2 }  }
    \\
    & \Omega(E) ={\pi\over \hbox{arcsh}\parton{\dis \pi \over \dis 2 \sqrt{|E|}} }
  \end{split}
 \label{eq_AVI_9}
\end{equation} 
for $-1/4<E<0 $. Notice that  $\Omega(E)>0$ and decreases with $E$ vanishing  for  $E\to 0$.
Therefore, both $d\Omega/dE$ and  $d\Omega/dJ$ are negative and diverge as $E\to 0$
and $J\to \Jsep$.
In  figure \ref{fig:fig_sep}  we  plot  the frequency $\Omega(E)$ as a function of $|E|\in[0,1/4]$
and of $|d\Omega/dJ|$ which enters in the evaluation of  the Lyapunov error.
%
\spazio
We compute now the period beyond the separatrix,
where $E>0$ in this case we  can still write
$2p^2= (x_+^2-x^2)(x^2-x_-)^2 $ with $x_+^2 = 1+\sqrt{1+4E}\ge 2 $ and $x_-^2 = 1-\sqrt{1+4E}<0$.
As a consequence in this case the inversion points are $\pm x_+$.
Notice that as the energy $E$ approaches zero from positive values we gave $x_+^2\simeq 2$ and
$|x_-^2|=-x_-^2\simeq 2E$.  The period is given by 
\begin{equation}
  \begin{split}
    T(E)& = 2\,\int_{-x+}^{x_+} \, {dx\over p(x)}=  \\ 
    & = 2\sqrt{2}\,\int_{-x_+}^{x_+}
    {dx\over \sqrt{(x_+^2-x^2)(x^2-x_-^2)} }=\\ 
    &  =  4\sqrt{2}\,\,\int_0^{x_+}{\,dx\over \sqrt{(x_+^2-x^2)(x^2+ |x_-^2|)} } 
  \end{split}
 \label{eq_AVI_10}
\end{equation} 
Changing the variable according to $x=x_+\,\sin\phi$ where $0\le \phi\le \pi/2$ we  have 
$dx/\sqrt{x_+^2-x^2}= d\phi$ and the integral becomes  
\begin{equation}
  \begin{split}
    T(E)& =  \,{ 4\sqrt{2}\over x_+}\,\int_0^{\pi/2}\,
    \parton{{|x_-^2|\over x_+^2} + \sin^2 \phi}^{-1/2}\,
    d\phi 
\end{split}
 \label{eq_AVI_11}
\end{equation} 
We evaluate the integral for $E$ close to 0  so that $x_+^2\simeq 2$ and $|x_-^2|\simeq 2E$ so that 
\begin{equation}
  \begin{split}
    T(E)&  \simeq 4 \int_0^{\pi/2}\, {d\phi\over \sqrt{E+ \sin^2 \phi} }\simeq 
    4 \int_0^{\pi/2}\, {d\phi\over \sqrt{E +  \phi^2} } = \\ \\ 
  & = 4 \,\hbox{arcsh}\parton{\pi\over 2\sqrt{E}}
  \end{split}
 \label{eq_AVI_12}
\end{equation} 
We have repalced $\sin\phi$ with $\phi$ since the leading
contribution come in  from small values of $\phi$  when $E$ is close to 0.
Asymptotically  for $E\to 0$ we have $T(E)\simeq 2 \log(1/E)$.
Summarizing for the frequency  and its derivative with respect to $J$ 
the following approximation holds for $|E|$ small but, it can  be safely extended to any value of $E$ in $[-1/4,+\infty[$
\begin{equation}
  \begin{split}
    \Omega(E)& = \begin{cases}  \pi \,\parqua{\hbox{arcsh}\parton{\dis \pi\over \dis 2\sqrt{|E|}} }^{-1}\qquad -1/4\le E <0 \\ \\ 
      {\dis \pi\over \dis 2}\,\, \parqua{ \hbox{arcsh}\parton{\dis \pi\over \dis 2\sqrt{E}}  }^{-1}\qquad \quad  E>0 \end{cases}  \\ \\ 
    {d\Omega\over dJ}& = \begin{cases}  - {\dis \Omega^3(E)\over \dis  2|E|\,\sqrt{4|E|+\pi^2} }   \qquad -1/4\le E <0 \\ \\
         { \dis \Omega^3(E)\over \dis E\,\sqrt{4E+\pi^2} }  \qquad  \quad  E>0 \end{cases}   
  \end{split}
 \label{eq_AVI_13}
\end{equation} 
Notice that for $E<0$ the function $\Omega(E)$ is decreasing to 0 wheres for $E>0$ it is increasing.
As a consequence $d\Omega/dE= -d\Omega/d|E|<0$ for $E<0$ whereas $d\Omega/dE>0$  for $E>0$. 
Since $\Omega >0$ for any $E$ we have $d\Omega/dJ<0$ for  $E<0$ and
$d\Omega/dJ>0$ for $E>0$.
\\
We notice that approacing $E=0$ from below  the $\Omega(E)\simeq 2\pi/\log(1/|E|)$
whereas approacing $E=0$ from above
$\Omega(E)\simeq \pi/\log(1/|E|)$. As a consequence the asymptotic behaviour near $E=0$ anf $J=\Jsep$ is given by
\begin{equation}
  \begin{split}
    {d\Omega\over dJ} \simeq \begin{cases}  -{ \dis (2\pi)^2 \over \dis |E|\, \log^3(1/|E|) }  \qquad E<0\quad J<\Jsep \\ \\
   \phantom{-} { \dis \pi^2 \over  \dis  E\,\log^3(1/E) } \qquad E>0\
    \end{cases}
  \end{split}
 \label{eq_AVI_14}
\end{equation} 
We recall that the acton is discontinuous for $E=0$   since $J\to \Jsep$ for $E\to 0-$ whereas $J\to 2\Jsep$
for $E\to 0+$.
\subsection{Inverting \texorpdfstring{$J=J(E)$}{TEXT}  }
In  general in order to obtain $E=E(J)$  we have to  invert $J=J(E)$ which is obtained by
integrating  $dE/dJ= \Omega(E)$. The result ca be easily obtained when the  approximation
to $\Omega(E)$ is made for $E\to 0$ namely when 
\begin{equation}
  \begin{split}
    {dE\over dJ} = \Omega(E)\simeq - {2\pi\over \log (\Esep-E) }
  \end{split}
 \label{eq_AVI_15}
\end{equation}  
which corresponds to (\ref{eq_AVI_7}) where $\Esep=0$ and $E<0$ so that $ \Esep-E=|E|$.
As a consequence  by approaching the separatrix we can write $H=E(J)$ where $E(J)$
is obtained by integrating (\ref{eq_AVI_15}) which gives
\begin{equation}
  \begin{split}
  (2\pi) ( \Jsep-J)= \Esep-E -(\Esep-E)\,\log(\Esep-E)
  \end{split}
 \label{eq_AVI_16}
\end{equation} 
By inverting this equation we obtain $E=E(J)$. For the Duffing oscillator $\Esep=0$ and
$-E=|E|$ so that  we have $(2\pi)(\Jsep-J)= |E|-|E| \log |E|$
In order to invert equation (\ref{eq_AVI_8}) we set $x=\Jsep-J$ and
$y=-E=|E|$  and   consider  the inversion of   $x= y-y\log y$  for $0<x<1$.
The function  $y-y\log y$  whose derivative is $-\log (y)$
is positive for $0\le y\le e$ and has a maximum for $y=1$. As a consequence
we have two inverses $y_1(x)\in ]0,1[$ and $y_2(x)\in ]1,e[$  for $0\le x\le 1$
and has no inverse  for $x>1$. 
The first  inverse $y_1(x)$   is the fixed point of the map
\begin{equation}
  \begin{split}
    M(y)= {x\over 1-\log(y)}
  \end{split}
 \label{eq_AVI_17}
\end{equation} 
since  the fixed point is attractive. Indeed  consider the  first derivative.
Evaluated at the fixed point the first derivative is
\begin{equation}
  \begin{split}
    M('y)= {x\over y}{1\over (1-y\log y)^2}= {1\over 1-\log y }<1   \qquad 0<y<1
  \end{split}
 \label{eq_AVI_18}
\end{equation} 
The map is contracting and the iterations converge.  The function $y_1(x)$ is monotonic increasing
and maps $[0,1]$ into $[0,1]$. We may consider the Newton's method which amounts
to find the zero in $y$ of the function $f(y) =x-y+y\log y$. The zero is the the fixed point of the 
the map
\begin{equation}
  \begin{split}
    N(y)= y-{f(y)\over f'(y)} = {y-x\over \log(y)}
  \end{split}
 \label{eq_AVI_19}
\end{equation} 
The fixed point is   attractive with quadratic convergence  for $0<y<1$.
As a consequence we determine $y_1(x)$  by iterating   the map
(\ref{eq_AVI_19}) by  choosing $y=x$
    to initialize it.

%
\section{Appendix VII.    Entropies}
The  entropy  is a measure of uncertainty   of a  physical state and  for an isolated system
it is an increasing function of time. The definition of entropy in the framework
of a mechanical microscopic description of the system was provided by Botzmann 
in its formulation  of statistical mechanics. The microstates correspond to single points in
phase space, the macrostates to subsets $A$ of positive measure $\mu(A)$  and 
the  dynamic evolution obeys   Hamilton's equations. The Liouville theorem implies that
the normalized Lebesgue measure  $\mu$ is the invariant probability measure $\mu$. Letting
$\Ecal$ denote the phase space and $m(A)$ the Lebesgue measure of anu $A\subset \Ecal)$
we have $\mu(A)=m(A)/m(\Ecal)$.
The first definition of entropy for a macrostate $A$ given by Boltzmann is  $\log(\mu(A))$.
This  definition is adequate only for equilibrium states due to the invariance of the measure
$\mu(S_t(A))=\mu(A)$. According to this definition the entropy of an initial  state $A$ and its
image $S_t(A)$ at any time $t>0$  are equal.
If the inital state $A$   is  a  ball of small radius and   small measure   it is appropriate
to say it is localized  since all its points, representing the microstates,  are close.
However if the system has positive and negative Lyapunov exponents the image of the ball
$S_t(\xbf)$ becomes a  long cylinder with a small base, because  the volume
cannot change. Since the cylinder cannot self intesect
and the phase space is a compact set the distance of any point from $S_t(A)$
becomes arbitrarily  small as $t$ increases.As a consequence the state $S_t(A)$
becomes delocalzed for $t$ large but since its entropy does not vary  this state
should have a  small uncertainty the same as for $t=0$.  
We know that the thermodynamic entropy is constant only if the system has reached the
thermodynamic equilibrium and only in this case Boltzmann first  definition  is appropriate. 
In order to circumvent this difficulty a second definition was  given  by
Boltzmann introducing an uncertainty  which affects each microstate so that it does no
longer correspond to a point but rather to a phase space cell of small volume.
\subsection{Boltzmann entropies}
To   investigate the  relaxation to equilibrium  Botzmann introduced 
partition of  of   phase space with small  cells (coarse-grained phase space),
whose size measures the uncertainty of  the microscopic states.
For  system of particles (atoms) the uncertainty is due do the finite accuracy with wich the 
space coordinates and the moment components of each can be determined. 
\def\xbar{\overline{x}}
A microstate is  represented by a point in phase space, a microstate with
uncertainty (noisy microstate) is represented by  a cell. Consider a phase space
given by the unit cube $\Ccal\subset \Reali^d$ and its partition into N cubic cells of side $a$
with $N\,a^d=1$. To the microstates $\xbf_i$ given by the centers of the cells partitioning $\Ccal$
we associate a noisy microstate $\xbf_i+\eps \xibf$ were $\xibf$ is a random vector
whose components  are uniformly distributed in $[-1/2,1/2[$.
\\
\\
The uncertainty of a micro-state introduced by noise,  amounts to  replacing  each  point   $\xbf_i$
with the  cell  $c_i$ of which it is the center.   The finite accuracy in the numerical
representation of real numbers has a comparable effect. In this case let $x\in [0,1[$ 
be a  real number  whose  binary representation (for the mantissa  and zero exponent)
is $x=0.b_1b_2\ldots  b_m\,\,b_{m+1}\ldots $. Its truncation to the first  $m$ bits i
$\xb=0.b_1b_2\ldots b_m=k/2^m$ where $0\le k\le 2^m-1$. All the reals having the same truncation
belong to the interval  $[k,k+1[\,/2^m$  where $k$ ranges between $0$ and $2^m-1$.
Let  $N=2^m$ and  $c_k$ for $k=1,\ldots N$ the cell $[k-1,k[\,/N$  whose center  $(k+{1/2})/N$.
The interval $[0,1[$ is partitioned into $N$ disjoint intervals  $c_i$ for $1\le i\le N$
    and to any macrosate $A$ of finite Lebesgue measure  we associate the conditional measures
    of the cells $c_i$ with respect to the set $A$
\begin{equation}
\begin{split}
  \mu_A(c_i)= {\mu(c_i\cap A) \over \mu(A)}
\end{split}
\label{eq_AVII_1}
\end{equation} 
The extension to a phase space given by a unit cube $\Ccal\subset \Reali^{2d}$ is
straightforward.
In this case letting $a=2^{-m}$ be the side of a cell its volume is $a^{2d}=2^{- 2d\,m}$
and  the total  number of cells is $N=1/a^{2d}$.
The weight $p_i\equiv \mu_A(c_i)$   is  the probabilities of the state $A$ to belong to the
cell $c_i$ and obviously $p_1+p_2+\ldots+p_N=1$. The second Boltzmann  entropy at time $t$ is
defined by
\begin{equation}
\begin{split}
  S_A(t) &  = -\sum_{i=1}^N \,p_i(t)\,\log p_i(t)  -\log N      \\  \\  
  p_i(t) &=  {\mu(c_i\cap S_t(A) ) \over \mu(A)} \qquad  A \subseteq \Ccal  \qquad 
\end{split}
\label{eq_AVII_2}
\end{equation} 
where  $f(u)=u\log u$  is  continuous by defining $f(0)=0$.
It is evident that  the conservation of measure $\mu(S_t(A)=\mu(A)$ implies that
the weights $p_i(t)$ satisfy the normalization condition 
$p_1(t)+p_2(t)+\ldots+p_d(t)=1$. However the weights  change  with time
$p_i(t)\not=p_i(0)$ and do not vary  only when the statistcal equilibrium is reached.
The weights for any   localized  state  $A= c_k$  with $\mu(A)=1/N$  are 
$p_i=\mu_A(c_i\cap A)= \delta_{ik}$   so that    $S_A =-\log N$  and this  is the  minimum
possible value of entropy. The same result is obtained with  the first definition of entropy.
If   $A=\Ccal$   then $p_1=p_2=\ldots=p_N=1/N$ and the   entropy  vanishes  $S_A=0$, just as with the first definition.
Consider now an initially localized state $A=c_k$  so that $\mu(A)=1/N$ and  $p_i(0)=\delta_{ik}$.  
Suppose that for $t$ large enough $S_t(c_k)$ has non zero intersection with all the cells.
The conservation of measure implies $\sum_{i=1}^N \,\mu(c_i\cap S_t(c_k))=\mu(c_k)=1/N$.
If we assume that the measures of  all the intersections is equal then we have 
$\mu(c_i\cap S_t(c_k))=1/N^2$ so that from $\mu(c_k)=1/N$ follows that 
\begin{equation}
\begin{split}
  p_i(t)=  {\mu_A(c_i\cap S_t(c_k))\over \mu(c_k)} ={1\over N}   
\end{split}
\label{eq_AVII_3}
\end{equation}
The entropy has changed with respect to the initial value reaching  $S_A(t)=0$.
To an initial localized state with minimum entropy $-\log N$ corresponds
as asymptotically delocalized state with   zero entropy which is the largest
possible value.
\\
The relaxation of $S_A(t)$ towards its  maximum value at equilibrium for $t\to \infty$
needs not to be monotonic since fluctuations   may occur.
The  numerical computation  of  the entropy  is straightforward for systems of low dimensionality,   since the 
weights  are evaluated    by choosing $M$ random points $\xbf_k$   in  $A$ uniformly
distributed, letting them evolve up to time $t$,   and  counting  the number the number  $m_i(t)$ of points 
in each  cubelet $c_i$  since   $p_i(t)=m_i(t)/M$  for $M\to \infty$.   High  computational  costs 
are face d  for systems of many interacting particles.
%
\begin{figure}[!ht]
\centerline{ \includegraphics[width=9 cm, height=4. cm]{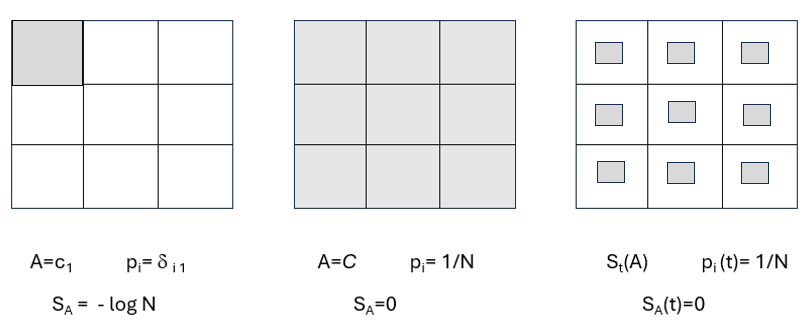}    }
\caption{Left frame: localized state $A=c_1$ for  $N=9$. Since $\mu(A)=1/N$
  and $p_i=\delta_{i1}$ the first and second  Boltzamann
  entropies  are  equal to $-\log N$. Center frame:  delocalized state $A=\Ccal=\cup_{i=1}^N c_i$.
  Since $\mu(A)=1$ and $p_i= 1/N$  the   first and second Boltzmann entropies are equal to 0.
  Right frame:  evolution $S_t(A)$  of an initially localized state $A=c_1$.  In this case 
  $\mu(c_i\cap S_t(A))=1/N^2$    and  $\mu(S_t(A))=1/N$ so that $p_i(t)=1/N$. The first   
  Boltzmann entropy is equal to  $-\log N$,  same  valueas  at $t=0$, 
  the second Boltzmann entropy is 0  the same  as for a delocalized state.   }
\label{fig:fig_AX_1} 
\end{figure}
\subsection{Shannon entropy}
The  second Boltzmann entropy $S_A$ of a set $A$  is simply related to the    Shannon  information entropy 
$H_A$ \cite{Shannon1948}  defined by 
\begin{equation}
\begin{split}
  H_A  = -\sum_{i=1}^N \,p_i\,\log_N p_i 
\end{split}
\label{eq_AVII_4}
\end{equation}   
Given an  alphabet with $N$ symbols $a_i$  (corresponding  to the cells $c_i$),
we  consider the ensemble of all possible strings of length $M$    and the subset $A$
of strings where the symbol $a_i$ appears $m_i$ times.
If  we increase $M$ keeping the ratios  $p_i=m_i/M$   fixed, then 
$A$  is   set of strings,  in which the symbol   $a_i$ occurs  with probability  $p_i$ when $M\to \infty$.
The Boltzmann entropy of a  state $A$ whose weghts are $p_i=\mu(c_i\cap A)/\mu(A)$
and $c_1,\ldots,c_n$ are the cells of a partition of phase space, is equal, up to a scaling factor,
to the Shannon entropy of any  very long  string $A$  randomly generated  with the symbols
$a_i$ having   probabilities $p_i$ 
\begin{equation}
\begin{split}
 & S_A =   -\sum_{i=1}^N \,p_i\,\log p_i -\log N =  \\ 
 & \;\; = \log N\,\Bigl(
  -\sum_{i=1}^N \,p_i\,\log_N p_i -1 \Bigr)    = \log N\,\,( H_A -1)  \\ 
 & H_A = {S_A\over \log N} +1 
\end{split}
\label{eq_AVII_5}
\end{equation}    
The Boltzmann entropy varies between $-\log N $ and $0$, whereas the Shannon entropy varies between
$0$ and 1.
\subsection{Gibbs  entropy}
Given  a dynamical system   with probability density $\rho(\xbf,t)$, which satisfies
the continuity equation, we have $\mu(S_t(A))= \int_{S_t(A)}\,\rho(\ybf,t) \,d\ybf$  and  the 
Gibbs entropy is defined  for any   $t>0$ 
\begin{equation}
\begin{split}
  S_A(t) =  -\int \,\rho(\ybf,t)\,\log \rho(\ybf,t)\,d\ybf  
\end{split}
\label{eq_AVII_6}
\end{equation}   
where $\rho(\xbf,0)=\rho_0(\xbf)$ is the initial assigned probability density. 
Volume preserving flows   preserve the  density and, given any initial density  $\rho_0(\xbf)$, 
we have $\rho(\ybf,t)=  \rho_0(\xbf)=\rho_0(S_{-t}(\ybf))$. In this case  the Gibbs
entropy  is just  the  second  Boltzmann entropy in the limit $N\to \infty$, 
when the initial condition is $\rho_0(\xbf)= \chi_A(\xbf) / m(A)$  and  $m$ is the Lebesgue measure.
Indeed in this case we have $\rho(\ybf,t)= \chi_A(S_{-t}(\ybf)) / m(A)= \chi_{S_t(A)}(\ybf) / m(A)$.
Choosing $A=c_i$,we find that the measure of a cell with respect to the density  $\rho(\ybf,t)$
is precisely the Boltzmann weight 
\begin{equation}
\begin{split}
  & \int_{c_i} \,\rho(\ybf,t)\,\,d\ybf = {1\over m(A) }\,\, \int \chi_{c_i}(\ybf) \chi_{S_t(A)}(\ybf)\,d\ybf = \\ \\
  & ={m(c_i\cap S_t(A))\over m(A)}  = {\mu(c_i\cap S_t(A))\over \mu(A)}  =  p_i(t)
\end{split}
\label{eq_AVII_7}
\end{equation}    
Finally, we can write the entropy using the partition of the  unit cube $\Ccal$ we have assumed to be our phase
space so that the invariant measure is the Lebesgue measure $\mu=m$. Using the mean value theorem 
\begin{equation}
\begin{split}
  S_A(t)& = - \sum_{i=1}^N\,\int_{c_i} \,\rho(\ybf,t)\,\log \rho (\ybf,t)\,d\ybf =  \\ \\
  & = - \sum_{i=1}^N\, p_i(t) \log (\ybf_i,t) \\ \\
    &  =-\sum_{i=1}^N \-p_i(t)  \log p_i(t) -\log N + O(1/N)
\end{split}
\label{eq_AVII_8}
\end{equation}  
where $\ybf_i$ is a point in the interior of $c_i$. The final result is obtained taking into account that  
$\log \rho(\ybf_i,t)\simeq  \,\log \bigl(p_i(t)/\mu(c_i)\,\bigr) $. 
where  $\mu(c_i)=1/N$ so that finally $\log \rho(\ybf_i,t)\simeq \log (N p_i(t)$
\\
\\
This definition of entropy applies
also to stochastic systems,  and in this case  the probability density $\rho(\ybf,t)$  satisfies the 
Fokker-Planck rather than the  continuity  equation. 
This definition of entropy is  very  convenient for  linear stochastic systems,
since  it can be expressed in analytic form.
If $d\ybf(t)/dt=\Aop(t)\,\ybf(t)  +\eps\xibf(t)$   then the solution is
$\ybf(t)=\Lop(t)\xbf + \Lop(t)\int_0^t \,\Lop^{-1}(s)\,\xibf(s)\,ds$ where $\Lop(t)$ is the fundamental matrix
satisfying the equation ${d\over dt} \Lop(t)= \Aop(t) \,\Lop(t)$  with initial condition
$\Lop(0)=\Iop$.
The distribution of $\ybf(t)$ is a multivariate Gaussian
$\rho(\ybf,t)=G\bigl(y-\Lop(t)\xbf,\eps^2\Sigma^2(t)\bigr)$
with mean value $\xbf(t)=\Lop(t)\xbf$ and  covariance matrix $\eps^2\Sigma^2(t)$   where 
\begin{equation}
\begin{split}
 \Sigma^2(t)= \Lop(t)\,\,\int_0^t \,\Lop^{-1}(s)\,\bigl(\Lop^{-1}(s)\bigr)^T\,ds \,\,\Lop^T(t) 
\end{split}
\label{eq_AVII_9}
\end{equation}   
 An analytic
expression  for the integral of $-\rho\log \rho$  is easily obtained according to 
\begin{equation}
  \begin{split}
  -\int \rho(\ybf,t)\,\log \rho(\ybf,t)\,d\ybf & =  {1\over 2}\,\log \parton{\det  \Sigma^2(t)} + c \\  
  c& ={d\over 2} \,\log(2\pi e\,\eps^2)
  \end{split}
\label{eq_AVII_10}
\end{equation}
When $\eps\to 0$ the  additive constant  diverges. As  consequence we define the Gibbs entropy by removing
this additive constant so that
\begin{equation}
 S(t) =  {1\over 2}\,\log \parton{\det  \Sigma^2(t)} 
\label{eq_AVII_11}
\end{equation}   
If the initial distribution is $\rho(\ybf,0)=\delta(\ybf-\xbf)$ corresponding to a
perfecty locaized state its entropy is $-\infty$.
\\
\\
Linear stochastic systems  cover only a  small  fraction of physically relevant systems. Given a nonlinear system
one has to rely on the definition (\ref{eq_AVII_5}) and numerical evaluations require expensive  Monte Carlo samplings.
The alternative we propose for nonlinear systems is to evaluate the entropy not  for the flow but rather for the tangent
flow  of   the backward-forward evolution up to time $t$.  In this case, we  recover a linear stochastic process
for  a vanishingly small  stochastic perturbation
The corresponding  entropies are simply related to the last of the  invariants 
introduced in the previous sections.
\subsection{The  reversibility error  entropy}
The deviation from the initial conditon we obtain when the the flow $S_{t',0}$ in the interval $[0,t]$
is stocastically perturbed by an additive noise in the vector fields  $\Phibf(\xbf,t')$  and  the system evolves 
bckwards in $[t,2t]$  under the action of the field  $-\Phi(\xbf,2t-t')$ whose flow is   $S_{2t-t',t}$. In the absence
of on noise  for $t'$ in the interval $[t,2t]$ we have 
\begin{equation}
  \xbf(t')= S_{2t-t',t} \xbf(t)=S_{2t-t',t} \circ S_{t,0}(\xbf) =S_{2t-t',0} (\xbf)
\label{eq_AVII_12}
\end{equation}   
and the orbits has the symmetry $\xbf(t')=\xbf(2t-t')$, which implies that at time $2t$ we are back
to the initial condition  $\xbf(2t)=\xbf$.
The linear response corresponding
to the zero amplitude limit limit of the noise is a stochastic vector
$\Xibf_{R}(\xbf,t)=\int_0^t \,\Lop^{-1}(\xbf,t')\,\xibf(t')\,dt'$ and the covariance matrix  is given by
\begin{equation}
  \Sigma^2_{BF}(\xbf,t) = \int_0^t \,\Bigl(\Lop^T(\xbf,t')\Lop(\xbf, t')\,\Bigr)^{-1}\,dt'
  \label{eq_AVII_13}
\end{equation}  
For Hamiltonian systems   the matrices  $\Lop$ and $\Lop^T\Lop$ are symplectic and the invariants
obtained by replacing in the integral $\bigl(\Lop^T(t')\Lop(t')\,\bigr)^{-1}$ with $\Lop^T(t')\Lop(t')^{-1}$
are equal to the invariants of  $\Sigma^2_{BF}(\xbf,t)$. In this case the entropy is given  by 
\begin{equation}
 S_{BF}(\xbf, t) =  {1\over 2}\,\log \parton{\det  \Sigma^2_{BF}(\xbf, t)} 
\label{eq_AVII_14}
\end{equation} 
The first $d$ Lyapunov exponents  are non negative the remaining $d$ being opposite  and  heuristic
arguments  suggest  that 
\begin{equation}
  \begin{split}
  \lim_{t\to \infty}  {1\over t} \, S_{BF}(\xbf,t) &  = \lim_{t\to \infty}
      {1\over 2t} \log \Bigl(\det \bigl( \Sigma^2_{BF}(\xbf,t)\,\bigr)\Bigr)   =\\ \\
      &  =\sum_{j=1}^d \,\lambda_j(\xbf)
  \end{split}
  \label{eq_AVII_15}
\end{equation}   
This is the Ruelle resut  for the Kolmogorov Sinai entropy.
For a generic system  whose Lyapunov exponents are $\lambda_1\ge \ldots\lambda_{n_+}\ge 0 \ge \lambda_{n_+ +1}\ge\ldots\ge\lambda_d$
heuristic arguments suggest that  we have the following limit 
\begin{equation}
   \begin{split}
 \lim_{t\to \infty}  {1\over t} \, S_{BF}(\xbf,t )  =\sum_{j=1}^{n_-} \, |\lambda_{d-j+1}(\xbf)|
    \end{split}
     \label{eq_AVII_16}
\end{equation}
where $n_-=d-n_+$. The sum over the absolute values of the negative lyapunov exponents occurs because the covariance matrix is
the integral over the inverse of  $\Lop^T\Lop$ which implies a change of sign in the
Lyapunov exponents. 
This conjecture can be checked  numerically.   For a linear autonomous system in $\Reali^2$  an
analytic proof has been worked out.

%
%

\def\diag{\hbox{diag}}
\section{Appendix  VIII. Local  random fidelity}
\spa
The random fidelity has been  introduced  for randomly perturbed maps \cite{Liverani2007}.
Given a  symplectic  map $M$ defined on an 
invariant compact manifold  $\Ecal$, we suppose to have unit Lebesgue measure $m(\Ecal)=1$, 
and a map $M_\eps$  obtained from $M$ by adding
a random perturbation of amplitude $\eps$  the fidelity for  an observable $f(\xbf)$
is defined by
\begin{equation}
  \begin{split}
    F_\eps(n) &=   \int _\Ecal\,\, \mean{\,f(M^n_\eps(\xbf))}\, f(M^n(\xbf))\,d\xbf  =   
\end{split}
 \label{eq_AVIII_1}
\end{equation} 
where we use the notation $d\xbf = dm(\xbf)$ for the Lebesgue measure.
The mean $\mean{\;\;}$ is taken with respect to
the stochastic process.  The random fidelity is  not stochastically stable,
since the limit $\eps\to 0$   followed by  $n\to \infty$  gives 
$\int_\Ecal \,f^2(\xbf)\,d\mu(\xbf)$ whereas interchanging the order the limit  
is $\bigl(\int_\Ecal \,f(\xbf)\,d\mu(\xbf)\bigr)^2$  if the correlations decay
is sufficiently rapid. The invariant measure $\mu$ is defined by
$\lim_{n\to \infty} \,\int_\Ecal f(M^n(\xbf))\,d\xbf =  \int_\Ecal f(\xbf)\,d\mu(\xbf)$
\\ \\
Given the flow $\xbf(t)= S_t(\xbf)$   one considers  the autocorrelation
$C_\eps(t)$ of  the noisy flow $\ybf(t)= S_{\eps, \,t}(\xbf)$. The fidelity  $F_\eps(t)$
is defined by the  correlation between the the flow and the noisy flow.
\begin{equation}
  \begin{split}   
    C_\eps(t) &=  \int _\Ecal\, \mean{\,f(S_{\eps, \,t}(\xbf))}\,f(\xbf)\, d\xbf  \\ \\
    F_\eps(t) &=  \int _\Ecal\, \mean{\,f(S_{\eps, \,t}(\xbf))}\,f(S_t(\xbf))\, d\xbf    
\end{split}
 \label{eq_AVIII_2}
\end{equation} 
If the measure of $\Ecal$ is not 1 the correlation and the fidelity are defined
multiplying the integral by $1/m(\Ecal)$.
\\
For an autonomous system we introduce  another fidelity  defined  as the autocorrelation
for the BF noisy  process 
\begin{equation}
  \begin{split}
    F_{\eps}^{BF}(t) &=  \int _\Ecal\, \mean{\,f(S_{-t}\circ S_{\eps,\,t}(\xbf))}\,f(\xbf)\,d\xbf 
\end{split}
 \label{eq_AVIII_3}
\end{equation} 
Notice that  $F_{\eps}^{BF}(t)$ differs from   $F_{\eps}(t)$ unless the
Lebesgue measure is invariant.
In the generic case of a non autonomous system we   considered the  BF  flow    $S_{t';t}^{BF}(\xbf)$
 generated by the field $\Phibf_{BF}(\xbf,t')$,  equal to   $\Phibf(\xbf,t')$  for $0\le t'\le t$ and 
to   $-\Phibf(\xbf,2t-t')$  for $t\le t'\le 2t$. The flow has the symmetry $S_{t';t}^{BF}(\xbf)=
S^{BF}_{2t-t';t}$
which implies  $S_{0;t}^{BF}(\xbf)=S_{2t;t}^{BF}(\xbf)=\xbf$.
By adding a white noise $\eps\,\xibf(t')$  for  $0\le t'\le t$ we obtain a   BF noisy  flow
$S_{\eps,\,\,t';t}^{BF}(\xbf)$ whose value at time $t'=2t$ we denote  $S_{\eps,\,t}^{BF}(\xbf)$.
The new  fidelity
is the autocorrelation for the BF noisy  flow $S_{\eps,\,t}^{BF}(\xbf)$
\begin{equation}
  \begin{split}
    F_\eps^{BF}(t) &= \, \int _\Ecal\, \mean{\,f(S_{\eps, \,\,t}^{BF}(\xbf))}\,f(\xbf)\,d\xbf   
\end{split}
 \label{eq_AVIII_4}
\end{equation} 
\subsection{Fidelity for a  linear system}
For an autonomous linear system $\Phi(\xbf)= \Aop \xbf$  the flow is 
$S_t(\xbf)= e^{\Aop t}\,\xbf $. The  F and BF noisy flows  $S_{\eps\,\,t}(\xbf)$ and
$S^{BF}_{\eps\,\,t}(\xbf)$  are given by
\begin{equation}
  \begin{split}
   S_{\eps\,\,t}(\xbf)& = e^{\Aop t} \xbf + \eps \int_0^t \,e^{\Aop(t-t')}\,\xibf(t') dt' \\ 
   S_{\eps\,\,t}^{BF}  &  = \xbf + \eps \int_0^t \,e^{-\Aop t'} \xibf(t')\,dt'
\end{split}
 \label{eq_AVIII_5}
\end{equation} 
The corresponding covariance matrices are given by $\eps^2\Sigma^2_F(t)$ and
$\eps^2\Sigma^2_{BF}(t)$
\begin{equation}
  \begin{split}
  \Sigma^2_F(t)& = \int_0^t \,e^{A t'}\, e^{A^T t'}\,dt' \\ 
  \Sigma^2_{BF}(t) & = \int_0^t\,( e^{A^T t'}\,e^{A t'})^{-1}\,dt'
\end{split}
 \label{eq_AVIII_6}
\end{equation} 
The distribution of $S_{\eps,\,t}(\xbf)$ is a Gaussian with a drift term since $S_{t}(\xbf)=e^{\Aop t}\,\xbf$,
whereas  the distribution of $S^{BF}_{\eps,\,t}(\xbf)$ has no drift term since $S_{t}^{R}(\xbf)=\xbf$.
We  introduce the functions    $u(\xbf,t)= \mean{S_{\eps,\,t}(\xbf)}$   and $v(\xbf, t)$ 
defined by 
\begin{equation}
  \begin{split}
  &   u(\xbf,t)\equiv\mean{f(S_{\eps,\,t})(\xbf)}  = \,\int_{\Ecal} \,f(\ybf) G_\Toro(\ybf-e^{\Aop t}\xbf,\,\eps^2\Sigma_F^2(t))
    \,d\ybf  \\ 
   &  v(\ybf,t)  = \,\int_{\Ecal}  G_\Toro(\ybf-e^{\Aop t}\xbf,\,\eps^2\Sigma_F^2(t)) \,f(\xbf) \,d\xbf
\end{split}
 \label{eq_AVIII_7}
\end{equation} 
Notice that  $u(\xbf,t)$ and $v(\xbf,t)$ satisfy the backward and forward Kolmogorov equation respectively.
The fidelity is given by 
\begin{equation}
  \begin{split}
  & F_\eps(t)= \,\int_{\Ecal} \,u(\xbf,t) \,f(e^{\Aop\,t}\,\xbf) \,d\xbf= 
  \,v(\ybf,t) \,d\ybf = \\ 
 & \quad =\int_\Ecal d\ybf  \int_\Ecal d\xbf  \,\,f(\ybf) G_\Toro(\ybf-e^{\Aop t}\xbf, \,\eps^2\Sigma_F^2(t)), f(e^{\Aop\,t}\,\xbf)
\end{split}
 \label{eq_AVIII_8}
\end{equation} 
For the BF  noisy  process the fidelity is defined by  
\begin{equation}
  \begin{split}  
   F_\eps^{BF}(t) & =\,\int_\Ecal \,d\xbf  \, u^{BF}(\xbf,t)\,f(\xbf) \,d\xbf = \\
   & = \int_\Ecal f(\ybf,t)\,v^{BF}(\ybf,t)\,d\ybf= \\ 
    & = \,\int_\Ecal d\ybf  \,\int_\Ecal d\xbf  \,\, f(\ybf) G_\Toro(\ybf-\xbf,\,\eps^2\Sigma^2_{R}(t)) f(\xbf)
\end{split}
 \label{eq_AVIII_9}
\end{equation} 
The fidelity $F^{BF}_\eps$ defined above is just the mean value with respect to the observable $f$
of a symmetric and positive operator.
The key advantage is that  in the absence of noise the  BF flow is the identity at any time
$S^{BF}_{t}(\xbf)=\xbf$. In view of the  extension to nonlinear flows 
we choose  $\Ecal$  to be  the torus $\Toro^d_\ell(\xbf_0)$ defined by
\def\halfell  {{ {\scriptstyle \ell \over \scriptstyle 2}}}
\begin{equation}
  \begin{split} \hskip -.5 truecm 
    \Toro^d_\ell(\xbf_0)\equiv   \begin{matrix} \otimes \\ {\scriptstyle j=1,d}\end{matrix} \, \Toro  
      \parqua{x_{0\,j}-\halfell, \,x_{0\,j}+\halfell}
\end{split}
 \label{eq_AVIII_10}
\end{equation} 
whose volume  is  $m( \Toro^d_\ell(\xbf_0))= \ell^d$.  In this case  the observable  we choose
is $f((\xbf-\xbf_0)/\ell)$,    where $f(\xbf)$ is a periodic function of period 1 in each
coordinate $x_j$  and cn  be expanded in a Fourier series 
\begin{equation}
  \begin{split} \hskip -.5 truecm 
  f(\xbf)= \sum_{\kbf} \,f_\kbf e^{2\pi\,i\, \kbf\cdot \xbf}
\end{split}
 \label{eq_AVIII_11}
\end{equation} 
Choosing $f\in C^m$ the coefficents decay as $|f_\kbf|\le C\,\Vert\kbf\Vert^{-m}$. 
The  flow $S_{-t}(\xbf)$ is discontinuous on the torus, nevertheless the   fidellity $F^R_\eps(t)$
can still be easily evaluated since the transition operator is diagonal.  As expected for a liear
system the result does not depend on $\xbf_0$ and depends on the ratio  $\eps/\ell$  of the noise amplitude
and torus size.   This allows to take the limits $\eps\to 0$ and $\ell \to 0$  
 \spa
    \subsection {The Gaussian distribution  on the torus}
    \spa
Starting from   the Gaussian distribution  $G(\ybf,\eps^2 \Sigma^2)$ on $\Reali^d$  the
Gaussian distribution on the Torus $\Toro^d $ is defined by 
\begin{equation}
  \begin{split}
    & G_\Toro\bigl(\ybf,\eps^2\Sigma^2(t)\bigr)= \sum_{\kbf\in \Interi^d}\,
    G_\Toro\bigl(\ybf +\kbf,\eps^2\Sigma_{}^2(t)\bigr)= \\
    & \qquad =\sum_{\kbf\in \Interi^d}   G_{\kbf}(t) \,e^{2\pi\,i\,\kbf\cdot \ybf} \\ \\   
\end{split}
 \label{eq_AVIII_12}
\end{equation}
where $G_\kbf(t)$ are are the coefficients of the Fourier expansion
of $G_\Toro(\ybf,\eps^2\Sigma^2(t))$ which is a
periodic function with period 1
in each coordinate. The   coefficents $G_\kbf(t)$ are given by the
characteristic function of the Gaussian
distribution. Indeed we have  
\begin{equation}
  \begin{split}
    G_\kbf (t)&= \int_{\Toro^d}\,e^{-2\pi\,i\,\kbf\cdot\ybf}\, G_\Toro\bigl(\ybf,\eps^2\Sigma(t)\bigr)\,d\ybf= \\
    &= \int_{\Toro^d}\,e^{-2\pi\,i\,\kbf\cdot\ybf}\, \sum_{\kbf\in \Interi^d}\,G\bigl(\ybf+\kbf,\,\eps^2\Sigma^2(t)\bigr)\,d\ybf = \\
    & =\sum_{\kbf\in \Interi^d}\,\int_{I_\kbf} e^{-2\pi\,i\,\kbf\cdot\ybf'}\, G\bigl(\ybf',\eps^2\Sigma^2(t)\bigr)\,d\ybf' =\\
    &= \int_{\Reali^d} \, e^{-2\pi\,i\,\kbf\cdot\xbf'} G\bigl(\xbf',\eps^2\Sigma^2(t)\bigr)\,d\xbf' =  \\ 
    & \qquad = e^{-{ 1\over 2}\, (2\pi)^2 \,\eps^2\,\kbf\cdot \Sigma^2(t)\kbf}
\end{split}
 \label{eq_AVIII_13}
\end{equation} 
%
%
%
%
where $I_{\kbf}=[k_1-{1\over 2},\,k_1+{1\over 2}]\times\cdots \times [k_d-{1\over 2},\,k_d+{1\over 2}]$.
As a consequence we can write the Gaussian on the torus as
\begin{equation}
  \begin{split}
    G_\Toro\bigl(\ybf,\eps^2\Sigma^2(t)\bigr)= \sum_{\kbf\in \Interi^d}\,e^{2\pi\,i\, \kbf\cdot \ybf}\,
    e^{-2\pi^2\,\eps^2 \kbf\,\cdot \Sigma^2(t)\kbf}
\end{split}
 \label{eq_AVIII_14}
\end{equation}
On the torus $\Toro_\ell^d =\Toro^d[-{\ell\over 2},\, {\ell\over 2}] $
the   transition probability  becomes 
\begin{equation}
  \begin{split}
  G_{\Toro_\ell^d}(\ybf) &  = \sum_\kbf\, G(\ybf+ \ell\kbf,\eps^2\Sigma^2)= \sum_{\kbf} \,e^{2\pi\,i\,\kbf\cdot \ybf/\ell} \, G_\kbf(t;\ell) \\ 
    G_\kbf(t;\ell) & =  {1\over \ell^d} \int_{\Reali^d}  \,e^{-2\pi\,i\, \kbf\cdot \ybf/\ell} G(\ybf,\eps^2\Sigma^2(t))\,d\ybf= \\
     & = {1\over \ell^d} e^{-{ 1\over 2}\, (2\pi)^2 \,\kbf\cdot \Sigma^2(t)\kbf\,\eps^2/\ell^2}
\end{split}
 \label{eq_AVIII_15}
\end{equation}
%
%
%
%
%
    \subsection{ Behaviour of  BF fidelity  for linear systems }
    \spazio
The evaluation of the classical random fidelity  defined on the torus $\Toro_\ell(\xbf_0)$ is difficult because 
the transition operator is not diagonal.
On the contrary the BF  random fidelity  can be easily obtained  and the result is 
\begin{equation}
  \begin{split}
    &  F_\eps^{BF}(t;\ell)= {1\over \ell^d }\int_{\Toro_\ell^d(\xbf_0)}\, \,u^{T}(\xbf,t)\, f\parton{\xbf-\xbf_0\over \ell} \,d\xbf = \\
  &\;\; =  {1\over \ell^d }\int_{\Toro_\ell^d(\xbf_0)}\,\,d\ybf d\xbf\, f\parton{\ybf-\xbf_0\over \ell}\,
  G_{\Toro_\ell^d}\bigl(\ybf-\xbf, \eps^2 \Sigma^2_{}(t)\bigr)\, \times  \\
    & \times f\parton{\xbf-\xbf_0\over \ell}    ={1\over \ell^{2d}} \int_{\Toro_\ell^d({\bf 0})}\,d\xbf'\,\, d\ybf'  \,f\parton{\xbf'\over \ell} f\parton{\ybf'\over \ell}     \times \\
 &  \qquad  \times \sum_\kbf\,\,e^{2\pi\,i\, \kbf\cdot (\ybf'-\xbf')/\ell}\,\, e^{-{ 1\over 2}\, (2\pi)^2 \,\kbf\cdot \Sigma^2(t)\kbf\,\eps^2/\ell^2}=  \\ \\
 &  \qquad   = \sum_\kbf \,|f_\kbf|^2 \,  e^{-{ 1\over 2}\, (2\pi)^2 \,\kbf\cdot \Sigma^2_{BF}(t)\kbf\,\eps^2/\ell^2} \phantom{\biggr)}
\end{split}
 \label{eq_AVIII_16}
\end{equation}
where we have set $\xbf'=\xbf-\xbf_0$ and $\ybf'=\ybf-\ybf_0$.
Due to linearity and translation invariance of the transition operator the result does not depend on $\xbf_0$
We may  consider the limit $\ell\to 0$ and $\eps\to 0$ keeping the ratio $\rho= \eps/\ell$ fixed.
For a linear system   this shows that  if we change the size $\ell$  of the torus and
the noise amplitude keeping their ratio constant the fidelity is not affected.
For a non-linear system $\Sigma^2_{BF}$ depends on the initial point $\xbf$ and such a dependence
prevents  an explicit computation of the integral which, in the linear case, leads to 
(\ref{eq_AVIII_16}) for the BF reverse process  fidelity.
However letting $\ell \to 0$  the fidelity  can be explicitly evaluated
and in addition the limit $\eps\to 0$  changes the linear approximation into the linear response
which is valid for any $t>0$ in the whole phase space.
\\
Notice that  setting $\rho=\eps/\ell$ we  denote the fidelity with $F_\rho^{BF}(t)$.
If we finally let $\rho\to 0$ which amounts to taking
first the $\eps\to 0$ limit keeping $\ell$ fixed the result is $\sum \,|f_\kbf|^2= \int f^2(\xbf)\,d\xbf$.
\\
\\
If $\Aop$ is a negative matrix then $e^{\Aop t}=\Wop\,e^{\Lambda t}\,\Wop^T$ where the entries of $\Lambda$ are 
$\lambda_j<0$ and $\Wop$ is orthogonal. In this case the fixed point $\xbf=0$ is stable and attractive.
The BF Reversibility error covariance matrix is
$\Sigma^2_{BF}= \Wop(e^{2|\Lambda|  t}-\Iop)(2|\Lambda|)^{-1} \Wop^T$ and the fidelity becomes 
\begin{equation}
  \begin{split}
   & F_\rho^{BF}(t) = \\
    & = \sum_\kbf \,|f_\kbf|^2  \exp\biggl(-2\pi^2 \,\rho^2\,
    \sum_{j=1}^d \,{e^{2|\lambda_j|\,t}  -1\over 2|\lambda_j|}\,(\kbf\cdot\wbf_j)^2\biggr)
\end{split}
 \label{eq_AVIII_17}   
\end{equation}   
The $t\to +\infty$ limit of the fidelity is  $f_{\bf 0}^2$
which corresponds to full de-correlation and the decay towards this limit is  super-exponential.
\\
\\
If $\Aop$ is a  positive  matrix  then $\lambda_j>0$ and the BF random fidelity is given by 
\begin{equation}
  \begin{split}
 & F_\rho^{BF}(t)= \\
 & = \sum_\kbf \,|f_\kbf|^2  \exp\biggl(-2\pi^2 \,\rho^2\,
    \sum_{j=1}^d \,{1-e^{-2\lambda_j\,t}\over 2\lambda_j}\,(\kbf\cdot\wbf_j)^2\biggr)
\end{split}
 \label{eq_AVIII_18}
\end{equation}   
In this case the asymptotic limit is
 $ F_\rho^{BF}=  \sum_\kbf \,|f_\kbf|^2  e^{\bigl( -\pi^2 \,\rho\,  \sum_{j=1}^d \,(\kbf\cdot\wbf_j)^2/\lambda_j\bigr)}$.
The result is intermediate between $f_{\bf 0}^2$ corresponding to full de-correlation and
 $ \sum_\kbf \,|f_\kbf|^2 $ corresponding to no de-correlation.  
 If all the $\lambda_j$ tend to zero   there is no deterministic evolution  $\Lop=\Iop$  and we have
 $\Sigma^2_{R}=\Iop\,t$ so that $ F_\rho^{R}(t)$  is given by
\begin{equation}
  \begin{split}
    F_\rho^{BF}(t)= \sum_\kbf \,|f_\kbf|^2  \exp\Bigl(-2\pi^2 \,\rho^2\, t \,\kbf\cdot \kbf \Bigr)
\end{split}
 \label{eq_AVIII_19}
\end{equation}  
and decays exponentially to the asymptotic limit $f_{\bf 0}^2$.  
Such a decay occurs  also when the linear system has even dimension $2d$  and is given by the product of $d$ rotations.
\spazio
If $\Aop$ is a symmetric matrix with positive and negative eigenvalues we introduce the projectors $\Pop_\pm$
into the subspaces of the positive and negative eigenvalues writing the matrix $\Aop$ as the sum
of two matrices $\Aop_{\pm}$  with positive and negative eigenvalues 
\begin{equation}
  \begin{split}
&   \Pop_+=\sum_{i=1}^{n_+}\, \wbf_i\,\wbf_i^T \qquad  \Pop_-=\sum_{i=n_+ +1}^{d}\, \wbf_i\,\wbf_i^T
\quad \Pop_++\Pop_-=\Iop\\ \\
& e^{\Aop_+t} =\sum_{i=1}^{n_+}\, e^{\lambda_i\,t}\,\wbf_i\,\wbf_i^T \qquad  e^{\Aop_- t}=\sum_{i=n_+ +1}^{d}  \,e^{\lambda_i\,t}\, \wbf_i\,\wbf_i^T
\end{split}
 \label{eq_AVIII_20}    
\end{equation}

The   BF  covariance matrix is written as the the sum of two matrices $\Sigma^2_{BF\,\,\pm}(t)$ defined by
\begin{equation}
  \begin{split}
&  \Sigma^2_{BF\,\,+}(t)= \sum_{i=1}^{n_+}\, {1-e^{-2\lambda_i\,t}\over 2\lambda_i}\,\wbf_i\,\wbf_i^T \\
&\Sigma^2_{BF\,\,-}(t)= \sum_{i=n_+ +1}^{d}  \,{e^{2|\lambda_i|\,t}-1\over 2|\lambda_i|}\, \wbf_i\,\wbf_i^T
\end{split}
 \label{eq_AVIII_21}   
\end{equation}
The evaluation of $F^{BF}_\rho(t)$ is simple since  the transition probability is diagonal
and we split the BF reversibility error  covariance matrix into matrices acting on the eigen-spaces
of positive and negative eigenvalues 
\begin{equation}
  \begin{split}
  F^{BF}_\rho(t) & = f_{\bf 0}^2 +   \,  \sum_{\kbf \not=0, } \,
 \exp\Bigl( -2\rho^2\pi^2  \kbf \,\cdot \bigl( \Sigma^2_{BF\,\,+}(t)\kbf \\
& \hskip 3. truecm  +\kbf \,\cdot \Sigma^2_{BF\,\,-}(t)\kbf\bigr)   \Bigr)\,|f_\kbf|^2
\end{split}
 \label{eq_AVIII_22}    
\end{equation}
The asymptotic limit is 
\begin{equation}
  \begin{split}
& \lim_{t\to +\infty }\, F^{R}_\rho(t) = \\
& = f_{\bf 0}^2 +\hskip -.3 truecm  \sum_{\kbf \not =0,\,P_-\kbf=0}\hskip -.3 truecm
\,|f_\kbf|^2 \,\,     \exp\Bigl( -2\pi^2\rho^2  \sum_{i=1}^{n_+}{ (\kbf\cdot \wbf_i)^2 \over 2\lambda_i} \Bigr)
\end{split}
 \label{eq_AVIII_23}    
\end{equation}  
and the convergence rate to the limit is exponential as $e^{-\lambda_{n_+} t}$ where $\lambda_{n_+}$ is the smallest of
the positive  Lyapunov exponents. If all the exponents are negative then  the limit is $f_{\bf 0}^2$ and
the convergence  is superexponetial. The convergence of the random BF fidelity to $f_{\bf 0}^2$
occurs only when the eigenvalues of $\Aop$ are have   negative or zero real part.  For linear Hamiltonian
systems this  condition is verified only when the eigenvalues of $\Aop$ are imaginary, namely the fixed point
is elliptic. When the fixed point is hyperbolic so that the eigenvalues of $\Aop$ are real and paiwise opposite
and the local random fidelity converges for $t\to +\infty$ to a limit different from $f_{\bf 0}^2$ 
according to (\ref{eq_AVIII_23}),  . The previous results are easily extended to invertible maps, such as  the 
symplectic maps. For the automorphism  of the torus $\Toro^2$  the BF random fidelity does not converge
to $f_{\bf 0}^2$ just as the Classical random fidelity, whose  proof can be found
in the appendix of \cite{Liverani2007}.
\subsection{The BF fidelity for nonlinear systems} 
\spa
For  a linear system we have considered the classical fidelity $F_\eps(t)$ and the local
random fidelity    $F_\rho^{R}(t)(t)$   for the BF noisy evolution the a torus  $\Toro_\ell(\xbf_0)$
where $\rho=\eps/\ell$ is kept constant when $\eps\to 0$ and $\ell\to 0$. The linearity renders
the  BF random fidelity   independent from the center $\xbf_0$ of the torus  and allows to
compute the limit for $t\to \infty$ and to  estimate the convergence rate.
This definition can be easily extended to non linear systems because the limit $\eps\to 0$
allows to use the linear response. The difference with respect to the linear case is that
covariance matrix $\Sigma^2_{BF}$ now depends on  the phase space point where it is computed.
If the  flow is generated by a nonlinear vector field $\Phi(\xbf,t)$  in  $\Reali^d$ the fidelities 
on the torus $\Toro_\ell^d(\xbf_0)$  are  defined by (\ref{eq_AVIII_16}). In this case $\Sigma^2_{BF}$ depends
not only on $t$ but also on $\xbf$ which varies on $\Toro_\ell^d(\xbf_0)$.
As a consequence the integral cannot  be evaluated analytically,  even when the  linear approximation
is justified.   The observable is this case is $f\bigl((\xbf-\xbf_0)/\ell\bigr)$
where $f(\xbf)$ is a periodic function of period 1 in each coordinate.
Since the fidelities  depend on $\rho=\eps/\ell$
taking the limit $\eps\to 0$ and $\ell\to 0$  while keeping $\rho$  constant has two advantages:
the result is  valid for any $t>0$ since  the limit  $\eps\to 0$ leads to the linear response,
the integral can be exactly computed and the fidelity  depends on  $\xbf_0$.
Letting  $S_{\eps\,\,t}(\xbf)$ be the solution for the noisy forward flow at time $t$, the  function $u(\xbf,t')
=\mean{f(S_{\eps\,\,t}(\xbf))}$,   satisfies the backward  Kolmogorov equation
\begin{equation}
  \begin{split}
     \derp{}{t'}u(\xbf,t') &= \Phibf(\xbf,t)\cdot \derp{}{\xbf}u(\xbf,t')   + {\eps^2\over 2}  \Delta_\xbf u(\xbf,t')
\end{split}
 \label{eq_AVIII_24}    
\end{equation}
for $0\le t'\le t$ and initial condition $u(\xbf,0)= f(\xbf)$. Periodic boundary
conditions allow to obtain a solution  defined on a torus.
\\
\\
To define the BF local  random fidelity  we consider the evolution
$ S^{BF}_{\eps\,\,t';t}(\xbf)$ defined as the solution
for a noisy forward flow up to time t followed by the backward flow up to time $2t$.
Introducing the field  $\Phibf_{FB}(\xbf,t'; t)$  and the flow $\ybf(t')\equiv S_{\eps\,\,t';t}(\xbf)$ defined by
\begin{equation}
  \begin{split}
   &\Phi_{BF}(\xbf,t';t) =\begin{cases}\quad  \Phi(\xbf,t') \quad \qquad\,\,\, 0\le t'\le t \\ \\
  -\Phi(\xbf,2t-t') \qquad t\le t'\le 2t \end{cases} \\ 
   & {d\over dt'} \ybf(t')=  \Phi_{BF}(\ybf(t'),t';t)+ \\ 
   & \eps\xibf(t') \vartheta(t-t'), \qquad 0\le t'\le 2t
\end{split}
 \label{eq_AVIII_25}    
\end{equation}
the function  $u(\xbf,t';t)= \mean{f((S_{\eps\,\,t';t}(\xbf))}$ satisfies the backward Kolmogorov equation
\begin{equation}
  \begin{split}
    & \derp{}{t'}u(\xbf,t';t) = \Phibf_{R}(\xbf,t'; t)\cdot \derp{}{\xbf}u(\xbf,t';t) +  \\
    & \hskip 3 truecm  + {\eps^2\over 2}\,\vartheta(t-t')
    \Delta u(\xbf,t';t)
\end{split}
 \label{eq_AVIII_26}   
\end{equation}
with initial condition $u(\xbf,0;t)=f\bigl((\xbf-\xbf_0)/\ell\bigr)$
and  periodic boundary conditions to obtain a solution
defined on  the torus $\Toro^d_\ell(\xbf_0)$.
The solution $u(\xbf,t)\equiv u(\xbf,2t;t)$  enters  in the definition of  the local random fidelity.
We  introduce the function  $v(\ybf,t';t)$  which satisfies the forward Kolmogorov equation
\begin{equation}
  \begin{split} 
    & \derp{}{t'} v(\ybf,t';t) +\div\bigl( \Phibf_{R}(\ybf,t';t)\,v(\ybf,t';t)\,\bigr) =  \\
    & \hskip 3 truecm =
         {\eps^2\over 2}\,\vartheta(t-t')\,\Delta v(\ybf,t';t)
\end{split}
 \label{eq_AVIII_27}    
\end{equation} 
with initial condition $f\bigl((\ybf-\xbf_0)/\ell\bigr)$  and periodic boundary conditions
 to obtain a solution defined on  the torus $\Toro^d_\ell(\xbf_0)$
Letting $v(\ybf,t)\equiv v(\ybf,2t;t)$ the BF local random fidelity is given by
\begin{equation}
  \begin{split} 
   F^{BF}_\eps(t;\,\xbf_0,\ell)& =\,\int_{\Toro^d_\ell(\xbf0)}\,d\ybf\,f\parton{\ybf-\xbf_0\over \ell} \,v(\ybf,t) = \\ 
   &= \int_{\Toro^d_\ell(\xbf_0)}\,d\xbf\, \,u(\xbf,t)\,f\parton{\xbf-\xbf_0\over \ell}
\end{split}
 \label{eq_AVIII_28}    
\end{equation} 
Having fixed $\ell$ the computation of the classical  fidelity $F_\eps$ or the new  fidelity $F^{BF}_\eps$
by varying $\xbf_0$ is computationally
expensive even for a low dimensionality such as $d=2$. 
\\
\\ 
In order to find an analytical solution   we must take the limit  $\eps\to 0$ which gives the linear
response valid for any $t>0$ and the limit $\ell\to 0$ which allows the integral to be evaluated,
while keeping constant the ratio $\rho=\eps/\ell$.
The key difference from the previous case is that the covariance matrix of BF error $\Sigma^2_{BF}$ depends on the point. The
local BF reverse   random fidelity is  obtained from  (\ref{eq_AVIII_16}) by   changing the
integration variables  according to  $\xbf'=(\xbf-\xbf_0)/\ell$ and $\ybf'=(\ybf-\xbf_0)/\ell$
and taking into account that the covariance matrix depends on $\xbf$  namely $\Sigma^2_{BF}(\xbf,t)$.
As a consequence after the change of coordinates the integral is over the unit torus $\Toro^d$
centered at the origin and  the fidelity    $F_\rho^{BF}(\xbf_0,t)$, which   depends 
on the ratio $\rho=\eps/\ell$,  $\xbf_0$ and $t$   is given by
\begin{equation}
  \begin{split} 
   \hskip -.3 truecm  F_\rho^{BF}(\xbf_0,\,t)& =
   \,\int_{\Toro^d\times \Toro^d} d\xbf'\,d\ybf'  \,f(\ybf')\,f(\xbf') \, \sum_\kbf \,e^{2\pi\,i\,\kbf \cdot(\ybf'-\xbf')} \times \\ \\
   & \times  \,e^{-2\pi^2\rho^2, \kbf\cdot \Sigma^2_{BF}(\xbf_0 + \ell \xbf',\,t)\,\kbf}
    \label{eq_AVIII_29}   
    \end{split}
\end{equation} 
Since the ratio $\eps/\ell$ is kept constant we can take the limit $\ell\to 0$ obtaining an
analytic result for the BF local random  fidelity which reads 
\begin{equation}
  \begin{split} 
    F_\rho^{R}(\xbf_0,\,t)= \sum_\kbf \,|f_\kbf|^2 \,  e^{-2 \pi^2\,\rho^2\,  \,\kbf\cdot \Sigma^2_{R}(\xbf_0,\, t)\kbf}
    \qquad\quad  \rho={\eps\over \ell}
    \label{eq_AVIII_30}   
    \end{split}
\end{equation} 
  where $\xbf_0$ is any point in phase space.  In the basin of attraction of a
  stable point the eigenvalues of $\Sigma^2_{BF}(\xbf,t)$ grow exponentially
  with $t$ and we expect a super-exponential
  decay of $F^{BF}_\rho(\xbf_0,t)$ to $f^2_{\bf 0}$.  On the stable manifold of an hyperbolic point the behaviour for
  $t\to +\infty$ is the same as for the linearized system. For an integrable system in the region
  delimited by a separatrix
  the eigenvalues of $\Sigma^2_{BF}$  grow according to a power law and   the fidelity decays exponentially
  towards its asymptotic value. 
\\
\\
  \subsection{Examples of local random fidelity}
  \spa
   We compute the   local random  fidelity  for the
  Duffing overdamped oscillator whose vector field is $\Phi(x)=x(1-x^2)$ and whose
  fundamental matrix is $L(x,t)= e^{-2t}(x^2-(1-x^2)e^{-2t})^{-3/2}$. The    variance $\sigma^2_{BF}(x,t)$
  (we change $x_0$ into $x$) is given by  
\begin{equation}
  \begin{split} 
    & \sigma^2_{BF}(x,t) =\int_0^t L^{-2}(x,\tau)\,d\tau= x^6\,{e^{4t}-1\over 4} +  \\
   &  + 3x^4(1-x^2)\times {e^{2t}-1\over 2} + \\
   &  + 3x^2(1-x^2)^2 \,t + (1-x^2)^3\,{1-e^{-2t}\over 2}
\end{split}
 \label{eq_AVIII_31}   
\end{equation}
The  local random fidelity in this case  for an observable
$f((x-x_0)/\ell)$ where $f(x)=f(x+1)$ is a periodic function
with Fourier coefficients $f_k$ is given by 
\begin{equation}
  \begin{split} 
  F^{BF}_\rho(x_0,t)= \sum_k \,|f_k|^2 e ^{-2\pi^2\,\rho^2 \,k^2\,\sigma^2_{BF}(x_0,t)} 
\end{split}
 \label{eq_AVIII_32}   
\end{equation}  
We remind that the Lyapunov exponent  is discontinuous since  $\lambda(0)=1$ and $\lambda(x)=-2$
for $x\not=0$. 
We notice that  also    $ \lambda_{BF}(x)=\lim_{t\to +\infty} \, (2t)^{-1}\,\log \sigma^2_{BF}(x,t)$
is  discontinuous.
Indeed $\lambda_{BF}(0)=0$  where the Lyapunov exponent is 1,  and  for any $x\not=0$ we
have  $\lambda_{BF}(x)=2$ where the Lyapunov exponent is  $-2$.
  The local random fidelity $F^{BF}_\rho(x,t)$  for $f(x)= 2\cos^2(\pi x)$  whose Fourier coefficients
  are $f_0=1$ and  $f_{\pm 1}=1/2$
  converges to $f_0^2=1$   superexponentially for $x\not =0$ but the convergence rate  slows down
  as we approach $x=0$.
  For $x=0$ the fidelity  decays exponentially to its limit value  $ 1+{1\over 2}\,e^{-\pi^2\,\rho^2}$.
  \\
  \\
  The second example  is  the Duffing oscillator whose Hamiltonian is   $H=p^2/2 -x^2/2+x^4/4$.
  In  angle action variables $(\theta,J)$  the
  Hamiltonian becomes $\hat H(J)$ defined for $0\le J\le J_s$  and  $J_s$ is the action of the separatrix
  where  the Hamiltonian has  a singularity, just as  $\Omega(J)=d\hat H/dJ$. 
  We recall that the matrix $\Lop^T\Lop$ and   and its eigenvalues $\mu^2_\pm$  are given by
\begin{equation}
  \begin{split}   
    \Lop^T\Lop& =\begin{pmatrix} 1 & \qquad  \Omega' t \\ \\ \Omega' t &  \qquad 1 +(\Omega' \,t)^2  \end{pmatrix}
    \\ \\
    \mu^2_+& = (2+{\Omega'}^2(J)\,t^2)\,\bigl(1+O(t^{-2}\bigr)\qquad \mu_-^2={1\over \mu_-^2}
\end{split}
 \label{eq_AVIII_33}   
\end{equation} 
As a consequence letting $\mu^2=\diag(\mu_+^2,\,\,\mu_-^2)$ the  matrix $\Lop^T\Lop$ asymptotically
can be written as 
\begin{equation}
  \begin{split} 
    \Lop^T\Lop= \Wop \,\mu^2\,  \Wop^T \quad
      \Wop={1\over \sqrt{1+(\Omega' t)^2}} \begin{pmatrix} 1 & - \Omega' t \\ \\ \Omega' t &  1   \end{pmatrix} 
\end{split}
 \label{eq_AVIII_34}    
\end{equation}
so that the  the asymptotic limits of the  eigengectors are  $\wbf_1(t)\to \ebf_2 \equiv (0,1)$ and    
$\wbf_2(t)\to -\ebf_1 \equiv (-1,0)$.  The largest variation corresponds to  displacements along
the $J$ axis.  The matrix $\Sigma^2_{BF}$ is obtained by integrating $\Lop^T\Lop)^{-1}$ in $[0,t]$ and reads 
 \def\scr{\scriptstyle}
\begin{equation}
  \begin{split} 
    \Sigma^2_{BF} = \begin{pmatrix} t+ {1\over 3}\,{\Omega'}^2 t^3  & - {1\over 2}\Omega' t^2 \\ \\
      - {1\over 2}\Omega' t^2  &  t   \end{pmatrix} 
\end{split}
 \label{eq_AVIII_35}   
\end{equation} 
The eigenvalues are  $\mu^2_{+}= {1\over 3}{\Omega'}^2t^3 + 2  t+ O(t^{-1})$ and $\mu^2_{-}={t\over 4}t+ O(t^{-1})$.
The  local random  fidelity is immediately obtained. We consider the  observable $f(\theta)= 2\cos^2(\pi\,\theta)$
whose Fourier coefficients are $f_{\bf 0}=1$ and $f_{\kbf}=1/2 $ for  $\kbf=(\pm 1,\,0)^T$ and
$f(J)= 2\cos^2(\pi\,J)$ whose Fourier coefficients correspond to  $f_{\bf 0}=1$ and $f_{\kbf}=1/2 $ for
$\kbf=(0,\,\pm 1)^T$.   As a consequence  the  local random  fidelity 
asymptotically reads
\begin{equation}
  \begin{split} 
    &f(\theta)  \qquad F^{BF}_\rho(J,t) = 1+ {1\over 2}\, \exp\parton{ - 2 \,\pi^2 \,\rho^2 \,\,\Bigl(
      {1\over 3}{\Omega'}^2 t^3 + t\Bigr)} \\ \\
    &f(J)  \qquad F^{BF}_\rho(J,t) = 1+ {1\over 2}\, \exp\parton{ - 2 \,\pi^2 \,\rho^2\, \,t } 
\end{split}
 \label{eq_AVIII_37}   
\end{equation} 
Notice that for a linear system $\hat H=\omega J$  we have  $\Omega'=0$  and in this case the decay is exponential
for whatever choice of the observable.
\\
\\
The results so far obtained allow to compute the local fidelity for a system with a limit cycle in $\Reali^2$ whose
vector field is $ \Phibf= \bigl(\omega p + x(1-x^2-p^2),\,\,-\omega x+ p(1-x^2-p^2)\bigr)$.
Introducing polar coordinates $x= r\cos \theta,\,p=-r\sin\theta$ the   vector field becomes
$\hat\Phibf= (\omega, \,\,r(1-r^2))$ so that $\theta(t)= \theta+\omega t$ and $r(t)$ is the solution of the
overdamped Duffing oscillator we have already considered, but restricted to the positive
$r$ axis. Since the equations  for $\theta$ and $r$
are decoupled we can apply the previous results. The covariance matrix  is diagonal
$\Sigma^2_{BF} = \diag\bigl( t,\, \sigma^2_{BF}(r,t)\bigr)$  where $\sigma^2_{BF}$ is given by
(\ref{eq_AVIII_31} )
with $x$ replaced by $r$. For an observable $f(\theta)$ the decay is exponential whereas
for any observable $f(r)$  the decay is super-exponential with a rate decreasing as we approach $r=0$.

\section { APPENDIX IX. Time reversal invariance}
    \spa
    For a Hamiltonian system in $\Reali^{2d}$ we denote with  $\xbf$ a point in phase space
    and   with $\Iop_R\,\xbf$  the point with inverted momentum coordinates 
\begin{equation}
  \begin{split}    
    \xbf =\begin{pmatrix} \qbf \\ \pbf\end{pmatrix} \qquad
    \Iop_R\,\, \xbf =\begin{pmatrix} \qbf \\ - \pbf\end{pmatrix} \qquad
    \Iop_R=\begin{pmatrix} \Iop & 0 \\ 0 & -\Iop\end{pmatrix}    
  \end{split}
\label{eq_AXI_1}
\end{equation}
Given the initial coordinates and moments  $\qbf_0,\,\pbf_0$  to which
corresponds  the phase space intial condition $\xbf_0$   the coordinates
and moments at time $t$ are  $\qbf(\qbf_0,\,\pbf_0,t),\,\pbf(\qbf_0,\,\pbf_0,t)\, $
to which corresponds the the phase space point we denote $\xbf=S_t(\xbf_0)$.
Consider now the coordinates and moments at time $-t$  defined by
$\qbf(\qbf_0,\,-\pbf_0,-t),\, -\pbf(\qbf_0,\,-\pbf_0,-t)\, $  to which
corresponds the phase space point  $ \Iop_R\,\,S_{-t}(\Iop_R\xbf_0)$.
\\
\\
The sytem is times reversal invariant if
\begin{equation}
  \begin{split}    
     & \Iop_R S_{-t}(\Iop_R \xbf_0) = S_t(\xbf_0) \\ \\
     & \Iop_R S_{t}(\Iop_R \xbf_0) = S_{-t}(\xbf_0) 
\end{split}
\label{eq_AXI_2}
\end{equation}
which in extended form reads
\begin{equation}
  \begin{split}        
   \qbf(\qbf_0,\,-\pbf_0,-t)& =\qbf(\qbf_0,\,\pbf_0,t)  \\ \\
   -\pbf(\qbf_0,\,-\pbf_0,-t)&=\pbf(\qbf_0,\,\pbf_0,t)
  \end{split}
\label{eq_AXI_3}
\end{equation}
The invariance condition is $H(\qbf,\pbf)=H(\qbf,-\pbf)$ and it 
is satisfied for $H(\qbf,\pbf)=\pbf^2/2+V(\qbf)$.  In this case  we prove
that the equations for  the coordinates and moments of the orbit for
negative times, we denote for brevity $\qbf(-t),- \pbf(-t)$,  and initial
condition $(\qbf_0,-\pbf_0)$  
\begin{equation}
  \begin{split}    
  & {d\over dt}\qbf(-t)= -{d\over d(-t)}\qbf(-t)  = -\pbf(-t)) \\  \\ 
  &  {d\over dt}(-\pbf(-t)){t}=  {d\over d (-t)}\pbf(-t)= \Fbf(\qbf(-t))
  \end{split}
\label{eq_AXI_4}
\end{equation} 
are the same as the  equations  satisfied  by $\qbf(t),\pbf(t)$
with initial condition $(\qbf_0,\pbf_0)$
\begin{equation} 
  {d\over dt}\qbf(t)= (\pbf(t)) \qquad   {d\over dt}\pbf(t)= \Fbf(\qbf(t)) 
\label{eq_AXI_5}
\end{equation}
In the previous equations  we have set $\Fbf(\qbf)=-\partial V(\qbf)/\partial \qbf$.
An immediate consequence of the time reversal invariance is that if the let
the system evolve up to time  $t$ and, after reversing the moments, we 
let it evolve up to time $2t$, and finally we invert the moments once more we
are back to the initial condition.
\begin{equation}
  \begin{split}    
   & \hskip -.3 truecm \Iop_R S_t(\Iop_R\,S_t(\xbf)) = S_{-t}\,\bigl (S_t(\xbf))\,\bigr)   =\xbf
  \end{split}
\label{eq_AXI_6}
\end{equation}      
\spa
Recalling that the fundamental matrix is the the tangent map of the flow,
 if the system is time reversal invariant we have from (\ref{eq_AXI_2}) 
\begin{equation}
  \begin{split}    
   \Lop(\xbf,-t) =DS_{-t}(\xbf) = \Iop_R \,\Lop(\Iop_R\xbf,t)\,\Iop_R
  \end{split}
\label{eq_AXI_7}
\end{equation}
\def\dt{{\Delta t}}
From (\ref{eq_AXI_7}) follows that   that the BF and FB Reversibility error
invariants are simply related if the system is Hamiltonian  and  time reversal invariant.
Denoting now with $ \Sigma^2_{BF}$ and $ \Sigma^2_{BF}$ the BF and FB Reversibility error
covaniance matrices  and taking into account that the  Lyapunov matrix $\Lop^T\Lop$
is symplectic we have 
\begin{equation}
  \begin{split}
    \Sigma^2_{BF}(\xbf,t) & =\Jop \,\int_0^t \,\Lop^T(\xbf,
  t')\,\Lop(\xbf,t')\,dt' \,\Jop^{-1} \\ \\
   \Sigma^2_{FB}(\xbf,t)  & =\Jop\, \int_0^t \Lop^T(\xbf,-t')\,\Lop(\xbf,-t')\,dt' \,\Jop^{-1}
  \end{split}
\label{eq_AXI_8}
\end{equation}
The time reversal invariance implies that 
\begin{equation}
  \begin{split}    
   \Lop^T(\xbf,-t')\,\Lop(\xbf,-t')= \Iop_R\, \Lop^T(\Iop_R\xbf,t')\,\Lop(\Iop_R \xbf,t')\,\Iop_R
  \end{split}
\label{eq_AXI_9}
\end{equation}
As a consequence taking into account $\Jop \Iop_R\Jop=\Iop_R$  and $\Iop_R=\Iop_R^{-1}$
\begin{equation}
  \begin{split}    
  \Sigma^2_{FB}(\xbf,t)=  \Iop_R \,\Sigma^2_{BF}(\Iop_R \xbf,t) \,\Iop_R
  \end{split}
\label{eq_AXI_10}
\end{equation}
The FB and BF reversibility error covariance matrices  are conjugated by a  a similarity tranformation
provided that we change $\xbf$ into $\Iop_R\xbf$. As a consequence their invariants are equal
provided that we change $\xbf$ into $\Iop_R \xbf$, that corresponds to  invert the moments.
\begin{equation}
  \begin{split}    
   I^{(k)}\bigl(\Sigma^2_{FB}(\xbf,t)\bigr)=   I^{(k)}\bigl(\Sigma^2_{BF}(\Iop_R \xbf,t)\bigr)
  \end{split}
\label{eq_AXI_11}
\end{equation}
\spazio
If $M_\dt(\xbf)$ is a symmetric symplectic integrator we have $M^{-1}_\dt(\xbf)= M_{-\dt}(\xbf)$
and if the Hamiltonian system  is time reversal invariant the map $M_\dt(\xbf)$
satisfies the following condition 
\begin{equation}
  \begin{split}    
      \Iop_R\,M_\dt(I_R\xbf)= M_{-\dt}(\xbf)\equiv M^{-1}_\dt(\xbf) 
\end{split}
\label{eq_AXI_12}
\end{equation}
For the tangent map $\Lop(\xbf)=DM_\dt(\xbf)$  of $M(\xbf)$ and
the tangent map $\Lop_n(\xbf)=DM_\dt^{\circ n}(\xbf)$ of its iterates
the time reversal invariance condition becomes 
\begin{equation}
  \begin{split}    
    \Lop_{-n}= \Iop_R\Lop_n(\Iop_R\xbf)\,\Iop_R
  \end{split}
\label{eq_AXI_13}
\end{equation}   
so that the BF and FB reversibility error covariance matrices are related by
\begin{equation}
  \begin{split}    
   \Sigma^2_{FB\,\,n}(\xbf) = \Iop_R  \,\Sigma^2_{BF\,\,n}(\Iop_R \xbf)\, \Iop_R
  \end{split}
\label{eq_AXI_14}
\end{equation}   
The BF and FB invariants are the equal  provided that we change $\xbf$ into $\Iop_R\xbf$,
that corresponds to change the sign of the moments.
 \section{Appendix X  A  symplectic 4D  map}
We consider  the following map $M$ proposed in \cite{Froeschle2000a, Froeschle2000b}
to exhibit the Arnold web using FLI
and by \cite{Faranda2012} to compare REM with other dynamical dynamical indicators.
The map is defined on the torus $\Toro^4$ and is obtaind by coupling two   standard maps.
the explicit form is given by
\def\mod{\hbox{mod}\,}
\begin{equation}
  \begin{split}
    p_{x\,\,n+1} & = p_{x\,\,n}-{\mu \over 2\pi}\, {\sin(2\pi\,x_n)\over C^2(x_n,y_n) } \;\;\mod 1 \qquad  \\
    p_{y\,\,n+1} & = p_{y\,\,n}-{\mu \over 2\pi}\, {\sin(2\pi\,y_n)\over C^2(x_n,y_n) } \;\;\mod 1 \\  \\
    x_{n+1}&= x_n+p_{x\,n+1}  \;\; \mod 1  \\
    y_{n+1}&= y_n+ p_{y\,n+1} \;\; \mod 1 \\ \\
     C(x,y)& = 4+\cos(2\pi\,x)+\cos(2\pi\,y)
  \end{split}
   \label{eq_AX_1}
\end{equation}
\pan
The torus $\Toro^4$ is the cube $[-1/2,1/2]^4$ with identified opposite faces.
The interpolating Hamiltonian is
\begin{equation}
  H= {p_x^2 + p_y^2\over 2} + {\mu\over (2\pi)^2}\, {1\over   C(x,y)}
  \label{eq_AX_2}
\end{equation}
The map is the  symplectic integrator of order one with unit time step.
For $|\mu|\ll 1 $ our map is a good integrator
and it is quasi-integrable. If we choose the initial conditions in one phase plane, this plane is invariant.
Choosing the phase space vector $\xbf=(x,p_x,y,p_y)^T$ the tangent map $DM$ is given by
\begin{equation}
  DM= \begin{pmatrix} 1+A_{xx}   & \; 1  &&  A_{xy} & \; 0 \\  
    A_{xx}  & \;  1  &&  A_{xy}  & \; 0  \\  \\
  A_{xy} & \; 0  &&  1+A_{yy}  & \;   1   \\
  A_{xy} & \; 0  &&  A_{yy}  &\;  1   \end{pmatrix}
  \label{eq_AX_3}
\end{equation}
where
\begin{equation}
\begin{split}
  A_{xx} & = -{\mu\over C^2(x,y) }\,\,\parton{\cos(2\pi x) +{2\over C(x,y)} \sin^2(2\pi x)}  \\ 
  A_{xy} & = -{2\mu\over C^3(x,y) }\, \sin(2\pi x)\,\,\sin(2\pi y)  \qquad \\
  A_{yy} & = -{\mu\over C^2(x,y) }\,\,\parton{\cos(2\pi y) +{2\over C(x,y)} \sin^2(2\pi y)}  \\
\end{split}
\label{eq_AX_4}
\end{equation}  
We may check   immediately that the matrix DM is symplectic, namely $DM \,\Jop\,(DM)^T= \Jop$
\\
\\
Another model extensively used in the literature, see for intance \cite{Skokos2001}
formula (11),   consists of two coupled standard maps  
\begin{equation}
  \begin{split}
  & p_{x\,n+1}= p_{x\, n}  -{\lambda\over 2\pi}\,\sin(2\pi\,x_n)  +{\mu\over 2\pi}\,\sin(2\pi(x_n-y_n)) \,\,\mod 1 \\ \\ 
    & p_{y\,n+1}= p_{y\, n}  -{\lambda\over 2\pi}\,\sin(2\pi\,y_n)  +{\mu\over 2\pi}\,\sin(2\pi(y_n-x_n))\,\,\mod 1 \\ \\
     & x_{n+1}= x_n +p_{x\,n} \,\,\mod 1 \qquad \qquad y_{n+1}= y_n +p_{y\,n} \,\,\mod 1 
  \end{split}
  \label{eq_AX_5}
\end{equation}

\end{document}